\documentclass[12pt]{article}
\pdfoutput=1
\usepackage{jheppub}
\usepackage{comment}
\usepackage{tikz}
\usepackage[vcentermath]{youngtab}
\usepackage{subfigure}

\newcommand\be{\begin{equation}}
\newcommand\ee{\end{equation}}

\newcommand\Tr{\mathrm{Tr}}

\allowdisplaybreaks

\preprint{KIAS-P22045}
\title{Dualities and flavored indices of M2-brane SCFTs}
\abstract{
We study various conjectural dual descriptions of 
a stack of M2-branes in M-theory including ADHM, ABJ(M), BLG, discrete gauge theories and quiver Chern-Simons (CS) theories and propose several new dualities of the M2-brane SCFTs by analyzing flavored supersymmetric indices in detail. 
The mapping of local operators, Coulomb, Higgs and mixed branch operators as well as global symmetries under the dualities are obtained from the precise matching of the indices. 
Furthermore, we find closed form expressions for the Coulomb limit of the indices of the $U(N)$ ADHM theory and the dual quiver CS theory for arbitrary $N$ and propose a refined generating function for plane partitions with trace $N$. 
For the quiver CS theories we also find an infinite-sum expression for the Higgs limit of the indices which is more useful than the original expression.
}
\author[a]{Hirotaka Hayashi,}
\emailAdd{h.hayashi@tokai.ac.jp}
\affiliation[a]{Department of Physics, School of Science, Tokai University, 4-1-1 Kitakaname, Hiratsuka-shi, Kanagawa 259-1292, Japan}
\author[b]{Tomoki Nosaka}
\emailAdd{nosaka@yukawa.kyoto-u.ac.jp}
\affiliation[b]{Kavli Institute for Theoretical Sciences, University of Chinese Academy of Sciences,\\
Beijing, China 100190}
\author[c]{and Tadashi Okazaki}
\emailAdd{tokazaki@kias.re.kr}
\affiliation[c]{
School of Physics, Korea Institute for Advanced Study,\\
85 Hoegi-ro, Cheongnyangri-dong, Dongdaemun-gu, Seoul 02455, Republic of Korea}
\begin{document}
\maketitle

\section{Introduction and summary}
%
The low-energy dynamics of M2-branes probing some backgrounds is described by certain three-dimensional (3d) superconformal field theories (SCFTs). 
For such M2-brane SCFTs, one finds various ultraviolet (UV) dual descriptions which flow to the same infrared (IR) fixed point. For example, the low energy dynamics of $N$ M2-branes probing the singularity of $\mathbb{C}^4/\mathbb{Z}_k$ can be captured by certain Chern-Simons matter theory called the Aharony-Bergman-Jafferis-Maldacena (ABJM) theory with gauge groups $U(N)_k \times U(N)_{-k}$ where the subscripts stand for the Chern-Simons level \cite{Aharony:2008ug}. One can generalize the configuration by introducing fractional M2-branes and/or replacing $\mathbb{C}^4/\mathbb{Z}_k$ with $\mathbb{C}^4/\widehat{D}_k$ where $\widehat{D}_k$ is the binary dihedral group of order $4k$. The low energy dynamics can be again described by a Chern-Simons matter theory with different gauge groups \cite{Hosomichi:2008jb, Aharony:2008gk}, which is referred to as the Aharony-Bergman-Jafferis (ABJ) theory. Another generalization is considering a background $(\mathbb{C}^2/\mathbb{Z}_p\times \mathbb{C}^2/\mathbb{Z}_q)/\mathbb{Z}_k$ with positive integers $p, q, k$, which generically yields a circular quiver Chern-Simons theory \cite{Imamura:2008nn,Imamura:2008dt}. A special case with $q = k = 1$ gives rise to the 3d $U(N)$ gauge theory with a hypermultiplet in the adjoint representation and $p$ hypermultiplets in the fundamental representation, which we call the $U(N)$ Atiyah-Drinfeld-Hitchin-Manin (ADHM) theory \cite{deBoer:1996mp,deBoer:1996ck}. The name comes from the fact that the Higgs branch of the ADHM theory describes the moduli space of $N$ $SU(p)$ instantons. 3d theories with other gauge group $G$ whose Higgs branches capture instanton moduli spaces may be also referred to as the $G$ ADHM theories. There is also another Lagrangian construction using a Lie-3 algebra for describing multiple M2-branes called the Bagger-Lambert-Gustavsson (BLG) theory \cite{Bagger:2006sk,Bagger:2007jr,Bagger:2007vi,Gustavsson:2007vu,Gustavsson:2008dy}. 

In this paper we study the flavored supersymmetric indices of the M2-brane SCFTs which at least have $\mathcal{N}=4$ supersymmetry, including ADHM theories, ABJ(M) theories, 
BLG theories,  
discrete gauge theories and quiver Chern-Simons theories.  
The 3d supersymmetric indices \cite{Bhattacharya:2008zy,Bhattacharya:2008bja,Kim:2009wb,Imamura:2011su,Kapustin:2011jm,Dimofte:2011py} are a powerful tool to study supersymmetric quantum field theories and their dualities. While the flavored indices of the ABJ(M) theory and BLG theory were computed in vast literature e.g. \cite{Kim:2009wb,Honda:2012ik,Agmon:2017lga,Beratto:2021xmn}, those of the $\mathcal{N}=4$ ADHM theories and their cousin have not yet been fully computed. \footnote{See \cite{Gang:2011xp} for the unflavored indices of specific ADHM theories.} 
We find precise agreement of flavored indices as strong evidence of the conjectural dualities of M2-brane SCFTs including ADHM theory and other descriptions by comparing a large number of terms by expanding the flavored indices. In addition, the flavored indices enable us to find the mapping of operators and  global symmetries under the dualities. We also explicitly give them in this paper. 

Also we find stringent evidence for new dualities of the M2-brane SCFTs
\begin{align}
\label{new_ABJdual}
&\textrm{$U(2)_{2}\times U(1)_{-2}$ ABJ $\otimes$ $U(1)_{1}\times U(1)_{-1}$ ABJM}
\nonumber\\
&\Leftrightarrow
\textrm{$U(2)_{1}\times U(2)_{-1}$ ABJM},
\end{align}
\begin{align}
\label{new_BLGdual}
&\textrm{$SU(2)_1 \times SU(2)_{-1}$ BLG}
\nonumber\\
&\Leftrightarrow 
\textrm{$U(2)_{2}\times U(1)_{-2}$ ABJ} \otimes 
\textrm{$U(1)_{2}\times U(1)_{-2}$ ABJM},
\end{align}
where some parameters of the factorized theories are identified and then the numbers of the parameters become the same between the dual theories.
The duality (\ref{new_ABJdual}) indicates that the $U(2)_{1}\times U(2)_{-1}$ ABJM theory can be factorized into the decoupled free sector isomorphic to the $U(1)_{1}\times U(1)_{-1}$ ABJM theory and the interacting sector described by the $U(2)_{1}\times U(1)_{-1}$ ABJ theory. The duality (\ref{new_BLGdual}) leads to an interpretation of the $SU(2)_1\times SU(2)_{-2}$ BLG theory as a product of M2-brane theories. We evaluate the flavored indices to find the precise  agreement. 

There exist limits of their fugacities \eqref{HS_lim} in which the $\mathcal{N}=4$ flavored indices reduce to the Hilbert series for the Coulomb and Higgs branches \cite{Razamat:2014pta}. We compute the Hilbert series of M2-brane SCFTs and find that the Hilbert series which counts the local operators on the $\mathcal{N}=4$ Coulomb branch precisely gives the Hilbert series for the geometry probed by M2-branes not only for the ADHM theory but also for highly supersymmetric $\mathcal{N}\ge 4$ Chern-Simons matter theories. We give several analytic and semi-analytic expressions of the Hilbert series for the supersymmetric Chern-Simons matter theories by taking the appropriate limit of the fugacities in the flavored indices. Besides, the flavored indices allow us to count the mixed branch operators which cannot be detected by the Hilbert series or unflavored indices. We concern ourselves with the analysis of the mixed branch operators in the $U(N)$ ADHM theory which consist of monopole operators dressed by the adjoint hypermultiplets.

For the $U(N)$ ADHM theory with $l$ fundamental hypermultiplets and the $U(N)_k\times U(N)_0^{\otimes (l-2)}\times U(N)_{-k}$ Chern-Simons matter theory which is conjectured to be dual to the ADHM theory when $k=1$, we also find closed form expressions of the fully dressed indices in the Coulomb limit.\footnote{
A different kind of Hilbert series of circular quiver Chern-Simons matter theories is also investigated in \cite{Cremonesi:2016nbo}.
}
For both of the two theories, our calculations are based on a special simplification which occurs in the grand canonical version of the indices, i.e.~the generating function of the indices in terms of rank $N$, which is reminiscent of the Fermi gas formalism for the $S^3$ partition functions \cite{Marino:2011eh}, correlation functions \cite{Gaiotto:2020vqj,Chester:2020jay,Hatsuda:2021oxa} and the four dimensional Schur indices \cite{Bourdier:2015wda,Bourdier:2015sga,Gaiotto:2020vqj,Hatsuda:2022xdv}.
Our result for the ADHM theory is the generalization of \cite{Gaiotto:2021xce} for $l\ge 1$, and when we take the unrefined limit our result also agrees with the closed form expressions for the corresponding Coulomb branch Hilbert series obtained in \cite{Okazaki:2022sxo}. 
Following the correspondence in \cite{Okazaki:2022sxo}, 
we conjecture that a refined generating function, that is a generating function for plane partitions of $n$ which has a trace $\tau(n)=N$ and the difference $\sum_{i>0}\tau_i(n)-\sum_{i<0}\tau_i(n)=M$ between the sum of the $i$-traces $\tau_i(n)$ with $i>0$ and the sum of those with $i<0$ is given by
\begin{align}
\label{conjecture_pp_conclude}
\sum_{n=1}^{\infty}\sum_{N=0}^{n}\sum_{M=-n}^{n} \alpha(n,N,M) \mathfrak{t}^n \kappa^{N} z^M
&=
    \prod_{m=0}^{\infty}\frac{1}{1-\kappa \mathfrak{t}^{2m+1}}
    \prod_{n=1}^{\infty}
    \prod_{\pm} \frac{1}{1-\kappa \mathfrak{t}^{2m+n+1}z^{\pm n}},
\end{align}
where the trace $\tau(n)=\sum_i n_{ii}$ is a sum of diagonal entries of the plane partitions and the $i$-trace $\tau_i(n)$ is a sum of entries in the $i$-th diagonal and $\alpha(n,N,M)$ is the number of plane partitions of $n$ with $\tau(n)=N$ and $\sum_{i>0}\tau_i-\sum_{i<0}\tau_i=M$. 
We find the closed form expression for the ADHM theory and that for the Chern-Simons matter theory with $k=1$ coincides with each other, giving a direct proof for the duality in the Coulomb limit.
By using the closed form expressions we can also write down the large $N$ expansion of the Coulomb limit of the fully dressed index, in a power series of $\mathfrak{t}^{N}$ ($\mathfrak{t}=t^{-1}q^{\frac{1}{4}}$) together with the explicit expressions for each coefficient of $\mathfrak{t}^{nN}$ \eqref{220429giantgraviton}.

For the $U(N)_k\times U(N)_0^{\otimes (l-2)}\times U(N)_{-k}$ Chern-Simons matter theory we also find that the integrations over holonomies can be performed explicitly in the Higgs limit, resulting in a new expression for the fully dressed index in the Higgs limit.
Although our final expression \eqref{220518_CSmatterHiggsdifficultsum} still contains infinite sums for the monopole charges which we could not perform explicitly for general $N$ and $l$, with our expression we can compute the Higgs limit of the index in the small $\mathfrak{t}$ ($\mathfrak{t}=tq^{\frac{1}{4}}$) expansion more easily than calculating the small $q$ expansion of the full index by using the original expression first and then taking the Higgs limit.

\subsection{Open problems}

\begin{itemize}
    \item There are variants which are not studied in this work, such as the $USp(2N)$ gauge theories with a rank 2 matter and an odd number of half-hypermultiplets, orthogonal gauge theories with gauge groups $SO(2N+\gamma)$, $Spin(2N+\gamma)$ and $Pin_{\pm}(2N+\gamma)$, Chern-Simons theory with affine D-type or affine E-type quiver \cite{Gulotta:2011vp} and quiver Chern-Simons theory with gauge group $SU(N)\times SU(N)$. We hope to report the analysis of such theories by evaluating the flavored indices. 
    \item The local operators in the M2-brane SCFTs form certain algebras. For the $U(N)$ ADHM theory with $l$ flavors the algebra formed by the Coulomb branch operators is the spherical part of the cyclotomic rational Cherednik algebra \cite{Kodera:2016faj}. It would be interesting to categorify the dualities of the M2-brane SCFTs as rigorous isomorphisms or equivalences of algebras or modules. 
    \item The BPS boundary conditions can realize the M5-branes on which M2-branes end. Such BPS boundary conditions are studied in the ABJM theory \cite{Berman:2009xd,Hosomichi:2014rqa,Okazaki:2015fiq}. The dualities of $\mathcal{N}=(0,4)$ boundary conditions and $\mathcal{N}=(2,2)$ boundary conditions in 3d $\mathcal{N}=4$ Abelian gauge theories are studied by evaluating the half-indices  \cite{Okazaki:2019bok,Okazaki:2020lfy} and by engineering them in brane setup \cite{Chung:2016pgt}. They will generalize the dualities of the M2-brane SCFTs associated with the 3d $\mathcal{N}=4$ ADHM theories. 
    \item The indices can be decorated by the BPS Wilson line and vortex line operators \cite{Drukker:2012sr}. The dualities of line operators in the M2-brane SCFTs may be studied by engineering the line operators in Type IIB setup \cite{Assel:2015oxa}. 
    \item The finite $N$ corrections of the gravity indices of the M2-brane SCFTs describing the M2-branes at the $A$-type singularity are investigated in \cite{Arai:2020uwd,Gaiotto:2021xce,Imamura:2022aua,Lee:2022vig}. 
    It would be nice to study the finite $N$ corrections for the other  M2-brane SCFTs by further analyzing our flavored indices.
    It would also be nice if we could find a gravitational interpretation of the giant graviton coefficients we obtained in the Coulomb limit \eqref{220429giantgraviton}.
\item The grand canonical index of the M2-brane SCFT describing a stack of $N$ M2-branes probing $\mathbb{C}^4$ is studied in  \cite{Gaiotto:2021xce}. Also the grand canonical index of its Coulomb limit is shown to be given by the generating functions for plane partitions \cite{Okazaki:2022sxo}. It would be interesting to study the grand canonical indices obtained from our flavored index of other M2-brane SCFTs with symplectic and orthogonal gauge groups and explore their combinatorial interpretation.
    \item It would be interesting to analyze the Higgs limit of the fully dressed indices of $U(N)_k\times U(N)_0^{\otimes (l-2)}\times U(N)_{-k}$ quiver Chern-Simons theory further.
For example, we may use \eqref{220518_CSmatterHiggsdifficultsum} to calculate the small $\mathfrak{t}$ expansion to very high order for various $l,k,N$ and try to guess a rational function which complete each series, as we do in \eqref{21modelHiggsN2closedconjecture}.
It would also be interesting to extend our analysis for the Coulomb/Higgs limit of the supersymmetric indices of $U(N)$ ADHM theory with $l$ and the $U(N)_k\times U(N)_0^{\otimes (l-2)}\times U(N)_{-k}$ quiver Chern-Simons theory to other theories of M2-branes, such as the $U(N)_k\times U(N)_0^{\otimes (p-1)}\times U(N)_{-k}\times U(N)_0^{\otimes (q-1)}$ quiver Chern-Simons theories and the ADHM theories/quiver Chern-Simons theories with orthogonal or symplectic gauge groups.

\item 
Higher-form symmetries of $\mathcal{N}\ge6$ quiver Chern-Simons matter theories including ABJ(M) and BLG theory are examined in  \cite{Tachikawa:2019dvq,Bergman:2020ifi,Beratto:2021xmn}. 
It would be nice to study higher-form symmetries in the proposed dual theories and explore further dualities of the $\mathcal{N}\ge4$ M2-brane SCFTs including the ADHM theory, the circular quiver CS theory and the discrete gauge theory. 
\end{itemize}

\subsection{Structure}
The organization of the paper is as follows. 
In section \ref{sec_brane} we review the three-dimensional  low-energy effective theories of D2-branes and M2-branes. 
We summarize the brane setup in Type IIA, Type IIB and M-theory and known dualities of these theories. 
In section \ref{sec_ADHMu} we study the $\mathcal{N}=4$ $U(N)$ ADHM theory which has $U(N)$ gauge group and a single adjoint hypermultiplet as well as fundamental hypermultiplets. By computing the indices we examine the local operators on the Coulomb, Higgs and mixed branches. We find the precise matching of the flavored indices with those of their mirror theories and derive the mapping of operators and symmetries under the mirror symmetry. 
In section \ref{ADHMCoulomballorder} we also derive the closed form expression for the Coulomb limit of the fully dressed indices of $U(N)$ ADHM theory with $l$ fundamental hypermultiplets. As a consequence, we propose a refined generating function for plane partitions. 
In section \ref{sec_sp} we investigate the $\mathcal{N}=4$ $USp(2N)$ gauge theories with a hypermultiplet transforming as (anti)symmetric representation and multiple fundamental half-hypermultiplets. The indices perfectly agree with those of their mirror theories. 
In section \ref{sec_O} we study the $\mathcal{N}=4$ gauge theories with orthogonal gauge groups, rank 2 matter fields and fundamental flavors. We give formulae of indices which allow us to get the  Hilbert series for the Coulomb and Higgs branches in appropriate limits and to check dualities of the orthogonal gauge theories. 
In section \ref{sec_abjm} we evaluate the flavored indices of  ABJ(M) theory. We show that their limits lead to the Hilbert series corresponding to the geometry probed by M2-branes. We confirm the proposed dualities with ADHM theory, discrete gauge theories as well as a new duality between $U(2)_{1}\times U(2)_{-1}$ ABJM theory and a product of $U(2)_{2}\times U(1)_{-2}$ ABJ theory and $U(1)_{1}\times U(1)_{-1}$ ABJM theory. 
In section \ref{sec_qCS} we study the $\mathcal{N}=4$ quiver Chern-Simons theories. The dualities between the ADHM theory with multiple flavors and the quiver Chern-Simons theories are confirmed as their flavored indices agree with each other.
We also derive the closed form expression for the fully dressed indices of these theories in the Coulomb limit in section \ref{sec_coulombCSmatterfermigas}, and reduce the Higgs limit of the indices into a simpler expression than the original expression \eqref{220413_CSmatterindexnewnotation} in section \ref{sec_Higgssimplerform}.
In section {\ref{sec_BLG}} we evaluate the flavored indices of the BLG theories with gauge groups $SU(2)\times SU(2)$ and $(SU(2)\times SU(2))/\mathbb{Z}_2$. The flavored indices reduce to the Hilbert series for $(\mathbb{C}^4\times \mathbb{C}^4)/\mathbb{D}_{m}$ where $\mathbb{D}_m$ is the dihedral group order $m$ in the limits. We also propose a new duality between the  $SU(2)_{1}\times SU(2)_{-1}$ BLG theory and a product of the $U(2)_{2}\times  U(1)_{-2}$ ABJ theory and the $U(1)_{2}\times U(1)_{-2}$ ABJM  theory and confirm it by finding the precise agreement of indices. 
In appendix \ref{app_notation} we introduce some notations for the supersymmetric indices.
In appendix \ref{app_auxdres} we consider a further generalization of the supersymmetric indices so that we can keep track which field components contribute to each term in the indices.

In this paper we evaluate the indices by expanding them with respect to $q$ at least up to $q^5$ for most of the examples except for the cases where we explicitly mention the orders we computed 
and show only several terms. 

\section{3d theories on probe M2-branes}
\label{sec_brane}
In this paper we consider 3d theories which can be engineered on M2-branes probing some backgrounds 
in M-theory. In this section we first review string theory construction of the 3d theories.

\subsection{Type IIA/M-theory construction}
\label{sec:typeIIAM}

Before considering M-theory configurations we start from type IIA string theory construction. The worldvolume theory on $N$ D2-branes yields the 3d $\mathcal{N}=8$ supersymmetric Yang-Mills theory with a gauge group $U(N)$. The supersymmetry can be reduced by half by introducing D6-branes. The brane configuration in the 10d spacetime of type IIA string theory is summarized in Table \ref{tb:typeIIA}.
\begin{table}[t]
\centering
\begin{tabular}{c|ccc|cccc|ccc}
&0&1&2&3&4&5&6&7&8&9\\
\hline
D2/O2 & $\times$ & $\times$ & $\times$ &&&&&&&\\
\hline
D6/O6 & $\times$ & $\times$ & $\times$ &$\times$&$\times$&$\times$&$\times$&&&\\
\hline
\end{tabular}
\caption{The configuration of branes and orientifolds in type IIA string theory.}
\label{tb:typeIIA}
\end{table}
The worldvolume theory on $N$ D2-branes in the presence of $l$ D6-branes give rise to 3d $\mathcal{N}=4$ $U(N)$ gauge theory with $l$ hypermultiplets in the fundamental representation and one hypermultiplet in the adjoint representation, which is called the $U(N)$ ADHM theory. 
The Coulomb branch of the 3d theory is realized when the D2-branes are apart from the D6-branes. It is parameterized by the position of the D2-branes together with vacuum expectation values (vevs) of scalars which are dual to 3d photons. The Higgs branch of the 3d theory is realized when the D2-branes are on top of the D6-branes 
and is give by the moduli space of $N$ $SU(l)$ instantons \cite{Douglas:1995bn, deBoer:1996mp, deBoer:1996ck}. 

We can add an O2-plane or an O6-plane into the configuration without further breaking the supersymmetry. The O2-plane and the O6-plane are placed parallel to the D2-branes and the D6-branes respectively, as shown in Table \ref{tb:typeIIA}. For both O2-plane and O6-plane, there are four types of orientifolds depending on the discrete torsion associated to the Neveu-Schwarz (NS) B-field and a Ramond-Ramond (R-R) field. The four types are denoted by $O2^-, \widetilde{O2}^-, O2^+, \widetilde{O2}^+$ and $O6^-, \widetilde{O6}^-, O6^+, \widetilde{O6}^+$. Note that in order to introduce an $\widetilde{O6}^{\pm}$-plane one needs to turn on an odd background cosmological constant in type IIA string theory \cite{Hyakutake:2000mr, deBoer:2001wca, Bergman:2001rp, Bertoldi:2002nn}. 
Since we are interested in configurations which can be lifted to M-theory we will not introduce an $\widetilde{O6}^{\pm}$-plane as their M-theory lift has not been known to our best knowledge. We will not also consider an $\widetilde{O2}^+$-plane since it will not give rise to a new theory in the setup we will focus on. 

Let us consider each case one by one. Introducing an $O2^-$-plane changes the gauge group $U(2N)$ into $O(2N)$. Similarly the adjoint hypermultiplet of $U(2N)$ becomes an adjoint  hypermultiplet of $O(2N)$. 
Then the 3d theory on $N$ D2-branes on top of an $O2^-$-plane with $l$ D6-branes gives 3d $\mathcal{N}=4$ $O(2N)$ gauge theory with $l$ hypermultiplets in the fundamental representation and one hypermultiplet in the adjoint (i.e. rank-two antisymmetric) representation. An $\widetilde{O2}^-$-plane can be effectively given by an $O2^-$-plane with a half D2-brane stuck at the orientifold as far as the D2-brane charge is concerned. The effective half D2-brane further alters the gauge group into $O(2N+1)$. Hence the 3d theory on $N$ D2-branes on top of an $\widetilde{O2}^-$-plane with $l$ D6-branes realizes 3d $\mathcal{N}=4$ $O(2N+1)$ gauge theory with $l$ hypermultiplets in the fundamental representation and one hypermultiplet in the adjoint (i.e. rank-two antisymmeric) representation. On the other hand an $O2^+$-plane 
changes the gauge group $U(2N)$ into $USp(2N)$. Then the 3d theory on $N$ D2-branes on top of an $O2^+$-plane 
with $l$ D6-branes gives 3d $\mathcal{N}=4$ $USp(2N)$ gauge theory with $l$ hypermultiplets in the fundamental representation and one hypermultiplet in the adjoint (i.e. rank-two symmetric) representation. 

In the cases of introducing an O6-plane, the orientifold action on the adjoint hypermultiplet is different from the action on the vector multiplets \cite{Witten:1995gx}. Then when the gauge group changes into $O(2N+\gamma)/USp(2N)$ $(\gamma=0 \text{ or } 1)$ the adjoint hypermultiplet becomes a hypermultiplet in the rank-two symmetric/antisymmetric representation respectively. When we consider $N$ D2-branes with an $O6^-$-plane and $l$ D6-branes, the worldvolume theory on the D2-branes is the 3d $\mathcal{N}=4$ $USp(2N)$ gauge theory with $l$ hypermultiplets or $2l$ half-hypermultiplets in the fundamental representation and a hypermultiplet in the rank-two antisymmetric representation. 
On the other hand an $O6^+$-plane 
changes the unitary gauge group into an orthogonal group. Then the 3d theory on $(2N+\gamma)$ half D2-branes with an $O6^+$-plane 
and $l$ D6-branes gives the 3d $\mathcal{N}=4$ $O(2N+\gamma)$ gauge theory with $l$ hypermultiplets in the fundamental representation and one hypermultiplet in the rank-two symmetric representation. When 
$\gamma=1$, one half D2-brane should be stuck at the orientifold.  
In these cases the Higgs branch moduli space is the moduli space of $N$ $G_l$ instantons where $G_l$ is the same as the flavor symmetry group associated with the fundamental hypermultiplets \cite{Witten:1995gx, Douglas:1995bn, deBoer:1996mp, Kapustin:1998fa, Hanany:1999sj, Hanany:2001iy}. 

\begin{table}[t]
\centering
\begin{tabular}{c|ccc|cccc|ccc|c}
&0&1&2&3&4&5&6&7&8&9&11\\
\hline
M2 & $\times$ & $\times$ & $\times$ &&&&&&&&\\
\hline
KK & $\times$ & $\times$ & $\times$ &$\times$&$\times$&$\times$&$\times$&&&&\\
\hline
\end{tabular}
\caption{The configuration of M2-branes and KK monopoles in M-theory. The $x_{11}$ is the direction along the M-theory circle. }
\label{tb:M}
\end{table}
It is possible to lift the type IIA configurations to M-theory by taking the strong string coupling limit. In M-theory $N$ D2-branes simply become $N$ M2-branes. On the other hand, D6-branes are geometrized and a D6-brane becomes a Kaluza-Klein (KK) monopole \cite{Townsend:1995kk}. The configuration in M-theory is summarized in Table \ref{tb:M} where $x_{11}$ is the direction along the M-theory circle. Then the transverse space of $l$ D6-branes is described by an $l$-center Taub-NUT space $TN_l$ in the $x_7, x_8, x_9, x_{11}$-direction . When $l$ D6-branes are on top of each other, the $l$ centers are at the same position which gives rises to an $A_{l-1}$ singularity. The transverse geometry around the singularity is described by an asymptotically locally Euclidean (ALE) space $X_{A_{l-1}} = \mathbb{C}^2/\mathbb{Z}_l$. Here $\mathbb{Z}_l$ is the cyclic group or order $l$. Therefore, $N$ D2-branes probing $l$ D6-branes in type IIA string theory become $N$ M2-branes probing the $A_{l-1}$ singularity of $\mathbb{C}^2/\mathbb{Z}_l$ in M-theory. In this picture the Coulomb branch of the 3d theory on the M2-branes can be explicitly seen as the geometry which M2-branes probe. When $N=1$ the Coulomb branch is $X_{A_{l-1}}$ itself and for general $N$ it is given by the $N$-th symmetric product,
\begin{align}\label{moduliADHM}
\mathcal{M}_C = \text{Sym}^NX_{A_{l-1}}.
\end{align}

Let us then consider the M-theory lift of the orientifolds. We start from an $O2$-plane. In the presence of an $O2$-plane the space in the $x_3, \cdots, x_9$-direction becomes an orbifold $\mathbb{R}^7/\mathbb{Z}_2$ where $\mathbb{Z}_2$ action is 
\be
(x_3, x_4, x_5, x_6, x_7, x_8, x_9) \to (-x_3, -x_4, -x_5, -x_6, -x_7, -x_8, -x_9),
\ee
with a sign flip for the R-R 1-form. An $O2$-plane at the origin of $\mathbb{R}^7/\mathbb{Z}_2$ is given by two $OM2$-planes which sit at the two fixed points of $(\mathbb{R}^7\times S^1)/\mathbb{Z}_2$ \cite{Sethi:1998zk, Berkooz:1998sn}. Here the $S^1$ is the M-theory circle on which M-theory is reduced to type IIA string theory. 
We can reparameterize the eight real coordinates of the $\mathbb{R}^7\times S^1$ by four complex coordinates $y_1, y_2, y_3, y_4$ of $\mathbb{C}^4$,
\begin{equation}\begin{split}
&x_3 = \text{Re}(y_1),\; x_4 = \text{Im}(y_1),\; x_5 = \text{Re}(y_2),\; x_6 = \text{Im}(y_2), \cr
&x_7 = \text{Re}(y_3y_4^{\ast}),\; x_8 = \text{Im}(y_3y_4^{\ast}),\; x_9 = |y_3|^2 - |y_4|^2,\; x_{11} = \frac{1}{2}(\text{arg}(y_3) + \text{arg}(y_4)).
\end{split}\end{equation}
Then the $\mathbb{Z}_2$ action on $(x_3, x_4, x_5, x_6, x_7, x_8, x_9, x_{11})$ can be realized by \cite{Mezei:2013gqa}
\begin{equation}\label{OM2Z2}
    (y_1, y_2, y_3, y_4) \to (-y_1, -y_2, iy_4^{\ast}, iy_3^{\ast}).
\end{equation}
There are two types of $OM2$-planes denoted by $OM2^{\pm}$-planes. On the quotient space, the $OM2^-$-plane has $-\frac{1}{16}$ units of M2-brane charge whereas $OM2^+$-plane has $\frac{3}{16}$ units of M2-brane charge. Their M2-brane charges suggest that an $O2^-$-plane splits into two $OM2^-$-planes and an $\widetilde{O2}^-$-plane splits into two $OM2^+$-planes. 
On the other hand an $O2^+$-plane splits into one $OM2^-$-plane at one fixed point and one $OM2^+$-plane at the other fixed point. 
Introducing $l$ physical D6-branes is realized by having $2l$ KK monopoles in the covering space. 
When the KK monopoles are on top of each other the space has an $A_{2l-1}$ singularity which is locally described by an ALE space $\mathbb{C}^2/\mathbb{Z}_{2l}$ in the $(y_3, y_4)$-direction where the $\mathbb{Z}_{2l}$ action is given by
\begin{equation}\label{Z2l}
    (y_3, y_4) \to e^{\frac{\pi i}{l}}(y_3, y_4).
\end{equation}
When the $2l$ KK monopoles are on top of the $OM2^{\pm}$-plane, the location of the KK monopoles with the $OM2^{\pm}$-plane develops a $D_{l+2}$ singularity since the combinations of the action \eqref{OM2Z2} and \eqref{Z2l} yield $X_{D_{l+2}} = \mathbb{C}^2/\widehat{D}_l$ \cite{Gang:2011xp,Mezei:2013gqa}. Here $\widehat{D}_{l}$ stands for the binary dihedral group of order $4l$, which is also known as dicyclic group.
Then the Coulomb branch of the 3d theory on the $N$ M2-branes is given by 
\begin{align}
\label{moduli_Dl+2}
\mathcal{M}_C = \text{Sym}^NX_{D_{l+2}}.
\end{align}

As for $O6$-planes, an $O6^-$-plane becomes a smooth geometry and the transverse space of the M-theory uplift of an $O6^-$-plane is given by the Atiyah-Hitchin space \cite{Seiberg:1996bs, Seiberg:1996nz, Sen:1997kz}. When $l\; (l \geq 3)$ physical D6-branes are on top of an $O6^-$-plane the configuration exhibits a $D_l$ singularity in M-theory at the location where the D6-branes and the $O6^-$-plane are placed. The transverse geometry around the singularity is described by $X_{D_l} = \mathbb{C}^2/\widehat{D}_{l-2}$ \cite{Seiberg:1996bs, Seiberg:1996nz, Sen:1997kz}. 
Then the Coulomb branch of the 3d theory on the $N$ M2-branes is given by
\begin{align}
\label{moduli_Dl}
\mathcal{M}_C = \text{Sym}^NX_{D_{l}}.
\end{align}
On the other hand an $O6^+$-plane is lifted to a frozen $D_4$ singularity \cite{Landsteiner:1997ei, Witten:1997bs}. A non-zero flux is turned on around the singularity. The flux prohibits a resolution and hence it is called a frozen singularity. When $l$ physical D6-branes are on top of an $O6^+$-plane the transverse geometry around the singularity is given by $X_{D_{l+4}} = \mathbb{C}^2/\widehat{D}_{l+2}$. Therefore the Coulomb branch of the 3d theory on the $N$ M2-branes is 
\begin{align}
\label{moduli_Dl+4}
\mathcal{M}_C = \text{Sym}^NX_{D_{l+4}}.
\end{align}

It is also possible to consider M-theory from the beginning to construct 3d theories. An interesting class of such examples arise from $N$ M2-branes probing a singularity of $\mathbb{C}^4/\mathbb{Z}_k\; (k=1. 2, \cdots)$ with the $\mathbb{Z}_k$ action given by 
\be\label{orbifold1}
(z_1, z_2, z_3, z_4) \to \left(e^{\frac{2\pi i}{k}}z_1, e^{\frac{2\pi i}{k}}z_2, e^{\frac{2\pi i}{k}}z_3, e^{\frac{2\pi i}{k}}z_4\right),
\ee 
where $z_1, z_2, z_3, z_4$ are the four complex coordinates of $\mathbb{C}^4$. The commutant of the orbifold action in $SO(8)$, which is the rotation group for $\mathbb{R}^8$, is $SU(4) \times U(1)$. In the 3d theory on the M2-branes $SU(4)$ serves as the R-symmetry. Hence the 3d theory generically has an $\mathcal{N}=6$ supersymmetry. In this case the $\mathcal{N}=4$ Coulomb branch is combined with the $\mathcal{N}=4$ Higgs branch and the total moduli sapce of the 3d theory on the $N$ M2-branes is given by
\begin{align}\label{moduliABJM}
\mathcal{M} = \text{Sym}^N\left(\mathbb{C}^4/\mathbb{Z}_k\right).
\end{align}
The supersymmetry is enhanced in the cases of $k=1, 2$. When $k=1$, the geometry which the M2-branes probe is simply $\mathbb{C}^4$ and the full $SO(8)$ symmetry remains. Then the theory possesses an $\mathcal{N}=8$ supersymmetry. When $k = 2$, the geometry is described by $\mathbb{C}^4/\mathbb{Z}_2$ with the orbifold action $z_I \to -z_I\; (I=1, 2, 3, 4)$. The orbifold action commutes with the whole $SO(8)$ and the supersymmetry is also enhanced to $\mathcal{N}=8$. The 3d theory realized on $N$ M2-branes probing a singularity of $\mathbb{C}^4/\mathbb{Z}_k$ can be described by a Lagrangian theory, called the ABJM theory, characterized by $U(N) \times U(N)$ gauge groups with a Chern-Simons term of level $k$ for one $U(N)$ and that of level $-k$ for the other $U(N)$ \cite{Aharony:2008ug}. The theory also has a hypermultipet in the bifundamental representation of $U(N) \times U(N)$ and a twisted hypermultiplet in the bifundamental representation of $U(N) \times U(N)$ in the $\mathcal{N}=4$ language. The relation between the M2-brane picture and the Lagrangian theory can be explicitly seen by considering a type IIB dual configuration, which will be discussed in section \ref{sec:typeIIB}. 

This class of theories can be generalized by deforming the rank of the gauge groups and/or changing the unitary gauge groups into $O\times USp$ \cite{Hosomichi:2008jb, Aharony:2008gk}. Such theories are referred to as the ABJ theories. The rank deformation can be achieved by introducing M5-branes wrapped on a vanishing 3-cycle in $\mathbb{C}^4/\mathbb{Z}_k$. The 3-cycle is a torsion cycle characterized by $H_3\left(S^7/\mathbb{Z}_k, \mathbb{Z}\right) = \mathbb{Z}_k$. Then wrapping $k$ M5-branes on it is equivalent to no M5-brane and we can wrap at most $(k-1)$ M5-branes. The M5-branes wrapped on the vanishing 3-cycle can be interpreted as fractional M2-branes. The presence of $L (< k)$ fractional M2-branes alters the gauge group $U(N)_k \times U(N)_{-k}$ into $U(N+L)_k \times U(N)_{-k}$ with the matter content unchanged. 
The subscripts of the gauge groups represent the CS levels associated with the gauge groups and we will use this notation throughout this paper. 
The amount of supersymmetry does not change since the fractional M2-branes preserve the same supersymmetry as that of M2-branes. The theory also has a duality 
given by \cite{Aharony:2008gk} 
\begin{equation}\label{ABJdual}
U(N + L)_k \times U(N)_{-k} \Leftrightarrow U(N)_k \times U(N+k-L)_{-k},
\end{equation}
which can be seen from a type IIB dual picture. 

The change of the unitary gauge groups into $O\times USp$ can be achieved by considering an M-theory background 
$\mathbb{C}^4/\widehat{D}_k\; (k=1, 2, \cdots)$ where the orbifold action is generated by
\begin{align}
(z_1, z_2, z_3, z_4) & \to \left(e^{\frac{\pi i}{k}} z_1, e^{\frac{\pi i}{k}} z_2, e^{\frac{\pi i}{k}} z_3, e^{\frac{\pi i}{k}} z_4\right),\\
(z_1, z_2, z_3, z_4) & \to (iz_2^{\ast}, -iz_1^{\ast}, iz_4^{\ast}, -iz_3^{\ast}).
\end{align}
The orbifold action can be embedded in $SU(2)$ and the commutant inside $SO(8)$ is $SO(5)$. Hence the theory has an $\mathcal{N}=5$ supersymmetry generically. The moduli space of the 3d theory on the $N$ M2-branes is given by
\begin{align}\label{moduliABJOSp}
\mathcal{M} = \text{Sym}^N\left(\mathbb{C}^4/\widehat{D}_k\right).
\end{align}
The geometry also has a vanishing 3-cycle which is characterized by $H_3\left(S^7/\widehat{D}_k, \mathbb{Z}\right) = \mathbb{Z}_{4k}$. Then we can also wrap M5-branes on the vanishing 3-cycle, leading to fractional M2-branes. 
When $N$ M2-branes probe the singularity of $\mathbb{C}^4/\widehat{D}_k$ with some fractional M2-branes the 3d theory on the M2-branes can be described again by a Chern-Simons matter theory \cite{Aharony:2008gk} and they are characterized by the following four types of gauge groups and the CS levels,
\begin{align}
O(2N+2L_1)_{2k} \times USp(2N)_{-k} 
, \label{OSpABJ1}\\
USp(2N + 2L_2)_k \times O(2N)_{-2k} 
, \label{OSpABJ2}\\
O(2N + 2L_3 + 1)_{2k} \times USp(2N)_{-k} 
,\label{OSpABJ3}\\
USp(2N + 2L_4)_k \times O(2N+1)_{-2k} 
.  \label{OSpABJ4}
\end{align} 
Each theory has a half-hypermultiplet in the bifundamental representation of $O \times USp$ and a twisted half-hypermultiplet in the bifundamental representation of $O \times USp$. 
The $L_1, L_2, L_3, L_4$  are 
restricted by $0 \leq L_1 \leq k+1$, $0\leq L_2 \leq k-1$, $0\leq L_3 \leq k$ and $0\leq L_4 \leq k$. 
The theories in \eqref{OSpABJ1}-\eqref{OSpABJ4} also have dual descriptions given by \cite{Aharony:2008gk}
\begin{align}
O(2N+2L_1)_{2k} \times USp(2N)_{-k}  &\Leftrightarrow O(2N+2(k-L_1+1))_{-2k} \times USp(2N)_{k}, \label{OSpABJdual1}\\
USp(2N + 2L_2)_k \times O(2N)_{-2k} &\Leftrightarrow USp(2N + 2(k-L_2-1))_{-k} \times O(2N)_{2k}, \label{OSpABJdual2}\\
O(2N + 2L_3 + 1)_{2k} \times USp(2N)_{-k} &\Leftrightarrow O(2N + 2(k-L_3) + 1)_{-2k} \times USp(2N)_{k}, \label{OSpABJdual3}\\
USp(2N + 2L_4)_k \times O(2N+1)_{-2k} &\Leftrightarrow USp(2N + 2(k-L_4))_{-k} \times O(2N+1)_{2k}.  \label{OSpABJdual4}
\end{align} 
%
The dualities imply that some of the theories of \eqref{OSpABJdual1}-\eqref{OSpABJdual4} at the boundary values of $L_i\; (i=1, 2, 3, 4)$ are equivalent with each other and we have $4k$ different choices of the $L_i$'s. 
When $k=1$ the supersymmetry is enhanced to $\mathcal{N}=6$ since the geometry becomes $\mathbb{C}^4/\mathbb{Z}_4$. This leads to orthosymplectic-unitary dualities given by \cite{Aharony:2008gk, Cheon:2012be}
\begin{align}
O(2N)_2 \times USp(2N)_{-1} &\Leftrightarrow U(N)_4 \times U(N)_{-4},\label{OSpABJdualkeq1no1}\\
O(2N+2)_2 \times USp(2N)_{-1} &\Leftrightarrow U(N+2)_4 \times U(N)_{-4},\label{OSpABJdualkeq1no2}\\
\left.\begin{array}{c}
O(2N+1)_2 \times USp(2N)_{-1}\\
O(2N+3)_2 \times USp(2N)_{-1}
\end{array}\right\}&\Leftrightarrow 
\left\{
\begin{array}{c}
U(N+1)_4 \times U(N)_{-4}\\
U(N+3)_4 \times U(N)_{-4}
\end{array}\right. . \label{OSpABJdualkeq1}
\end{align}
Due to the duality \eqref{OSpABJdual3} or \eqref{ABJdual}, the two theories on each side of \eqref{OSpABJdualkeq1} are related by the parity transformation. Furthermore, gauging one-form symmetries of the theories lead to another dualities \cite{Beratto:2021xmn}
\begin{align}
SO(2N)_2\times USp(2N)_{-1} &\Leftrightarrow \left[U(N)_4 \times U(N)_{-4}\right]/\mathbb{Z}_2, \label{BMS1}\\
\left[SO(2N)_2\times USp(2N)_{-1}\right]/\mathbb{Z}_2 &\Leftrightarrow \left[U(N)_4 \times U(N)_{-4}\right]/\mathbb{Z}_4,\label{BMS2}\\
SO(2N+1)_2\times USp(2N)_{-1} &\Leftrightarrow \left[U(N+1)_4 \times U(N)_{-4}\right]/\mathbb{Z}_2,\label{BMS3}\\
SO(2N+2)_2\times USp(2N)_{-1} &\Leftrightarrow \left[U(N+2)_4 \times U(N)_{-4}\right]/\mathbb{Z}_2.\label{BMS4}
\end{align}
The special cases with $N=1$ for \eqref{BMS1} and \eqref{BMS2} give \cite{Beratto:2021xmn}
\begin{align}
SO(2)_2\times USp(2)_{-1} &\Leftrightarrow \left[U(1)_4 \times U(1)_{-4}\right]/\mathbb{Z}_2 \Leftrightarrow U(1)_2 \times U(1)_{-2}, \label{BMS1no2}\\
\left[SO(2)_2\times USp(2)_{-1}\right]/\mathbb{Z}_2 &\Leftrightarrow \left[U(1)_4 \times U(1)_{-4}\right]/\mathbb{Z}_4 \Leftrightarrow U(1)_1 \times U(1)_{-1}. \label{BMS2no2}
\end{align}

Another generalization is to consider a M-theory background $\left(\mathbb{C}^2/\mathbb{Z}_p \times \mathbb{C}^2/\mathbb{Z}_q\right)/\mathbb{Z}_k$ where $p, q, k$ are positive integers. The action of $\mathbb{Z}_p$ is given by $(z_1, z_2) \to \left(e^{\frac{2\pi i}{p}}z_1, e^{-\frac{2\pi i}{p}}z_2\right)$ while the action of $\mathbb{Z}_q$ is given by$(z_3, z_4) \to \left(e^{\frac{2\pi i}{q}}z_3, e^{-\frac{2\pi i}{q}}z_4\right)$. The $\mathbb{Z}_k$ acts on the all four complex coordinates and it is given by \eqref{orbifold1}. The whole orbifold action can be summarized as
\be
(z_1, z_2, z_3, z_4) \to \left(e^{\frac{2\pi i}{kp}}z_1, e^{-\frac{2\pi i}{kp}}z_2, e^{\frac{2\pi i}{kq}}z_3, e^{-\frac{2\pi i}{kq}}z_4\right).
\ee
The orbifold action can be embedded in $SU(2) \times SU(2)$ and the commutant inside $SO(8)$ is $SO(4)$. Hence the 3d theory on M2-branes probing the singularity has an $\mathcal{N}=4$ supersymmetry generically. The moduli space of the 3d theory on the $N$ M2-branes is given by 
\begin{align}
\mathcal{M} = \text{Sym}^N\left(\left(\mathbb{C}^2/\mathbb{Z}_p \times \mathbb{C}^2/\mathbb{Z}_q\right)/\mathbb{Z}_k\right).
\end{align}
It turns out that the moduli space can be achieved by a Chern-Simons matter theory characterized by a circular quiver theory with the following gauge groups and the CS level \cite{Imamura:2008dt, Imamura:2008nn},
\begin{align}\label{circulargauge}
U(N)_k\times \underbrace{U(N)_0 \times \cdots U(N)_0}_{p-1} \times U(N)_{-k} \times \underbrace{U(N)_0 \times \cdots U(N)_0}_{q-1}.
\end{align}
The theory has twisted hypermultiplets and hypermultiplets, both of which are in the bifundamental representation of $U(N) \times U(N)$ that are next to each other.  
When one of the two $U(N)$'s comes from the $(p-1)$ $U(N)$'s of \eqref{circulargauge} then the bifundamental matter is a twisted hypermultiplet. When it comes from the $(q-1)$ $U(N)$'s of \eqref{circulargauge} then the bifundamental matter is a hypermultiplet. 
We can further introduce fractional M2-branes into the configuration. 
The presence of the fractional M2-branes can change the gauge group \eqref{circulargauge} into
\begin{equation}
\begin{split}
&U(N + L_1)_k \times\underbrace{U(N+L_2)_0 \times \cdots U(N+L_p)_0}_{p-1} \times \cr
&\hspace{4cm} \times U(N + L_{p+1})_{-k}\times \underbrace{U(N+L_{p+2})_0 \times \cdots U(N + L_{p+q})_0}_{q-1},
\end{split}
\end{equation}
where $0\leq L_1, L_{p+1} \leq k-1$.
The theory also has various dual descriptions, which can be seen in a dual type IIB picture. 

When $p=2, q=1, k=2$, the moduli space becomes 
\begin{equation}
\text{Sym}^N\left(\left(\mathbb{C}^2/\mathbb{Z}_2 \times \mathbb{C}^2 \right)/\mathbb{Z}_2\right).
\end{equation}
The overall $\mathbb{Z}_2$ quotient may 
imply the presence an $OM2$-plane. Furthermore the singularity for the first $\mathbb{C}^2$ in the case of $N=1$ is given by $\mathbb{C}^2/\mathbb{Z}_4$, which is isomorphic to $\mathbb{C}^2/\widehat{D}_1$. Hence the configuration is equivalent to $N$ M2-branes probing an $OM2$-plane on top of 2 KK monopoles. When no fractional M2-brane is introduced the configuration corresponds to the one with an $OM2^-$-plane. Hence this suggests a duality \cite{Gang:2011xp}
\begin{align}\label{OCSdual}
&\text{3d $O(2N)$ gauge theory with one adjoint hyper and a fundamental hyper}\cr 
&\qquad\Leftrightarrow \quad \text{3d $U(N)_2 \times U(N)_0 \times U(N)_{-2}$ Chern-Simons matter theory}.
\end{align}
We can also introduce fractional M2-branes. Since $k=2$, we can have $0\leq L_1, L_3 \leq 1$. It turns out that this case leads to another duality \cite{Gang:2011xp},
\begin{align}\label{USpCSdual}
&\text{3d $USp(2N)$ gauge theory with one adjoint hyper and a fundamental hyper}\cr
&\qquad\Leftrightarrow\quad \text{3d $U(N)_2 \times U(N)_0 \times U(N+1)_{-2}$ Chern-Simons matter theory}.\cr
\end{align}

There is yet another Lagrangian construction called the BLG theories which describe 3d theories on multiplet M2-branes using a Lie 3-algebra \cite{Bagger:2006sk, Bagger:2007jr, Bagger:2007vi, Gustavsson:2007vu, Gustavsson:2008dy}. When we preserve $\mathcal{N}=8$ supersymmetry, there are two families of the BLG theories, which are characterized by the gauge groups $G=SU(2)_k \times SU(2)_{-k}$ or $G=(SU(2)_k \times SU(2)_{-k})/\mathbb{Z}_2$ \cite{VanRaamsdonk:2008ft, Aharony:2008ug,Lambert:2010ji}. From the Lagrangian description one can calculate the moduli space for each case and it is given by 
\begin{equation}\label{moduliBLG1}
\mathcal{M}= \left(\mathbb{C}^4 \times \mathbb{C}^4\right)/\mathbb{D}_{4k},
\end{equation}
for $G=SU(2)_k \times SU(2)_{-k}$ and 
\begin{equation}\label{moduliBLG2}
\mathcal{M}= \left(\mathbb{C}^4 \times \mathbb{C}^4\right)/\mathbb{D}_{2k},
\end{equation}
for $G=(SU(2)_k \times SU(2)_{-k})/\mathbb{Z}_2$  \cite{Lambert:2008et,Distler:2008mk} where $\mathbb{D}_k$ is the dihedral group of order $k$. For special values of $k$ the moduli spaces \eqref{moduliBLG1} and \eqref{moduliBLG2} agree with the $N=2$ case of \eqref{moduliABJM}. Then the BLG theory has an interpretation of a 3d theory on two M2-branes probing an $A$-type singularity. Since $\mathbb{D}_2 \cong \mathbb{Z}_2$, we have
\begin{align}
\mathcal{M}_{(SU(2)_1 \times SU(2)_{-1})/\mathbb{Z}_2} = \text{Sym}^2\left(\mathbb{C}^4\right),
\end{align}
and this implies a duality \cite{Lambert:2010ji, Bashkirov:2011pt}
\begin{equation} \label{BLGdual1}
(SU(2)_1 \times SU(2)_{-1})/\mathbb{Z}_2 \text{ BLG} \Leftrightarrow U(2)_1 \times U(2)_{-1} \text{ ABJM}.
\end{equation}
The other two cases are $SU(2)_2 \times SU(2)_{-2}$ and $(SU(2)_4 \times SU(2)_{-4})/\mathbb{Z}_2$ and their moduli spaces can be identified with $\text{Sym}^2\left(\mathbb{C}^4/\mathbb{Z}_2\right)$ with discrete torsion turned on for the latter case. Hence we have dualities \cite{Lambert:2010ji, Bashkirov:2011pt}
\begin{align}
SU(2)_2 \times SU(2)_{-2}\text{ BLG} \Leftrightarrow U(2)_2\times U(2)_{-2} \text{ ABJM}, \label{BLGdual2}\\
(SU(2)_4 \times SU(2)_{-4})/\mathbb{Z}_2 \text{ BLG} \Leftrightarrow U(3)_2\times U(2)_{-2} \text{ ABJ}. \label{BLGdual3}
\end{align}
There is also another type of duality given by \cite{Agmon:2017lga}
\be
(SU(2)_3 \times SU(2)_{-3})/\mathbb{Z}_2 \text{ BLG} \otimes U(1)_1\times U(1)_{-1} \text{ ABJM} \Leftrightarrow U(3)_1\times U(3)_{-1} \text{ ABJM}. \label{ACPdual}
\ee

\subsection{Type IIB construction}
\label{sec:typeIIB}
In fact 
most of the theories considered in the previous subsection can be also realized as 
low energy effective theories on D3-branes in type IIB string theory compactified on $S^1$.
These brane setups consist of D3-branes wrapped on $S^1$ segmented by the NS5-branes or the bound states of an NS5-brane and $k$ D5-branes (which we call $(1,k)$ 5-brane) and D5-branes.
Here each five-brane is extended in the directions indicated in Table \ref{220327_table_ADHM} and Table \ref{220327_table_N4circularquiverSCCS}.
In a brane configuration consisting only of D3-branes, NS5-branes and D5-branes, each segment of D3-branes corresponds to a gauge node.
The open strings ending on single segment correspond to an ${\cal N}=4$ vector multiplet, while the open strings between the D3-brane segment and somewhere else corresponds to an ${\cal N}=4$ hypermultiplet.
We can also consider 
NS5-branes and D5-branes extended in the different directions, which we shall call ${\widetilde {\rm NS5}}$-brane and ${\widetilde {\rm D5}}$-brane, as indicated in Table \ref{220327_table_ADHM}.
Brane setups consisting only of D3-branes, ${\widetilde {\rm NS}}$5-branes and ${\widetilde {\rm D5}}$-branes also realizes ${\cal N}=4$ quiver gauge theories, where the open strings ending on single segment correspond to an ${\cal N}=4$ twisted vector multiplet, while the open strings between the D3-brane segment and somewhere else corresponds to an ${\cal N}=4$ twisted hypermultiplet.

\begin{table}
\begin{center}
\begin{tabular}{c|ccc|c|ccc|ccc}
&0&1&2&3&4&5&6&7&8&9\\ \hline
D3(
\begin{tikzpicture}[scale=0.05]
\draw (0,2)--(12,2);
\end{tikzpicture}
)&$\times$&$\times$&$\times$&$\times$&&&&&&\\ \hline
NS5(
\begin{tikzpicture}[scale=0.05]
\draw (6,0)--(6,8);
\end{tikzpicture}
)&$\times$&$\times$&$\times$&&$\times$&$\times$&$\times$\\ \hline
D5(
\begin{tikzpicture}[scale=0.05]
\draw [blue] (8,8)--(4,6);
\end{tikzpicture}
)&$\times$&$\times$&$\times$&&&&&$\times$&$\times$&$\times$\\ \hline
${\widetilde {\rm NS5}}$(
\begin{tikzpicture}[scale=0.05]
\draw (8,8)--(4,6);
\end{tikzpicture}
)&$\times$&$\times$&$\times$&&&&&$\times$&$\times$&$\times$\\ \hline
${\widetilde {\rm D5}}$(
\begin{tikzpicture}[scale=0.05]
\draw [blue] (6,0)--(6,8);
\end{tikzpicture}
)&$\times$&$\times$&$\times$&&$\times$&$\times$&$\times$\\ \hline
\end{tabular}
\vspace{1cm}\\
\begin{tabular}{c|c|c}
brane configuration&supermultiplet&quiver\\ \hline
\begin{tikzpicture}[scale=0.1]
\draw (0,0)--(0,8);
\draw (12,0)--(12,8);
\draw (0,2)--(12,2);
\draw [red, domain=0:180, samples=100] plot({6+cos(\x)*(2+0.04*cos(20*\x))},{2+sin(\x)*(2+0.04*cos(20*\x))});
\end{tikzpicture}
&${\cal N}=4$ vector multiplet&
\scalebox{0.7}{
\begin{tikzpicture}[scale=0.1]
\draw [domain=0:360, samples=100, fill=yellow!20] plot({6+5.7*cos(\x)},{4+5.7*sin(\x)});
\node [below] at (6,7.7) {$U(N)$};
\end{tikzpicture}
}
\\ \hline
\begin{tikzpicture}[scale=0.1]
\draw (6,0)--(6,8);
\draw (0,2)--(12,2);
\draw [red, domain=0:180, samples=100] plot({6+cos(\x)*(2+0.04*cos(20*\x))},{2+sin(\x)*(2+0.04*cos(20*\x))});
\end{tikzpicture}
& ${\cal N}=4$ bifundamental hypermultiplet&
\begin{tikzpicture}[scale=0.1]
\draw [domain=0:180, samples=100, fill=yellow!20] plot({0+4*cos(\x-90)},{4+4*sin(\x-90)});
\draw [domain=0:180, samples=100, fill=yellow!20] plot({22+4*cos(\x+90)},{4+4*sin(\x+90)});
\draw (4,4)--(18,4);
\node [below] at (11,4) {$(X,Y)$};
\end{tikzpicture}
\\ \hline
\begin{tikzpicture}[scale=0.1]
\draw (0,0)--(0,8);
\draw (12,0)--(12,8);
\draw [blue] (8,8)--(4,6);
\draw (0,2)--(12,2);
\draw [red, domain=0:180, samples=100] plot({6+0.04*cos(20*\x)},{2+5*\x/(180)});
\end{tikzpicture}
& ${\cal N}=4$ fundamental hypermultiplet&
\begin{tikzpicture}[scale=0.1]
\draw [domain=0:180, samples=100, fill=yellow!20] plot({0+4*cos(\x-90)},{0+4*sin(\x-90)});
\draw (4,0)--(18,0);
\draw [fill=cyan!10] (18,-3)--(24,-3)--(24,3)--(18,3)--(18,-3);
\node [below] at (11,0) {$(I,J)$};
\end{tikzpicture}
\\ \hline
\begin{tikzpicture}[scale=0.1]
\draw (2,3)--(-2,1);
\draw (14,3)--(10,1);
\draw (0,2)--(12,2);
\draw [red, domain=0:180, samples=100] plot({6+cos(\x)*(2+0.04*cos(20*\x))},{2+sin(\x)*(2+0.04*cos(20*\x))});
\end{tikzpicture}
&${\cal N}=4$ twisted vector multiplet&
\scalebox{0.7}{
\begin{tikzpicture}[scale=0.1]
\draw [domain=0:360, samples=100, fill=orange!20] plot({6+5.7*cos(\x)},{4+5.7*sin(\x)});
\node [below] at (6,7.7) {$U(N)$};
\end{tikzpicture}
}
\\ \hline
\begin{tikzpicture}[scale=0.1]
\draw (8,3)--(4,1);
\draw (0,2)--(12,2);
\draw [red, domain=0:180, samples=100] plot({6+cos(\x)*(2+0.04*cos(20*\x))},{2+sin(\x)*(2+0.04*cos(20*\x))});
\end{tikzpicture}
& ${\cal N}=4$ bifundamental twisted hypermultiplet&
\begin{tikzpicture}[scale=0.1]
\draw [domain=0:180, samples=100, fill=orange!20] plot({0+4*cos(\x-90)},{4+4*sin(\x-90)});
\draw [domain=0:180, samples=100, fill=orange!20] plot({22+4*cos(\x+90)},{4+4*sin(\x+90)});
\draw (4,4)--(18,4);
\node [below] at (11,4) {$({\widetilde X},{\widetilde Y})$};
\end{tikzpicture}
\\ \hline
\begin{tikzpicture}[scale=0.1]
\draw (2,3)--(-2,1);
\draw (14,3)--(10,1);
\draw [blue] (4,0)--(4,8);
\draw (0,2)--(3.5,2);
\draw (4.5,2)--(12,2);
\draw [red, domain=0:180, samples=100] plot({6-2*\x/(180)+0.04*cos(20*\x)},{2-\x/(180)});
\end{tikzpicture}
& ${\cal N}=4$ fundamental twisted hypermultiplet&
\begin{tikzpicture}[scale=0.1]
\draw [domain=0:180, samples=100, fill=orange!20] plot({0+4*cos(\x-90)},{0+4*sin(\x-90)});
\draw (4,0)--(18,0);
\draw [fill=cyan!10] (18,-3)--(24,-3)--(24,3)--(18,3)--(18,-3);
\node [below] at (11,0) {$({\widetilde I},{\widetilde J})$};
\end{tikzpicture}
\\ \hline
\end{tabular}
\caption{Top: directions of the branes in the configuration realizing 3d quiver gauge theories; 
bottom: supermultiplets corresponding to the open string (red line) ending on D3-branes in various situations.
}
\label{220327_table_ADHM}
\end{center}
\end{table}
On the other hand, 
circular quiver superconformal Chern-Simons matter theories are realized by the brane configurations consisting of D3-branes, NS5-branes and $(1,k)$ 5-branes \cite{Bergman:1999na,Kitao:1998mf}.
The ${\cal N}=4$ supersymmetry is realized by assigning the supermultiplets to the open strings ending on each D3-brane segment appropriately depending on the type of 5-branes involved \cite{Hosomichi:2008jd,Imamura:2008dt}, as summarized in Table \ref{220327_table_N4circularquiverSCCS}.\footnote{
Here we assume that D3-branes are wrapped on the direction compactified on $S^1$ and also that there is at least one NS5-brane in each configuration.
}
\begin{table}
\begin{center}
\begin{tabular}{c|ccc|c|ccc|ccc}
&0&1&2&3&4&5&6&7&8&9\\ \hline
D3(
\begin{tikzpicture}[scale=0.05]
\draw (0,2)--(12,2);
\end{tikzpicture}
)&$\times$&$\times$&$\times$&$\times$&&&&&&\\ \hline
NS5(
\begin{tikzpicture}[scale=0.05]
\draw (6,0)--(6,8);
\end{tikzpicture}
)&$\times$&$\times$&$\times$&&$\times$&$\times$&$\times$\\ \hline
$(1,k)$5(
\begin{tikzpicture}[scale=0.05]
\draw [dashed] (6,0)--(6,8);
\end{tikzpicture}
)&$\times$&$\times$&$\times$&&\multicolumn{6}{|c}{$(47)_k,(58)_k,(69)_k$}\\ \hline
\end{tabular}
\vspace{1cm}\\
\begin{tabular}{c|c|c}
brane configuration&supermultiplet&quiver\\ \hline
\begin{tikzpicture}[scale=0.1]
\draw (0,0)--(0,8);
\draw (12,0)--(12,8);
\draw (0,2)--(12,2);
\draw [red, domain=0:180, samples=100] plot({6+cos(\x)*(2+0.04*cos(20*\x))},{2+sin(\x)*(2+0.04*cos(20*\x))});
\end{tikzpicture}
&${\cal N}=4$ vector multiplet
&\scalebox{0.7}{
\begin{tikzpicture}[scale=0.1]
\draw [domain=0:360, samples=100, fill=yellow!20] plot({6+5.7*cos(\x)},{4+5.7*sin(\x)});
\node [below] at (6,7.7) {$U(N)$};
\end{tikzpicture}
}
\\ \hline
\begin{tikzpicture}[scale=0.1]
\draw (0,0)--(0,8);
\draw [dashed] (12,0)--(12,8);
\draw (0,2)--(12,2);
\draw [red, domain=0:180, samples=100] plot({6+cos(\x)*(2+0.04*cos(20*\x))},{2+sin(\x)*(2+0.04*cos(20*\x))});
\end{tikzpicture}
& ${\cal N}=2$ vector multiplet with CS level $k$
&\scalebox{0.5}{
\begin{tikzpicture}[scale=0.1]
\draw [domain=0:360, samples=100, fill=red!10] plot({6+8*cos(\x)},{4+8*sin(\x)});
\node [below] at (6,7.7) {$U(N)_k$};
\end{tikzpicture}
}
\\ \hline
\begin{tikzpicture}[scale=0.1]
\draw [dashed] (0,0)--(0,8);
\draw (12,0)--(12,8);
\draw (0,2)--(12,2);
\draw [red, domain=0:180, samples=100] plot({6+cos(\x)*(2+0.04*cos(20*\x))},{2+sin(\x)*(2+0.04*cos(20*\x))});
\end{tikzpicture}
& ${\cal N}=2$ vector multiplet with CS level $-k$
&\scalebox{0.5}{
\begin{tikzpicture}[scale=0.1]
\draw [domain=0:360, samples=100, fill=green!10] plot({6+8*cos(\x)},{4+8*sin(\x)});
\node [below] at (6,7.7) {$U(N)_{-k}$};
\end{tikzpicture}
}
\\ \hline
\begin{tikzpicture}[scale=0.1]
\draw [dashed] (0,0)--(0,8);
\draw [dashed] (12,0)--(12,8);
\draw (0,2)--(12,2);
\draw [red, domain=0:180, samples=100] plot({6+cos(\x)*(2+0.04*cos(20*\x))},{2+sin(\x)*(2+0.04*cos(20*\x))});
\end{tikzpicture}
& ${\cal N}=4$ twisted vector multiplet
&\scalebox{0.7}{
\begin{tikzpicture}[scale=0.1]
\draw [domain=0:360, samples=100, fill=orange!20] plot({6+5.7*cos(\x)},{4+5.7*sin(\x)});
\node [below] at (6,7.7) {$U(N)$};
\end{tikzpicture}
}
\\ \hline
\begin{tikzpicture}[scale=0.1]
\draw (6,0)--(6,8);
\draw (0,2)--(12,2);
\draw [red, domain=0:180, samples=100] plot({6+cos(\x)*(2+0.04*cos(20*\x))},{2+sin(\x)*(2+0.04*cos(20*\x))});
\end{tikzpicture}
& ${\cal N}=4$ bifundamental hypermultiplet
&\begin{tikzpicture}[scale=0.1]
\draw [domain=0:180, samples=100] plot({0+4*cos(\x-90)},{4+4*sin(\x-90)});
\draw [domain=0:180, samples=100] plot({22+4*cos(\x+90)},{4+4*sin(\x+90)});
\draw (4,4)--(18,4);
\node [below] at (11,4) {$(H,{\widetilde H})$};
\end{tikzpicture}
\\ \hline
\begin{tikzpicture}[scale=0.1]
\draw [dashed] (6,0)--(6,8);
\draw (0,2)--(12,2);
\draw [red, domain=0:180, samples=100] plot({6+cos(\x)*(2+0.04*cos(20*\x))},{2+sin(\x)*(2+0.04*cos(20*\x))});
\end{tikzpicture}
& ${\cal N}=4$ bifundamental twisted hypermultiplet
&\begin{tikzpicture}[scale=0.1]
\draw [domain=0:180, samples=100] plot({0+4*cos(\x-90)},{4+4*sin(\x-90)});
\draw [domain=0:180, samples=100] plot({22+4*cos(\x+90)},{4+4*sin(\x+90)});
\draw (4,4)--(18,4);
\node [below] at (11,4) {$(T,{\widetilde T})$};
\end{tikzpicture}
\\ \hline
\end{tabular}
\caption{Top: directions of the branes in the configuration realizing 
${\cal N}=4$ superconformal Chern-Simons matter theories, where $(ab)_k$ stands for the direction in $ab$-plane with angle $\arctan k$ from $a$-axis;
bottom: supermultiplets corresponding to the open string (red line) ending on D3-branes in various situations.
In \cite{Imamura:2008dt} the ${\cal N}=4$ vector multiplet and the ${\cal N}=4$ twisted vector multiplet are referred to as the auxiliary vector multiplet.
}
\label{220327_table_N4circularquiverSCCS}
\end{center}
\end{table}

These type IIB brane configurations are related to the configuration of M2-branes as follows.
By taking the T-duality in the $x_3$-direction we obtain the type IIA brane configuration, where D3-branes are transformed into D2-branes, D5-branes are transformed into D6-branens while NS5-branes become a 
KK monopole along $x_3$ (a $(1,k)$5-brane is treated as an NS5-brane and $k$ D5-branes).
By further uplifting the type IIA configuration to the M-theory with a new $S^1$ direction $x_{11}$, D2-branes become M2-branes while D6-branes become a KK monople along $x_{11}$.
Hence each of the ${\cal N}=4$ theories realized by a IIB brane configuration with $N$ D3-branes wrapped on $S^1$ and the five-branes can be interpreted as the theory of $N$ M2-branes in M-theory probing some singularity of $\mathbb{C}^4/\Gamma$.
The detail of the singularity can be read off from the KK monopole background.
For example, for the $U(N)$ ADHM theory with $l$ flavors, which is realized by the type IIB brane configuration with one NS5-brane and $l$ D5-branes, the singularity is $\mathbb{C}^2/\mathbb{Z}_l\times \mathbb{C}^2$.
For the superconformal Chern-Simons matter theory realized by $p$ NS5-branes and $q$ $(1,k)5$-branes, the singularity is $(\mathbb{C}^2/\mathbb{Z}_p \times \mathbb{C}^2/\mathbb{Z}_q)/\mathbb{Z}_k$ where $\mathbb{Z}_k$ acts on $(z_1,z_2)\in\mathbb{C}^2/\mathbb{Z}_p,(z_3,z_4)\in\mathbb{C}^2/\mathbb{Z}_q$ as $(z_1,z_2,z_3,z_4)\rightarrow (e^{\frac{2\pi i}{kp}}z_1, e^{-\frac{2\pi i}{kp}}z_2, e^{\frac{2\pi i}{kq}}z_3, e^{-\frac{2\pi i}{kq}}z_4)$ \cite{Imamura:2008nn}.

In the brane setup with NS5-branes and $(1,k)$ 5-branes we can also realize non-uniform ranks $N_1,N_2,\cdots$ by introducing fractional D3-branes stretched between each pair of an NS5-brane and a $(1,k)$ 5-brane.
See Table \ref{220327_table_N4circularquiverSCCS}.
Under the M-theory uplift these fractional D3-branes become fractional M2-branes which are trapped on top of the singularity $\mathbb{C}^4/\Gamma$ and cannot move, hence the fractional D3-branes do not affect the structure of the singularity \cite{Aharony:2008gk}.

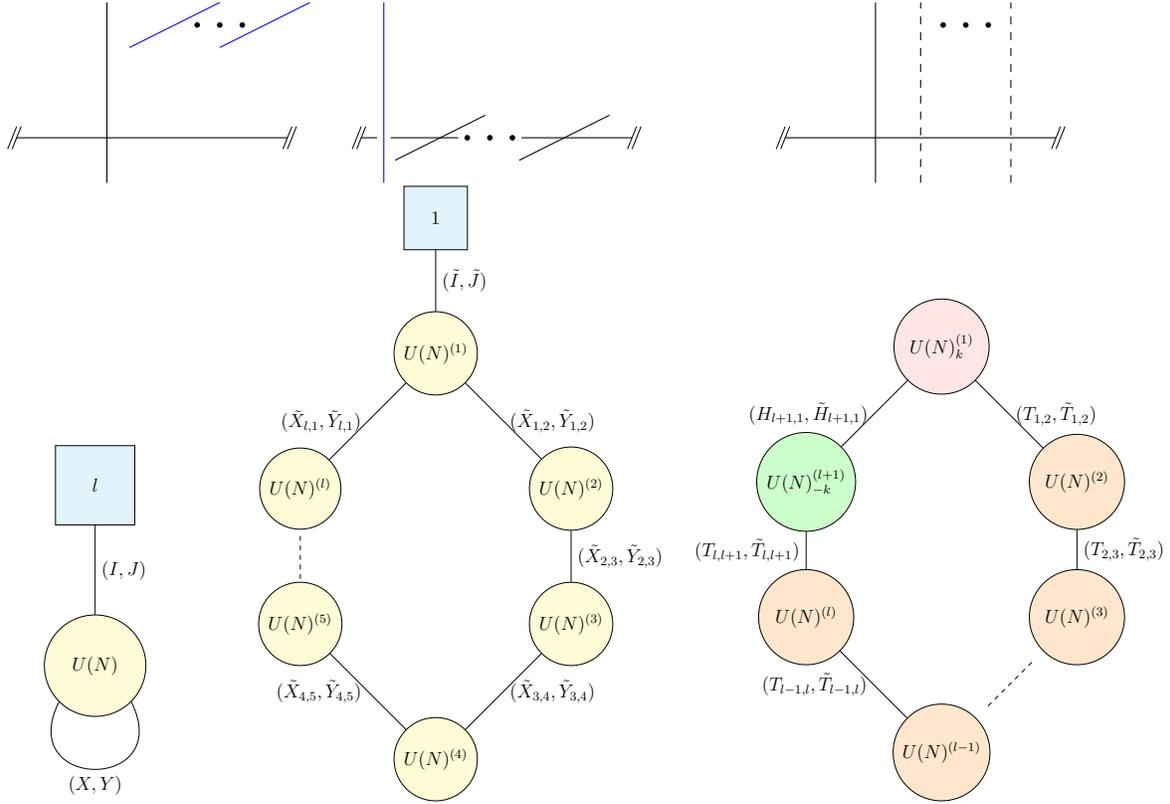
\begin{figure}
\begin{tikzpicture}[scale=0.3]
\draw (4,0)--(4,8);
\draw (0,2)--(12,2);
\draw [blue] (9,8)--(5,6);
\filldraw (8,7) circle (0.1);
\filldraw (9,7) circle (0.1);
\filldraw (10,7) circle (0.1);
\draw [blue] (13,8)--(9,6);
\draw [very thin] (0.2,2.5)--(-0.2,1.5);
\draw [very thin] (0,2.5)--(-0.4,1.5);
\draw [very thin] (12+0.2,2.5)--(12-0.2,1.5);
\draw [very thin] (12+0.4,2.5)--(12-0,1.5);
\end{tikzpicture}
\quad
\begin{tikzpicture}[scale=0.3]
\draw [blue] (1,0)--(1,8);
\draw (0,2)--(0.7,2);
\draw (1.3,2)--(4.3,2);
\draw (7.1,2)--(12,2);
\draw (5.5,3)--(1.5,1);
\filldraw (4.7,2) circle (0.1);
\filldraw (5.7,2) circle (0.1);
\filldraw (6.7,2) circle (0.1);
\draw (11,3)--(7,1);
\draw [very thin] (0.2,2.5)--(-0.2,1.5);
\draw [very thin] (0,2.5)--(-0.4,1.5);
\draw [very thin] (12+0.2,2.5)--(12-0.2,1.5);
\draw [very thin] (12+0.4,2.5)--(12-0,1.5);
\end{tikzpicture}
\quad
\quad
\quad
\quad
\begin{tikzpicture}[scale=0.3]
\draw (4,0)--(4,8);
\draw (0,2)--(12,2);
\draw [dashed] (6,0)--(6,8);
\filldraw (7,7) circle (0.1);
\filldraw (8,7) circle (0.1);
\filldraw (9,7) circle (0.1);
\draw [dashed] (10,0)--(10,8);
\draw [very thin] (0.2,2.5)--(-0.2,1.5);
\draw [very thin] (0,2.5)--(-0.4,1.5);
\draw [very thin] (12+0.2,2.5)--(12-0.2,1.5);
\draw [very thin] (12+0.4,2.5)--(12-0,1.5);
\end{tikzpicture}\\
\scalebox{0.6}{
\begin{tikzpicture}
\path 
(-2,0) node[circle, minimum size=64, fill=yellow!20, draw](AG1) {$U(N)$}
(-2,4) node[minimum size=50, fill=cyan!10, draw](AF1) {$l$};
\draw (AG1) -- node[right]{$(I,J)$}(AF1);
\draw (AG1.south west) .. controls ++(-1, -2) and ++(1, -2) .. node[below]{$(X,Y)$} (AG1.south east);
\end{tikzpicture}
}
\quad\quad
\scalebox{0.6}{
\begin{tikzpicture}
\path 
(7,2) node[circle, minimum size=48, fill=yellow!20, draw](BG6) {$U(N)^{(l)}$}
(7,-1) node[circle, minimum size=48, fill=yellow!20, draw](BG5) {$U(N)^{(5)}$}
(10,-4) node[circle, minimum size=48, fill=yellow!20, draw](BG4) {$U(N)^{(4)}$}
(13,-1) node[circle, minimum size=48, fill=yellow!20, draw](BG3) {$U(N)^{(3)}$}
(13,2) node[circle, minimum size=48, fill=yellow!20, draw](BG2) {$U(N)^{(2)}$}
(10,5) node[circle, minimum size=48, fill=yellow!20, draw](BG1) {$U(N)^{(1)}$}
(10,8) node[minimum size=40, fill=cyan!10, draw](BF1) {$1$};
\draw (BG6) -- node[left]{$(\tilde{X}_{l,1},\tilde{Y}_{l,1})$} (BG1);
\draw (BG1) -- node[right]{$(\tilde{X}_{1,2},\tilde{Y}_{1,2})$} (BG2);
\draw (BG2) -- node[right]{$(\tilde{X}_{2,3},\tilde{Y}_{2,3})$} (BG3);
\draw (BG3) -- node[right]{$(\tilde{X}_{3,4},\tilde{Y}_{3,4})$} (BG4);
\draw (BG4) -- node[left]{$(\tilde{X}_{4,5},\tilde{Y}_{4,5})$} (BG5);
\draw (BG1) -- node[right]{$(\tilde{I},\tilde{J})$}(BF1);
\draw[dashed,-]  (7,-0) to(7,1);
\end{tikzpicture}
}
\scalebox{0.6}{
\begin{tikzpicture}
\path 
(10,5) node[circle, minimum size=60, fill=red!10, draw](BG1) {$U(N)_k^{(1)}$}
(13,2) node[circle, minimum size=60, fill=orange!20, draw](BG2) {$U(N)^{(2)}$}
(13,-1) node[circle, minimum size=60, fill=orange!20, draw](BG3) {$U(N)^{(3)}$}
(10,-4) node[circle, minimum size=60, fill=orange!20, draw](BG4) {$U(N)^{(l-1)}$}
(7,-1) node[circle, minimum size=60, fill=orange!20, draw](BG5) {$U(N)^{(l)}$}
(7,2) node[circle, minimum size=48, fill=green!20, draw](BG6) {$U(N)_{-k}^{(l+1)}$};
\draw (BG1) -- node[right]{$(T_{1,2},\tilde{T}_{1,2})$} (BG2);
\draw (BG2) -- node[right]{$(T_{2,3},\tilde{T}_{2,3})$} (BG3);
\draw[dashed,-]  (12,-2) to(11,-3);
\draw (BG4) -- node[left]{$(T_{l-1,l},\tilde{T}_{l-1,l})$} (BG5);
\draw (BG5) -- node[left]{$(T_{l,l+1},\tilde{T}_{l,l+1})$} (BG6);
\draw (BG6) -- node[left]{$(H_{l+1,1},\tilde{H}_{l+1,1})$} (BG1);
\end{tikzpicture}
}
\caption{
Top: type IIB brane configuration related by $SL(2,\mathbb{Z})$ transformations;
bottom: $U(N)$ ADHM theory with $l$ flavors, $U(N)^{\otimes l}$ necklace quiver theory with one flavor and $U(N)_k\times U(N)^{\otimes (l-1)}\times U(N)_{-k}$ quiver superconformal Chern-Simons matter theory with $k=1$, each of which are realized by the three brane configuration on top of the quiver diagram.
}
\label{fig:ADHM_mirror}
\end{figure}

The type IIB brane configurations are also useful to predict the dualities of the ${\cal N}=4$ theories.
The first example is the duality induced by the $SL(2,\mathbb{Z})$ transformations of the 5-brane charges.
Let $\tau=\chi+i/g_s$ be the Type IIB coupling where $\chi=C_0$ is the axion (R-R scalar) 
and $g_s=e^{\Phi}$ is the string coupling, i.e. the expectation value of the dilaton $\Phi$ (NS-NS scalar). 
The $SL(2,\mathbb{Z})$ S-duality in Type IIB string theory act on $\tau$ as
\begin{align}
\label{S_tau}
\tau&\rightarrow 
\frac{a\tau +b}{c\tau+d},
\end{align}
with $a,b,c,d,\in \mathbb{Z}$ and $ad-bc=1$. 
A $(p,q)$ 5-brane with $p$ units of NS-NS charge and $q$ units of R-R charge transforms as 
\begin{align}
\label{S_pq}
(p\quad q)&\rightarrow (p\quad q)
\left(
\begin{matrix}
a&b\\
c&d\\
\end{matrix}
\right). 
\end{align}
The action of $SL(2,\mathbb{Z})$ S-duality can be specified by the action of two generators 
\begin{align}
\label{S_gene}
S&=\left(
\begin{matrix}
0&1\\
-1&0\\
\end{matrix}
\right),& 
T&=\left(
\begin{matrix}
1&1\\
0&1\\
\end{matrix}
\right). 
\end{align}
The $S$ transformation that swaps 
an NS5-brane with a D5-brane conjectures that the $U(N)$ ADHM theory with $l$ flavors is mirror to the 
$U(N)^{\otimes l}$ necklace quiver theory \cite{deBoer:1996mp, Porrati:1996xi, deBoer:1996ck}. The brane configuration and the quiver diagram for the $U(N)$ ADHM theory are depicted in the leftmost column in Figure \ref{fig:ADHM_mirror}. The notation of the brane setup and the quiver diagram are explained in Table \ref{220327_table_ADHM}. 
The mirror theory has the $\mathcal{N}=4$ twisted vector multiplets of gauge groups $\prod_{I=1}^{l}U(N)^{(I)}$, 
the twisted hypermultiplets $(\tilde{X}_{I,I+1}, \tilde{Y}_{I,I+1})$ transforming as the bifundamental representation 
under the $I$-th factor $U(N)^{(I)}$ and the $(I+1)$-th factor $U(N)^{(I+1)}$ of the gauge groups where $I=1,\cdots, l$ and $l+1=1$.  
Also it has a single twisted hypermultiplet $(\tilde{I},\tilde{J})$ transforming as the fundamental representation under the first factor $U(1)^{(1)}$ of the gauge groups. 
The brane setup and the quiver diagram of the mirror necklace quiver theory are displayed in the center picture in Figure \ref{fig:ADHM_mirror}. 
On the other hand, if we perform the $STS$ transformation, 
an NS5-brane turns into 
an NS5-brane while 
a D5-brane turns into 
a $(1,1)$ 5-brane.
This proposes the duality between the $U(N)$ ADHM theory with $l$ flavors and an ${\cal N}=4$ circular quiver Chern-Simons matter theory with $l+1$ nodes which consist of an ${\cal N}=2$ $U(N)^{(1)}$ vector multiplet with 
the CS level $k=1$, $l-1$ $U(N)^{(a)}$ twisted vectormultiplet ($a=2,3,\cdots,l$), an ${\cal N}=2$ $U(N)^{(l+1)}$ vector multiplet with 
the CS level $k=-1$, $l$ twisted bifundamental hypermultiplets $(T_{a,a+1},{\widetilde T}_{a,a+1})$ ($a=1,2,,\cdots,l$) and a bifundamental hypermultiplet $(H_{l+1,1},{\widetilde H}_{l+1,1})$. The brane configuration and the quiver diagram of the theory are given in 
the rightmost picture in Figure \ref{fig:ADHM_mirror}. In this case with CS levels, the notation of the brane setup and the quiver diagram in Table \ref{220327_table_N4circularquiverSCCS} is used.

\begin{figure}
\begin{center}
\scalebox{0.7}{
\begin{tikzpicture}[scale=0.3]
\filldraw (-3,2) circle (0.1);
\filldraw (-2,2) circle (0.1);
\filldraw (-1,2) circle (0.1);
\node [below] at (2,2) {$N_1$};
\node [below] at (6,2) {$N_2$};
\node [below] at (10,2) {$N_3$};
\draw (4,0)--(4,8);
\draw (0,2)--(12,2);
\draw [dashed] (8,0)--(8,8);
\filldraw (13,2) circle (0.1);
\filldraw (14,2) circle (0.1);
\filldraw (15,2) circle (0.1);
\draw (17,4) -- (18,4.5);
\draw (17,4) -- (18,3.5);
\draw (17,4) -- (21,4);
\draw (21,4) -- (20,4.5);
\draw (21,4) -- (20,3.5);
\filldraw (23,2) circle (0.1);
\filldraw (24,2) circle (0.1);
\filldraw (25,2) circle (0.1);
\node [below] at (28,2) {$N_1$};
\node [below] at (36,2) {$N_1+N_3-N_2+k$};
\node [below] at (44,2) {$N_3$};
\draw [dashed] (30,0)--(30,8);
\draw (26,2)--(46,2);
\draw (42,0)--(42,8);
\filldraw (47,2) circle (0.1);
\filldraw (48,2) circle (0.1);
\filldraw (49,2) circle (0.1);
\end{tikzpicture}
}
\end{center}
\caption{
The change of the number of D3-branes due to the Hanany-Witten brane creation/annihilation effect.
}
\label{fig_HWduality}
\end{figure}
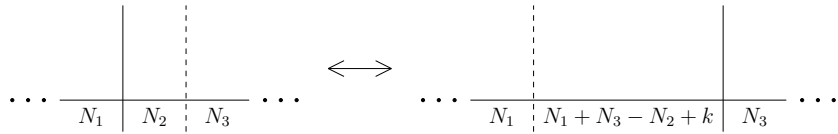

One can also move the five-branes along the $x_3$-direction and create/annihilate D3-branes on each segment according to the Hanany-Witten effect \cite{Hanany:1996ie}, which transforms the brane configuration with $M$ fractional D3-branes into a different configuration with $k-M$ fractional D3-branes (see Figure \ref{fig_HWduality}).
The ${\cal N}=4$ theories realized by the two configurations before and after this transformation are suggested to be dual to each other.
The duality of the $U(N+L)_k\times U(N)_{-2k}$ ABJ theory \eqref{ABJdual} is also obtained from this effect.

In section \ref{sec:typeIIAM}, we also constructed 3d theories with other gauge groups. Some of the theories can be also realized by introducing an orientifold in the type IIB configuration. We can introduce either an $O3$-plane along the D3-branes or an $O5$-plane along the D5-branes without further breaking supersymmetry. In the type IIA brane setup we considered an $O2$-plane or an $O6$-plane. While an $O6$-plane is T-dual to two $O5$-planes, we need two $O2$-planes to obtain an $O3$-plane. Hence we focus on the type IIB dual descriptions of the type IIA construction for the theories realized on D2-branes with an $O6$-plane.

\begin{figure}
\centering
\scalebox{0.5}{
\begin{tikzpicture}
\path 
(-2,-2) node[circle, minimum size=64, fill=yellow!20, draw](AG1) {$USp(2N)$}
(-2,1) node[minimum size=50, fill=cyan!10, draw](AF1) {$SO(2l)$};
\draw (AG1) -- node[right]{$(I,J)$}(AF1);
\draw (AG1.south west) .. controls ++(-1, -2) and ++(1, -2) .. node[below]{$(X,Y)$} (AG1.south east);
\path 
(10,5) node[circle, minimum size=64, fill=yellow!20, draw](BG1) {$U(N)^{(1)}$}
(4,5) node[circle, minimum size=64, fill=yellow!20, draw](BG2) {$U(N)^{(2)}$}
(7,2) node[circle, minimum size=64, fill=yellow!20, draw](BG3) {$U(2N)^{(5)}$}
(7,-1) node[circle, minimum size=64, fill=yellow!20, draw](BG4) {$U(2N)^{(6)}$}
(7,-4) node[circle, minimum size=64, fill=yellow!20, draw](BG5) {$U(2N)^{(l+1)}$}
(10,-7) node[circle, minimum size=64, fill=yellow!20, draw](BG6) {$U(N)^{(3)}$}
(4,-7) node[circle, minimum size=64, fill=yellow!20, draw](BG7) {$U(N)^{(4)}$}
(10,8) node[minimum size=50, fill=cyan!10, draw](BF1) {$1$};
\draw (BG3) -- node[right]{$(\tilde{X}_{1,5},\tilde{Y}_{1,5})$} (BG1);
\draw (BG3) -- node[left]{$(\tilde{X}_{2,5},\tilde{Y}_{2,5})$} (BG2);
\draw (BG3) -- node[right]{$(\tilde{X}_{5,6},\tilde{Y}_{5,6})$} (BG4);
\draw (BG5) -- node[right]{$(\tilde{X}_{l+1,3},\tilde{Y}_{l+1,3})$} (BG6);
\draw (BG5) -- node[left]{$(\tilde{X}_{l+1,4},\tilde{Y}_{l+1,4})$} (BG7);
\draw (BG1) -- node[right]{$(\tilde{I},\tilde{J})$}(BF1);
\draw[dashed,-]  (7,-2.8) to(7,-2.2);
\end{tikzpicture}
}
\caption{
Left: The 3d $USp(2N)$ ADHM theory with one antisymmetric hyper $(X,Y)$ and $2l$ half-hypers $(I,J)$. Right: The 3d $U(N)^{\otimes 4}\times U(2N)^{\otimes l-3}$ 
quiver theory with one flavor $(\tilde{I},\tilde{J})$. This theory is mirror dual of the theory on the left. } \label{fig:USp_mirror}
\end{figure}
First we consider 
the 3d $\mathcal{N}=4$ $USp(2N)$ gauge theory with $l$ flavors and a antisymmetric hypermultiplet. The field content 
can be summarized as a quiver diagram given by the left figure in Figure \ref{fig:USp_mirror}. 
The brane configuration which realizes the 3d theory is depicted in \eqref{fig:IIBUSp}. 
\begin{equation}\label{fig:IIBUSp}
 \begin{scriptsize}
\begin{tikzpicture}[scale=0.5]
\draw[thick] (-2.5,0)--(2.5,0);
\draw[blue, dotted, thick] (2,-1/4)--(3,1/4) {};
\draw[blue, dotted, thick] (-3,-1/4)--(-2,1/4) {};
\node[label=left:{$O5^-$}]  at (-2.5,0) {};
\node[label=right:{$O5^-$}]  at (2.5,0) {};
\draw[thick] (2.5,-1)--(2.5,1);
\draw[blue] (-2,1-1/2)--(-1,3/2-1/2);
\draw[blue] (-2+1/2,1-1/2)--(-1+1/2,3/2-1/2);
\node[label=right:{$\cdots$}] at (-1,1+1/4-1/2) {};
\draw[blue] (1/2,1-1/2)--(3/2,3/2-1/2);
\node[label=above:{$\overbrace{\hspace{2cm}}$}]  at (0,1+1/8-1/2) {};
\node[label=above:{$l$}]  at (0,1+3/4) {};
\node[label=below:{$2N$ D3}]  at (0,0) {};
\end{tikzpicture}
\end{scriptsize}
\end{equation}
We use a dotted diagonal line for representing an $O5$-plane. In \eqref{fig:IIBUSp} the two dotted diagonal lines are two $O5^-$-planes. The presence of the two $O5^-$-planes makes the horizontal direction periodic. 
When one end of an open string is on the D3-branes, the other end can cross the NS5-brane and then it ends on the D3-branes. Such an open string yields the antisymmetric hypermultiplet. The mirror dual of the theory is obtained by the S-dual of the configuration. Then a D5-brane on top of an $O5^-$-plane changes into an $ON^0$-plane \cite{Kutasov:1995te, Sen:1996na}. The brane setup after the S-duality becomes
\begin{equation}\label{fig:IIBUSpmirror}
 \begin{scriptsize}
\begin{tikzpicture}[scale=0.5]
\draw[thick] (-2.5,0)--(2.5,0);
\node[label=left:{$ON^0$}]  at (-2.5,0) {};
\node[label=right:{$ON^0$}]  at (2.5,0) {};
\draw[dotted, thick] (2.5,-1)--(2.5,1);
\draw[dotted, thick] (-2.5,-1)--(-2.5,1);
\draw[thick] (-2+1/4,1)--(-2+1/4,-1);
\draw[thick] (-2+3/4,1)--(-2+3/4,-1);
\node[label=right:{$\cdots$}] at (-1,1/2) {};
\draw[thick] (2-1/4,1)--(2-1/4,-1);
\node[label=above:{$\overbrace{\hspace{1.8cm}}$}]  at (0,1/2+1/8) {};
\node[label=above:{$l-2$}]  at (0,1+1/4) {};
\node[label=below:{$2N$ D3}]  at (0,-1/2) {};
\draw[blue] (2.2,1/2)--(2.8,1);
\end{tikzpicture}
\end{scriptsize}
\end{equation}
where the vertical dotted lines represent the $ON^0$-planes. Then the 3d theory realized on the D3-branes is a quiver theory whose quiver shape is given by the Dynkin diagram of $\widehat{\mathfrak{so}(2l)}$ with one flavor attached to an end node \cite{Kapustin:1998fa, Hanany:1999sj}. The extra flavor comes from the D5-brane in \eqref{fig:IIBUSpmirror}. The quiver diagram of the theory is given by the right diagram in Figure \ref{fig:USp_mirror}. This theory is the mirror dual of the $USp(2N)$ gauge theory with $l$ flavors and a antisymmetric hypermultiplet \cite{deBoer:1996mp, Porrati:1996xi, Kapustin:1998fa}.

On the other hand, the 3d $\mathcal{N}=4$ $O(2N+\gamma)$ gauge theory with $l$ fundamental hypermultiplets and a symmetric hypermultiplet is realized by brane construction with two $O5^+$-planes. The configuration can be depicted as  
\begin{equation}\label{fig:IIBO}
 \begin{scriptsize}
\begin{tikzpicture}[scale=0.5]
\draw[thick] (-2.5,0)--(2.5,0);
\draw[blue, dotted, thick] (2,-1/4)--(3,1/4) {};
\draw[blue, dotted, thick] (-3,-1/4)--(-2,1/4) {};
\node[label=left:{$O5^+$}]  at (-2.5,0) {};
\node[label=right:{$O5^+$}]  at (2.5,0) {};
\draw[thick] (2.5,-1)--(2.5,1);
\draw[blue] (-2,1-1/2)--(-1,3/2-1/2);
\draw[blue] (-2+1/2,1-1/2)--(-1+1/2,3/2-1/2);
\node[label=right:{$\cdots$}] at (-1,1+1/4-1/2) {};
\draw[blue] (1/2,1-1/2)--(3/2,3/2-1/2);
\node[label=above:{$\overbrace{\hspace{2cm}}$}]  at (0,1+1/8-1/2) {};
\node[label=above:{$l$}]  at (0,1+3/4-1/2) {};
\node[label=below:{$(2N+\gamma)$ D3}]  at (0,0) {};
\end{tikzpicture}
\end{scriptsize}
\end{equation}
Although it is also possible to consider the S-dual of the configurations 
\eqref{fig:IIBO} a conventional Lagrangian description of the theories on the D3-branes has not been known. However the Coulomb branches of the mirror theories can be extracted by non-simply laced quiver theories \cite{Cremonesi:2014xha}. 

\begin{figure}%
\centering
\subfigure[]{\label{fig:IIBABJ1}
\begin{scriptsize}
\begin{tikzpicture}[scale=0.5]
\draw (0,0) circle[x radius=3.5, y radius=0.5];
\draw[dotted, thick] (0,-1/8) circle[x radius=3.5, y radius=0.5];
\draw[thick] (3.5,-1)--(3.5,1);
\draw[dashed, thick] (-3.5,-1)--(-3.5,1);
\node[label=below:{$O3^{+} + N$ D3}]  at (0,-1/2) {};
\node[label=above:{$O3^{-} + (N+L_1)$ D3}]  at (0,1/2) {};
\end{tikzpicture}
\end{scriptsize}}\hspace{2cm}
\subfigure[]{\label{fig:IIBABJ2}
\begin{scriptsize}
\begin{tikzpicture}[scale=0.5]
\draw (0,0) circle[x radius=3.5, y radius=0.5];
\draw[dotted, thick] (0,-1/8) circle[x radius=3.5, y radius=0.5];
\draw[thick] (3.5,-1)--(3.5,1);
\draw[dashed, thick] (-3.5,-1)--(-3.5,1);
\node[label=below:{$O3^{-} + N$ D3}]  at (0,-1/2) {};
\node[label=above:{$O3^{+} + (N+L_2)$ D3}]  at (0,1/2) {};
\end{tikzpicture}
\end{scriptsize}}\\
\subfigure[]{\label{fig:IIBABJ3}
\begin{scriptsize}
\begin{tikzpicture}[scale=0.5]
\draw (0,0) circle[x radius=3.5, y radius=0.5];
\draw[dotted, thick] (0,-1/8) circle[x radius=3.5, y radius=0.5];
\draw[thick] (3.5,-1)--(3.5,1);
\draw[dashed, thick] (-3.5,-1)--(-3.5,1);
\node[label=below:{$\widetilde{O3}^{+} + N$ D3}]  at (0,-1/2) {};
\node[label=above:{$\widetilde{O3}^{-} + (N+L_3)$ D3}]  at (0,1/2) {};
\end{tikzpicture}
\end{scriptsize}}\hspace{2cm}
\subfigure[]{\label{fig:IIBABJ4}
\begin{scriptsize}
\begin{tikzpicture}[scale=0.5]
\draw (0,0) circle[x radius=3.5, y radius=0.5];
\draw[dotted, thick] (0,-1/8) circle[x radius=3.5, y radius=0.5];
\draw[thick] (3.5,-1)--(3.5,1);
\draw[dashed, thick] (-3.5,-1)--(-3.5,1);
\node[label=below:{$\widetilde{O3}^{-} + N$ D3}]  at (0,-1/2) {};
\node[label=above:{$\widetilde{O3}^{+} + (N+L_4)$ D3}]  at (0,1/2) {};
\end{tikzpicture}
\end{scriptsize}}
\caption{The brane configurations which realize the $O\times USp$ ABJ theories. 
The configurations in 
\ref{fig:IIBABJ1}, \ref{fig:IIBABJ2}, \ref{fig:IIBABJ3} and \ref{fig:IIBABJ4} give rise to \eqref{OSpABJ1}, \eqref{OSpABJ2}, \eqref{OSpABJ3} and \eqref{OSpABJ4} respectively. The vertical dashed line in each figure is a $(1, 2k)$ 5-brane. 
}
\label{fig:IIBABJ}
\end{figure}
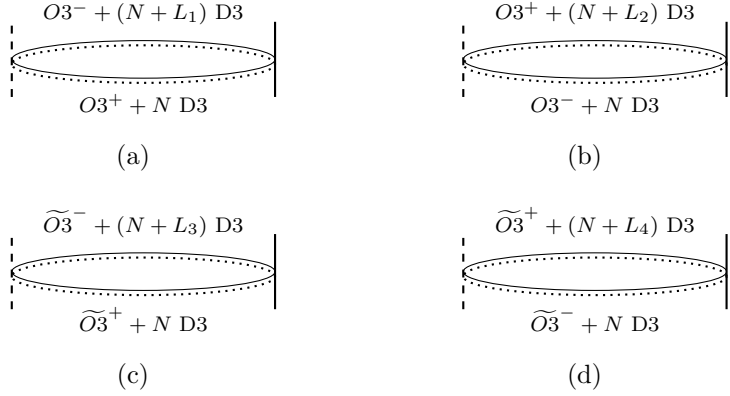

An $O3$-plane can be introduced to the type IIB brane setup for the ABJM theory, In the original ABJM setup the configuration contains an NS5-brane and a $(1, k)$ 5-brane, When an $O3^{\pm}/\widetilde{O3}^{\pm}$-plane crosses an NS5-brane it changes into respectively an $O3^{\mp}/\widetilde{O3}^{\mp}$-plane. On the other hand when an $O3^{\pm}/\widetilde{O3}^{\pm}$-plane crosses a D5-brane it becomes an $\widetilde{O3}^{\pm}/{O3}^{\pm}$-plane respectively. Hence a consistent setup requires a pair of an NS5-brane and a $(1, 2k)$ 5-brane in the setup with an $O3$-plane. Then there are four configurations with an O3-plane along D3-branes and they are summarized in Figure \ref{fig:IIBABJ}. 
The brane configurations in Figure \ref{fig:IIBABJ1}, \ref{fig:IIBABJ2}, \ref{fig:IIBABJ3} and \ref{fig:IIBABJ4} realize the 3d theories which are written in respectively \eqref{OSpABJ1}, \eqref{OSpABJ2}, \eqref{OSpABJ3} and \eqref{OSpABJ4}. The equivalence of the theories \eqref{OSpABJdual1}, \eqref{OSpABJdual2}, \eqref{OSpABJdual3} and \eqref{OSpABJdual4} 
may be seen from the Hanany-Witten effect when the NS5-brane is exchanged with the $(1, 2k)$ 5-brane \cite{Aharony:2008gk}.

\section{$U(N)$ ADHM theory with $l$ flavors}
\label{sec_ADHMu}
We start from the $U(N)$ ADHM theory, that is the low-energy effective theory of $N$ coincident M2-branes probing the $A_{l-1}$ singilarity. 
As reviewed in section \ref{sec_brane}, it is a 3d $\mathcal{N}=4$ supersymmetric gauge theory with $U(N)$ gauge group 
and one adjoint hypermultiplet $(X,Y)$ and $l$ fundamental hypermultiplets $(I_{\alpha},J_{\alpha})$, $\alpha=1,\cdots, l$. 

\subsection{Moduli space and local operators}
The moduli space of supersymmetric vacua of the gauge theory 
is determined by the following equations:
\begin{align}
    \label{adhm_bps1}
    &[\phi,\phi]=0,\\
    \label{adhm_bps4}
    &[X,X^{\dag}]+[Y,Y^{\dag}]+JJ^{\dag}-I^{\dag}I=0,\\
    \label{adhm_bps2}
    &[\phi,X]=0,\qquad [\phi,Y]=0, \qquad \phi J=0,\qquad  I\phi=0,\\
    \label{adhm_bps3}
    &[X,Y]+JI=0,
\end{align}
where $\phi$ is the the adjoint scalar field in the $\mathcal{N}=4$ vector multiplet and we split it into a real component $\sigma$ and a complex component $\varphi$. 
The first two equations (\ref{adhm_bps1}) and (\ref{adhm_bps4}) are the D-term equations and the 
equations (\ref{adhm_bps2}) and (\ref{adhm_bps3}) are the F-term equations. 
The equations (\ref{adhm_bps2}) and (\ref{adhm_bps3}) are deformed when one turns on mass parameters $m$, $m_{\textrm{adj}}$ and FI parameters $\zeta$ 
\begin{align}
    \label{adhm_bps2a}
    &[\phi,X]=m_{\textrm{adj}} X,\qquad [\phi,Y]=m_{\textrm{adj}} Y, \qquad \phi J=Jm,\qquad  I\phi=mI,\\
    \label{adhm_bps3a}
    &[X,Y]+JI=\zeta. 
\end{align}

\subsubsection{Coulomb branch}
By setting the hypermultiplet scalar fields $(X,Y)$ and $(I_{\alpha},J_{\alpha})$ to zero, 
we obtain the Coulomb branch which is parametrized by the local operators constructed from the monopole operators dressed by the vector multiplet scalar field $\varphi$. 
A solution to the equation (\ref{adhm_bps1}) is given by 
\begin{align}
\label{uN_phi_C}
\varphi=\mathrm{diag}(\varphi_1, \cdots, \varphi_N),
\end{align}
and the gauge group is broken to $U(1)^{N}$. 
The Coulomb branch receives the non-perturbative quantum corrections from the monopole operators. 
The bare monopole $v^{\{m_i\}}$ for the $U(N)$ ADHM theory with $l$ flavors carries the GNO charge as an integer vector $\vec{m}$ $=$ $(m_1,\cdots, m_N)$. 
It has the conformal dimension \cite{Gaiotto:2008ak}
\begin{align}
\label{uN_l_monoD}
\Delta(m_i)&=
\frac{l}{2}\sum_{i=1}^N |m_i|. 
\end{align}

First consider the Abelian case with $N=1$. 
The Coulomb branch operators are not independent as they obey a chiral ring relation which determines the OPE
\begin{align}
v^{+}v^{-}\sim \varphi^l. 
\end{align}
This is consistent with the dimension (\ref{uN_l_monoD}) of monopole. 

We can parametrize $\varphi$ and $v^{\pm1}$ as $\varphi=z_1 z_2$, $v^+=z_1^l$ and $v^-=z_2^l$. Since $(z_1, z_2)$ is identified with $(e^{\frac{2\pi i}{l}} z_1, e^{-\frac{2\pi i}{l}}z_2)$, the Coulomb branch operators describe the ALE singularity $X_{A_{l-1}}=\mathbb{C}^2/\mathbb{Z}_l$. 
More generally, the Coulomb branch of the non-Abelian $U(N)$ ADHM theory with $l$ flavors is given by the $N$-th symmetric product (\ref{moduliADHM}) of the ALE space $X_{A_{l-1}}$
\begin{align}
\label{Mc_u}
    \mathrm{Sym}^N X_{A_{l-1}}&=\mathrm{Sym}^{N} (\mathbb{C}^2/\mathbb{Z}_l),
\end{align}
whose dimension is $\dim_{\mathbb{C}}\mathcal{M}_C=2 N$. 
It has singularities coming from the $A_{l-1}$ singularity and from the quotient singularity of the symmetric group. 
The adjoint mass parameter $m$ resolves the quotient singularity, which results in a Hilbert scheme of $N$ points of ALE space of $A_{l-1}$-type. 
The fundamental mass parameters $m_{\alpha}$ $\alpha=1,\cdots, l$ resolve the $A_{l-1}$ singularity. 
It gives the resolved ALE space $\widetilde{X}_{A_{l-1}}$ of $A_{l-1}$-type.  

\subsubsection{Higgs branch}
When the vector multiplet scalar is turned off, we find 
the Higgs branch that is parametrized by the half-BPS local operators constructed from two types of hypermultiplet scalar fields $(X,Y)$ and $(I,J)$. 
They obey the equations (\ref{adhm_bps4}) and (\ref{adhm_bps3}). 

When the theory has a single flavor, that is $l=1$, 
the equations (\ref{adhm_bps4}) and  (\ref{adhm_bps3}) implies that the fundamental hypermultiplet $(I,J)$ vanishes and that 
the adjoint hypermultiplet $(X,Y)$ can be diagonalized
\begin{align}
    X&=\mathrm{diag}(X_1,\cdots, X_N),& 
    Y&=\mathrm{diag}(Y_1,\cdots, Y_N). 
\end{align}
The Higgs branch is given by $N$ copies of $\mathbb{C}^2$ parametrized by $(X_i,Y_i)$, $i=1,\cdots, N$ divided by the residual permutation symmetry $S_N$, which is the $N$-th symmetric product of $\mathbb{C}^2$. 

For $l\ge 2$ the equations (\ref{adhm_bps4}) and (\ref{adhm_bps3}) are identified with the ADHM equations for the $N$ $SU(l)$ instantons on $\mathbb{R}^4$ \cite{Donaldson:1984tm}. Hence The Higgs branch of the $U(N)$ ADHM theory with $l$ flavors is the moduli space of $SU(l)$ $N$-instantons. 
It has dimension  $\dim_{\mathbb{C}}\mathcal{M}_H=2Nl$

The gauge invariant operators can be described by closed words with the form $\Tr X^l Y^m$ and open words with the form $J X^l Y^m I$. 
For the closed words the multi-traces at level $n\le N$ give the gauge invariant basis 
which are one-to-one correspondence with the $p(n)$ conjugacy classes of the permutation group $S_n$ 
where $p(n)$ is the number of partition of $n$. 

\subsubsection{Mixed branch}
On the mixed branch in the moduli space, 
both scalar fields in the hypermultiplet and the vector multiplet do not vanish so that the bare monopole can be also dressed by the adjoint scalar fields $(X,Y)$ in the hypermultiplet. 
Consider the configuration where the vector multiplet scalar fields takes the form 
\begin{align}
    \varphi&=\mathrm{diag}(\underbrace{\varphi_1,\cdots,\varphi_1}_{N_1},\cdots,\underbrace{\varphi_n,\cdots,\varphi_n}_{N_n},\underbrace{0,\cdots,0}_{N_0}),
\end{align}
where $\sum_{i=0}^{n} N_i=N$. 
By fixing the gauge for the action of the Weyl group of $U(N)$, one can write the GNO charge as
\begin{align}
\label{uN_GNO}
(\underbrace{m_1,\cdots, ,m_1}_{N_1}, 
\cdots, \underbrace{m_{n'},\cdots, ,m_{n'}}_{N_{n'}}, \underbrace{m_{n'+1},\cdots,m_{n'+1}}_{N_{n'+1}},\cdots,\underbrace{m_{n'+m'},\cdots,m_{n'+m'}}_{N_{n'+m'}}\underbrace{0,\cdots,0}_{N_0}),
\end{align}
where $m_1> m_2>\cdots>m_{n'}>0$ and $m_{n'+1}<m_{n'+2}<\cdots <m_{n'+m'}<0$ with $n'+m'=n$.
The magnetic flux for the bare monopole with the GNO charge (\ref{uN_GNO}) 
breaks the $U(N)$ gauge group down to the residual gauge group $H_{\{m_i\}}=\prod_{j=1}^n U(N_j)$. 
Consequently, the adjoint scalar field takes the block-diagonal form so that the bare monopole operator with the GNO charge (\ref{uN_GNO}) will be dressed by 
\begin{align}
X&=
\left(
\begin{matrix}
X^{(1)}_{N_1\times N_1}&&& \\
&X^{(2)}_{N_2\times N_2}&&& \\
&&\ddots&& \\
&&&X^{(n)}_{N_n\times N_n}& \\
&&&&X^{(0)}_{N_0\times N_0} \\
\end{matrix}
\right),
\end{align}
which obey the F-term constraint (\ref{adhm_bps3}). 
Here the $U(N_i)$ adjoint scalar field $X^{(i)}_{N_i\times N_i}$ shows up for each factor $U(N_i)$ in $U(N)$. 
Therefore general monopole operators in the ADHM theory are dressed by a collection of adjoint scalar fields $\varphi$, $X$ and $Y$. In the following we check that such dressed monopoles contribute to the indices. 

\subsection{Indices}
The index of the $U(N)$ ADHM theory with an adjoint hyper and $l$ fundamental hypers is given by
\begin{align}
\label{uN_l_index}
&I^{\textrm{$U(N)$ ADHM$-[l]$}}(t,x,y_{\alpha},z;q)
\nonumber\\
&=\frac{1}{N!} \frac{(q^{\frac12}t^2;q)_{\infty}^N}{(q^{1/2}t^{-2};q)_{\infty}^N}
\sum_{m_1,\cdots,m_N\in \mathbb{Z}}
\oint\prod_{i=1}^N \frac{ds_i}{2\pi is_i}
\prod_{i<j}
(1-q^{\frac{|m_i-m_j|}{2}}s_i^{\pm}s_j^{\mp})
\frac{(q^{\frac{1+|m_i-m_j|}{2}}t^2s_i^{\mp}s_j^{\pm};q)_{\infty}}{(q^{\frac{1+|m_i-m_j|}{2}}t^{-2}s_i^{\pm}s_j^{\mp};q)_{\infty}}
\nonumber\\
&\times 
\frac{(q^{\frac34}t^{-1}x^{\mp};q)_{\infty}^N}{(q^{\frac14}tx^{\pm};q)_{\infty}^N}
\prod_{i<j}
\frac{(q^{\frac34+\frac{|m_i-m_j|}{2}}t^{-1}s_i^{\mp}s_j^{\pm}x^{\mp};q)_{\infty}}
{(q^{\frac14+\frac{|m_i-m_j|}{2}}ts_i^{\pm}s_j^{\mp}x^{\pm};q)_{\infty}}
\frac{(q^{\frac34+\frac{|m_i-m_j|}{2}}t^{-1}s_i^{\mp}s_j^{\pm}x^{\pm};q)_{\infty}}
{(q^{\frac14+\frac{|m_i-m_j|}{2}}ts_i^{\pm}s_j^{\mp}x^{\mp};q)_{\infty}}
\nonumber\\
&\times 
\prod_{i=1}^N
\prod_{\alpha=1}^{l}
\frac{(q^{\frac34+\frac{|m_{i}|}{2}}t^{-1}s_{i}^{\mp}y_{\alpha}^{\mp};q)_{\infty}}
{(q^{\frac14+\frac{|m_i|}{2}}ts_{i}^{\pm}y_{\alpha}^{\pm};q)_{\infty}}
q^{\sum_{i=1}\frac{l |m_i|}{4}}t^{-l \sum_{i=1}^N|m_i|}z^{l \sum_{i=1}^N m_i}. 
\end{align}
Here $x$ is the fugacity for a flavor symmetry of the adjoint hyper,  $y_{\alpha}$ are the fugacities for an $SU(l)$ flavor symmetry of the  fundamental hypers obeying $\prod_{\alpha}y_{\alpha}=1$ 
and $z$ is the fugacity for a topological symmetry. 

\subsubsection{$U(1)$ ADHM with one flavor ($N=1$, $l=1$)}
The simplest example is the case with $N=1$ and $l=1$. 
The theory describes a single M2-brane moving in a flat space $\mathbb{C}^4$. 
It contains the BPS bare monopole operators $v^m$ of the GNO charges $m\in \mathbb{Z}$ and the dimensions $\Delta(m)=|m|/2$. 

We find the index 
\begin{align}
\label{u1_1_findex}
&I^{\textrm{$U(1)$ ADHM$-[1]$}}(t,x,z;q)
\nonumber\\
&=1+
\Bigl[ (\underbrace{x}_{X}+\underbrace{x^{-1}}_{Y})t
+(\underbrace{z}_{v^{1}}+\underbrace{z^{-1}}_{v^{-1}})t^{-1} \Bigr]q^{1/4}
\nonumber\\
&+\Bigl[
\underbrace{xz}_{v^{1}X}
+\underbrace{x^{-1}z^{-1}}_{v^{-1}Y}
+\underbrace{xz^{-1}}_{v^{-1}X}
+\underbrace{x^{-1}z}_{v^{1}Y}
+(\underbrace{1}_{XY}
+\underbrace{x^2}_{X^2}
+\underbrace{x^{-2}}_{Y^2})t^2
+(\underbrace{1}_{\varphi}
+\underbrace{z^2}_{v^{2}}
+\underbrace{z^{-2}}_{v^{-2}} )t^{-2}
\Bigr]q^{1/2}
\nonumber\\
&
+\Biggl[
(\underbrace{x^2 z}_{v^{1}X^{2}}
+\underbrace{x^{-2}z^{-1}}_{v^{-1}Y^2}
+\underbrace{x^{-2}z}_{v^{1}Y^2}
+\underbrace{x^2z^{-1}}_{v^{-1}X^2}
)t
+
(\underbrace{xz^2}_{v^2 X}
+\underbrace{x^{-1}z^{-2}}_{v^{-2}Y}
+\underbrace{xz^{-2}}_{v^{-2}X}x
+\underbrace{x^{-1}z^2}_{v^2 Y}
)t^{-1}
\nonumber\\
&
+
(\underbrace{x}_{X^2 Y}
+\underbrace{x^{-1}}_{XY^2}
+\underbrace{x^{3}}_{X^3}
+\underbrace{x^{-3}}_{Y^3} )t^{3}
+
(\underbrace{z}_{v^1 \varphi}
+\underbrace{z^{-1}}_{v^{-1}\varphi}
+\underbrace{z^{3}}_{v^3}
+\underbrace{z^{-3}}_{v^{-3}} ) t^{-3}
\Biggr]q^{3/4}+\cdots.
\end{align}
The index (\ref{u1_1_findex}) has no contributions from the fundamental hyper. 

Notice that the $U(1)$ ADHM theory with one flavor can be viewed as a theory of SQED$_1$, or equivalently a $U(1)$ gauge theory coupled to a hypermultiplet of gauge charge one (or equivalently a free twisted hypermultiplet) and a free hypermultiplet. 
Thus the index (\ref{u1_1_findex}) has the closed form
\begin{align}
\label{u1_1_findex2}
&I^{\textrm{$U(1)$ ADHM$-[1]$}}(t,x,z;q)
=I^{\textrm{SQED}_{1}}(t,z;q)\times I^{\textrm{HM}}(t,x;q)
\nonumber\\
&=
\frac{(q^{\frac34}tz^{\mp};q)_{\infty}}
{(q^{\frac14}t^{-1}z^{\pm};q)_{\infty}}
\frac{(q^{\frac34}t^{-1}x^{\mp};q)_{\infty}}
{(q^{\frac14}tx^{\pm};q)_{\infty}}. 
\end{align}

When we turn off the global fugacities $x$ and $z$, we get the simplified indices
\begin{align}
\label{u1_1_index}
&I^{\textrm{$U(1)$ ADHM$-[1]$}}(t,x=1,z=1;q)
\nonumber\\
&=1+(2t+2t^{-1})q^{1/4}+(4+3t^2+3t^{-2})q^{1/2}
+(4t^3+4t+4t^{-1}+4t^{-3})q^{3/2}
\nonumber\\
&+
(1+5t^4+4t^{2}+4t^{-2}+5^{-4})q
+
(6t^5+4t^3+4t^{-3}+6t^{-5})q^{5/4}
+\cdots.
\end{align}
The flavored index generally admits two limits of the fugacities in which the Coulomb and Higgs branch operators are counted respectively. They are referred to as the Coulomb and Higgs limits (see \cite{Razamat:2014pta} and Appendix \ref{app_notation}).
The Coulomb limit and Higgs limit \eqref{HS_lim} of the index (\ref{u1_1_index}) are equal. 
They are given by
\begin{align}
\label{HS_C2}
\mathcal{I}^{\textrm{$U(1)$ ADHM$-[1]$}(C)}(\mathfrak{t})&=\mathcal{I}^{\textrm{$U(1)$ ADHM$-[1]$}(H)}(\mathfrak{t})
=\frac{1}{(1-\mathfrak{t})^2}, 
\end{align}
which simply counts two bosonic generators parametrizing $\mathbb{C}^2\subset \mathbb{C}^4$. 
As argued in \cite{Okazaki:2022sxo}, 
the Coulomb and Higgs branch operators correspond to the plane partition with trace $1$. 

\subsubsection{$U(2)$ ADHM with one flavor ($N=2, l=1$)}
Now consider the non-Abelian example where the BPS local operators include single-trace operators 
as well as multi-trace operators. 

When $N=2$ and $l=1$, the theory captures a stack of two M2-branes propagating in flat space. 
The monopole operator $v^{m_1, m_2}$ has the dimension $\Delta(m_i)=\sum_{i=1}^2 |m_i|/2$. 

The ADHM index for $N=2$ and $l=1$ is given by\footnote{
The flavored indices for $N=2,3$ and $l=1$ are also analyzed in \cite{Beratto:2020qyk} to study the enhancement of the supersymmetry using the technology developed in \cite{Garozzo:2019ejm}, where the the Coulomb branch operators contributing to the indices are also identified.
}
\begin{align}
\label{u2_1_findex}
&I^{\textrm{$U(2)$ ADHM$-[1]$}}(t,x,z;q)
\nonumber\\
&=1+
\Biggl[ (\underbrace{x}_{\Tr X}+\underbrace{x^{-1}}_{\Tr Y})t
+(\underbrace{z}_{v^{1,0}}+\underbrace{z^{-1}}_{v^{-1,0}})t^{-1} \Biggr]q^{1/4}
+\Biggl[
\underbrace{2xz}_{\begin{smallmatrix}v^{1,0}X^{(1)}, \\ v^{1,0}X^{(2)}\end{smallmatrix}}
+\underbrace{2x^{-1}z^{-1}}_{\begin{smallmatrix}v^{-1,0}Y^{(1)},\\ v^{-1,0}Y^{(2)}\end{smallmatrix}}
+\underbrace{2xz^{-1}}_{\begin{smallmatrix} v^{-1,0}X^{(1)},\\ v^{-1,0}X^{(2)} \end{smallmatrix}}
+\underbrace{2x^{-1}z}_{\begin{smallmatrix}v^{1,0}Y^{(1)},\\ v^{1,0}Y^{(2)} \end{smallmatrix}} 
\nonumber\\
&
+(\underbrace{2}_{\begin{smallmatrix}\Tr XY,\\\Tr X \Tr Y \end{smallmatrix}}
+\underbrace{2x^2}_{\begin{smallmatrix} \Tr X^2,\\ (\Tr X)^2\end{smallmatrix}}
+\underbrace{2x^{-2}}_{\begin{smallmatrix}\Tr Y^2\\ (\Tr Y)^2\end{smallmatrix}})t^2
+(\underbrace{2}_{\begin{smallmatrix}\Tr\varphi,\\v^{1,-1}\end{smallmatrix}}
+\underbrace{2z^2}_{\begin{smallmatrix}v^{2,0},\\v^{1,1}\end{smallmatrix}}
+\underbrace{2z^{-2}}_{\begin{smallmatrix}v^{-2,0},\\v^{-1,-1}\end{smallmatrix}})t^{-2}
\Biggr]q^{1/2}
\nonumber\\
&
+
(\underbrace{3x^2 z}_{\begin{smallmatrix}v^{1,0}X^{(1)2},\\ v^{1,0}X^{(2)2},\\v^{1,0}X^{(1)}X^{(2)} \end{smallmatrix}}
+\underbrace{3x^{-2}z^{-1}}_{\begin{smallmatrix}v^{-1,0}Y^{(1)2},\\ v^{-1,0}Y^{(2)2},\\v^{-1,0}Y^{(1)}Y^{(2)} \end{smallmatrix}}
+\underbrace{3x^{-2}z}_{\begin{smallmatrix}v^{1,0}Y^{(1)2},\\ v^{1,0}Y^{(2)2},\\v^{1,0}Y^{(1)}Y^{(2)} \end{smallmatrix}}
+\underbrace{3x^2z^{-1}}_{\begin{smallmatrix}v^{-1,0}X^{(1)2},\\ v^{-1,0}X^{(2)2},\\v^{-1,0}X^{(1)}X^{(2)} \end{smallmatrix}}
+\underbrace{3z}_{
\begin{smallmatrix}v^{1,0}X^{(1)} Y^{(1)},\\ v^{1,0}X^{(2)} Y^{(2)}, \\ v^{1,0}X^{(2)}Y^{(1)},\\v^{1,0}X^{(1)} Y^{(2)}, \\
v^{1,0}\psi_{\varphi^{(1)}}, \\
v^{1,0}\psi_{\varphi^{(2)}}, \\
v^{1,0} J^{(2)}I^{(2)}, \\
\end{smallmatrix}}
+\underbrace{3z^{-1}}_{
\begin{smallmatrix}v^{-1,0}X^{(1)} Y^{(1)},\\v^{-1,0}X^{(2)} Y^{(2)},\\ v^{-1,0}X^{(2)}Y^{(1)},\\v^{-1,0}X^{(1)} Y^{(2)}, \\
v^{-1,0}\psi_{\varphi^{(1)}}, \\
v^{-1,0}\psi_{\varphi^{(2)}}, \\
v^{-1,0} J^{(2)}I^{(2)}, \\
\end{smallmatrix}}
)t
\nonumber\\
&
+
(\underbrace{3x z^2}_{\begin{smallmatrix}v^{2,0}X^{(1)},\\ v^{2,0}X^{(2)},\\v^{1,1}\Tr X \end{smallmatrix}}
+\underbrace{3x^{-1}z^{-2}}_{\begin{smallmatrix}v^{-2,0}Y^{(1)},\\ v^{-2,0}Y^{(2)},\\v^{-1,-1}\Tr Y \end{smallmatrix}}
+\underbrace{3xz^{-2}}_{\begin{smallmatrix}v^{-2,0}X^{(1)},\\ v^{2,0}X^{(2)},\\v^{-1,-1}\Tr X \end{smallmatrix}}
+\underbrace{3x^{-1}z^2}_{\begin{smallmatrix}v^{2,0}Y^{(1)},\\ v^{2,0}Y^{(2)},\\v^{1,1}\Tr Y \end{smallmatrix}}
+\underbrace{3x}_{\begin{smallmatrix}
v^{1,-1}X^{(1)},\\ 
v^{1,-1}X^{(2)},\\ 
\Tr (\varphi X),\\
\Tr (\varphi) \Tr(X),\\
\Tr \psi_{X}\\
\end{smallmatrix}}
+\underbrace{3x^{-1}}_{
\begin{smallmatrix}
v^{1,-1}Y^{(1)},\\ 
v^{1,-1}Y^{(2)},\\ 
\Tr (\varphi Y), \\ 
\Tr (\varphi) \Tr (Y), \\
\Tr \psi_{Y} \\
\end{smallmatrix}}
)t^{-1}
\nonumber\\
&
+
(
\underbrace{3x}_{\begin{smallmatrix}\Tr X^2 Y,\\ \Tr X\Tr XY,\\ \Tr X^2 \Tr Y \end{smallmatrix}}
+\underbrace{3x^{-1}}_{\begin{smallmatrix}\Tr X Y^2,\\ \Tr XY\Tr Y,\\ \Tr X \Tr Y^2 \end{smallmatrix}}
+\underbrace{2x^{3}}_{\begin{smallmatrix}\Tr X^3,\\ \Tr X^2 \Tr X \end{smallmatrix}}
+\underbrace{2x^{-3}}_{\begin{smallmatrix}\Tr Y^3,\\ \Tr Y^2 \Tr Y \end{smallmatrix}}
)t^{3}
+
(\underbrace{3z}_{\begin{smallmatrix} v^{2,-1},\\ v^{1,0}\varphi^{(1)},\\ v^{1,0}\varphi^{(2)} \end{smallmatrix}}
+\underbrace{3z^{-1}}_{\begin{smallmatrix} v^{-2,1},\\ v^{-1,0}\varphi^{(1)},\\ v^{-1,0}\varphi^{(2)} \end{smallmatrix}}
+\underbrace{2z^{3}}_{\begin{smallmatrix} v^{3,0},\\ v^{2,1} \end{smallmatrix}}
+\underbrace{2z^{-3}}_{\begin{smallmatrix} v^{-3,0},\\ v^{-2,-1} \end{smallmatrix}}
)t^{-3}
\Biggr]q^{3/4}+\cdots.
\end{align}
Again the equations (\ref{adhm_bps4}) and (\ref{adhm_bps3}) imply that 
the fundamental hypermultiplet scalar fields cannot get a non-trivial vev so that the index has no contribution from the fundamental hyper. 
The Higgs branch operators are constructed as closed words of the form $\Tr X^l Y^m$ and their double-trace operators. 

We observe that 
on the mixed branch for the non-Abelian ADHM theory there exist more operators corresponding to the terms $q^{3/4}tz$, $q^{3/4}tz^{-1}$, $q^{3/4}t^{-1}x$ and $q^{3/4}t^{-1}x^{-1}$ than those for the Abelian ADHM theory. 

The first two terms $q^{3/4}tz$ and $q^{3/4}tz^{-1}$ 
are associated with the monopole operator $v^{\pm,0}$. 
The magnetic flux for the monopole operator $v^{\pm,0}$ breaks down the $U(2)$ gauge group down to $U(1)\times U(1)$ where 
the vacuum equations (\ref{adhm_bps1})-(\ref{adhm_bps4}) has a solution 
\begin{align}
\label{u2_mixed1}
    \varphi&=\left(
    \begin{array}{cc}
    \varphi^{(1)}&0\\
    0&0\\
    \end{array}
    \right),& 
    X&=\left(
    \begin{array}{cc}
    X^{(1)}&0\\
    0&X^{(2)}\\
    \end{array}
    \right),& 
    Y&=\left(
    \begin{array}{cc}
    Y^{(1)}&0\\
    0&Y^{(2)}\\
    \end{array}
    \right),\nonumber \\
    J&=\left(\begin{array}{c}
    0\\
    J^{(2)}\\
    \end{array}
    \right),& 
    I&=(0\quad I^{(2)}),& 
    \zeta&=\left(
    \begin{array}{cc}
    0&0\\
    0&\zeta^{(2)}\\
    \end{array}
    \right),
\end{align}
on the mixed branch.  
The configuration (\ref{u2_mixed1}) admits four monopole operators 
\begin{align}
v^{\pm,0}X^{(1)}Y^{(1)}, \qquad 
v^{\pm,0}X^{(1)}Y^{(2)}, \qquad 
v^{\pm,0}X^{(2)}Y^{(1)}, \qquad 
v^{\pm,0}X^{(2)}Y^{(2)},
\end{align}
dressed by the adjoint hypermultiplet scalars $(X,Y)$, 
two monopole operators 
\begin{align}
v^{\pm,0}\psi_{\varphi^{(1)}}, \qquad 
v^{\pm,0}\psi_{\varphi^{(2)}},
\end{align}
dressed by the adjoint fermions 
and a single monopole 
\begin{align}
    v^{\pm,0}J^{(2)}I^{(2)},
\end{align}
dressed by the fundamental hypermultiplet scalars. 
The terms $q^{3/4}tz$ and $q^{3/4}tz^{-1}$ in (\ref{u2_1_findex}) count these monopole operators as $4+(-2)+1=3$ contributions. 

The terms $q^{3/4}t^{-1}x$ involves two dressed monopole operators 
\begin{align}
\label{u2_1_mix_mono}
v^{1,-1}X^{(1)},\qquad 
v^{1,-1}X^{(2)}, 
\end{align}
two bosonic operators
\begin{align}
    \Tr \varphi \Tr X,\qquad \Tr(\varphi X)
\end{align}
and a single fermionic operator
\begin{align}
    \Tr \psi_{X}. 
\end{align}
For the Abelian case there is a single bosonic operator $\varphi X$ since the dressed monopole operators (\ref{u2_1_mix_mono}) do not exist and the double-trace operator is not available. Therefore the term $q^{3/4}t^{-1}x$ does not show up in the index (\ref{u1_1_findex}). 
The absence of the term $q^{3/4}t^{-1}x^{-1}$ is similarly argued by replacing $X$ with $Y$. 
The indices with auxiliary fugacities are shown in \eqref{ADHMl1N2auxdres} in appendix \ref{app_auxdres}.

The index (\ref{u2_1_findex}) is simplified by turning off the fugacities $x$, $z$ for the global symmetries: 
\begin{align}
\label{u2_1_index}
&I^{\textrm{$U(2)$ ADHM$-[1]$}}(t,x=1,z=1;q)
\nonumber\\
&=1+(2t+2t^{-1})q^{1/4}+(8+6t^2+6t^{-2})q^{1/2}
+(10t^3+18t+18t^{-1}+10t^{-3})q^{3/4}
\nonumber\\
&+(37+19t^4+32t^2+32t^{-2}+19t^{-4})q
+\cdots.
\end{align}
The difference from the $U(1)$ ADHM index (\ref{u1_1_index}) appears from the power $q^{1/2}$. 
This reflects the fact that the $U(2)$ ADHM theory has gauge invariant double-trace operators. 
In the Coulomb and Higgs limit the index (\ref{u2_1_index}) reduces to 
\begin{align}
\label{HS_Sym2C2}
\mathcal{I}^{\textrm{$U(2)$ ADHM$-[1]$}(C)}(\mathfrak{t})
&=\mathcal{I}^{\textrm{$U(2)$ ADHM$-[1]$}(H)}(\mathfrak{t})
=\frac{1+\mathfrak{t}^2}{(1+\mathfrak{t})^2 (1-\mathfrak{t})^4}. 
\end{align}
This describes the symmetric product $\mathrm{Sym}^2 (\mathbb{C}^2)$ 
which are identified with the Coulomb and Higgs branches. 
The function (\ref{HS_Sym2C2}) counts the plane partitions 
of trace $2$ which corresponds to the pairs of column-strict plane partitions of the same shape $\lambda$ whose weight is $|\lambda|=\sum_i \lambda_i$ $=$ $2$ \cite{Okazaki:2022sxo}. 

\subsubsection{$U(3)$ ADHM with one flavor ($N=3, l=1$)}
Next consider the case with higher rank gauge group. 
For $N=3$ and $l=1$ we find the ADHM index 
\begin{align}
&I^{\textrm{$U(3)$ ADHM$-[1]$}}(t,x,z;q)
\nonumber\\
&=1+
\Biggl[ (\underbrace{x}_{\Tr X}+\underbrace{x^{-1}}_{\Tr Y})t
+(\underbrace{z}_{v^{1,0,0}}+\underbrace{z^{-1}}_{v^{-1,0,0}})t^{-1} 
\Biggr]q^{1/4}
+\Biggl[
\underbrace{2xz}_{\begin{smallmatrix}v^{1,0,0}X^{(1)}, \\ v^{1,0,0} \Tr X^{(2)}\end{smallmatrix}}
+\underbrace{2x^{-1}z^{-1}}_{\begin{smallmatrix}v^{-1,0,0}Y^{(1)},\\ v^{-1,0,0}\Tr Y^{(2)}\end{smallmatrix}}
\nonumber\\
&
+\underbrace{2xz^{-1}}_{\begin{smallmatrix} v^{-1,0,0}X^{(1)},\\ v^{-1,0,0}\Tr X^{(2)} \end{smallmatrix}}
+\underbrace{2x^{-1}z}_{\begin{smallmatrix}v^{1,0,0}Y^{(1)},\\ v^{1,0,0}\Tr Y^{(2)} \end{smallmatrix}} 
+(\underbrace{2}_{\begin{smallmatrix}\Tr XY,\\\Tr X \Tr Y \end{smallmatrix}}
+\underbrace{2x^2}_{\begin{smallmatrix} \Tr X^2,\\ (\Tr X)^2\end{smallmatrix}}
+\underbrace{2x^{-2}}_{\begin{smallmatrix}\Tr Y^2\\ (\Tr Y)^2\end{smallmatrix}})t^2
+(\underbrace{2}_{\begin{smallmatrix}\varphi,\\v^{1,-1,0}\end{smallmatrix}}
+\underbrace{2z^2}_{\begin{smallmatrix}v^{2,0,0},\\v^{1,1,0}\end{smallmatrix}}
+\underbrace{2z^{-2}}_{\begin{smallmatrix}v^{-2,0,0},\\v^{-1,-1,0}\end{smallmatrix}})t^{-2}
\Biggr]q^{1/2} 
\nonumber\\
&
+\Biggl[
(\underbrace{4x^2 z}_{\begin{smallmatrix}v^{1,0,0}X^{(1)2},\\ v^{1,0,0} (\Tr X^{(2)})^2,\\ v^{1,0,0}\Tr (X^{(2)2}) \\v^{1,0,0}X^{(1)}\Tr X^{(2)} \end{smallmatrix}}
+\underbrace{4x^{-2}z^{-1}}_{\begin{smallmatrix}v^{-1,0,0}Y^{(1)2},\\ v^{-1,0,0} (\Tr Y^{(2))^2},\\ v^{-1,0,0}\Tr (Y^{(2)2}), \\ v^{-1,0,0}Y^{(1)}\Tr Y^{(2)} \end{smallmatrix}}
+\underbrace{4x^{-2}z}_{\begin{smallmatrix}v^{1,0,0}Y^{(1)2},\\ v^{1,0,0} (\Tr Y^{(2))^2},\\ v^{1,0,0}\Tr (Y^{(2)2}), \\ v^{1,0,0}Y^{(1)}\Tr Y^{(2)} \end{smallmatrix}}
+\underbrace{4x^2z^{-1}}_{\begin{smallmatrix}v^{-1,0,0}X^{(1)2},\\ v^{-1,0,0} (\Tr X^{(2)})^2,\\ v^{-1,0,0}\Tr (X^{(2)2}) \\v^{-1,0,0}X^{(1)}\Tr X^{(2)} \end{smallmatrix}}
\nonumber
\end{align}
\begin{align}
&+\underbrace{4z}_{\begin{smallmatrix}
v^{1,0,0}X^{(1)} Y^{(1)},\\ 
v^{1,0,0} X^{(1)}\Tr Y^{(2)}, \\
v^{1,0,0}\Tr X^{(2)} X^{(1)}, \\
v^{1,0,0} \Tr X^{(2)} \Tr Y^{(2)},\\ 
v^{1,0,0} \Tr (X^{(2)} Y^{(2)}),\\ 
v^{1,0,0}\psi_{\varphi^{(1)}},\\
v^{1,0,0}\Tr \psi_{\varphi^{(2)}},\\
v^{1,0,0}J^{(2)}I^{(2)},\\
\end{smallmatrix}}
+\underbrace{4z^{-1}}_{\begin{smallmatrix}
v^{-1,0,0}X^{(1)} Y^{(1)},\\ 
v^{-1,0,0} X^{(1)} \Tr Y^{(2)},\\ 
v^{-1,0,0} \Tr X^{(2)} Y^{(1)}, \\
v^{-1,0,0}\Tr X^{(2)} \Tr Y^{(2)},\\
v^{-1,0,0}\Tr (X^{(2)}Y^{(2)}),\\
v^{-1,0,0}\psi_{\varphi^{(1)}},\\
v^{-1,0,0}\Tr \psi_{\varphi^{(2)}},\\
v^{-1,0,0}J^{(2)}I^{(2)},\\
\end{smallmatrix}}
)t
+
(\underbrace{4z^2 x}_{\begin{smallmatrix}v^{2,0,0}X^{(1)},\\ v^{2,0,0}\Tr X^{(2)},\\ v^{1,1,0} \Tr X^{(1)},\\ v^{1,1,0} X^{(2)} \end{smallmatrix}}
+\underbrace{4x^{-1}z^{-2}}_{\begin{smallmatrix}v^{-2,0,0}Y^{(1)},\\ v^{-2,0,0}\Tr Y^{(2)},\\ v^{-1,-1,0} \Tr Y^{(1)},\\ v^{-1,-1,0} Y^{(2)} \end{smallmatrix}}
\nonumber\\
&
+\underbrace{4xz^{-2}}_{\begin{smallmatrix}v^{-2,0,0}X^{(1)},\\ v^{-2,0,0}\Tr X^{(2)},\\ v^{-1,-1,0} \Tr X^{(1)},\\ v^{-1,-1,0} X^{(2)} \end{smallmatrix}}
+\underbrace{4x^{-1}z^2}_{\begin{smallmatrix}v^{2,0,0}Y^{(1)},\\ v^{2,0,0}\Tr Y^{(2)},\\ v^{1,1,0} \Tr Y^{(1)},\\ v^{1,1,0} Y^{(2)} \end{smallmatrix}}
+\underbrace{4x}_{\begin{smallmatrix}
v^{1,-1,0}X^{(1)},\\ 
v^{1,-1,0} X^{(2)},\\ 
v^{1,-1,0} X^{(3)},\\ 
\Tr (\varphi X),\\
\Tr \varphi \Tr X,\\
\Tr \psi_{X},\\
\end{smallmatrix}}
+\underbrace{4x^{-1}}_{\begin{smallmatrix}
v^{1,-1,0}Y^{(1)},\\ 
v^{1,-1,0} Y^{(2)},\\ 
v^{1,-1,0} Y^{(3)},\\ 
\Tr (\varphi Y),\\ 
\Tr \varphi \Tr Y,\\ 
\Tr \psi_{Y},\\
\end{smallmatrix}}
)t^{-1}
\nonumber
\end{align}
\begin{align}
\label{u3_1_findex}
&
+
(
\underbrace{4x}_{\begin{smallmatrix}\Tr X^2 Y,\\ \Tr X\Tr XY,\\ \Tr X^2 \Tr Y,\\ (\Tr X)^2 \Tr Y \end{smallmatrix}}
+\underbrace{4x^{-1}}_{\begin{smallmatrix}\Tr Y^2 X,\\ \Tr Y\Tr XY,\\ \Tr Y^2 \Tr X,\\ (\Tr Y)^2 \Tr X \end{smallmatrix}}
+\underbrace{3x^{3}}_{\begin{smallmatrix}\Tr X^3,\\ \Tr X^2 \Tr X,\\ (\Tr X)^3 \end{smallmatrix}}
+\underbrace{3x^{-3}}_{\begin{smallmatrix}\Tr Y^3,\\ \Tr Y^2 \Tr Y,\\ (\Tr Y)^3 \end{smallmatrix}}
)t^{3}
\nonumber\\
&
+
(\underbrace{4z}_{\begin{smallmatrix} v^{2,-1,0},\\ v^{1,1,-1},\\ v^{1,0,0}\varphi^{(1)},\\ v^{1,0,0} \Tr \varphi^{(2)} \end{smallmatrix}}
+\underbrace{4z^{-1}}_{\begin{smallmatrix} v^{-2,1,0},\\ v^{-1,0}\varphi^{(1)},\\ v^{-1,0}\varphi^{(2)} \end{smallmatrix}}
+\underbrace{3z^{3}}_{\begin{smallmatrix} v^{3,0,0},\\ v^{2,1,0},\\ v^{1,1,1} \end{smallmatrix}}
+\underbrace{3z^{-3}}_{\begin{smallmatrix} v^{-3,0,0},\\ v^{-2,-1,0},\\v^{-1,-1,-1} \end{smallmatrix}}
)t^{-3}
\Biggr]q^{3/4}
+\cdots.
\end{align}
The listed terms in the expansion (\ref{u3_1_findex}) generally appear in the index of the $U(N)$ ADHM theory with one flavor for $N\ge 3$. 
The finite $N$ correction in the large $N$ limit typically shows up from the term with $q^{(N+1)/4}$ 
as the $U(N)$ ADHM theory does not contain $(N+1)$-trace operators as gauge invariant operators. 

From (\ref{u3_1_findex}) we see that 
for $N\ge3$ there appear more operators corresponding to the terms 
$q^{3/4}tz$, $q^{3/4}tz^{-1}$, $q^{3/4}t^{-1}x$ and $q^{3/4}t^{-1}x^{-1}$ 
on the mixed branch. 
The first two terms are the contributions from the dressed monopole operators involving $v^{\pm,0,0,\cdots,0}$ for which the gauge group is broken to $U(1)\times U(N-1)$. 
The vacuum equations (\ref{adhm_bps1})-(\ref{adhm_bps4}) admit a solution 
of the same form as (\ref{u2_mixed1}) on the mixed branch. 
However, when $N\ge3$, $X^{(2)}$ and $Y^{(2)}$ are the $(N-1)\times (N-1)$ matrices and $J^{(2)}$ and $I^{(2)}$ are the $(N-1)$-vectors so that the monopole $v^{\pm,0,0,\cdots,0}$ can be dressed by two distinguished operators $\Tr X^{(2)}\Tr Y^{(2)}$ and $\Tr (X^{(2)}Y^{(2)})$. 
This leads to an additional operator that shows up in each of the terms $q^{3/4}tz$ and $q^{3/4}tz^{-1}$. 

The terms $q^{3/4}t^{-1}x$ and $q^{3/4}t^{-1}x^{-1}$ contain the monopole operator $v^{1,-1,0,\cdots, 0}$ for which the gauge group is broken to $U(1)\times U(1)\times U(N-2)$. While for $N=2$ the adjoint scalar $X$ or $Y$ split into two parts, for $N\ge 3$ there are three parts. So the monopole can be dressed by three distinct scalar fields $X^{(i)}$ (resp. $Y^{(i)}$), $i=1,2,3$ and there appears an additional contribution in the term  $q^{3/4}t^{-1}x$ or $q^{3/4}t^{-1}x^{-1}$. 
The indices with auxiliary fugacities are shown in \eqref{ADHMl1N3auxdres} in appendix \ref{app_auxdres}.

When the fugacities $x$ and $z$ are set to $1$, the flavored index (\ref{u3_1_findex}) becomes 
\begin{align}
\label{u3_1_index}
&I^{\textrm{$U(3)$ ADHM$-[1]$}}(t,x=1,z=1;q)
\nonumber\\
&=1+(2t+2t^{-1})q^{1/4}+(8+6t^2+6t^{-2})q^{1/2}
+(14t^3+24t+24t^{-1}+14t^{-3})q^{3/4}
\nonumber\\
&+(71+28t^4+56t^2+56t^{-2}+28t^{-4})q
+\cdots.
\end{align}
The difference of the $U(3)$ ADHM index (\ref{u3_1_index}) from the $U(2)$ ADHM index (\ref{u2_1_index}) begins with the power $q^{3/4}$ 
as the $U(3)$ ADHM theory contains gauge invariant triple-trace operators unlike the $U(2)$ ADHM theory. 
The Coulomb and Higgs limits of the index (\ref{u3_1_index}) are
\begin{align}
\label{HS_Sym3C2}
\mathcal{I}^{\textrm{$U(3)$ ADHM$-[1]$}(C)}(\mathfrak{t})&=\mathcal{I}^{\textrm{$U(3)$ ADHM$-[1]$}(H)}(\mathfrak{t})
=\frac{1+\mathfrak{t}^2+2\mathfrak{t}^3+\mathfrak{t}^4+\mathfrak{t}^6}{(1+\mathfrak{t})^2 (1+\mathfrak{t}+\mathfrak{t}^2)^2 (1-\mathfrak{t})^6}. 
\end{align}
This is the Hilbert series for the symmetric product $\mathrm{Sym}^3 (\mathbb{C}^2)$. 
Again the function (\ref{HS_Sym3C2}) plays a role of 
the generating function for the plane partitions with trace $3$  \cite{Okazaki:2022sxo}. 

\subsubsection{$U(1)$ ADHM with two flavors ($N=1$, $l=2$)}
Next example is the case with multiple flavors with $l>1$. 
Unlike the case with one flavor describing M2-branes in a flat space, the theory describes the M2-branes probing $\mathbb{C}^2\times (\mathbb{C}^2/\mathbb{Z}_l)$. 

For $N=1$ and $l=2$ the monopole operator $v^m$ of the GNO charge $m$ has the dimension $\Delta(m)=|m|$. 
The basic monopoles of $m=\pm 1$ have the dimension one, which is consistent with the OPE $v^{1}v^{-1}\sim \varphi^2$. 
While the hypermultiplet scalar fields $(X,Y)$ parametrize $\mathbb{C}^2$, 
the monopole operators $v^{\pm1}$ and the vector multiplet scalar $\varphi$ obeying the chiral ring relation parametrize $\mathbb{C}^2/\mathbb{Z}_2$. 

The flavored index is computed as 
\begin{align}
\label{u1_2_findex}
&I^{\textrm{$U(1)$ ADHM$-[2]$}}(t,x,y_{\alpha},z;q)
\nonumber\\
&=1+(\underbrace{x}_{X}+\underbrace{x^{-1}}_{Y})tq^{1/4}
+\Biggl[
(\underbrace{2}_{\begin{smallmatrix}XY,\\J_1 I_1 \end{smallmatrix}}
+\underbrace{x^2}_{X^2}
+\underbrace{x^{-2}}_{Y^2}
+\underbrace{\frac{y_1}{y_2}}_{J_2 I_1}
+\underbrace{\frac{y_2}{y_1}}_{J_1 I_2}
)t^2
+(\underbrace{1}_{\varphi}
+\underbrace{z^2}_{v^{1}}
+\underbrace{z^{-2}}_{v^{-1}} )t^{-2}
\Biggr]q^{1/2}
\nonumber\\
&
+\Biggl[
(\underbrace{xz^2}_{v^1 X}
+\underbrace{x^{-1}z^{-2}}_{v^{-1}Y}
+\underbrace{xz^{-2}}_{v^{-1}X}
+\underbrace{x^{-1}z^2}_{v^1 Y}
)t^{-1} 
\nonumber\\
&
+
(\underbrace{2x}_{\begin{smallmatrix}X^2 Y,\\ J_1 X I_1\end{smallmatrix}}
+\underbrace{2x^{-1}}_{\begin{smallmatrix}XY^2,\\ J_1 Y I_1 \end{smallmatrix}}
+\underbrace{x^{3}}_{X^3}
+\underbrace{x^{-3}}_{Y^3}
+\underbrace{\frac{xy_1}{y_2}}_{J_2 X I_1}
+\underbrace{\frac{y_2}{xy_1}}_{J_1 Y I_2}
+\underbrace{\frac{xy_2}{y_1}}_{J_1 X I_2}
+\underbrace{\frac{y_1}{xy_2}}_{J_2 Y I_1}
)t^{3}
\Biggr]q^{3/4}+\cdots.
\end{align}
In this case, the three terms $XY$, $J_1 I_1$ and $J_2I_2$ are not independent due to the F-term relation (\ref{adhm_bps3}) so that only two of them, e.g. $XY$ and $J_1 I_1$ show up in the expansion. 

When the fugacities for the global symmetries are turned off, the flavored index (\ref{u1_2_findex}) reduces to
\begin{align}
\label{u1_2_index}
&I^{\textrm{$U(1)$ ADHM$-[2]$}}(t,x=1,y_{\alpha}=1,z=1;q)
\nonumber\\
&=1+2tq^{1/4}+(6t^2+3t^{-2})q^{1/2}
+(10t^3+4t^{-1})q^{3/4}
+(-2+19t^4+5t^{-4})q
\nonumber\\
&
+(28t^5-12t+4t^{-3})q^{5/4}
+(44t^6-26t^2+7t^{-6})q^{3/2}
+\cdots.
\end{align}
The Coulomb branch limit of the index (\ref{u1_2_index}) is
\begin{align}
\label{HS_C2Z2}
\mathcal{I}^{\textrm{$U(1)$ ADHM$-[2]$}(C)}(\mathfrak{t})
&=\frac{1+\mathfrak{t}^2}{(1-\mathfrak{t}^2)^2}
=\frac{1+\mathfrak{t}^2}{(1+\mathfrak{t})^2 (1-\mathfrak{t})^2},
\end{align}
that describes the singularity $\mathbb{C}^2/\mathbb{Z}_2$. 
It is the geometry probed by a single M2-brane.  
The Higgs limit coincides with 
the Hilbert series (\ref{HS_Sym2C2}) that corresponds to $\mathrm{Sym}^2(\mathbb{C}^2)$. 

\subsubsection{$U(1)$ ADHM with three flavors ($N=1$, $l=3$)}
Let us then consider the case with three flavors. The refined index of the $U(1)$ ADHM with three flavors is
\begin{align}
\label{u1_3_findex}
&I^{\textrm{$U(1)$ ADHM$-[3]$}}(t,x,y_{\alpha},z;q)
\nonumber\\
&=1+(\underbrace{x}_{X}+\underbrace{x^{-1}}_{Y})tq^{1/4}
+\Biggl[
(\underbrace{3}_{\begin{smallmatrix}XY,\\J_1 I_1,\\J_2 I_2 \end{smallmatrix}}
+\underbrace{x^2}_{X^2}
+\underbrace{x^{-2}}_{Y^2}
+\sum_{\alpha\neq \beta}^3\underbrace{\frac{y_\alpha}{y_\beta}}_{J_\beta I_\alpha}
)t^2
+\underbrace{t^{-2}}_{\varphi}
\Biggr]q^{1/2}
\nonumber\\
&
+\Biggl[
(\underbrace{3x}_{\begin{smallmatrix}X^2 Y,\\ J_1 X I_1,\\ J_2XI_2 \end{smallmatrix}}
+\underbrace{3x^{-1}}_{\begin{smallmatrix}XY^2,\\ J_1 Y I_1,\\J_2YI_2 \end{smallmatrix}}
+\underbrace{x^{3}}_{X^3}
+\underbrace{x^{-3}}_{Y^3}
+\sum_{\alpha\neq \beta}\underbrace{\frac{xy_\alpha}{y_\beta}}_{J_\beta X I_\alpha}
+\sum_{\alpha\neq \beta}\underbrace{\frac{y_\alpha}{xy_\beta}}_{J_\beta Y I_\alpha}
)t^{3}
+\Bigl(
\underbrace{z^3}_{v^1}
+\underbrace{z^{-3}}_{v^{-1}}
\Bigr)
\Biggr]q^{3/4}+\cdots.
\end{align}
Keeping fugacity $t$ and setting other global fugacities to $1$, we get
\begin{align}
\label{u1_3_index}
&I^{\textrm{$U(1)$ ADHM$-[3]$}}(t,x=1,y_{\alpha}=1,z=1;q)
\nonumber\\
&=1+2tq^{1/4}+(11t^2+t^{-2})q^{1/2}
+(20t^3+2t^{-3})q^{3/4}
+(-11+56t^4+4t^{-2}+t^{-4})q
\nonumber\\
&+(92t^5-36t+4t^{-1}+2t^{-5})q^{5/4}
+(4+192t^6-107t^2+3t^{-6})q^{3/2}+\cdots
\end{align}
We have the Coulomb limit 
\begin{align}
\label{HS_C2Z3}
\mathcal{I}^{\textrm{$U(1)$ ADHM$-[3]$}(C)}(\mathfrak{t})
&=\frac{1-\mathfrak{t}+\mathfrak{t}^2}{(1+\mathfrak{t}+\mathfrak{t}^2) (1-\mathfrak{t})^2}
=\frac{1+\mathfrak{t}^3}{(1+\mathfrak{t}) (1+\mathfrak{t}+\mathfrak{t}^2) (1-\mathfrak{t})^2}
\end{align}
that describes the geometry $\mathbb{C}^2/\mathbb{Z}_3$. 
This is the expected geometry probed by an M2-brane. 
In the Higgs branch limit we get
\begin{align}
\mathcal{I}^{\textrm{$U(1)$ ADHM$-[3]$}(H)}(\mathfrak{t})
&=\frac{1+4\mathfrak{t}^2+\mathfrak{t}^4}
{(1+\mathfrak{t})^4 (1-\mathfrak{t})^6}.
\end{align}

\subsubsection{$U(1)$ ADHM with $l$ flavors ($N=1$, $l\ge 4$)}
We also show the expansion of the index for $N=1$ and $l=4$ as we will see various dual descriptions of the $U(1)$ ADHM with four flavors in the following discussion. 
It is given by
\begin{align}
\label{u1_4_findex}
&I^{U(1)\text{ADHM-}[4]}(t,x,y_\alpha,z;q)\nonumber \\
&=
1+(x+x^{-1})tq^{\frac{1}{4}}
+\biggl[t^{-2}+\Bigl(4+x^2+x^{-2}+\sum_{\alpha\neq\beta}\frac{y_\alpha}{y_\beta}\Bigr)t^2\biggr]q^{\frac{1}{2}}
+\biggl(x^{-3}+4x^{-1}+4x+x^3\nonumber \\
&\quad +(x+x^{-1})\sum_{\alpha\neq\beta}\frac{y_\alpha}{y_\beta}\biggr)t^3q^{\frac{3}{4}}
+\cdots.
\end{align}
For $x=y_\alpha=z=1$ we have
\begin{align}
\label{u1_4_index}
&I^{\textrm{$U(1)$ ADHM$-[4]$}}(t,x=1,y_{\alpha}=1,z=1;q)
\nonumber\\
&=1+2tq^{1/4}+(18t^2+t^{-2})q^{1/2}+34t^3q^{3/4}
+(-18+134t^4+3t^{-4})q
\nonumber\\
&+(234t^5-64t+4t^{-3})q^{5/4}
+(634t^6-283t^2+4t^{-2}+3t^{-6})q^{3/2}+\cdots.
\end{align}
The Coulomb limit is 
\begin{align}
\label{HS_C2Z4}
\mathcal{I}^{\textrm{$U(1)$ ADHM$-[4]$}(C)}(\mathfrak{t})
&=\frac{1+\mathfrak{t}^4}{(1+\mathfrak{t}^2) (1-\mathfrak{t}^2)^2},
\end{align}
which describes the singularity $\mathbb{C}^2/\mathbb{Z}_4$. 
Again this is identified with the geometry probed by an M2-brane. 
In the Higgs limit the index (\ref{u1_4_index}) reduces to 
\begin{align}
\label{HS_1_su4inst}
\mathcal{I}^{\textrm{$U(1)$ ADHM$-[4]$}(H)}(\mathfrak{t})
&=\frac{1+9\mathfrak{t}^2+9\mathfrak{t}^4+\mathfrak{t}^6}{(1+\mathfrak{t})^6 (1-\mathfrak{t})^8}. 
\end{align}

More generally for general $l$ flavors  we can get the Coulomb and Higgs limits of the index. 
In the Coulomb limit we find
\begin{align}
\label{HS_C2Zl}
\mathcal{I}^{\textrm{$U(1)$ ADHM$-[l]$}(C)}(\mathfrak{t})&=
\frac{1+\mathfrak{t}^l}
{(1+\mathfrak{t}) (1+\mathfrak{t}+\mathfrak{t}^2+\cdots+\mathfrak{t}^{l-1}) (1-\mathfrak{t})^2}. 
\end{align}
As expected this describes the $A_{l-1}$ singularity $\mathbb{C}^2/\mathbb{Z}_{l}$ \cite{Cremonesi:2013lqa}. 
The Higgs limit is given by
\begin{align}
\label{HS_1_sulinst}
\mathcal{I}^{\textrm{$U(1)$ ADHM$-[l]$}(H)}(\mathfrak{t})&=
\frac{\sum_{m=0}^{l-1} \left(\begin{smallmatrix} l-1\\m \end{smallmatrix}\right)^2 \mathfrak{t}^{2m}}
{(1+\mathfrak{t})^{2(l-1)} (1-\mathfrak{t})^{2l}}
=
\frac{{}_{2}F_{1} (1-l,1-l;1;\mathfrak{t}^2)}
{(1+\mathfrak{t})^{2(l-1)} (1-\mathfrak{t})^{2l}},
\end{align}
where ${}_{2}F_{1}(a,b;c;z)$ is the hypergeometric function of the first kind. 
The order of pole at $\mathfrak{t}=1$ in (\ref{HS_1_sulinst}) is $2l$ which is equivalent to the dimension of the Higgs branch for the $U(1)$ ADHM with $l$ flavors. 
This is reproduces the formula in \cite{Benvenuti:2010pq}. 

\subsubsection{$U(2)$ ADHM with two flavors ($N=2$, $l=2$)}
As the simplest example of the non-Abelian ADHM theory with multi flavors, we consider the case with $N=2$ and $l=2$. 
The theory has the monopole operator of dimension $\Delta(m_i)=|m_1|+|m_2|$. 

We obtain the index 
\begin{align}
\label{u2_2_findex}
&I^{\textrm{$U(2)$ ADHM$-[2]$}}(t,x,y_{\alpha},z;q)
\nonumber\\
&=1+(\underbrace{x}_{\Tr X}+\underbrace{x^{-1}}_{\Tr Y})tq^{1/4}
+\Biggl[
(\underbrace{3}_{\begin{smallmatrix}\Tr XY,\\ \Tr X\Tr Y,\\J_1 I_1 \end{smallmatrix}}
+\underbrace{2x^2}_{\begin{smallmatrix}\Tr X^2,\\ (\Tr X)^2 \end{smallmatrix}}
+\underbrace{2x^{-2}}_{\begin{smallmatrix}\Tr Y^2,\\ (\Tr Y)^2\end{smallmatrix}}
+\underbrace{\frac{y_1}{y_2}}_{J_2 I_1}
+\underbrace{\frac{y_2}{y_1}}_{J_1 I_2}
)t^2
+(\underbrace{1}_{\Tr \varphi}
\nonumber\\
&
+\underbrace{z^2}_{v^{1,0}}
+\underbrace{z^{-2}}_{v^{-1,0}} )t^{-2}
\Biggr]q^{1/2}
+\Biggl[
(
\underbrace{x}_{\Tr(\varphi X)}
+\underbrace{x^{-1}}_{\Tr (\varphi Y)}
+\underbrace{2xz^2}_{\begin{smallmatrix}v^{1,0} X^{(1)},\\ v^{1,0} X^{(2)}\\ \end{smallmatrix}}
+\underbrace{2x^{-1}z^{-2}}_{\begin{smallmatrix}v^{-1,0}Y^{(1)},\\ v^{-1,0}Y^{(2)} \end{smallmatrix}}
+\underbrace{2xz^{-2}}_{\begin{smallmatrix} v^{-1,0}X^{(1)},\\ v^{-1,0}X^{(2)} \end{smallmatrix}}
+\underbrace{2x^{-1}z^2x}_{\begin{smallmatrix} v^{1,0} Y^{(1)},\\ v^{1,0}Y^{(2)} \end{smallmatrix}}
)t^{-1} 
\nonumber\\
&
+
(\underbrace{5x}_{\begin{smallmatrix}\Tr X^2 Y,\\ \Tr X^2 \Tr Y,\\ \Tr XY \Tr X,\\ J_1 I_1 \Tr X,\\ J_1 X I_1\end{smallmatrix}}
+\underbrace{5x^{-1}}_{\begin{smallmatrix}\Tr XY^2,\\ \Tr Y^2 \Tr X, \\ \Tr XY\Tr Y, \\ J_1 I_1 \Tr Y, \\ J_1 Y I_1 \end{smallmatrix}}
+\underbrace{2x^{3}}_{\begin{smallmatrix}\Tr X^3,\\ \Tr X \Tr X^2 \end{smallmatrix}}
+\underbrace{2x^{-3}}_{\begin{smallmatrix}\Tr Y^3,\\ \Tr Y \Tr Y^2 \end{smallmatrix}}
+\underbrace{\frac{2xy_1}{y_2}}_{\begin{smallmatrix}J_2 I_1 \Tr X,\\ J_2 X I_1,\\ \end{smallmatrix}}
+\underbrace{\frac{2y_2}{xy_1}}_{\begin{smallmatrix}J_1 I_2 \Tr Y,\\ J_1 Y I_2\end{smallmatrix}}
+\underbrace{\frac{2xy_2}{y_1}}_{\begin{smallmatrix}J_1 I_2 \Tr X,\\ J_1 X I_2\end{smallmatrix}}
+\underbrace{\frac{2y_1}{xy_2}}_{\begin{smallmatrix}J_2 I_1 \Tr Y,\\ J_2 Y I_1\end{smallmatrix}}
)t^{3}
\Biggr]q^{3/4}+\cdots.
\end{align}
Again the F-term constraint (\ref{adhm_bps3}) gets rid of one of the open or closed words. 

The simplified index is 
\begin{align}
\label{u2_2_index}
&I^{\textrm{$U(2)$ ADHM$-[2]$}}(t,x=1,y_{\alpha}=1,z=1;q)
\nonumber\\
&=1+2tq^{1/4}+(9t^2+3t^{-2})q^{1/2}+(22t^3+10t^{-1})q^{3/4}
+(25+55t^4+11t^{-4})q
\nonumber\\
&+(116t^5+46t+26t^{-3})q^{5/4}
+(242t^6+60t^2+44t^{-2}+22t^{-6})q^{3/2}+\cdots.
\end{align}
The Coulomb branch limit of the index (\ref{u2_2_index}) is 
\begin{align}
\label{HS_Sym2C2Z2}
\mathcal{I}^{\textrm{$U(2)$ ADHM$-[2]$}(C)}(\mathfrak{t})&=
\frac{1+\mathfrak{t}^2+4\mathfrak{t}^4+\mathfrak{t}^6+\mathfrak{t}^8}
{(1+\mathfrak{t}^2)^2 (1-\mathfrak{t}^2)^4}.
\end{align}
This is the Coulomb branch Hilbert series describing the $\mathrm{Sym}^2 (\mathbb{C}^2/\mathbb{Z}_2)$. 
The Higgs limit of the index (\ref{u2_2_index}) gives 
\begin{align}
\label{HS_2_su2inst}
\mathcal{I}^{\textrm{$U(2)$ ADHM$-[2]$}(H)}(\mathfrak{t})&=
\frac{
1+\mathfrak{t}+3\mathfrak{t}^2+6\mathfrak{t}^3+8\mathfrak{t}^4+6\mathfrak{t}^5
+8\mathfrak{t}^6+6\mathfrak{t}^7+3\mathfrak{t}^8+\mathfrak{t}^9+\mathfrak{t}^{10}
}
{
(1+\mathfrak{t})^4 (1+\mathfrak{t}+\mathfrak{t}^2)^3 (1-\mathfrak{t})^8
}.
\end{align}
This is the Hilbert sereis for the two $SU(2)$ instanton moduli space \cite{Hanany:2012dm,Bourget:2017tmt}. 
\footnote{Note that the expression (\ref{HS_Sym2C2Z2}) is different from eq.(4.24) in \cite{Hanany:2012dm} by the overall factor $(1-\mathfrak{t})^{-2}$. }

\subsubsection{$U(2)$ ADHM with four flavors ($N=2$, $l=4$)}
Another interesting example is the case with $N=2$ and $l=4$, i.e. the $U(2)$ ADHM theory with four flavors. 
It has the monopole operator of dimension $\Delta(m_i)=2(|m_1|+|m_2|)$. 

The flavored index is given by 
\begin{align}
\label{u2_4_findex}
&I^{\textrm{$U(2)$ ADHM$-[4]$}}(t,x,y_{\alpha},z;q)
\nonumber\\
&=1+(\underbrace{x}_{\Tr X}+\underbrace{x^{-1}}_{\Tr Y})tq^{1/4}
+\Biggl[
(\underbrace{5}_{\begin{smallmatrix}\Tr XY,\\ \Tr X\Tr Y,\\J_1 I_1,\\J_2 I_2 \end{smallmatrix}}
+\underbrace{2x^2}_{\begin{smallmatrix}\Tr X^2,\\ (\Tr X)^2 \end{smallmatrix}}
+\underbrace{2x^{-2}}_{\begin{smallmatrix}\Tr Y^2,\\ (\Tr Y)^2\end{smallmatrix}}
+\sum_{\alpha\neq \beta}\underbrace{\frac{y_\alpha}{y_\beta}}_{J_{\beta} I_{\alpha}}
)t^2
+\underbrace{t^{-2}}_{\Tr \varphi}
\Biggr]q^{1/2}
\nonumber\\
&
+\Biggl[
(
\underbrace{x}_{\Tr(\varphi X)}
+\underbrace{x^{-1}}_{\Tr (\varphi Y)}
)t^{-1} 
+
(\underbrace{9x}_{\begin{smallmatrix}\Tr X^2 Y,\\ \Tr X^2 \Tr Y,\\ \Tr XY \Tr X,\\ J_{\alpha} I_{\alpha} \Tr X,\\ J_{\alpha} X I_{\alpha}\\ \end{smallmatrix}}
+\underbrace{9x^{-1}}_{\begin{smallmatrix}\Tr XY^2,\\ \Tr Y^2 \Tr X, \\ \Tr XY\Tr Y, \\ J_{\alpha} I_{\alpha} \Tr Y, \\ J_{\alpha} Y I_{\alpha}\\ \end{smallmatrix}}
\nonumber\\
&
+\underbrace{2x^{3}}_{\begin{smallmatrix}\Tr X^3,\\ \Tr X \Tr X^2 \end{smallmatrix}}
+\underbrace{2x^{-3}}_{\begin{smallmatrix}\Tr Y^3,\\ \Tr Y \Tr Y^2 \end{smallmatrix}}
+\sum_{\alpha\neq \beta}\underbrace{\frac{2xy_\alpha}{y_\beta}}_{\begin{smallmatrix}J_{\alpha} I_{\beta} \Tr Y,\\ J_{\beta} Y I_{\alpha}\end{smallmatrix}}
+\sum_{\alpha\neq \beta}\underbrace{\frac{2y_\alpha}{xy_\beta}}_{\begin{smallmatrix}J_{\beta} I_{\alpha} \Tr Y,\\ J_{\beta} Y I_{\alpha}\end{smallmatrix}}
)t^{3}
\Biggr]q^{3/4}+\cdots.
\end{align}

When we turn off the global fugacities $x$, $y_{\alpha}$ and $z$, we find
\begin{align}
\label{u2_4_index}
&I^{\textrm{$U(2)$ ADHM$-[4]$}}(t,x=1,y_{\alpha}=1,z=1;q)
\nonumber\\
&=1+2tq^{1/4}+(21t^2+t^{-2})q^{1/2}+(70t^3+t^{-1})q^{3/4}
+(1+289t^4+4t^{-4})q
\nonumber\\
&+(946t^5-34t+10t^{-3})q^{5/4}
+(2961t^6-335t^2+48t^{-2}+6t^{-6})q^{3/2}+\cdots.
\end{align}
The Coulomb limit of the index (\ref{u2_4_index}) is
\begin{align}
\label{HS_Sym2C2Z4}
\mathcal{I}^{\textrm{$U(2)$ ADHM$-[4]$}(C)}(\mathfrak{t})&=
\frac{1-\mathfrak{t}^2+2\mathfrak{t}^4+2\mathfrak{t}^8-\mathfrak{t}^{10}+\mathfrak{t}^{12}}{(1+\mathfrak{t}^4) (1+\mathfrak{t}^2)^2 (1-\mathfrak{t}^2)^4}. 
\end{align}
This is the Hilbert series of $\mathrm{Sym}^2 (\mathbb{C}^2/\mathbb{Z}_4)$ corresponding to the Coulomb branch. 
The index (\ref{u2_4_index}) reduces to 
\begin{align}
\label{HS_2_su4inst}
    \mathcal{I}^{\textrm{$U(2)$ ADHM$-[4]$}(H)}(\mathfrak{t})&=
    \frac{1}{(1-\mathfrak{t})^{16}(1+\mathfrak{t})(1+\mathfrak{t}+\mathfrak{t}^2)^2}
    \nonumber\\
    &\times 
    (1+\mathfrak{t}+11\mathfrak{t}^2+34\mathfrak{t}^3+
    88\mathfrak{t}^4+216\mathfrak{t}^5
    +473\mathfrak{t}^6+797\mathfrak{t}^7
    \nonumber\\
    &
    +1243\mathfrak{t}^8
    +1738\mathfrak{t}^9+2080\mathfrak{t}^{10}
    +2152\mathfrak{t}^{11}+\textrm{palindrome}+\mathfrak{t}^{22})
\end{align}
in the Higgs limit. 
This is the Hilbert series for the moduli space of two $SU(4)$ instantons \cite{Hanany:2012dm}. \footnote{Again the expression (\ref{HS_2_su4inst}) is different from eq.(3.21) in \cite{Hanany:2012dm} by the overall factor $(1-\mathfrak{t})^{-2}$. }

\subsubsection{Mirror symmetry}
\label{sec_umirror}
The $U(N)$ ADHM theory with $l$ flavors has the mirror description given in the leftmost and center figures in Figure \ref{fig:ADHM_mirror}. 
We have confirmed that the ADHM index (\ref{uN_l_index}) perfectly agrees with the following index  
\begin{align}
\label{muN_l_index}
&I^{\textrm{$U(N)^{\otimes l}$mADHM$-[1]$}}(t,x,y_{\alpha},z;q)=
\left( \frac{1}{N!} \frac{(q^{\frac12}t^{-2};q)_{\infty}^N}{(q^{1/2}t;q)_{\infty}^N} \right)^{l}
\sum_{m_1^{(1)},\cdots,m_N^{(l)} \in \mathbb{Z}}
\oint \prod_{I=1}^{l} \prod_{i=1}^N \frac{ds_i^{(I)}}{2\pi is_i^{(I)}}
\nonumber\\
&\times 
\prod_{i<j}
(1-q^{\frac{|m_i^{(I)}-m_j^{(I)}|}{2}}s_i^{(I)\pm}s_j^{(I)\mp})
\frac{(q^{\frac{1+|m_i^{(I)}-m_j^{(I)}|}{2}}t^{-2}s_i^{(I)\mp}s_j^{(I)\pm};q)_{\infty}}{(q^{\frac{1+|m_i^{(I)}-m_j^{(I)}|}{2}}t^{2}s_i^{(I)\pm}s_j^{(I)\mp};q)_{\infty}}
\nonumber\\
&\times \prod_{I=1}^{l} \prod_{i,j=1}^{N}
\frac{(q^{\frac34+\frac{|m_i^{(I)}-m_j^{(I+1)}|}{2}}t s_i^{(I)\mp}s_j^{(I+1)\pm}z^{\mp};q)_{\infty}}
{(q^{\frac14+\frac{|m_i^{(I)}-m_j^{(I+1)}|}{2}}t^{-1}s_i^{(I)\pm}s_j^{(I+1)\mp}z^{\pm};q)_{\infty}}
\frac{(q^{\frac34+\frac{|m_i^{(l)}-m_j^{(1)}|}{2}}t s_i^{(l)\mp}s_j^{(1)\pm}z^{\pm};q)_{\infty}}
{(q^{\frac14+\frac{|m_i^{(l)}-m_j^{(1)}|}{2}}t^{-1}s_i^{(l)\pm}s_j^{(1)\mp}x^{\mp};q)_{\infty}}
\nonumber\\
&\times 
\prod_{i=1}^N\frac{(q^{\frac34+\frac{|m_{i}^{(1)}|}{2}}t s_{i}^{(1)\mp};q)_{\infty}}
{(q^{\frac14+\frac{|m_i^{(1)}|}{2}}t^{-1}s_{i}^{(1)\pm};q)_{\infty}}
\nonumber\\
&\times 
q^{\frac14 \left(\sum_{i=1}^N |m_i^{(1)}|+\sum_{I=1}^{l-1}\sum_{i,j}|m_i^{(I)}-m_j^{(I+1)}|+\sum_{i,j}|m_i^{(l)}-m_j^{(1)}| \right) - \frac12 \sum_{I=1}^{l}\sum_{i<j} |m_{i}^{(I)}-m_{j}^{(I)}| } 
\nonumber\\
&\times 
t^{ \sum_{i=1}^N |m_i^{(1)}|+\sum_{I=1}^{l-1}\sum_{i,j}|m_i^{(I)}-m_j^{(I+1)}|+\sum_{i,j}|m_i^{(l)}-m_j^{(1)}|  - 2 \sum_{I=1}^{l}\sum_{i<j} |m_{i}^{(I)}-m_{j}^{(I)}| }
\nonumber\\
&\times 
x^{\sum_{i=1}^N m_i^{(1)}}
\prod_{\alpha=1}^{l}
\left( \frac{y_\alpha}{y_{\alpha+1}} \right)^{\sum_{i=1}^{N}m_{i}^{(\alpha)}}. 
\end{align}
of the mirror $U(N)^{\otimes l}$ necklace quiver gauge theory with one flavor. 

In particular, the ADHM theories with $l=1$ are self-mirror 
where their indices are invariant under the transformation
\begin{align}
\label{mirror_transf}
t&\rightarrow t^{-1},& 
x&\rightarrow z,& 
z&\rightarrow x,
\end{align}
as explicitly checked from the previous computations, e.g. (\ref{u1_1_findex}), (\ref{u2_1_findex}) and (\ref{u3_1_findex}). 

From the equality of the indices (\ref{uN_l_index}) and (\ref{muN_l_index}) we find the following fugacity map: 
\begin{align}
\label{ADHM_mirror_fug}
\begin{array}{c|c}
\textrm{$U(N)$ ADHM with $l$ flavors}&\textrm{$U(N)^{\otimes l}$ necklace quiver with one flavor} \\ \hline 
\textrm{$z$ (topological sym.)}&\textrm{$\tilde{x}$ (flavor sym. for $(\tilde{X},\tilde{Y})$)} \\  
&\tilde{x}=z \\ \hline
\textrm{$x$ (flavor sym. for $(X,Y)$)}&\textrm{$\tilde{z}^{(\alpha)}$ (topological sym.)} \\ 
\textrm{$y_{\alpha}$ (flavor sym. for $(I,J)$)}&\tilde{z}^{(1)}=x\frac{y_1}{y_2} \\
&\tilde{z}^{(\alpha)}=\frac{y_{\alpha}}{y_{\alpha+1}}, \quad \alpha=2,\cdots, l \\
\end{array}
\end{align}
where $\tilde{x}$ and $\tilde{z}^{(\alpha)}$ are the fugacities coupled to 
the flavor symmetry for the bifundamental hypers $(\tilde{X},\tilde{Y})$ 
and the topological symmetry for the $\alpha$-th factor of gauge node. 

Also we obtain the operator mapping under the mirror symmetry. 
For the Abelian case we find 
\begin{align}
\label{ADHM_mirrormap1}
\begin{array}{c|c}
\textrm{$U(1)$ ADHM with $l$ flavors}&\textrm{$U(1)^{\otimes l}$ necklace quiver with one flavor} \\ \hline 
X^m&v^{m;m;\cdots;m}\\
Y^m&v^{-m;-m;\cdots;-m} \\
XY,\quad \sum_{\alpha\neq l} J_{\alpha}I_{\alpha}&\tilde{\varphi}^{(\alpha)} \\
\sum_{\begin{smallmatrix}\alpha< \beta,\\ \alpha\neq 1\end{smallmatrix}}J_{\alpha}I_{\beta}
&v^{0;\cdots;0;m^{(\alpha)}=-1;-1;\cdots;m^{(\beta-1)}=-1;0;\cdots;0} \\
J_{1}I_{\alpha>1}
&v^{0;\cdots;0;m^{(\alpha)}=1;1,\cdots;1} \\
\sum_{\begin{smallmatrix}\alpha> \beta,\\ \beta\neq 1 \end{smallmatrix}}J_{\alpha}I_{\beta}
&v^{0;\cdots;0;m^{(\beta)}=1;1;\cdots;m^{(\alpha-1)}=1;0;\cdots;0} \\
J_{\alpha>1}I_{1}
&v^{0;\cdots;0;m^{(\alpha)}=-1;-1,\cdots;-1} \\ \hline 
v^{m}& (\tilde{X}_{1,2}\tilde{X}_{2,3}\cdots \tilde{X}_{l,1})^m \\
v^{-m}& (\tilde{Y}_{1,2}\tilde{Y}_{2,3}\cdots \tilde{Y}_{l,1})^m \\
\varphi&\tilde{J}\tilde{I} \\
\end{array}.
\end{align}
For the non-Abelian case we get
\begin{align}
\label{ADHM_mirrormap2}
\begin{array}{c|c}
\textrm{$U(N)$ ADHM with $l$ flavors}&\textrm{$U(N)^{\otimes l}$ necklace quiver with one flavor} \\ \hline 
\Tr X&v^{1,0,\cdots,0;\cdots;1,0,\cdots,0} \\
\Tr Y&v^{-1,0,\cdots,0;\cdots;-1,0,\cdots,0} \\
\Tr XY,\quad \Tr X \Tr Y, \quad \sum_{\alpha\neq l} J_{\alpha}I_{\alpha}
& v^{1,-1,0,\cdots; 1,-1,0,\cdots; \cdots; 1,-1,0,\cdots},\quad \Tr \tilde{\varphi}^{(\alpha)} \\
\sum_{\begin{smallmatrix}\alpha< \beta,\\ \alpha\neq 1\end{smallmatrix}}J_{\alpha}I_{\beta}
&v^{0;\cdots;0;m^{(\alpha)}_1=-1,0,\cdots,0;\cdots;m^{(\beta-1)}_1=-1,0,\cdots,0;0;\cdots;0} \\
J_{1}I_{\alpha>1}
&v^{0;\cdots;0;m^{(\alpha)}_1=1,0\cdots,0;\cdots;1,0,\cdots,0} \\
\sum_{\begin{smallmatrix}\alpha> \beta,\\ \beta\neq 1 \end{smallmatrix}}J_{\alpha}I_{\beta}
&v^{0;\cdots;0;m^{(\beta)}_1=1,0\cdots,0;\cdots;m^{(\alpha-1)}_1=1,0,\cdots,0;0;\cdots;0} \\
J_{\alpha>1}I_{1}
&v^{0;\cdots;0;m^{(\alpha)}_1=-1,0,\cdots,0;\cdots;-1,0,\cdots,0} \\ 
\hline 
v^{1,0,\cdots,0}&\Tr \tilde{X}_{1,2}\tilde{X}_{2,3}\cdots \tilde{X}_{l,1} \\
v^{-1,0,\cdots,0}&\Tr \tilde{Y}_{1,2}\tilde{Y}_{2,3}\cdots \tilde{Y}_{l,1} \\
\Tr \varphi&\tilde{J}\tilde{I} \\
\end{array}.
\end{align}


\subsection{Closed form expression for the Coulomb limit with general $N,l$}
\label{ADHMCoulomballorder}

In fact, it is possible to obtain a closed form expression for the Coulomb limit \eqref{HS_lim} of the supersymmetric index of the $U(N)$ ADHM theory.
Here we assume $|\mathfrak{t}|<1$ and that all the other remaining fugacities has absolute value $1$.

First of all, since the overall factor $q^{\frac{l}{4}\sum_i|m_i|}t^{-l\sum_i|m_i|}=\mathfrak{t}^{l\sum_i|m_i|}$ is not in a negative power of $q$, we can take the Coulomb limit \eqref{HS_lim} separately in each factor within the summation and the integration of \eqref{uN_l_index}:
\begin{align}
{\cal I}^{U(N) \text{ADHM-}[l]\,(C)}&=
\lim_{\substack{\mathfrak{t}=q^{\frac{1}{4}}t^{-1}\text{: fixed}\\ q\rightarrow 0}}
I^{U(N) \text{ADHM-}[l]}\nonumber \\
&=\frac{1}{N!}\sum_{m_i\in\mathbb{Z}}\prod_{i=1}^N\frac{ds_i}{2\pi is_i}
\prod_{i\neq j}^N \lim_{\substack{\mathfrak{t}=q^{\frac{1}{4}}t^{-1}\text{: fixed}\\ q\rightarrow 0}}\Bigl(1-q^{\frac{|m_i-m_j|}{2}}\frac{s_i}{s_j}\Bigr)\nonumber \\
&\quad \prod_{i,j}^N \lim_{\substack{\mathfrak{t}=q^{\frac{1}{4}}t^{-1}\text{: fixed}\\ q\rightarrow 0}}\frac{(q^{\frac{1}{2}+\frac{|m_i-m_j|}{2}}t^2\frac{s_i}{s_j};q)_\infty}{(q^{\frac{1}{2}+\frac{|m_i-m_j|}{2}}t^{-2}\frac{s_i}{s_j};q)_\infty}
\mathfrak{t}^{l\sum_i^N|m_i|}
z^{l\sum_im_i}.
\label{220425ADHMCoulomb}
\end{align}
To further proceed we notice that each factor under the limit is $1$ unless $m_i=m_j$.
This motivate us to label each choice of the monopole charges $m_i\in\mathbb{Z}^N$ by $\nu_m$, the number of $i$ where $m_i=m$, with which the Coulomb limit of the index \eqref{220425ADHMCoulomb} can be written as
\begin{align}
{\cal I}^{U(N) \text{ADHM-}[l]\,(C)}=\sum_{\substack{\nu_m=0\\ (\sum_m\nu_m=N)}}^\infty \prod_{m=-\infty}^\infty\frac{1}{\nu_m!}\prod_{i=1}^{\nu_m}\frac{ds_i}{2\pi is_i}\prod_{i\neq j}^{\nu_m}\Bigl(1-\frac{s_i}{s_j}\Bigr)\prod_{i,j}^{\nu_m}\frac{1}{1-\mathfrak{t}^2\frac{s_i}{s_j}}\mathfrak{t}^{l|m|\nu_m}z^{lm\nu_m}.
\end{align}
The constraint on the summation over $\nu_m$ can be removed by considering the grand canonical sum
\begin{align}
\Xi(\kappa)=\sum_{N=0}^\infty \kappa^N{\cal I}^{U(N) \text{ADHM-}[l]\,(C)}
=\prod_{m=-\infty}^\infty {\widetilde \Xi}(\kappa\mathfrak{t}^{l|m|}z^{lm},\mathfrak{t}),
\label{220429Xi}
\end{align}
where
\begin{align}
{\widetilde \Xi}(\kappa,\mathfrak{t})=\sum_{\nu=0}^\infty\frac{\kappa^\nu}{\nu!}\prod_{i=1}^\nu\oint\frac{ds_i}{2\pi is_i}\prod_{i\neq j}^\nu\Bigl(1-\frac{s_i}{s_j}\Bigr)\prod_{i,j}^\nu\frac{1}{1-\mathfrak{t}^2\frac{s_i}{s_j}}.
\label{220518_Z2inHiggsstart}
\end{align}
By using the Cauchy determinant formula
\begin{align}
\frac{\prod_{i<j}^N(x_i-x_j)(y_i-y_j)}{\prod_{i,j}^N(x_i+y_j)}=\det_{i,j}\frac{1}{x_i+y_j},
\end{align}
we can rewrite ${\widetilde \Xi}$ as
\begin{align}
{\widetilde \Xi}(\kappa,\mathfrak{t})=\sum_{\nu=0}^\infty \kappa^\nu\Omega_\nu(\mathfrak{t}),
\end{align}
where
\begin{align}
\Omega_\nu(\mathfrak{t})=\frac{\mathfrak{t}^{-\nu(\nu-1)}}{\nu!}\prod_{i=1}^\nu\oint\frac{ds_i}{2\pi i}\det_{i,j}\frac{1}{s_i-\mathfrak{t}^2s_j}.
\label{220429Omeganu}
\end{align}
By using $\Omega_\nu(\mathfrak{t})$ we can write the grand canonical sum $\Xi(\kappa)$ \eqref{220429Xi} as
\begin{align}
\log\Xi(\kappa)=\sum_{\nu=1}^\infty \kappa^\nu\frac{1-\mathfrak{t}^{2l\nu}}{\prod_\pm (1-\mathfrak{t}^{l\nu}z^{\pm l\nu})}\sum_{n=1}^\nu\frac{(-1)^{n-1}}{n}\sum_{\substack{\nu_1,\cdots,\nu_n\ge1\\ (\sum_i\nu_i=\nu)}}\prod_{i=1}^n\Omega_{\nu_i}(\mathfrak{t}).
\label{220429_logXi}
\end{align}
Here $\Omega_\nu(\mathfrak{t})$ \eqref{220429Omeganu}
can be obtained by the following generating function
\begin{align}
\sum_{\nu=0}^\infty\kappa^\nu \mathfrak{t}^{\nu(\nu-1)}\Omega_\nu(\mathfrak{t})
&=\text{Det}(1+\kappa\rho)\nonumber \\
&=\text{exp}\text{Tr}\log (1+\kappa\rho)\nonumber \\
&=\text{exp}\Bigl[\sum_{n=1}^\infty\frac{(-1)^{n-1}}{n}\kappa^n\frac{1}{1-\mathfrak{t}^{2n}}\Bigr],
\end{align}
where $\rho(s,s')=\frac{1}{s-\mathfrak{t}^2s'}$, operator product is defined as $(\rho\circ \rho)(s,s')=\oint\frac{ds''}{2\pi i}\rho(s,s'')\rho(s'',s')$ and in the third line we have used $\text{Tr}\rho^n=\oint\frac{ds}{2\pi i}(\rho^{\circ n})(s,s)=\frac{1}{1-\mathfrak{t}^{2n}}$ which is obtained by evaluating the integrations explicitly.
Expanding the right-hand side we observe
\begin{align}
\Omega_\nu(\mathfrak{t})=\prod_{n=1}^\nu\frac{1}{1-\mathfrak{t}^{2n}}.
\label{220518_Z2inHiggsend}
\end{align}
Using this result we further observe
\begin{align}
\sum_{n=1}^\nu\frac{(-1)^{n-1}}{n}\sum_{\substack{\nu_1,\cdots,\nu_n=1\\ (\sum_i\nu_i=\nu)}}^\infty\prod_{i=1}^n\Omega_{\nu_i}(\mathfrak{t})
=\frac{1}{\nu(1-\mathfrak{t}^{2\nu})}.
\end{align}
Hence \eqref{220429_logXi} simplifies as
\begin{align}
\Xi(\kappa)=e^{\sum_{\nu=1}^\infty\kappa^\nu A_\nu(z,\mathfrak{t})},\quad A_\nu(z,\mathfrak{t})=\frac{1-\mathfrak{t}^{2l\nu}}{\nu(1-\mathfrak{t}^{2\nu})\prod_{\pm}(1-\mathfrak{t}^{l\nu}z^{\pm l\nu})}.
\label{220513_usefulforsmallN}
\end{align}
After a few manipulation we can also write this as
\begin{align}
\Xi(\kappa)=\prod_{m=0}^\infty\frac{1}{1-\mathfrak{t}^{2m}\kappa}\prod_{n=1}^\infty\prod_\pm\frac{1}{1-\mathfrak{t}^{2m+nl}z^{\pm ln}\kappa}.
\label{220513_productform}
\end{align}
Note that this result is consistent with the result for $l=1$ \cite{Gaiotto:2021xce}, and also with the results for the Hilbert series obtained in \cite{Okazaki:2022sxo} if we set $z=1$. 

Here we propose a combinatorial interpretation of the formula (\ref{220513_productform}). Let $\pi=\{n_{ij}\}$ be a plane partition. The sum of all the entries is called the norm $n=\sum_{ij} n_{ij}$ of $\pi$ and the sum $\tau_i(\pi)$ $=$ $\sum_{k} \tau_{k k+i}$ of the $i$-th diagonal entries is referred to as the $i$-trace of $\pi$. We  write the $0$-trace as $\tau(\pi)$ and simply call it the trace of  $\pi$, i.e. $\tau(\pi)=\sum_{i}n_{ii}$. 

According to the correspondence in \cite{Okazaki:2022sxo} the local operators with scaling dimension $\Delta$ and flavor charge $M$ in the M2-brane SCFT parametrizing the geometry $\mathbb{C}^2$ probed by $N$ M2-branes correspond to plane partitions of $n=2\Delta$ with trace $\tau(\pi)=N$ and the difference $\sum_{i>0}\tau_i(\pi) -\sum_{i<0}\tau_i(\pi)$ $=$ $M$ of the sums of the $i$-traces.  Therefore we conjecture that 
\begin{align}
\label{conjecture_pp}
\sum_{n=1}^{\infty}\sum_{N=0}^{n}\sum_{M=-n}^{n} \alpha(n,N,M) q^n \kappa^{N} z^M
&=
    \prod_{m=0}^{\infty}\frac{1}{1-\kappa q^{2m+1}}
    \prod_{n=1}^{\infty}
    \prod_{\pm} \frac{1}{1-\kappa q^{2m+n+1}z^{\pm n}}
\end{align}
where $\alpha(n,N,M)$ is the number of plane partitions of $n=2\Delta$ with trace $\tau(\pi)=N$ and $\sum_{i>0}\tau_i(\pi) -\sum_{i<0}\tau_i(\pi)$ $=$ $M$. 

Also we can obtain ${\cal I}^{U(N) \text{ADHM-}[l]\,(C)}$ by inverting \eqref{220429Xi}.
For small $N$, this can be done explicitly by using \eqref{220513_usefulforsmallN} as
\begin{align}
{\cal I}^{U(N)\text{ ADHM-}[l](C)}&=\sum_{n=1}^N\frac{1}{n!}\sum_{\substack{\nu_1,\cdots,\nu_n\ge 1\\ (\sum_{i=1}^n\nu_i=N)}}\prod_{i=1}^nA_{\nu_i}(z,\mathfrak{t}),\nonumber \\
{\cal I}^{U(1)\text{ ADHM-}[l]\,(C)}&=\frac{1-\mathfrak{t}^{2l}}{(1-\mathfrak{t}^2)\prod_{\pm}(1-\mathfrak{t}^lz^{\pm l})},\nonumber \\
{\cal I}^{U(2)\text{ ADHM-}[l]\,(C)}&=
\frac{
(1-\mathfrak{t}^{2l})(1+(z^l+z^{-l})\mathfrak{t}^{2+l}+\mathfrak{t}^{2l}-\mathfrak{t}^{2l+2}-(z^l+z^{-l})\mathfrak{t}^{3l}-\mathfrak{t}^{4l+2})
}{
(1-\mathfrak{t}^2)
(1-\mathfrak{t}^4)
\prod_{\pm}
(1-z^{\pm l}\mathfrak{t}^l)
(1-z^{\pm 2l}\mathfrak{t}^{2l})
},\nonumber \\
{\cal I}^{U(3)\text{ ADHM-}[l]\,(C)}&=
\frac{
(1-\mathfrak{t}^{2l})
}{(1-\mathfrak{t}^2)(1-\mathfrak{t}^4)(1-\mathfrak{t}^6)\prod_{\pm}
(1-z^{\pm l}\mathfrak{t}^l)
(1-z^{\pm 2l}\mathfrak{t}^{2l})
(1-z^{\pm 3l}\mathfrak{t}^{3l})
}\nonumber \\
&\quad \times \Bigl[1 
+ (z^l+z^{-l})(\mathfrak{t}^{l+2}+ \mathfrak{t}^{l+4})
+ \mathfrak{t}^{2l} 
+ (z^{2l}+z^{-2l})(\mathfrak{t}^{2l+2} + \mathfrak{t}^{2l+4})
+ \mathfrak{t}^{2l+6}\nonumber \\
&\quad\quad + (z^{3l}+z^{-3l})\mathfrak{t}^{3l+6}
+(-z^{2l}-z^{-2l})\mathfrak{t}^{4l}
- 2 (\mathfrak{t}^{4l+2} +\mathfrak{t}^{4l+4})\nonumber \\
&\quad\quad + (1-z^{2l}-z^{-2l})\mathfrak{t}^{4l+6}
+ (-z^l-z^{-l})\mathfrak{t}^{5l}\nonumber \\
&\quad\quad +(-z^{3l}-z^l-z^{-l}-z^{-3l})(\mathfrak{t}^{5l+2}+\mathfrak{t}^{5l+4})
+(-z^l-z^{-l})\mathfrak{t}^{5l+6}\nonumber \\
&\quad\quad + (1-z^{2l}-z^{-2l})\mathfrak{t}^{6l} 
- 2 (\mathfrak{t}^{6l+2} +\mathfrak{t}^{6l+4})
+(-z^{2l}-z^{-2l}) \mathfrak{t}^{6l+6}\nonumber \\
&\quad\quad + (z^{3l}+z^{-3l})\mathfrak{t}^{7l}
+ \mathfrak{t}^{8l} 
+(z^{2l}+z^{-2l})(\mathfrak{t}^{8l+2}+\mathfrak{t}^{8l+4})
+ \mathfrak{t}^{8l+6}\nonumber \\
&\quad\quad + (z^l+z^{-l})(\mathfrak{t}^{9l+2}+\mathfrak{t}^{9l+4})
+ \mathfrak{t}^{10l+6}
\Bigr],\nonumber \\
&\vdots.
\end{align}
After setting $z$ to $1$, these results precisely reproduce the Coulomb branch Hilbert series written in \eqref{HS_C2Zl} ($N=1$),\eqref{HS_Sym2C2},\eqref{HS_Sym2C2Z2},\eqref{HS_Sym2C2Z4} ($N=2$) and \eqref{HS_Sym3C2} ($N=3$).

For general $N$ we can use the product expression \eqref{220513_productform} together with the following relation
\begin{align}
{\cal I}^{U(N) \text{ADHM-}[l]\,(C)}=\oint\frac{d\kappa}{2\pi i\kappa}\kappa^{-N}\Xi(\kappa).
\end{align}
By evaluating the integration by collecting the residues of the poles at $|\kappa|>1$ we obtain
\begin{align}
&{\cal I}^{U(N) \text{ADHM-}[l]\,(C)}\nonumber \\
&=
\sum_{m=0}^\infty\mathfrak{t}^{2mN}\prod_{\substack{m'\ge 0\\ (m'\neq m)}}\frac{1}{1-\mathfrak{t}^{2(m'-m)}}\prod_{m'\ge 0,n'\ge 1,\sigma'=\pm}\frac{1}{1-\mathfrak{t}^{2(m'-m)+n'l}z^{\sigma' n'l}}\nonumber \\
&\quad +\sum_{m=0}^\infty\sum_{n=1}^\infty\sum_{\sigma=\pm}(\mathfrak{t}^{2m+nl}z^{\sigma nl})^N\prod_{m'=0}^\infty\frac{1}{1-\mathfrak{t}^{2(m'-m)-nl}z^{-\sigma nl}}\nonumber \\
&\quad \quad \quad \quad \quad\quad\quad\quad\times \prod_{\substack{m'\ge 0,n'\ge 1,\sigma'=\pm\\ ((m',n',\sigma')\neq (m,n,\sigma))}}\frac{1}{1-\mathfrak{t}^{2(m'-m)+(n'-n)l}z^{l(\sigma'n'-\sigma n)}}.
\label{220429giantgraviton}
\end{align}
Note that \eqref{220429giantgraviton} gives the explicit expression for ${\cal I}^{U(\infty) \text{ADHM-}[l]\,(C)}$
\begin{align}
{\cal I}^{U(\infty) \text{ADHM-}[l]\,(C)}=\prod_{m\ge 1}\frac{1}{1-\mathfrak{t}^{2m}}\prod_{m\ge 0,n\ge 1,\pm}\frac{1}{1-\mathfrak{t}^{2m+nl}z^{\pm nl}},
\end{align}
which is consistent with the results obtained in \cite{Okazaki:2022sxo} for $l=1,2$ and $z=1$, as well as the explicit coefficients of all order giant graviton expansion.

\section{$USp(2N)$ gauge theories of M2-branes}
\label{sec_sp}
Let us study 3d $\mathcal{N}=4$ supersymmetric gauge theories with a $USp(2N)$ gauge group which can describe $N$ M2-branes probing a $D$-type singularity. 
As reviewed in section \ref{sec:typeIIAM}, there are two types of hypermultiplets $(X,Y)$ transforming as rank 2 tensor representations; either a symmetric (i.e. adjoint) or an antisymmetric under the gauge group 
as well as $2l$ 
half-hypermultiplets $(I,J)$ transforming as the fundamental representation. The quiver diagram for the antisymmetric case is depicted in the left figure in Figure \ref{fig:USp_mirror}.

\subsection{Moduli space and local operators}

\subsubsection{Coulomb branch}
On the Coulomb branch the vevs of the (half-)hypermultiplet scalar fields are turned off and the equation (\ref{adhm_bps1}) can be solved by 
\begin{align}
\label{sp_phi_C}
\varphi&=\mathrm{diag}(\varphi_1,-\varphi_1,\varphi_2,-\varphi_2,\cdots,\varphi_N,-\varphi_N)
\end{align}
so that the gauge group is broken to $U(1)^N$. 
The monopole in the $USp(2N)$ gauge theory has the GNO charge labeled by $N$ integers $(m_1,\cdots, m_N)$ as points in the weight lattice of the Langlands dual group $SO(2N+1)$. 
The monopole operator $v^{\{m_i\}}$ has the dimension
\begin{align}
\label{sp_l_monoD}
\Delta(m_i)&=
(l-2+2\epsilon+\delta/2)\sum_{i=1}^N |m_i|
\end{align}
where 
\begin{align}
\epsilon&=
\begin{cases}
0&\textrm{for antisym. hyper}\cr
1&\textrm{for sym. hyper}\cr
\end{cases}, 
& 
\delta&=
\begin{cases}
0&\textrm{for $2l$ fund. half-hypers}\cr
1&\textrm{for $(2l+1)$ fund. half-hypers}\cr
\end{cases}.
\end{align}
The good UV theories in the classification of \cite{Gaiotto:2008ak} can be obtained when one has at least two (resp. six) fundamental half-hypers for the theory with a symmetric (resp. antisymmetric) hyper. In the expression \eqref{sp_l_monoD} we include the cases with odd number of half-hypermultiplets but we will focus on the cases with $\delta=0$ in the following, where the theories are expected to describe the $N$ M2-branes probing a $D$-type singularity. 

For $N=1$, i.e. $USp(2)\cong SU(2)$ and $\delta=0$ the Coulomb branch operators describe the singularity $X_{D_{l+2\epsilon}}$ $=$ $\mathbb{C}^2/\widehat{D}_{l-2+2\epsilon}$ 
\footnote{A quotient singularity $X_{\mathfrak{g}}=\mathbb{C}^2/\Gamma$ with $\Gamma=\widehat{D}_{n-2}$ of order $4(n-2)$ corresponds to $\mathfrak{g}=D_{n}$. } 
where the dicyclic group $\widehat{D}_{l-2+2\epsilon}$ is generated by the rotation associated to the chiral ring relation consistent with the OPE
\begin{align}
v^{+}v^{-}\sim \varphi^{2(l-2+2\epsilon)}
\end{align}
and by the reflection $\varphi\rightarrow -\varphi$, $v^{\pm}\leftrightarrow v^{\mp}$ corresponding to the $\mathbb{Z}_2$ Weyl group of $SU(2)$. 
For $N>1$ the Coulomb branch is identified with the $N$-th symmetric product (\ref{moduli_Dl}) or (\ref{moduli_Dl+2}) of the ALE space $X_{D_{l+2\epsilon}}$
\begin{align}
\label{Mc_sp}
\mathcal{M}_C&=
\mathrm{Sym}^{N}X_{D_{l+2\epsilon}}
=\mathrm{Sym}^{N}(\mathbb{C}^2/\widehat{D}_{l-2+2\epsilon})
\end{align}
whose dimension is $\dim_{\mathbb{C}}\mathcal{M}_C$ $=$ $2N$. 

\subsubsection{Higgs branch}
When the vector multiplet scalars vanish, one finds the Higgs branch that is parametrized by the hypermultiplet scalar fields $(X,Y)$ and the half-hypermultiplet scalar fields $(I,J)$ where the gauge group is completely broken. 

The Higgs branch of the $USp(2N)$ gauge theory with either an adjoint or antisymmetric hyper 
and $2l+\delta$ fundamental half-hypers has dimension 
\begin{align}
\label{dim_H_sp}
\dim_{\mathbb{C}}\mathcal{M}_H=2N(l+\delta/2-1+2\epsilon). 
\end{align}

In the case with an antisymmetric hyper, i.e. $\epsilon=0$, 
the equations (\ref{adhm_bps4}) and (\ref{adhm_bps3}) are the ADHM equations for the $N$ $SO(2l+\delta)$ instantons on $\mathbb{R}^4$ \cite{Donaldson:1984tm}. 
So the Higgs branch is identified with the moduli space of $SO(2l+\delta)$ $N$-instantons. 

\subsubsection{Mixed branch}
The vacuum equations (\ref{adhm_bps1})-(\ref{adhm_bps3}) can be also solved when both of the vector multiplet scalar and the (half-)hypermultiplet scalar fields are non-zero. 
The equation (\ref{adhm_bps1}) can be solved by the configuration 
\begin{align}
    \varphi&=
    \left(
    \begin{matrix}
    1&0\\
    0&-1\\
    \end{matrix}
    \right)
    \otimes 
    \mathrm{diag}
    (
    \underbrace{\varphi_1, \cdots, \varphi_1}_{N_1}, 
    \cdots, 
    \underbrace{\varphi_n, \cdots, \varphi_n}_{N_n},
    \underbrace{0,\cdots,0}_{N_0}
    ),
\end{align}
where $2N_0$ of $2N$ components of scalar fields in (\ref{sp_phi_C}) vanish so that $USp(2N_0)$ gauge group is restored. 
The monopole operators can be dressed by the rank-2 tensor matter fields 
\begin{align}
    X&=
    \left(
    \begin{matrix}
    1&0\\
    0&-1\\
    \end{matrix}
    \right)
    \otimes 
    \mathrm{diag}
    (
    X_1,\cdots,X_{N_0}
    ),\nonumber\\
        Y&=
    \left(
    \begin{matrix}
    1&0\\
    0&-1\\
    \end{matrix}
    \right)
    \otimes 
    \mathrm{diag}
    (
    Y_1,\cdots,Y_{N_0}
    ), 
\end{align}
which solve the remaining vacuum equations (\ref{adhm_bps4}), (\ref{adhm_bps2}) and (\ref{adhm_bps3}) when one turns on the FI parameter.  
Such dressed monopole operators form the gauge invariant half-BPS local operators on the mixed branch as they are distinguished from the Coulomb and Higgs branch operators. 

\subsection{Indices}
The supersymmetric index of the $USp(2N)$ gauge theory with a single symmetric or an antisymmetric hyper and $2l$ fundamental hypers can be calculated as
\begin{align}
\label{sp_index}
&I^{\textrm{$USp(2N)+$(a)sym. hyper$-[2l+\delta]$}}(t,x,y_{\alpha};q)
\nonumber\\
&=\frac{1}{2^N N!}
\frac{(q^{\frac12}t^2;q)_{\infty}^{N}}
{(q^{\frac12}t^{-2};q)_{\infty}^{N}}
\sum_{m_1,\cdots, m_N\in \mathbb{Z}}
\oint \prod_{i=1}^{N}\frac{ds_i}{2\pi is_{i}}
\nonumber\\
&\times 
\prod_{i=1}^N (1-q^{|m_i|} s_i^{\pm 2})
\prod_{i< j}
(1-q^{\frac{|m_i-m_j|}{2}} s_i^{\pm}s_j^{\mp})
(1-q^{\frac{|m_i+m_j|}{2}} s_i^{\pm}s_j^{\pm})
\nonumber\\
&\times 
\prod_{i=1}^{N}
\frac{(q^{\frac12+|m_i|}t^{2}s_i^{\mp 2};q)_{\infty}}
{(q^{\frac12+|m_i|}t^{-2}s_i^{\pm 2};q)_{\infty}}
\prod_{i< j}
\frac{(q^{\frac12+\frac{|m_i-m_j|}{2}}t^{2}s_i^{\mp}s_j^{\pm};q)_{\infty}}
{(q^{\frac12+\frac{|m_i-m_j|}{2}}t^{-2}s_i^{\pm}s_j^{\mp};q)_{\infty}}
\prod_{i< j}
\frac{(q^{\frac12+\frac{|m_i+m_j|}{2}}t^{2}s_i^{\mp}s_j^{\mp};q)_{\infty}}
{(q^{\frac12+\frac{|m_i+m_j|}{2}}t^{-2}s_i^{\pm}s_j^{\pm};q)_{\infty}}
\nonumber\\
&\times 
\frac{(q^{\frac34}t^{-1} x^{\mp};q)_{\infty}^N}
{(q^{\frac14}t x^{\pm};q)_{\infty}^N}
\prod_{i< j}
\frac{(q^{\frac34+\frac{|m_i-m_j|}{2}}t^{-1}s_i^{\mp}s_j^{\pm}x^{\mp};q)_{\infty}}
{(q^{\frac14+\frac{|m_i-m_j|}{2}}ts_i^{\pm}s_j^{\mp}x^{\pm};q)_{\infty}}
\frac{(q^{\frac34+\frac{|m_i-m_j|}{2}}t^{-1}s_i^{\mp}s_j^{\pm}x^{\pm};q)_{\infty}}
{(q^{\frac14+\frac{|m_i-m_j|}{2}}ts_i^{\pm}s_j^{\mp}x^{\mp};q)_{\infty}}
\nonumber\\
&\times 
\prod_{i< j}
\frac{(q^{\frac34+\frac{|m_i+m_j|}{2}}t^{-1}s_i^{\mp}s_j^{\mp}x^{\mp};q)_{\infty}}
{(q^{\frac14+\frac{|m_i+m_j|}{2}}ts_i^{\pm}s_j^{\pm}x^{\pm};q)_{\infty}}
\frac{(q^{\frac34+\frac{|m_i+m_j|}{2}}t^{-1}s_i^{\mp}s_j^{\mp}x^{\pm};q)_{\infty}}
{(q^{\frac14+\frac{|m_i+m_j|}{2}}ts_i^{\pm}s_j^{\pm}x^{\mp};q)_{\infty}}
\nonumber\\
&\times 
\prod_{i=1}^{N}
\prod_{\alpha=1}^{2l+\delta}
\frac{(q^{\frac34+\frac{|m_i|}{2}}t^{-1}s_i^{\mp} y_{\alpha}^{-1};q)_{\infty}}
{(q^{\frac14+\frac{|m_i|}{2}}ts_i^{\pm} y_{\alpha};q)_{\infty}}
\Biggl[
\prod_{i=1}^{N}
\frac{(q^{\frac34+|m_i|}t^{-1}s_i^{\mp 2}x^{\mp};q)_{\infty}}
{(q^{\frac14+|m_i|}t s_i^{\pm 2}x^{\pm};q)_{\infty}}
\frac{(q^{\frac34+|m_i|}t^{-1}s_i^{\mp 2}x^{\pm};q)_{\infty}}
{(q^{\frac14+|m_i|}t s_i^{\pm 2}x^{\mp};q)_{\infty}}
\Biggr]^{\epsilon}
\nonumber\\
&\times 
q^{\frac{l-2+2\epsilon+\delta/2}{2}\sum_{i=1}^N |m_i|}
t^{-2(l-2+2\epsilon+\delta/2)\sum_{i=1}^N |m_i|}.
\end{align}
Here the flavor fugacities $y_{\alpha}$ obey the $SO(2l+\delta)$ conditions $y_{\alpha+l}=y_{\alpha}^{-1}$ for $\alpha=1,\cdots, l$ 
and $y_{2l+1}=1$. 
In the Coulomb limit (\ref{HS_lim}) the $USp(2N)$ index (\ref{sp_index}) for the case with an antisymmetric hyper $(\epsilon=0)$ and $2l$ fundamental hypers $(\delta=0)$ becomes the Coulomb branch Hilbert series studied in \cite{Cremonesi:2013lqa}, which corresponds to the geometry $\mathbb{C}^2/\hat{D}_{l-2}$ probed by the M2-branes. 
In the Higgs limit (\ref{HS_lim}) it becomes the Higgs branch Hilbert series studied in \cite{Benvenuti:2010pq}. 

\subsubsection{$USp(2)$ with 1 sym. and 2 fund. ($N=1$, $\epsilon=1$, $l=1$)}
We start with the $USp(2)\cong SU(2)$ gauge theory with a symmetric hypermultiplet. 
The dimension (\ref{sp_l_monoD}) of monopole operator can be positive when $l\ge 1$ or $\delta\ge0$. 
For $l=1$ with two half-hypers we find the flavored index 
\begin{align}
\label{Sp2_S_2_findex}
&I^{\textrm{$USp(2)+$sym. hyper$-[2]$}}(t,x,y;q)
\nonumber\\
&=1+\Bigl(
(2+x^2+x^{-2})t^2+t^{-2}
\Bigr)q^{1/2}
+\Bigl(
(xy^2+x^{-1}y^{-2}+xy^{-2}+x^{-1}y^2)t^3
\nonumber\\
&+(2x+2x^{-1})t^{-1}
\Bigr)q^{3/4}
+\Bigl(
-3+(3+x^4+x^{-4}+2x^2+2x^{-2})t^4+3t^{-4}
\Bigr)q+
\nonumber\\
&+\Bigl(
(x^3y^2+x^{-3}y^{-2}+x^3y^{-2}+x^{-3}y^2+2xy^2+2x^{-1}y^{-2}+2xy^{-2}+2x^{-1}y^2)t^5
\nonumber\\
&+(2x^3+2x^{-3}-xy^2-x^{-1}y^{-2}-xy^{-2}-x^{-1}y^2)t
\Bigr)q^{5/4}+\cdots.
\end{align}
For $x=1$ and $y=1$ it becomes
\begin{align}
\label{Sp2_S_2_index}
&I^{\textrm{$USp(2)+$sym. hyper$-[2]$}}(t,x=1,y=1;q)
\nonumber\\
&=1+
\Bigl(4t^2+t^{-2}\Bigr)q^{1/2}
+\Bigl(4t^3+4t^{-1}\Bigr)q^{3/4}
+\Bigl(-3+9t^4+3t^{-4} \Bigr)q
+12t^5q^{5/4}
\nonumber\\
&+\Bigl(-12+22t^6+3t^{-2}+3t^{-6} \Bigr)q^{3/2}
+\Bigl(-20t^3+24t^7-4t^{-1}+4t^{-5} \Bigr)q^{7/4}+\cdots.
\end{align}
Its Coulomb limit (\ref{HS_lim}) yields the Hilbert series
that coincides with (\ref{HS_C2Z4}). 
This is consistent with the Coulomb branch (\ref{Mc_sp}), that is 
$\mathbb{C}^2/\widehat{D}_1$ $=$ $\mathbb{C}^2/\mathbb{Z}_4$. 
Also we find the Higgs limit
\begin{align}
\label{HS_Sp2_S_2_H}
\mathcal{I}^{\textrm{$USp(2)+$sym. hyper$-[2]$}(H)}(\mathfrak{t})&=
\frac{
1+2\mathfrak{t}^2+2\mathfrak{t}^3+2\mathfrak{t}^4+\mathfrak{t}^6
}
{(1+\mathfrak{t})^2 (1+\mathfrak{t}+\mathfrak{t}^2)^2 (1-\mathfrak{t})^4}. 
\end{align}
The order $4$ of the pole at $\mathfrak{t}=1$ in (\ref{HS_Sp2_S_2_H}) is the complex dimension (\ref{dim_H_sp}) of the Higgs branch. 

\subsubsection{$USp(2)$ with 1 antisym. and 6 fund. ($N=1$, $\epsilon=0$, $l=3$)}
For the $USp(2)$ ADHM theory with an antisymmetric hypermultiplet 
the dimension (\ref{sp_l_monoD}) of monopole operator can be positive when $l\ge 3$. 

So the simplest example of the good theory is realized when $l=3$, that is the $USp(2)$ gauge theory with an antisymmetric hyper and six fundamental half-hypers. 
We find that the flavored index precisely agrees with 
the flavored index (\ref{u1_4_findex}) of the $U(1)$ ADHM theory with four flavors. 
This can be understood as a duality associated with the 
$\widehat{\mathfrak{so}(6)} = \widehat{\mathfrak{su}(4)}$ quiver \cite{Kapustin:1998fa} (see Section \ref{sec_spmirror}). 
This is consistent with the statement that 
the theory has the Coulomb branch $\mathbb{C}^2/\widehat{D}_1$ and that 
the Higgs branch is the moduli space of a single $SO(6)\cong SU(4)$ instanton. 

\subsubsection{$USp(2)$ with 1 antisym. and 8 fund. ($N=1$, $\epsilon=0$, $l=4$)}
Next example is the $USp(2)$ gauge theory with an antisymmetric hyper and eight half-hypers. 
In this case we get the flavored index 
\begin{align}
\label{Sp2_A_8_findex}
&I^{\textrm{$USp(2)+$asym. hyper$-[8]$}}(t,x,y_{\alpha};q)
\nonumber\\
&=1+(x+x^{-1})tq^{1/4}
+\Bigl(
5+x^{2}+5x^{-2}+\sum_{\alpha<\beta}^4(y_{\alpha}y_{\beta}+y_{\alpha}^{-1}y_{\beta}^{-1}+y_{\alpha}y_{\beta}^{-1}+y_{\alpha}^{-1}y_{\beta})
\Bigr)t^2q^{1/2}
\nonumber\\
&+\Bigl(
(x^3+x^{-3}+5x+5x^{-1}+(x+x^{-1})\sum_{\alpha<\beta}^4(y_{\alpha}y_{\beta}+y_{\alpha}^{-1}y_{\beta}^{-1}+y_{\alpha}y_{\beta}^{-1}+y_{\alpha}^{-1}y_{\beta}))t^3
\nonumber\\
&-(x+x^{-1})t^{-1} \Bigr)q^{3/4}+\cdots
\end{align}
For $x=y_{\alpha}=1$ it becomes
\begin{align}
\label{Sp2_A_8_index}
&I^{\textrm{$USp(2)+$asym. hyper$-[8]$}}(t,x=1,y_{\alpha}=1;q)
\nonumber\\
&=1+2tq^{1/4}
+31t^2q^{1/2}
+\Bigl( -2t^{-1}+60t^3 \Bigr)q^{3/4}
+\Bigl(
-33+2t^{-4}+389t^4
\Bigr)q
\nonumber\\
&
+
\Bigl(
4t^{-3}-118t+718t^5
\Bigr)q^{5/4}
+
\Bigl(
t^{-6}+4t^{-2}-852t^2+2972t^6
\Bigr)q^{3/2}
+\cdots.
\end{align}
The Coulomb limit of the index (\ref{Sp2_A_8_index}) is 
\begin{align}
\label{HS_C2D2}
\mathcal{I}^{\textrm{$USp(2)+$asym. hyper$-[8]$}(C)}(\mathfrak{t})
&=\frac{1-\mathfrak{t}^2+\mathfrak{t}^4}{(1+\mathfrak{t}^2) (1-\mathfrak{t}^2)^2}
=\frac{1+\mathfrak{t}^6}{(1+\mathfrak{t}^2) (1+\mathfrak{t}^2) (1-\mathfrak{t}^2)^2}.
\end{align}
This is the Hilbert series for the $\mathbb{C}^2/\widehat{D}_{2}$ $=$ $\mathbb{C}^2/\mathbb{Q}_8$ 
where $\mathbb{Q}_8$ is the quaternion group of order $8$. 
This is compatible with the expectation that 
the theory describes an M2-brane probing $\mathbb{C}^2\times (\mathbb{C}^2/\mathbb{Q}_8)$. 
The Higgs limit of the index (\ref{Sp2_A_8_index}) is 
\begin{align}
\label{HS_1_so8inst}
\mathcal{I}^{\textrm{$USp(2)+$asym. hyper$-[8]$}(H)}(\mathfrak{t})&=
\frac{(1+\mathfrak{t}^2)(1+17\mathfrak{t}^2+48\mathfrak{t}^4+17\mathfrak{t}^6+\mathfrak{t}^8)}
{(1+\mathfrak{t})^{10} (1-\mathfrak{t})^{12}}.
\end{align}
This reproduces the Hilbert series of the moduli space of one $SO(8)$ instanton \cite{Benvenuti:2010pq}. 

\subsubsection{$USp(4)$ with 1 sym. and 2 fund. ($N=2$, $\epsilon=1$, $l=1$)}
The flavored index of the $USp(4)$ gauge theory with an adjoint hyper and two half-hypers is evaluated as 
\begin{align}
\label{Sp4_S_2_findex}
&I^{\textrm{$USp(4)+$sym. hyper$-[2]$}}(t,x,y;q)
=1+\Bigl(
(2+x^2+x^{-2})t^2+t^{-2}
\Bigr)q^{1/2}
\nonumber\\
&+\Bigl(
(xy^2+x^{-1}y^{-2}+xy^{-2}+x^{-1}y^2)t^3
+(2x+2x^{-1})t
\Bigr)q^{3/4}
\nonumber\\
&
+\Bigl(
-1+x^2+x^{-2}
+(6+2x^4+2x^{-4}+4x^2+4x^{-2})t^4
+4t^{-4}
\Bigr)q+\cdots. 
\end{align}
When $x=1$ and $y=1$ it reduces to
\begin{align}
    \label{Sp4_S_2_index}
&I^{\textrm{$USp(4)+$sym. hyper$-[2]$}}(t,x=1,y_{\alpha}=1;q)
\nonumber\\
&=1+(4t^2+t^{-2})q^{1/2}+(4t^3+t^{-1})q^{3/4}
+(1+18t^4+4t^{-4})q
\nonumber\\
&+(24t^5+20t+4t^{-3})q^{5/4}
+(58t^6+9t^2+23t^{-2}+6t^{-6})q^{3/2}+\cdots
\end{align}
The Coulomb limit of the index (\ref{Sp4_S_2_index}) coincides with (\ref{HS_Sym2C2Z4}) as the Coulomb branch of the theory is  $\mathrm{Sym}^2(\mathbb{C}^2/\widehat{D}_1)$ $\cong$ $\mathrm{Sym}^2(\mathbb{C}^2/\mathbb{Z}_4)$. 
As stated in \eqref{USpCSdual}, the theory has a dual description as the $U(2)_{2}\times U(2)_{0}\times U(3)_{-2}$ quiver Chern-Simons theory. 

\subsubsection{$USp(4)$ with 1 antisym. and 6 fund. ($N=2$, $\epsilon=0$, $l=3$)}
We find that the flavored index of the $USp(4)$ ADHM theory with six fundamental half-hypers coincides with the flavored index (\ref{u2_4_findex}) 
of the $U(2)$ ADHM theory with four flavors for $z=1$. 
This again supports a special duality corresponding to the 
$\widehat{\mathfrak{so}(6)} = \widehat{\mathfrak{su}(4)}$ quiver proposed in  \cite{Kapustin:1998fa}. 

\subsubsection{Mirror symmetry}
\label{sec_spmirror}
As reviewed in section \ref{sec:typeIIB}, 
the $USp(2N)$ gauge theory with an antisymmetric hyper and $2l$ half-hypers has a conjectural mirror theory which is 
the quiver gauge theory with $U(N)^{\otimes 4}\times U(2N)^{\otimes l-3}$ gauge group and matter content which are encoded by the 
$\widehat{\mathfrak{so}(l)}$ affine Dynkin diagram as in Figure \ref{fig:USp_mirror}  \cite{deBoer:1996mp}. 

The dimension of the monopole operator for the mirror theory is given by
\begin{align}
\Delta(m_i^{(I)})&=
-\sum_{I=1}^{4}\sum_{i<j}^N |m_i^{(I)}-m_j^{(I)}|
-\sum_{I=5}^{l+1}\sum_{i<j}^{2N}|m_i^{(I)}-m_{j}^{(I)}|
+\frac12 \sum_{i=1}^N |m_i^{(1)}|
\nonumber\\
&+\frac12 \sum_{I=1}^2 \sum_{i,j}|m_i^{(I)}-m_j^{(5)}|
+\frac12 \sum_{I=5}^{l}\sum_{i,j}|m_i^{(I)}-m_j^{(I+1)}|
+\frac12 \sum_{I=3}^4 \sum_{i,j}|m_i^{(I)}-m_j^{(l+1)}|,
\end{align}
where $\{m_i^{(I)}\}$ is the GNO charge for the $I$-th gauge node with
\begin{align}
i=
\begin{cases}
1,\cdots, N&\textrm{for $I=1,2,3,4$}\cr
1,\cdots, 2N&\textrm{otherwise}\cr
\end{cases}
\end{align}

The index for the conjectural mirror theory takes the form
\begin{align}
\label{msp_index}
&I^{\textrm{$U(N)^{\otimes 4}\times U(2N)^{\otimes l-3}$}}(t,z_{I};q)=
\frac{1}{(N!)^4} \frac{1}{(2N!)^{l-3}}
\sum_{m_i^{(I)}\in \mathbb{Z}}
\prod_{I=1}^{4}
\oint \prod_{i=1}^N \frac{ds_i^{(I)}}{2\pi is_i^{(I)}}
\prod_{I=5}^{l+1}
\oint \prod_{i=1}^{2N} \frac{ds_i^{(I)}}{2\pi is_i^{(I)}}
\nonumber\\
&\times 
\prod_{i<j}(1-q^{\frac{|m_i^{(I)}-m_j^{(I)}|}{2}} {s_i^{(I)}}^{\pm} s_j^{(I)\mp})
\frac{
(q^{\frac12+\frac{|m_i^{(I)}-m_j^{(I)}|}{2}}t^{-2}s_i^{(I)\mp}s_j^{(I)\pm};q)_{\infty}
}
{
(q^{\frac12+\frac{|m_i^{(I)}-m_j^{(I)}|}{2}}t^{2}s_i^{(I)\pm}s_j^{(I)\mp};q)_{\infty}
}
\nonumber\\
&\times 
\prod_{I=1}^2 \prod_{i,j}
\frac{
(q^{\frac34+\frac{|m_i^{(I)}-m_j^{(5)}|}{2}}t s_i^{(I)\mp}s_j^{(5)\pm};q)_{\infty}
}
{
(q^{\frac14+\frac{|m_i^{(I)}-m_j^{(5)}|}{2}}t^{-1} s_i^{(I)\pm}s_j^{(5)\mp};q)_{\infty}
}
\prod_{I=5}^{l} \prod_{i,j}
\frac{
(q^{\frac34+\frac{|m_i^{(I)}-m_j^{(I+1)}|}{2}}t s_i^{(I)\mp}s_j^{(I+1)\pm};q)_{\infty}
}
{
(q^{\frac14+\frac{|m_i^{(I)}-m_j^{(I+1)}|}{2}}t^{-1} s_i^{(I)\pm}s_j^{(I+1)\mp};q)_{\infty}
}
\nonumber\\
&\times 
\prod_{I=3}^4 \prod_{i,j}
\frac{
(q^{\frac34+\frac{|m_i^{(I)}-m_j^{(l+1)}|}{2}}t s_i^{(I)\mp}s_j^{(l+1)\pm};q)_{\infty}
}
{
(q^{\frac14+\frac{|m_i^{(I)}-m_j^{(l+1)}|}{2}}t^{-1} s_i^{(I)\pm}s_j^{(l+1)\mp};q)_{\infty}
}
\prod_{i=1}^N 
\frac{(q^{\frac34+\frac{|m_i^{(1)}|}{2}}t s_{i}^{(1)\mp};q)_{\infty}}
{(q^{\frac14+\frac{|m_i^{(1)}|}{2}}t^{-1} s_{i}^{(1)\pm};q)_{\infty}}
\nonumber\\
&\times 
(q^{-\frac12}t^{-2})^{
\sum_{I=1}^{4}\sum_{i<j}^N |m_i^{(I)}-m_j^{(I)}|
+\sum_{I=5}^{l+1}\sum_{i<j}^{2N}|m_i^{(I)}-m_{j}^{(I)}|
 }
 \nonumber\\
&(q^{\frac14}t)^{(\sum_{i=1}^N |m_i^{(1)}|
+\sum_{I=1}^2 \sum_{i,j}|m_i^{(I)}-m_j^{(5)}|
+\sum_{I=5}^{l}\sum_{i,j}|m_i^{(I)}-m_j^{(I+1)}|
+\sum_{I=3}^4 \sum_{i,j}|m_i^{(I)}-m_j^{(l+1)}|
)
 }
\prod_{I=1}^{l+1} z_{I}^{\sum_{i} m_i^{(I)}}.
\end{align}
For $N=1$ and $l=4$  we confirm that the index (\ref{msp_index}) for the mirror theory precisely agrees with the index (\ref{Sp2_A_8_findex}) for the $USp(2)$ ADHM theory with an antisymmetric hyper and eight fundamental half-hypers under the following mapping of fugacities:
\begin{align}
z_1=xy_1^{-1}y_2^{-1}, \qquad  
z_2=y_1y_2^{-1}, \qquad
z_3=y_3y_4^{-1}, \qquad 
z_4=y_3y_4, \qquad
z_5=y_2 y_3^{-1}.
\end{align}
For $l=5$, we found that the index \eqref{msp_index} agrees with that of the $USp(2)$ ADHM theory with the following parameter identification:
\begin{align}
z_1=xy_1^{-1}y_2^{-1}, \quad  
z_2=y_1y_2^{-1}, \quad
z_3=y_4y_5^{-1}, \quad 
z_4=y_4y_5, \quad
z_5=y_2 y_3^{-1},\quad
z_6=y_3 y_4^{-1}.
\end{align}
From these results we conjecture the following parameter identification for general $l\ge 4$:
\begin{align}
&z_1=xy_1^{-1}y_2^{-1}, \quad  
z_2=y_1y_2^{-1}, \quad
z_3=y_{l-1}y_l^{-1}, \quad 
z_4=y_{l-1}y_l, \quad
z_5=y_2 y_3^{-1},\nonumber \\
&z_6=y_3 y_4^{-1},\quad
\cdots,\quad
z_{l+1}=y_{l-2} y_{l-1}^{-1}.
\end{align}

\begin{figure}
\centering
\scalebox{0.7}{
\begin{tikzpicture}
\path 
(-2,0) node[circle, minimum size=64, fill=green!10, draw](AG1) {$USp(2N)$}
(-2,4) node[minimum size=50, fill=cyan!10, draw](AF1) {$6$};
\draw (AG1) -- node[right]{$(I,J)$}(AF1);
\draw (AG1.south west) .. controls ++(-1, -2) and ++(1, -2) .. node[below]{$(X,Y)$} (AG1.south east);
\path 
(-2,-10) node[circle, minimum size=64, fill=yellow!20, draw](AG1) {$U(N)$}
(-2,-6) node[minimum size=50, fill=cyan!10, draw](AF1) {$4$};
\draw (AG1) -- node[right]{$(I,J)$}(AF1);
\draw (AG1.south west) .. controls ++(-1, -2) and ++(1, -2) .. node[below]{$(X,Y)$} (AG1.south east);
\path 
(6,-3) node[circle, minimum size=48, fill=yellow!20, draw](BG1) {$U(N)^{(1)}$}
(11,-3) node[circle, minimum size=48, fill=yellow!20, draw](BG2) {$U(N)^{(2)}$}
(6,-7) node[circle, minimum size=48, fill=yellow!20, draw](BG4) {$U(N)^{(4)}$}
(11,-7) node[circle, minimum size=48, fill=yellow!20, draw](BG3) {$U(N)^{(3)}$}
(6,0) node[minimum size=40, fill=cyan!10, draw](BF1) {$1$};
\draw (BG1) -- node[above]{$(\tilde{X}_{1,2},\tilde{Y}_{1,2})$} (BG2);
\draw (BG1) -- node[right]{$(\tilde{X}_{1,4},\tilde{Y}_{1,4})$} (BG4);
\draw (BG2) -- node[left]{$(\tilde{X}_{2,3},\tilde{Y}_{2,3})$} (BG3);
\draw (BG3) -- node[below]{$(\tilde{X}_{3,4},\tilde{Y}_{3,4})$} (BG4);
\draw (BG1) -- node[right]{$(\tilde{I},\tilde{J})$}(BF1);
\end{tikzpicture}
}
\caption{Mirror symmetry of the $USp(2N)$ ADHM theory with one antisymmetric hyper $(X,Y)$ and $6N$ half-hypers $(I,J)$ 
and the $U(N)^{\otimes 4}$ 
$\widehat{\mathfrak{so}(6)}=\widehat{\mathfrak{su}(4)}$ quiver theory with one flavor $(\tilde{I},\tilde{J})$. It is also dual to the $U(N)$ ADHM theory with four flavors.} \label{fig:USp_mirror2}
\end{figure}
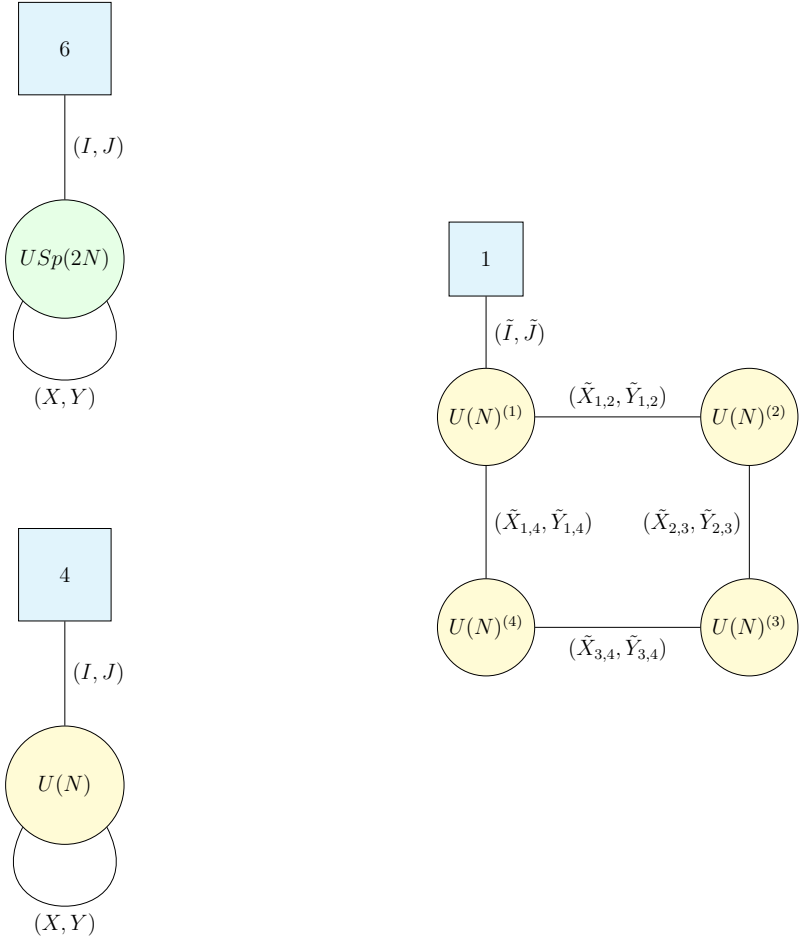

For $l=3$, the mirror theory is identical to the $U(N)^{\otimes 4}$ gauge theory 
which is mirror to the $U(N)$ ADHM theory with four flavors 
corresponding to the 
$\widehat{\mathfrak{so}(6)}=\widehat{\mathfrak{su}(4)}$ quiver theory depicted in 
Figure \ref{fig:USp_mirror2}). 
In this case, the index (\ref{msp_index}) is equal to the index (\ref{muN_l_index}). 
This 
confirms the duality \cite{Kapustin:1998fa} between 
the $USp(2N)$ ADHM theory with six fundamental half-hypers and 
the $U(N)$ ADHM theory with four hypers. 

\section{$O(N)$ gauge theories of M2-branes}
\label{sec_O}
Let us study 3d $\mathcal{N}=4$ supersymmetric gauge theories with orthogonal gauge group which can describe M2-branes. 
As reviewed in section \ref{sec:typeIIAM}, the theories have rank-2 tensor matter, either an antisymmetric (i.e. adjoint) or a symmetric hypermultiplet $(X,Y)$ and $l$ fundamental hypermultiplets $(I,J)$. 

\subsection{Moduli space and local operators}

\subsubsection{Coulomb branch}
Setting the hypermultiplet scalar fields to zero, we obtain the Coulomb branch. For the $SO(2N+\gamma)$ gauge theory with $\gamma$ $=$ $1$ or $0$ the equation (\ref{adhm_bps1}) can be solved by skew-diagonal configuration
\begin{align}
\label{o_phi_C}
    \varphi&=
    \left(
    \begin{matrix}
    0&\varphi_1& &&\\
    -\varphi_1&0& &&\\
    &&\ddots &&\\
    &&&0&\varphi_{N}\\
    &&&-\varphi_{N}&0\\
    \end{matrix}
    \right). 
\end{align}
This breaks the gauge group down to $U(1)^N$. 
For $\gamma=1$ there is an additional row and a column of zeroes in (\ref{o_phi_C}). 
The Coulomb branch receives perturbative and non-perturbative quantum corrections. 
The monopole operators for $SO(2N+\gamma)$ gauge theories   
carry the GNO charge labeled by integers $(m_1, \cdots, m_N)$ in the weight lattice of the Langlands dual group $USp(2N)$. 
When the theory contains a rank-2 hyper and fundamental hypers,  
it has the monopole operator whose dimension is 
\begin{align}
\Delta(m_i)&=
(l+2\epsilon)\sum_{i=1}^{N}|m_i|,
\end{align}
where 
\begin{align}
\epsilon&=
\begin{cases}
0&\textrm{for antisym. hyper}\cr
1&\textrm{for sym. hyper}\cr
\end{cases}.
\end{align}
According to the classification in \cite{Gaiotto:2008ak} the theory with a symmetric (resp. antisymmetric) hyper 
is good for $l\ge 0$ (resp. $l\ge1$). 

The Lie algebra $\mathfrak{so}(2N+\gamma)$ admits several gauge theories of distinct gauge groups 
including $O(2N+\gamma)_{\pm}$, $Pin(2N+\gamma)_{\pm}$, $SO(2N+\gamma)$ and $Spin(2N+\gamma)$. 
The $SO(2N+\gamma)$ gauge group has two zero-form global symmetries, the charge conjugation symmetry $\mathbb{Z}_2^{\mathcal{C}}$ and the magnetic symmetry $\mathbb{Z}_2^{\mathcal{M}}$. 
The other gauge groups can be obtained by gauging these global symmetries. 

In particular, 3d $\mathcal{N}=4$ $O(2N+\gamma)_{+}$ gauge theories with a rank-2 hyper and fundamental hypers are expected to describe multiple M2-branes at a $D$-type singularities. 
So we will mainly focus on this case. 

For $N=1$, i.e. $O(2+\gamma)$ gauge theories with a rank-2 tensor hyper and $l$ fundamental hypers the Coulomb branch is the quotient singularities $X_{D_{l+2\epsilon+2}}=\mathbb{C}^2/\widehat{D}_{l+2\epsilon}$. 
For higher rank gauge groups the Coulomb branch is given by the $N$-th symmetric product (\ref{moduli_Dl+2}) or (\ref{moduli_Dl+4}) of the ALE space $X_{D_{l+2\epsilon+2}}$
\begin{align}
\label{Mc_O}
\mathcal{M}_C&=\mathrm{Sym}^{N}X_{D_{l+2\epsilon+2}}
=\mathrm{Sym}^N(\mathbb{C}^2/\widehat{D}_{l+2\epsilon})
\end{align}
of the singularity 
whose dimension is $\dim_{\mathbb{C}}\mathcal{M}_C$ $=$ $2N$. 

\subsubsection{Higgs branch}
The vacuum equations (\ref{adhm_bps1})-(\ref{adhm_bps3}) can be solved by setting the vector multiplet scalar field to zero, 
for which we find the Higgs branch parametrized by the hypermultiplet scalar fields. 

For $l\ge N$ the orthogonal gauge theories, e.g. $SO(2N+\gamma)$ gauge theory can admit baryonic operators 
of the form 
\begin{align}
    \epsilon^{a_1\cdots a_{2N+\gamma}}J_{a_1}\cdots J_{a_{2N+\gamma}}, \qquad 
    \epsilon_{a_1\cdots a_{2N+\gamma}}I^{a_1}\cdots I^{a_{2N+\gamma}}. 
\end{align}

For the $O(2N+\gamma)$ theory with a symmetric hyper and $l$ fundamental hypers the equations (\ref{adhm_bps1})-(\ref{adhm_bps3}) are the ADHM equations for the $(2N+\gamma)$ $USp(2l)$ instantons on $\mathbb{R}^4$ \cite{Donaldson:1984tm} so that the Higgs branch is identified with the moduli space of the $(2N+\gamma)$ $USp(2l)$ instantons. 

\subsubsection{Mixed branch}
There exist solutions to the equations (\ref{adhm_bps2})-(\ref{adhm_bps3}) where 
both of the vector multiplet scalar and the hypermultiplet scalars take non-zero values. 
The monopole operators may be dressed by the rank-2 tensor matter fields $(X,Y)$ which solve the remaining vacuum equations (\ref{adhm_bps4}), (\ref{adhm_bps2}) and (\ref{adhm_bps3}) in the presence of the FI parameter. 
They form the gauge invariant half-BPS local operators which are distinguished from the Coulomb and Higgs branch operators. 

\subsection{Indices}
The supersymmetric index of 3d gauge theories with orthogonal gauge groups depends on the global structure of the gauge group  \cite{Aharony:2013kma}. 
\footnote{
The indices of 3d gauge theories with gauge group $O(2N+\gamma)_{+}$ are computed in \cite{Hwang:2011qt,Hwang:2011ht}. 
}
All the indices can be obtained from the $SO(2N+\gamma)$ indices with discrete fugacities $\zeta$ and $\chi$ 
for the $\mathbb{Z}_2^{\mathcal{M}}$ and $\mathbb{Z}_2^{\mathcal{C}}$ global symmetries. 

For $\chi=1$ or $\gamma=1$ 
the $O(2N+\gamma)$ holonomy can take the following form:  
\begin{align}
\mathrm{diag}(s_1,s_1^{-1},\cdots,s_N,s_N^{-1}),\qquad \textrm{or} \qquad 
    \mathrm{diag}(s_1,s_1^{-1},\cdots, s_N,s_N^{-1},\chi). 
\end{align}
Accordingly, the index takes the form
\begin{align}
\label{SO_index1}
&I^{\textrm{$SO(2N+\gamma)+$(a)sym. hyper$-[2l]$}}(t,x,y_{\alpha};\zeta;\chi;q)
=\frac{1}{2^{N+\gamma-1}N!}\frac{(q^{\frac12}t^2;q)_{\infty}^N}
{(q^{\frac12}t^{-2};q)_{\infty}^N}
\nonumber\\
&\times 
\sum_{m_1,\cdots, m_n\in \mathbb{Z}}\oint \prod_{i=1}^N \frac{ds_i}{2\pi is_i} 
\prod_{i=1}^N 
(1-\chi q^{\frac{|m_i|}{2}}s_i^{\pm})^{\gamma} 
\prod_{i<j}
(1-q^{\frac{|m_i -m_j|}{2}} s_i^{\pm}s_j^{\mp})
(1-q^{\frac{|m_i +m_j|}{2}} s_i^{\pm}s_j^{\pm})
\nonumber\\
&\times 
\left[\prod_{i=1}^N 
\frac{(\chi q^{\frac12+\frac{|m_i|}{2}}t^2 s_i^{\mp};q)_{\infty}}
{(\chi q^{\frac12+\frac{|m_i|}{2}}t^{-2} s_i^{\pm};q)_{\infty}}
\right]^{\gamma}
\prod_{i<j} 
\frac{(q^{\frac12+\frac{|m_i-m_j|}{2}}t^2 s_i^{\mp}s_j^{\pm};q)_{\infty}}
{(q^{\frac12+\frac{|m_i-m_j|}{2}}t^{-2} s_i^{\pm}s_j^{\mp};q)_{\infty}}
\frac{(q^{\frac12+\frac{|m_i+m_j|}{2}}t^2 s_i^{\mp}s_j^{\mp};q)_{\infty}}
{(q^{\frac12+\frac{|m_i+m_j|}{2}}t^{-2} s_i^{\pm}s_j^{\pm};q)_{\infty}}
\nonumber\\
&\times 
\frac{(q^{\frac34}t^{-1}x^{\mp};q)_{\infty}^{N+\gamma}}
{(q^{\frac14}tx^{\pm};q)_{\infty}^{N+\gamma}}
\prod_{i<j}
\frac{(q^{\frac34+\frac{|m_i-m_j|}{2}} t^{-1}s_i^{\mp} s_j^{\pm}x^{\mp};q)_{\infty}}
{(q^{\frac14+\frac{|m_i-m_j|}{2}} ts_i^{\pm} s_j^{\mp}x^{\pm};q)_{\infty}}
\frac{(q^{\frac34+\frac{|m_i-m_j|}{2}} t^{-1}s_i^{\mp} s_j^{\pm}x^{\pm};q)_{\infty}}
{(q^{\frac14+\frac{|m_i-m_j|}{2}} ts_i^{\pm} s_j^{\mp}x^{\mp};q)_{\infty}}
\nonumber\\
&\times 
\prod_{i<j}
\frac{(q^{\frac34+\frac{|m_i+m_j|}{2}} t^{-1}s_i^{\mp} s_j^{\mp}x^{\mp};q)_{\infty}}
{(q^{\frac14+\frac{|m_i+m_j|}{2}} ts_i^{\pm} s_j^{\pm}x^{\pm};q)_{\infty}}
\frac{(q^{\frac34+\frac{|m_i+m_j|}{2}} t^{-1}s_i^{\mp} s_j^{\mp}x^{\pm};q)_{\infty}}
{(q^{\frac14+\frac{|m_i+m_j|}{2}} ts_i^{\pm} s_j^{\pm}x^{\mp};q)_{\infty}}
\nonumber\\
&\times 
\left[
\prod_{i=1}^N
\frac{(\chi q^{\frac34+\frac{|m_i|}{2}} t^{-1}s_i^{\mp}x^{\mp};q)_{\infty}}
{(\chi q^{\frac14+\frac{|m_i|}{2}} ts_i^{\pm}x^{\pm};q)_{\infty}}
\frac{(\chi q^{\frac34+\frac{|m_i|}{2}} t^{-1}s_i^{\mp}x^{\pm};q)_{\infty}}
{(\chi q^{\frac14+\frac{|m_i|}{2}} ts_i^{\pm}x^{\mp};q)_{\infty}}
\right]^{\gamma}
\nonumber\\
&\times 
\left[
\prod_{i=1}^{N}
\frac{(q^{\frac34+|m_i|}t^{-1}s_i^{\mp 2}x^{\mp};q)_{\infty}}
{(q^{\frac34+|m_i|}t s_i^{\pm 2}x^{\pm};q)_{\infty}}
\frac{(q^{\frac34+|m_i|}t^{-1}s_i^{\mp 2}x^{\pm};q)_{\infty}}
{(q^{\frac34+|m_i|}t s_i^{\pm 2}x^{\mp};q)_{\infty}}
\right]^{\epsilon}
\nonumber\\
&\times 
\prod_{i=1}^N
\prod_{\alpha=1}^{2l}
\frac{(q^{\frac34+\frac{|m_i|}{2}} t^{-1}s_i^{\mp}y_{\alpha}^{-1};q)_{\infty}}
{(q^{\frac14+\frac{|m_i|}{2}} ts_i^{\pm}y_{\alpha};q)_{\infty}}
\left[
\frac{(\chi q^{\frac34}t^{-1}y_{\alpha}^{-1};q)_{\infty}}
{(\chi q^{\frac14}ty_{\alpha};q)_{\infty}}
\right]^{\gamma}
\nonumber\\
&\times 
q^{\frac{l+2\epsilon}{2}\sum_{i=1}^{N}|m_i|}
t^{-2 (l+2\epsilon)\sum_{i=1}^{N}|m_i|}
\zeta^{\sum_{i=1}^N m_i}. 
\end{align}
Again the flavor fugacities $y_{\alpha}$ satisfy the $USp(2l)$ condition $y_{l+\alpha}=y_{\alpha}^{-1}$. 

When $\chi=-1$ and $\gamma=0$, one can set the $O(2N)$ holonomy to 
\begin{align}
    \mathrm{diag}(s_1, s_1^{-1},\cdots, s_{N-1},s_{N-1}^{-1},1,-1)
\end{align}
so that the gauge fugacity $s_N$ is simply replaced with $\pm1$ and the magnetic flux $m_{N}$ is set to zero. 
The formula of the index for $\chi=-1$ and $\gamma=0$ is 
\begin{align}
\label{SO_index2}
&I^{\textrm{$SO(2N)+$(a)sym. hyper$-[2l]$}}(t,x,y_{\alpha};\zeta;\chi=-;q)
\nonumber\\
&=
\frac{1}{2^{N-1}(N-1)!}
\frac{(-q^{\frac12}t^2;q)_{\infty}}
{(-q^{\frac12}t^{-2};q)_{\infty}}
\frac{(q^{\frac12}t^2;q)_{\infty}^{N-1}}
{(q^{\frac12}t^{-2};q)_{\infty}^{N-1}}
\sum_{m_1,\cdots, m_{N-1}\in \mathbb{Z}}
\oint \prod_{i=1}^{N-1}
\frac{ds_i}{2\pi is_i}
\nonumber\\
&\times 
\prod_{i=1}^{N-1}
(1-q^{|m_i|}s_i^{\pm2})
\prod_{i<j}
(1-q^{\frac{|m_i -m_j|}{2}} s_i^{\pm}s_j^{\mp})
(1-q^{\frac{|m_i +m_j|}{2}} s_i^{\pm}s_j^{\pm})
\nonumber\\
&\times 
\prod_{i=1}^{N-1}
\frac{(q^{\frac12+\frac{|m_i|}{2}}t^2 s_i^{\mp};q)_{\infty}}
{(q^{\frac12+\frac{|m_i|}{2}}t^{-2}s_i^{\pm};q)_{\infty}}
\frac{(-q^{\frac12+\frac{|m_i|}{2}}t^2 s_i^{\mp};q)_{\infty}}
{(-q^{\frac12+\frac{|m_i|}{2}}t^{-2}s_i^{\pm};q)_{\infty}}
\nonumber\\
&\times 
\prod_{i<j} 
\frac{(q^{\frac12+\frac{|m_i-m_j|}{2}}t^2 s_i^{\mp}s_j^{\pm};q)_{\infty}}
{(q^{\frac12+\frac{|m_i-m_j|}{2}}t^{-2} s_i^{\pm}s_j^{\mp};q)_{\infty}}
\frac{(q^{\frac12+\frac{|m_i+m_j|}{2}}t^2 s_i^{\mp}s_j^{\mp};q)_{\infty}}
{(q^{\frac12+\frac{|m_i+m_j|}{2}}t^{-2} s_i^{\pm}s_j^{\pm};q)_{\infty}}
\nonumber\\
&\times 
\frac{(-q^{\frac34}t^{-1}x^{\mp};q)_{\infty}}
{(-q^{\frac14}tx^{\pm};q)_{\infty}}
\frac{(q^{\frac34}t^{-1}x^{\mp};q)_{\infty}^{N-1}}
{(q^{\frac14}tx^{\pm};q)_{\infty}^{N-1}}
\prod_{i=1}^{N-1}
\frac{(q^{\frac34+\frac{|m_i|}{2}}t^{-1}s_i^{\mp}x^{\mp};q)_{\infty}}
{(q^{\frac14+\frac{|m_i|}{2}}ts_i^{\pm}x^{\pm};q)_{\infty}}
\frac{(q^{\frac34+\frac{|m_i|}{2}}t^{-1}s_i^{\mp}x^{\pm};q)_{\infty}}
{(q^{\frac14+\frac{|m_i|}{2}}ts_i^{\pm}x^{\mp};q)_{\infty}}
\nonumber\\
&\times 
\prod_{i=1}^{N-1}
\frac{(-q^{\frac34+\frac{|m_i|}{2}}t^{-1}s_i^{\mp}x^{\mp};q)_{\infty}}
{(-q^{\frac14+\frac{|m_i|}{2}}ts_i^{\pm}x^{\pm};q)_{\infty}}
\frac{(-q^{\frac34+\frac{|m_i|}{2}}t^{-1}s_i^{\mp}x^{\pm};q)_{\infty}}
{(-q^{\frac14+\frac{|m_i|}{2}}ts_i^{\pm}x^{\mp};q)_{\infty}}
\nonumber\\
&\times 
\prod_{i<j}
\frac{(q^{\frac34+\frac{|m_i-m_j|}{2}} t^{-1}s_i^{\mp} s_j^{\pm}x^{\mp};q)_{\infty}}
{(q^{\frac14+\frac{|m_i-m_j|}{2}} ts_i^{\pm} s_j^{\mp}x^{\pm};q)_{\infty}}
\frac{(q^{\frac34+\frac{|m_i-m_j|}{2}} t^{-1}s_i^{\mp} s_j^{\pm}x^{\pm};q)_{\infty}}
{(q^{\frac14+\frac{|m_i-m_j|}{2}} ts_i^{\pm} s_j^{\mp}x^{\mp};q)_{\infty}}
\nonumber\\
&\times 
\prod_{i<j}
\frac{(q^{\frac34+\frac{|m_i+m_j|}{2}} t^{-1}s_i^{\mp} s_j^{\mp}x^{\mp};q)_{\infty}}
{(q^{\frac14+\frac{|m_i+m_j|}{2}} ts_i^{\pm} s_j^{\pm}x^{\pm};q)_{\infty}}
\frac{(q^{\frac34+\frac{|m_i+m_j|}{2}} t^{-1}s_i^{\mp} s_j^{\mp}x^{\pm};q)_{\infty}}
{(q^{\frac14+\frac{|m_i+m_j|}{2}} ts_i^{\pm} s_j^{\pm}x^{\mp};q)_{\infty}}
\nonumber\\
&\times 
\left[
\prod_{i=1}^{N-1}
\frac{(q^{\frac34+|m_i|}t^{-1}s_i^{\mp 2}x^{\mp};q)_{\infty}}
{(q^{\frac34+|m_i|}t s_i^{\pm 2}x^{\pm};q)_{\infty}}
\frac{(q^{\frac34+|m_i|}t^{-1}s_i^{\mp 2}x^{\pm};q)_{\infty}}
{(q^{\frac34+|m_i|}t s_i^{\pm 2}x^{\mp};q)_{\infty}}
\frac{(q^{\frac34}t^{-1}x^{\mp};q)_{\infty}^2}
{(q^{\frac14}tx^{\pm};q)_{\infty}^2}
\right]^{\epsilon}
\nonumber\\
&\times 
\prod_{i=1}^N
\prod_{\alpha=1}^{2l}
\frac{(q^{\frac34+\frac{|m_i|}{2}} t^{-1}s_i^{\mp}y_{\alpha}^{-1};q)_{\infty}}
{(q^{\frac14+\frac{|m_i|}{2}} ts_i^{\pm}y_{\alpha};q)_{\infty}}
\frac{(\pm q^{\frac34}t^{-1}y_{\alpha}^{-1};q)_{\infty}}
{(\pm q^{\frac14}ty_{\alpha};q)_{\infty}}
\nonumber\\
&\times 
q^{\frac{(l+2\epsilon)}{2} \sum_{i=1}^{N-1}|m_i|}
t^{-2 (l+2\epsilon) \sum_{i=1}^{N-1}|m_i|}
\zeta^{\sum_{i=1}^{N-1} m_i}. 
\end{align}
For simplicity we often use the shorthand notation $I(t;\zeta;\chi;q)=I(\zeta;\chi)$ etc. in the following analysis. 

The index of the $O(2N+\gamma)$ gauge theory can be obtained by gauging the $\mathbb{Z}_2^{\mathcal{C}}$ charge conjugation symmetry
\begin{align}
I^{O(2N+\gamma)}(\zeta,\chi')
&=\frac12 
\left(
I^{SO(2N+\gamma)}(\zeta,+)
+\chi'
I^{SO(2N+\gamma)}(\zeta,-)
\right),
\end{align}
where $\chi'$ is $+$ or $-$. 
Since the integral formulae of the indices are more subtle than the previous theories with gauge groups, $U(N)$ and $USp(2N)$, we will  present explicit expressions for several examples. 

\subsubsection{$O(1)$ with 1 fund. ($N=0$, $\gamma=1$, $l=1$)}
\label{sec_o1}
The $SO(1)$ gauge theory is a free theory with matter fields.  
The index is not sensitive to the value of $\zeta$. 
For example, when the theory has a single hyper the index reads
\begin{align}
\label{SO1++_index}
&I^{\textrm{$SO(1)-[1]$}}(t;\zeta=\pm;\chi=+;q)
=\frac{(q^{\frac34}t^{-1};q)_{\infty}^2}{(q^{\frac14}t;q)_{\infty}^2},\\
\label{SO1+-_index}
&I^{\textrm{$SO(1)-[1]$}}(t;\zeta=\pm;\chi=-;q)
=\frac{(-q^{\frac34}t^{-1};q)_{\infty}^2}{(-q^{\frac14}t;q)_{\infty}^2}.
\end{align}
By gauging the global $\mathbb{Z}_2^{\mathcal{C}}$ we get the index for the $O(1)_{+}$ gauge theory
\begin{align}
\label{O1++_index}
&I^{\textrm{$O(1)_{+}-[1]$}}(\zeta,\chi')
=\frac12 \left(
I^{\textrm{$SO(1)-[1]$}}(\zeta,+)+\chi'
I^{\textrm{$SO(1)-[1]$}}(\zeta,-)
\right). 
\end{align}
While the $O(1)_{+}=\mathbb{Z}_2$ gauge theory has no Coulomb branch, the Higgs branch is $\mathbb{C}^2/\mathbb{Z}_2$. 
In fact, we find that the Higgs limit of the index (\ref{O1++_index}) agrees with (\ref{HS_C2Z2}) for $\mathbb{C}^2/\mathbb{Z}_2$. 

In section \ref{sec_Zk} we will see discrete gauge theories of the M2-brane 
which generalize the $O(1)_+=\mathbb{Z}_2$ gauge theories. 

\subsubsection{$O(2)$ with 1 antisym. and 1 fund. ($N=1$, $\gamma=0$, $\epsilon=0$, $l=1$)}
The $O(2)$ gauge theory with an adjoint hyper and one flavor has a conjectural 
dual theory, a $U(1)_2\times U(1)\times U(1)_{-2}$ circular quiver Chern-Simons matter theory as in \eqref{OCSdual}. 
Thus we give a full flavored index of this theory. 
The index of the $SO(2)$ gauge theory with an adjoint hyper and one flavor can be expressed as
\begin{align}
\label{SO2z+_A_2_index}
&I^{\textrm{$SO(2)+$asym. hyper$-[2]$}}(\zeta;\chi=+;x,y)
\nonumber\\
&=
\frac{(q^{\frac12}t^2;q)_{\infty}}
{(q^{\frac12}t^{-2};q)_{\infty}}
\frac{(q^{\frac34}t^{-1}x^{\mp};q)_{\infty}}
{(q^{\frac14}tx^{\pm};q)_{\infty}}
\sum_{m\in \mathbb{Z}}\oint 
\frac{(q^{\frac34+\frac{|m|}{2}}t^{-1}s^{\mp}y^{\mp};q)_{\infty}}
{(q^{\frac14+\frac{|m|}{2}} ts^{\pm}y^{\pm};q)_{\infty}}
\frac{(q^{\frac34+\frac{|m|}{2}}t^{-1}s^{\mp}y^{\pm};q)_{\infty}}
{(q^{\frac14+\frac{|m|}{2}} ts^{\pm}y^{\mp};q)_{\infty}}
\nonumber\\
&\times 
q^{\frac{|m|}{2}}t^{-2|m|} \zeta^{m}, \\
\label{SO2z-_A_2_index}
&I^{\textrm{$SO(2)+$asym. hyper$-[2]$}}(\zeta;\chi=-;x,y)
\nonumber\\
&=
\frac{(-q^{\frac12}t^{2};q)_{\infty}}
{(-q^{\frac12}t^{-2};q)_{\infty}}
\frac{(-q^{\frac34}t^{-1}x^{\mp};q)_{\infty}}
{(-q^{\frac14}tx^{\pm};q)_{\infty}}
\frac{(q^{\frac34}t^{-1}y^{\pm};q)_{\infty}}
{(q^{\frac14}ty^{\pm};q)_{\infty}^2}
\frac{(-q^{\frac34}t^{-1}y^{\mp};q)_{\infty}}
{(-q^{\frac14}ty^{\pm};q)_{\infty}}. 
\end{align}
One can evaluate the indices as
\begin{align}
\label{SO2++_A_2_sub}
&I^{\textrm{$SO(2)+$asym. hyper$-[2]$}}(\zeta=+;\chi=+;x,y)
=I^{\textrm{$U(1)$ ADHM$-[2]$}}(t,x,y;q), \\
\label{SO2+-_A_2}
&I^{\textrm{$SO(2)+$asym. hyper$-[2]$}}(\zeta=+;\chi=-;x,y)
\nonumber\\
&=1-(x+x^{-1})tq^{1/4}
+\Bigl((2+x^2+x^{-2}+y^2+y^{-2})t^2-t^2 \Bigr)q^{1/2}+\cdots,
\end{align}
where $I^{\textrm{$U(1)$ ADHM$-[2]$}}(t;q)$ is the index (\ref{u1_2_findex}) of the $U(1)$ ADHM with two flavors. 
When the fugacities $x$ and $y$ are turned off, the index (\ref{SO2+-_A_2}) becomes 
\begin{align}
\label{SO2+-_A_2_sub}
&I^{\textrm{$SO(2)+$asym. hyper$-[2]$}}(\zeta=+;\chi=-)
\nonumber\\
&=1-2tq^{1/4}+(6t^2-t^{-2})q^{1/2}
+(-10t^3+4t^{-1})q^{3/4}
\nonumber\\
&+(-10+19t^4+t^{-4})q
+(-28t^5+20t+4t^{-3})q^{5/4}
\nonumber\\
&+(44t^6-34t^2+8t^{-2}-t^{-6})q^{3/2}+\cdots.
\end{align}

By gauging the $\mathbb{Z}_2^{\mathcal{C}}$ symmetry we obtain the index of the $O(2)_{+}$ gauge theory 
with an antisymmetric hyper and a fundamental hyper:
\begin{align}
 \label{O2++_A_2_indexFull}
&I^{\textrm{$O(2)_{+}+$asym. hyper$-[2]$}}(\zeta=+;\chi'=+;x,y)
\nonumber\\
&=
\frac12 \Bigl[ 
I^{\textrm{$SO(2)+$asym. hyper$-[2]$}}(\zeta=+;\chi=+;x,y)+ 
I^{\textrm{$SO(2)+$asym. hyper$-[2]$}}(\zeta=+;\chi=-;x,y)
\Bigr]
\nonumber\\
&=
1+\Bigl(
(2+x^2+x^{-2}+y^2+y^{-2})t^2+t^{-2}
\Bigr)q^{1/2}
+
2(x+x^{-1})t^{-1}q^{3/4}
+\Bigl(
-4-y^2-y^{-2}
\nonumber\\
&+
(3+2x^2+2x^{-2}+2y^2+2y^{-2}+x^4+x^{-4}+x^2y^2+x^2y^{-2}+x^{-2}y^2+x^{-2}y^{-2})t^4
+3t^{-4}
\Bigr)q
\nonumber\\
&+2(x^3+x^{-3})tq^{5/4}+\cdots.
\end{align}
When $t=1$, we have 
\begin{align}
\label{O2++_A_2_index}
&I^{\textrm{$O(2)_{+}+$asym. hyper$-[2]$}}(\zeta=+;\chi'=+)
\nonumber\\
&=1+(6t^2+t^{-2})q^{1/2}+4t^{-1}q^{3/4}+(-6+19t^4+3t^{-4})q+4tq^{5/4}
\nonumber\\
&+(44t^6-30t^2+4t^{-2}+3t^{-6})q^{3/2}
+(4t^3+4t^{-5})q^{7/4}
\nonumber\\
&+(24+85t^8-70t^4-4t^{-4}+5t^{-8})q^2+\cdots.
\end{align}
As expected from (\ref{Mc_O}), 
in the Coulomb limit the index (\ref{O2++_A_2_index}) agrees with the Hilbert series (\ref{HS_C2Z4}) for $\mathbb{C}^2/\widehat{D}_1$ $\cong$ $\mathbb{C}^2/\mathbb{Z}_4$. 
On the other hand, the Higgs limit is 
\begin{align}
\label{HS_O2++_A_2}
\mathcal{I}^{\textrm{$O(2)_{+}+$asym. hyper$-[2]$}(H)}(\mathfrak{t})&=\frac{1+2\mathfrak{t}^2+\mathfrak{t}^4}{(1-\mathfrak{t}^2)^4}. 
\end{align}

\subsubsection{$O(2)$ with 1 sym. and 1 fund. ($N=1$, $\gamma=0$, $\epsilon=1$, $l=1$)}
The indices of the $SO(2)$ gauge theory with a symmetric hyper and a fundamental hyper read
\begin{align}
\label{SO2z+_S_2_index}
&I^{\textrm{$SO(2)+$sym. hyper$-[2]$}}(\zeta;\chi=+)
\nonumber\\
&=
\frac{(q^{\frac12}t^2;q)_{\infty}}
{(q^{\frac12}t^{-2};q)_{\infty}}
\frac{(q^{\frac34}t^{-1};q)_{\infty}^2}
{(q^{\frac14}t;q)_{\infty}^2}
\sum_{m\in \mathbb{Z}}\oint 
\frac{(q^{\frac34+\frac{|m|}{2}}t^{-1}s^{\mp2};q)_{\infty}^2}
{(q^{\frac14+\frac{|m|}{2}} ts^{\pm2};q)_{\infty}^2}
\frac{(q^{\frac34+\frac{|m|}{2}}t^{-1}s^{\mp};q)_{\infty}^2}
{(q^{\frac14+\frac{|m|}{2}} ts^{\pm};q)_{\infty}^2}
q^{\frac{3|m|}{2}}t^{-6|m|} \zeta^{m}, \\
\label{SO2z-_S_2_index}
&I^{\textrm{$SO(2)+$sym. hyper$-[2]$}}(\zeta;\chi=-)
\nonumber\\
&=
\frac{(-q^{\frac12}t^{2};q)_{\infty}}
{(-q^{\frac12}t^{-2};q)_{\infty}}
\frac{(-q^{\frac34}t^{-1};q)_{\infty}^2}
{(-q^{\frac14}t;q)_{\infty}^2}
\frac{(q^{\frac34}t^{-1};q)_{\infty}^4}{(q^{\frac14}t;q)^4}
\frac{(q^{\frac34}t^{-1};q)_{\infty}^2}
{(q^{\frac14}t;q)_{\infty}^2}
\frac{(-q^{\frac34}t^{-1};q)_{\infty}^2}
{(-q^{\frac14}t;q)_{\infty}^2}. 
\end{align}
For $(\zeta,\chi)=(+,+)$ and $(+,-)$ we find
\begin{align}
\label{SO2++_S_2_sub}
&I^{\textrm{$SO(2)+$sym. hyper$-[2]$}}(\zeta=+;\chi=+)
\nonumber\\
&=1+2tq^{1/4}+(10t^2+t^{-2})q^{1/2}+30t^3q^{3/4}+(-10+76t^4+t^{-4})q
\nonumber\\
&+(178t^5-48t)q^{5/4}+(380t^6-165t^2+3t^{-6})q^{3/2}
+\cdots,\\
\label{SO2+-_S_2_sub}
&I^{\textrm{$SO(2)+$sym. hyper$-[2]$}}(\zeta=+;\chi=-)
\nonumber\\
&=1+2tq^{1/4}+(8t^2-t^{-2})q^{1/2}+(14t^3-4t^{-1})q^{3/4}+(-12+34t^4+t^{-4})q
\nonumber\\
&+(54t^5-28t+4t^{-3})q^{5/4}
+(104t^6-57t^2+8t^{-2}-t^{-6})q^{3/2}+\cdots.
\end{align}

After gauging the $\mathbb{Z}_2^{\mathcal{C}}$, 
we find the flavored index of the $O(2)_{+}$ gauge theory with a symmetric hyper and a fundamental hyper 
\begin{align}
\label{O2++_S_2_findex}
&I^{\textrm{$O(2)+$sym. hyper$-[2]$}}(\zeta=+;\chi'=+;x,y)
\nonumber\\
&=
\frac12 \Bigl[ 
I^{\textrm{$SO(2)+$sym. hyper$-[2]$}}(\zeta=+;\chi=+;x,y)+ 
I^{\textrm{$SO(2)+$sym. hyper$-[2]$}}(\zeta=+;\chi=-;x,y)
\Bigr]
\nonumber\\
&=1+(x+x^{-1})tq^{1/4}
+\Bigl(
3+2x^2+2x^{-2}+y^2+y^{-2}
\Bigr)t^2q^{1/2}
\nonumber\\
&+
\Bigl(
(2x^3+2x^{-3}+5x+5x^{-1}+2xy^2+2x^{-1}y^{-2}+2xy^{-2}+2x^{-1}y^2)t^3
\nonumber\\
&-(x+x^{-1})t^{-1}
\Bigr)q^{3/4}+\cdots. 
\end{align}
Turning off $x$ and $y$, we get
\begin{align}
\label{O2++_S_2_sub}
&I^{\textrm{$O(2)+$sym. hyper$-[2]$}}(\zeta=+;\chi'=+)
\nonumber\\
&=1+2tq^{1/4}+9t^2q^{1/2}+(22t^3-2t^{-1})q^{3/4}+(-11+55t^4+t^{-4})q
\nonumber\\
&+(116t^5-38t+2t^{-3})q^{5/4}+(242t^6-111t^2+4t^{-2}+t^{-6})q^{3/2}+\cdots.
\end{align}
The Coulomb limit of the index (\ref{O2++_S_2_sub}) is
\begin{align}
\label{HS_C2D3}
\mathcal{I}^{\textrm{$O(2)+$sym. hyper$-[2]$}(C)}(\mathfrak{t})&=\frac{1+\mathfrak{t}^8}{1-\mathfrak{t}^4-\mathfrak{t}^6+\mathfrak{t}^{10}}
=\frac{1+\mathfrak{t}^8}{(1+\mathfrak{t}^2) (1+\mathfrak{t}^2+\mathfrak{t}^4) (1-\mathfrak{t}^2)^2}.
\end{align}
This is the Hilbert series of $\mathbb{C}^2/\widehat{D}_{3}$. 
The Higgs limit of the index (\ref{O2++_S_2_sub}) coincides with (\ref{HS_2_su2inst}). 
It is consistent with the fact that the Higgs branch of the theory is the two $USp(2)$ instanton moduli space. 

\subsubsection{$O(2)$ with 1 sym. and 2 fund. ($N=1$, $\gamma=0$, $\epsilon=1$, $l=2$)}
In this case the indices take the similar form as (\ref{SO2z+_S_2_index}), (\ref{SO2z-_S_2_index}) and (\ref{O2++_S_2_sub}). 
The flavored index is evaluated as
\begin{align}
\label{O2++_S_4_findex}
&I^{\textrm{$O(2)+$sym. hyper$-[4]$}}(\zeta=+;\chi'=+;x,y_{\alpha})
\nonumber\\
&=1+(x+x^{-1})tq^{1/4}
+\Bigl(
4+2x^2+2x^{-2}+\sum_{\alpha=1}^2(y_{\alpha}^2+y_{\alpha}^{-2})
+\sum_{\alpha<\beta}(y_{\alpha}^{\pm}y_{\beta}^{\pm}
+y_{\alpha}^{\pm}y_{\beta}^{\mp})
\Bigr)t^2q^{1/2}
\nonumber\\
&+\Bigl(
(2x^3+2x^{-3}+7x+7x^{-1}+2(x+x^{-1})
\sum_{\alpha=1}^2 (y_{\alpha}^2+y_{\alpha}^{-2})
+2(x+x^{-1})\sum_{\alpha<\beta}(y_{\alpha}^{\pm}y_{\beta}^{\pm}+y_{\alpha}^{\pm}y_{\beta}^{\mp})
)t^3
\nonumber\\
&-(x+x^{-1})t^{-1}
\Bigr)q^{3/4}+\cdots. 
\end{align}
For $x=y_{\alpha}=1$, it reduces to 
\begin{align}
\label{O2++_S_4_sub}
&I^{\textrm{$O(2)+$sym. hyper$-[4]$}}(\zeta=+;\chi'=+)
\nonumber\\
&=1+2tq^{1/4}+16t^2q^{1/2}+(50t^3-2t^{-1})q^{3/4}+(-18+174t^4+t^{-4})q
\nonumber\\
&+(498t^5-90t+2t^{-3})q^{5/4}
+(1359t^6-399t^2+4t^{-2})q^{3/2}+\cdots.
\end{align}
In the Coulomb limit the index (\ref{O2++_S_4_sub}) becomes
\begin{align}
\label{HS_C2D4}
\mathcal{I}^{\textrm{$O(2)+$sym. hyper$-[4]$}(C)}(\mathfrak{t})&=
\frac{1-\mathfrak{t}^{20}}{(1-\mathfrak{t}^4) (1-\mathfrak{t}^8) (1-\mathfrak{t}^{10})}
=\frac{1+\mathfrak{t}^{10}}{(1+\mathfrak{t}^2) (1+\mathfrak{t}^2+\mathfrak{t}^4+\mathfrak{t}^6) (1-\mathfrak{t}^2)^2}.
\end{align}
This is the Hilbert series for $\mathbb{C}^2/\widehat{D}_{4}$. 
The Higgs branch limit of the index (\ref{O2++_S_4_sub}) is
\begin{align}
\label{HS_2_sp4inst}
\mathcal{I}^{\textrm{$O(2)+$sym. hyper$-[4]$}(H)}(\mathfrak{t})&=
\frac{1}{(1+\mathfrak{t})^6(1+\mathfrak{t}+\mathfrak{t}^2)^5(1-\mathfrak{t})^{12}}
\nonumber\\
&\times 
(1+\mathfrak{t}+8\mathfrak{t}^2+23\mathfrak{t}^3+50\mathfrak{t}^4+95\mathfrak{t}^5+177\mathfrak{t}^6
+222\mathfrak{t}^7+236\mathfrak{t}^8
\nonumber\\
&+222\mathfrak{t}^{9}+177\mathfrak{t}^{10}+95\mathfrak{t}^{11}
+50\mathfrak{t}^{12}+23\mathfrak{t}^{13}+8\mathfrak{t}^{14}+\mathfrak{t}^{15}+\mathfrak{t}^{16}
). 
\end{align}
This is the Hilbert series of the moduli space of two $USp(4)$ instantons \cite{Hanany:2012dm}. 

\subsubsection{$O(3)$ with 1 sym. and 1 fund. ($N=1$, $\gamma=1$ $\epsilon=1$, $l=1$)}
The indices of the $SO(3)$ gauge theory with a symmetric hyper one flavor is
\begin{align}
\label{SO3++_A_2_index}
&I^{\textrm{$SO(3)+$sym. hyper$-[2]$}}(\zeta=+;\chi=+)
\nonumber\\
&
=\frac12 \frac{(q^{\frac12}t^2;q)_{\infty}}{(q^{\frac12}t^{-2};q)_{\infty}} 
\sum_{m\in \mathbb{Z}}\oint \frac{ds}{2\pi is}(1-q^{\frac{|m|}{2}} s^{\pm})
\frac{(q^{\frac12}t^2s^{\mp};q)_{\infty}}{(q^{\frac12} t^{-2}s^{\pm};q)_{\infty}}
\nonumber\\
&\times 
\frac{(q^{\frac34}t^{-1};q)_{\infty}^4}{(q^{\frac14}t;q)_{\infty}^4}
\frac{(q^{\frac34+|m|}t^{-1}s^{\mp2};q)_{\infty}^2}{(q^{\frac14+|m|}ts^{\pm 2};q)_{\infty}^2}
\frac{(q^{\frac34+\frac{|m|}{2}}t^{-1}s^{\mp};q)_{\infty}^2}{(q^{\frac14+\frac{|m|}{2}}ts^{\pm};q)_{\infty}^2}
\nonumber\\
&\times 
\frac{(q^{\frac34+\frac{|m|}{2}}t^{-1}s^{\mp};q)_{\infty}^2}{(q^{\frac14+\frac{|m|}{2}}ts^{\pm};q)_{\infty}^2}
\frac{(q^{\frac34}t^{-1};q)_{\infty}^2}{(q^{\frac14}t;q)_{\infty}^2}
q^{\frac32 |m|} t^{-6|m|}, 
\\
\label{SO3+-_A_2_index}
&I^{\textrm{$SO(3)+$sym. hyper$-[2]$}}(\zeta=+;\chi=-)
\nonumber\\
&=\frac12 \frac{(q^{\frac12}t^2;q)_{\infty}}{(q^{\frac12}t^{-2};q)_{\infty}} 
\sum_{m\in \mathbb{Z}}\oint \frac{ds}{2\pi is}(1+q^{\frac{|m|}{2}} s^{\pm})
\frac{(-q^{\frac12}t^2s^{\mp};q)_{\infty}}{(-q^{\frac12} t^{-2}s^{\pm};q)_{\infty}}
\nonumber\\
&\times 
\frac{(q^{\frac34}t^{-1};q)_{\infty}^4}{(q^{\frac14}t;q)_{\infty}^4}
\frac{(q^{\frac34+|m|}t^{-1}s^{\mp2};q)_{\infty}^2}{(q^{\frac14+|m|}ts^{\pm 2};q)_{\infty}^2}
\frac{(-q^{\frac34+\frac{|m|}{2}}t^{-1}s^{\mp};q)_{\infty}^2}{(-q^{\frac14+\frac{|m|}{2}}ts^{\pm};q)_{\infty}^2}
\nonumber\\
&\times 
\frac{(q^{\frac34+\frac{|m|}{2}}t^{-1}s^{\mp};q)_{\infty}^2}{(q^{\frac14+\frac{|m|}{2}}ts^{\pm};q)_{\infty}^2}
\frac{(-q^{\frac34}t^{-1};q)_{\infty}^2}{(-q^{\frac14}t;q)_{\infty}^2}
q^{\frac32 |m|} t^{-6|m|}. 
\end{align}
We find the flavored index for the $O(3)_{+}$ gauge theory by gauging the $\mathbb{Z}_2^{\mathcal{C}}$:
\begin{align}
\label{O3++_A_2_findex}
&I^{\textrm{$O(3)+$sym. hyper$-[2]$}}(\zeta=+;\chi'=+;x,y)
\nonumber\\
&=1+(x+x^{-1})tq^{1/4}
+\Bigl(
3+2x^2+2x^{-2}+y^2+y^{-2}
\Bigr)t^2q^{1/2}
\nonumber\\
&+\Bigl(
(3x^3+3x^{-3}+6x+6x^{-1}+2(xy^2+x^{-1}y^{-2}+xy^{-2}+x^{-1}y^2))t^3
\nonumber\\
&-(x+x^{-1})t^{-1}
\Bigr)q^{3/4}+\cdots. 
\end{align}
By setting $x=y=1$, one finds
\begin{align}
\label{O3++_A_2_index}
&I^{\textrm{$O(3)+$sym. hyper$-[2]$}}(\zeta=+;\chi'=+)
\nonumber\\
&=1+2tq^{1/4}+9t^2q^{1/2}+(26t^3-2t^{-1})q^{3/4}+(-11+73t^4+t^{-4})q
\nonumber\\
&+(178t^5-42t+4t^{-3})q^{5/4}+(430t^6-140t^2+14t^{-2}+t^{-6})q^{3/2}+\cdots.
\end{align}
The Coulomb limit of the index (\ref{O3++_A_2_index}) agrees with the Hilbert series (\ref{HS_C2D3}) for the $\mathbb{C}^2/\widehat{D}_3$. 
We have checked that in the Higgs limit the index (\ref{O3++_A_2_index}) coincides with the Higgs branch Hilbert series of the $U(3)$ ADHM theory with two flavors. 
This is consistent with the fact that both theories have the same Higgs branch which is identical to the moduli space of three $USp(2)$ instantons. 

\subsubsection{$O(3)$ with 1 sym. and 2 fund. ($N=1$, $\gamma=1$ $\epsilon=1$, $l=2$)}
The flavored index of the $O(3)$ gauge theory with a symmetric hyper and two flavors is evaluated as
\begin{align}
\label{O3++_A_4_findex}
&I^{\textrm{$O(3)+$sym. hyper$-[4]$}}(\zeta=+;\chi=+;x,y)
\nonumber\\
&=1+(x+x^{-1})tq^{1/4}
+\Bigl(
4+2x^2+2x^{-2}+\sum_{\alpha}(y_{\alpha}^2+y_{\alpha}^{-2})
+\sum_{\alpha<\beta}(y_{\alpha}^{\pm}y_{\beta}^{\pm}+ y_{\alpha}^{\pm} y_{\beta}^{\mp})
\Bigr)t^2 q^{1/2}
\nonumber\\
&+\Bigl(
(3x^3+3x^{-3}+8x+8x^{-1}+2(x+x^{-1})\sum_{\alpha}(y_{\alpha}^2+y_{\alpha}^{-2})+2(x+x^{-1})\sum_{\alpha<\beta}(y_{\alpha}^{\pm}y_{\beta}^{\pm}+y_{\alpha}^{\pm} y_{\beta}^{\mp})
)t^3
\nonumber\\
&-(x+x^{-1})t^{-1}
\Bigr)q^{3/4}+\cdots. 
\end{align}
For $x=y_{\alpha}=1$, it becomes 
\begin{align}
\label{O3++_A_4_index}
&I^{\textrm{$O(3)+$sym. hyper$-[4]$}}(\zeta=+;\chi=+)
\nonumber\\
&=1+2tq^{1/4}+16t^2q^{1/2}+(54t^3-2t^{-1})q^{3/4}+(-18+213t^4+t^{-4})q
\nonumber\\
&+(618t^5-84t+4t^{-3})q^{5/4}+(2193t^6-414t^2+21t^{-2})q^{3/2}+\cdots.
\end{align}
The Coulomb limit of the index (\ref{O3++_A_4_index}) is equal to the Hilbert series (\ref{HS_C2D4}) for the $\mathbb{C}^2/\widehat{D}_4$. 
The Higgs limit of the index (\ref{O3++_A_4_index}) is 
\begin{align}
\label{HS_3_sp4inst}
&\mathcal{I}^{\textrm{$O(3)+$sym. hyper$-[4]$}(H)}(\mathfrak{t})=
\frac{1}{(1+\mathfrak{t})^{10} (1+\mathfrak{t}^2)^5 (1+\mathfrak{t}+\mathfrak{t}^2)^6 (1-\mathfrak{t})^{18}}
\nonumber\\
&\times 
(1+8\mathfrak{t}^2+18\mathfrak{t}^3+61\mathfrak{t}^4+142\mathfrak{t}^5+388\mathfrak{t}^6+792\mathfrak{t}^7+1691\mathfrak{t}^8+2996\mathfrak{t}^9+5255\mathfrak{t}^{10}
\nonumber\\
&+7994\mathfrak{t}^{11}
+11713\mathfrak{t}^{12}+15134\mathfrak{t}^{13}+18773\mathfrak{t}^{14}
+20796\mathfrak{t}^{15}+21980\mathfrak{t}^{16}+\textrm{palindrome}+\mathfrak{t}^{32}
).
\end{align}
This describes the Hilbert series for the moduli space of three $USp(4)$ instantons \cite{Cremonesi:2014xha}. 
\footnote{Note that we have the additional factor $(1-\mathfrak{t})^{-2}$ in the denominator compared to that in \cite{Cremonesi:2014xha}. }

\subsubsection{$O(4)$ with 1 antisym. and 1 fund. ($N=2$, $\gamma=0$ $\epsilon=0$, $l=1$)}
Let us study higher rank orthogonal gauge theories. 
In order to see the duality \eqref{OCSdual} between 
the $O(4)$ gauge theory with an adjoint hyper and one flavor and 
the quiver Chern-Simons theory, 
we compute the relevant full indices. 
The indices of the $SO(4)$ gauge theory with an antisymmetric hyper and one flavor can be obtained from the integrals
\begin{align}
    \label{SO4++_A_2_index}
&I^{\textrm{$SO(4)+$asym. hyper$-[2]$}}(\zeta=+;\chi=+;x,y)
\nonumber\\
&=\frac14 \frac{(q^{\frac12}t^2;q)_{\infty}^2}{(q^{\frac12}t^{-2};q)_{\infty}^2} \sum_{m_1, m_2\in \mathbb{Z}}
\oint \prod_{i=1}^2 \frac{ds_i}{2\pi is_i} 
(1-q^{\frac{|m_1-m_2|}{2}}s_1^{\pm} s_2^{\mp})
(1-q^{\frac{|m_1+m_2|}{2}}s_1^{\pm} s_2^{\pm})
\nonumber\\
&\times 
\frac{(q^{\frac12+\frac{|m_1-m_2|}{2}} t^2 s_1^{\mp}s_2^{\pm};q)_{\infty} }
{(q^{\frac12+\frac{|m_1-m_2|}{2}} t^{-2} s_1^{\pm}s_2^{\mp};q)_{\infty} }
\frac{(q^{\frac12+\frac{|m_1+m_2|}{2}} t^2 s_1^{\mp}s_2^{\mp};q)_{\infty} }
{(q^{\frac12+\frac{|m_1+m_2|}{2}} t^{-2} s_1^{\pm}s_2^{\pm};q)_{\infty} }
\nonumber\\
&\times 
\frac{(q^{\frac34}t^{-1}x^{\mp};q)_{\infty}^2}
{(q^{\frac14}tx^{\pm};q)_{\infty}^2}
\prod_{i=1}^2
\frac{(q^{\frac34+\frac{|m_1-m_2|}{2}} t^{-1} s_1^{\mp}s_2^{\pm}x^{\mp};q)_{\infty}}
{(q^{\frac14+\frac{|m_1-m_2|}{2}} t s_1^{\pm}s_2^{\mp}x^{\pm};q)_{\infty}}
\frac{(q^{\frac34+\frac{|m_1-m_2|}{2}} t^{-1} s_1^{\mp}s_2^{\pm}x^{\pm};q)_{\infty}}
{(q^{\frac14+\frac{|m_1-m_2|}{2}} t s_1^{\pm}s_2^{\mp}x^{\mp};q)_{\infty}}
\nonumber\\
&\times 
\frac{(q^{\frac34+\frac{|m_1+m_2|}{2}} t^{-1} s_1^{\mp}s_2^{\mp}x^{\mp};q)_{\infty}}
{(q^{\frac14+\frac{|m_1+m_2|}{2}} t s_1^{\pm}s_2^{\pm}x^{\pm};q)_{\infty}}
\frac{(q^{\frac34+\frac{|m_1+m_2|}{2}} t^{-1} s_1^{\mp}s_2^{\mp}x^{\pm};q)_{\infty}}
{(q^{\frac14+\frac{|m_1+m_2|}{2}} t s_1^{\pm}s_2^{\pm}x^{\mp};q)_{\infty}}
\nonumber\\
&\times 
\prod_{i=1}^2
\frac{(q^{\frac34+\frac{|m_i|}{2}} t^{-1} s_i^{\mp}y^{\mp};q)_{\infty}}
{(q^{\frac14+\frac{|m_i|}{2}} t s_i^{\pm}y^{\pm};q)_{\infty}}
\frac{(q^{\frac34+\frac{|m_i|}{2}} t^{-1} s_i^{\mp}y^{\pm};q)_{\infty}}
{(q^{\frac14+\frac{|m_i|}{2}} t s_i^{\pm}y^{\mp};q)_{\infty}}
q^{\frac12 \sum_{i=1}^2 |m_i|} t^{-2\sum_{i=1}^2 |m_i|}, 
\\
\label{SO4+-_A_2_index}
&I^{\textrm{$SO(4)+$asym. hyper$-[2]$}}(\zeta=+;\chi=-;x,y)
\nonumber\\
&=
\frac12 
\frac{(\pm q^{\frac12}t^2;q)_{\infty}}{(\pm q^{\frac12}t^{-2};q)_{\infty}}
\sum_{m\in \mathbb{Z}}\oint \frac{ds}{2\pi is} (1-q^{|m|}s^{\mp 2})
\frac{(q^{\frac12+\frac{|m|}{2}} t^2s^{\mp};q)_{\infty}}
{(q^{\frac12+\frac{|m|}{2}} t^{-2}s^{\pm};q)_{\infty}}
\frac{(-q^{\frac12+\frac{|m|}{2}} t^2s^{\mp};q)_{\infty}}
{(-q^{\frac12+\frac{|m|}{2}} t^{-2}s^{\pm};q)_{\infty}}
\nonumber\\
&\times 
\frac{(-q^{\frac34}t^{-1}x^{\mp};q)_{\infty}}{(-q^{\frac14}tx^{\pm};q)_{\infty}}
\frac{(-q^{\frac34}t^{-1}x^{\mp};q)_{\infty}}{(-q^{\frac14}tx^{\pm};q)_{\infty}}
\frac{(q^{\frac34+\frac{|m|}{2}} t^{-1}s^{\mp}x^{\pm};q)_{\infty}}
{(q^{\frac14+\frac{|m|}{2}} t s^{\pm}x^{\pm};q)_{\infty}}
\frac{(q^{\frac34+\frac{|m|}{2}} t^{-1}s^{\mp}x^{\mp};q)_{\infty}}
{(q^{\frac14+\frac{|m|}{2}} t s^{\pm}x^{\mp};q)_{\infty}}
\nonumber\\
&\times 
\frac{(-q^{\frac34+\frac{|m|}{2}} t^{-1}s^{\pm}x^{\mp};q)_{\infty}}
{(-q^{\frac14+\frac{|m|}{2}} t s^{\pm}x^{\pm};q)_{\infty}}
\frac{(-q^{\frac34+\frac{|m|}{2}} t^{-1}s^{\pm}x^{\pm};q)_{\infty}}
{(-q^{\frac14+\frac{|m|}{2}} t s^{\pm}x^{\mp};q)_{\infty}}
\nonumber\\
&\times 
\frac{(q^{\frac34+\frac{|m|}{2}} t^{-1}s^{\mp}y^{\mp};q)_{\infty}}
{(q^{\frac14+\frac{|m|}{2}} t s^{\pm}y^{\pm};q)_{\infty}}
\frac{(q^{\frac34+\frac{|m|}{2}} t^{-1}s^{\mp}y^{\pm};q)_{\infty}}
{(q^{\frac14+\frac{|m|}{2}} t s^{\pm}y^{\mp};q)_{\infty}}
\frac{(\pm q^{\frac34}t^{-1}y^{\mp};q)_{\infty}}{(\pm q^{\frac14}ty^{\pm};q)_{\infty}}
q^{\frac12 |m|} t^{-2|m|}. 
\end{align}
By gauging the $\mathbb{Z}_2^{\mathcal{C}}$ we find the index of the $O(4)_{+}$ gauge theory with an antisymmetric hyper and one flavor
\begin{align}
\label{O4++_A_2_index}
    &I^{\textrm{$O(4)_{+}+$asym. hyper$-[2]$}}(\zeta=+;\chi'=+,x,y)
    \nonumber\\
    &=\frac12
\left[I^{\textrm{$SO(4)+$asym. hyper$-[2]$}}(\zeta=+;\chi=+,x,y)
+I^{\textrm{$SO(4)+$asym. hyper$-[2]$}}(\zeta=+;\chi=-,x,y)
\right]
\nonumber\\
&=1+\Bigl( 
(2+x^2+x^{-2}+y^2+y^{-2})t^2+t^{-2}
\Bigr)q^{1/2}
+2(x+x^{-1})t^{-1}q^{3/4}
\nonumber\\
&+
\Bigl(
-1+x^2+x^{-2}
+(7+4x^{2}+4x^{-2}+3y^2+3y^{-2}
\nonumber\\
&+
+2x^{4}+2x^{-4}+2x^2y^2+2x^{-2}y^{-2}+2x^2y^{-2}+2x^{-2}y^2+y^4+y^{-4})t^4
+4t^{-4}
\Bigr)q+\cdots.
    \end{align}
For $x=y=1$ the index (\ref{O4++_A_2_index}) reduces to
\begin{align}
&I^{\textrm{$O(4)_{+}+$asym. hyper$-[2]$}}(\zeta=+;\chi'=+)
    \nonumber\\
    &=
    1+(6t^2+t^{-2})q^{1/2}+4t^{-1}q^{3/4}+(1+35t^4+4t^{-4})q+(28t+4t^{-3})q^{5/4}
    \nonumber\\
    &+(131t^6-22t^2+26t^{-2}+6t^{-6})q^{3/2}+\cdots.
\end{align}
The index (\ref{O4++_A_2_index}) in fact coincides with the index of the circular $U(2)_{2}\times U(2)\times U(2)_{-2}$ quiver Chern-Simons theory which will be discussed in section \ref{sec_qCS}. 
In the Coulomb limit the index (\ref{O4++_A_2_index}) becomes the Hilbert series (\ref{HS_Sym2C2Z4}) for $\mathrm{Sym}^2(\mathbb{C}^2/\widehat{D}_1)$ $\cong$ $\mathrm{Sym}^2(\mathbb{C}^2/\mathbb{Z}_4)$. 
In the Higgs limit we find the Higgs branch Hilbert series 
\begin{align}
\label{O4asymhyp1Higgslimit}
    \mathcal{I}^{\textrm{$O(4)_{+}+$asym. hyper$-[2]$}(H)}(\mathfrak{t})&=
    \frac{1 
+ 2 \mathfrak{t}^{2}
+ 13 \mathfrak{t}^{4}
+ 15 \mathfrak{t}^{6}
+ 28 \mathfrak{t}^{8}
+ 15 \mathfrak{t}^{10}
+ 13 \mathfrak{t}^{12}
+ 2 \mathfrak{t}^{14}
+ \mathfrak{t}^{16}
}{
(1 - \mathfrak{t}^2)^4(1 - \mathfrak{t}^4)^4
}. 
\end{align}

\subsubsection{$O(4)$ with 1 sym. and 1 fund. ($N=2$, $\gamma=0$ $\epsilon=1$, $l=1$)}
The indices for the $SO(4)$ gauge theory with a symmetric hyper and one flavor take the form
\begin{align}
\label{SO4++_S_2_index}
&I^{\textrm{$SO(4)+$sym. hyper$-[2]$}}(\zeta=+;\chi=+)
\nonumber\\
&
=\frac14 \frac{(q^{\frac12}t^2;q)_{\infty}^2}{(q^{\frac12}t^{-2};q)_{\infty}^2} \sum_{m_1, m_2\in \mathbb{Z}}
\oint \prod_{i=1}^2 \frac{ds_i}{2\pi is_i} 
(1-q^{\frac{|m_1-m_2|}{2}}s_1^{\pm} s_2^{\mp})
(1-q^{\frac{|m_1+m_2|}{2}}s_1^{\pm} s_2^{\pm})
\nonumber\\
&\times 
\frac{(q^{\frac12+\frac{|m_1-m_2|}{2}} t^2 s_1^{\mp}s_2^{\pm};q)_{\infty} }
{(q^{\frac12+\frac{|m_1-m_2|}{2}} t^{-2} s_1^{\pm}s_2^{\mp};q)_{\infty} }
\frac{(q^{\frac12+\frac{|m_1+m_2|}{2}} t^2 s_1^{\mp}s_2^{\mp};q)_{\infty} }
{(q^{\frac12+\frac{|m_1+m_2|}{2}} t^{-2} s_1^{\pm}s_2^{\pm};q)_{\infty} }
\nonumber\\
&\times 
\frac{(q^{\frac34}t^{-1};q)_{\infty}^4}
{(q^{\frac14}t;q)_{\infty}^4}
\prod_{i=1}^2
\frac{(q^{\frac34+|m_i|} t^{-1} s_i^{\mp 2};q)_{\infty}^2}
{(q^{\frac14+|m_i|} t s_i^{\pm 2};q)_{\infty}^2}
\frac{(q^{\frac34+\frac{|m_1-m_2|}{2}} t^{-1} s_1^{\mp}s_2^{\pm};q)_{\infty}^2}
{(q^{\frac14+\frac{|m_1-m_2|}{2}} t s_1^{\pm}s_2^{\mp};q)_{\infty}^2}
\frac{(q^{\frac34+\frac{|m_1+m_2|}{2}} t^{-1} s_1^{\mp}s_2^{\mp};q)_{\infty}^2}
{(q^{\frac14+\frac{|m_1+m_2|}{2}} t s_1^{\pm}s_2^{\pm};q)_{\infty}^2}
\nonumber\\
&\times 
\prod_{i=1}^2
\frac{(q^{\frac34+\frac{|m_i|}{2}} t^{-1} s_i^{\mp};q)_{\infty}^2}
{(q^{\frac14+\frac{|m_i|}{2}} t s_i^{\pm};q)_{\infty}^2}
q^{\frac32 \sum_{i=1}^2 |m_i|} t^{-6\sum_{i=1}^2 |m_i|}, 
\\
\label{SO4+-_S_2_index}
&I^{\textrm{$SO(4)+$sym. hyper$-[2]$}}(\zeta=+;\chi=-)
\nonumber\\
&=
\frac12 
\frac{(\pm q^{\frac12}t^2;q)_{\infty}}{(\pm q^{\frac12}t^{-2};q)_{\infty}}
\sum_{m\in \mathbb{Z}}\oint \frac{ds}{2\pi is} (1-q^{|m|}s^{\mp 2})
\frac{(q^{\frac12+\frac{|m|}{2}} t^2s^{\mp};q)_{\infty}}
{(q^{\frac12+\frac{|m|}{2}} t^{-2}s^{\pm};q)_{\infty}}
\frac{(-q^{\frac12+\frac{|m|}{2}} t^2s^{\mp};q)_{\infty}}
{(-q^{\frac12+\frac{|m|}{2}} t^{-2}s^{\pm};q)_{\infty}}
\nonumber\\
&\times 
\frac{(\pm q^{\frac34}t^{-1};q)_{\infty}^2}{(\pm q^{\frac14}t;q)_{\infty}^2}
\frac{(q^{\frac34+\frac{|m|}{2}} t^{-1}s^{\mp};q)_{\infty}^2}
{(q^{\frac14+\frac{|m|}{2}} t s^{\pm};q)_{\infty}^2}
\frac{(-q^{\frac34+\frac{|m|}{2}} t^{-1}s^{\mp};q)_{\infty}^2}
{(-q^{\frac14+\frac{|m|}{2}} t s^{\pm};q)_{\infty}^2}
\frac{(q^{\frac34+|m|}t^{-1}s^{\mp 2};q)_{\infty}^2}
{(q^{\frac14+|m|}t s^{\pm 2};q)_{\infty}^2}
\frac{(q^{\frac34}t^{-1};q)_{\infty}^4}{(q^{\frac14}t;q)_{\infty}^4}
\nonumber\\
&\times 
\frac{(q^{\frac34+\frac{|m|}{2}} t^{-1}s^{\mp};q)_{\infty}^2}
{(q^{\frac14+\frac{|m|}{2}} t s^{\pm};q)_{\infty}^2}
\frac{(\pm q^{\frac34}t^{-1};q)_{\infty}^2}{(\pm q^{\frac14}t;q)_{\infty}^2}
q^{\frac32 |m|} t^{-6|m|}.
\end{align}
The flavored index of the $O(4)_{+}$ gauge theory with a symmetric hyper and a fundamental hyper is
\begin{align}
\label{O4++_S_2_findex}
&I^{\textrm{$O(4)+$sym. hyper$-[2]$}}(\zeta=+;\chi'=+;x,y)
\nonumber\\
&=1+(x+x^{-1})tq^{1/4}
+\Bigl(
3+2x^2+2x^{-2}+y^2+y^{-2}
\Bigr)t^2q^{1/2}
\nonumber\\
&+\Bigl(
(3x^3+3x^{-3}+6x+6x^{-1}+2(xy^2+x^{-1}y^{-2}+xy^{-2}+x^{-1}y))t^3
\nonumber\\
&-(x+x^{-1})t^{-1}
\Bigr)q^{3/4}+\cdots. 
\end{align}
By setting the fugacities $x$ and $y$ to unity, we get
\begin{align}
\label{O4++_S_2_index}
&I^{\textrm{$O(4)+$sym. hyper$-[2]$}}(\zeta=+;\chi'=+)
\nonumber\\
&=1+2tq^{\frac14}+9t^2q+(26t^3-2t^{-1})q^{\frac34}+(-11+78t^4+t^{-4})q+(202t^5-42t+4t^{-3})q^{\frac54}
\nonumber\\
&+(518t^6-145t^2+17t^{-2}+t^{-6})q^{\frac32}
+(1228t^7-452t^3+64t^{-1})q^{\frac74}+\cdots.
\end{align}
In the Coulomb limit the index (\ref{O4++_S_2_index}) is equal to the Hilbert series 
\begin{align}
    \mathcal{I}^{\textrm{$O(4)+$sym. hyper$-[2]$}(C)}(\mathfrak{t})&=
    \frac{1+\mathfrak{t}^8+\mathfrak{t}^{10}+\mathfrak{t}^{12}+\mathfrak{t}^{14}+\mathfrak{t}^{16}+\mathfrak{t}^{18}+\mathfrak{t}^{26}}{(1+\mathfrak{t}^4)(1+\mathfrak{t}^6)(1-\mathfrak{t}^6)^2(1-\mathfrak{t}^4)^2},
\end{align}
which describes $\mathrm{Sym}^2 (\mathbb{C}^2/\widehat{D}_3)$. 
In the Higgs limit the index (\ref{O4++_S_2_index}) becomes
\begin{align}
\label{HS_4_sp2inst}
&\mathcal{I}^{\textrm{$O(4)+$sym. hyper$-[2]$}(H)}(\mathfrak{t})
=\frac{1}{(1+\mathfrak{t})^8 (1+\mathfrak{t}^2)^4 (1+\mathfrak{t}+\mathfrak{t}^2)^4 (1+\mathfrak{t}+\mathfrak{t}^2+\mathfrak{t}^3+\mathfrak{t}^4)^3 (1-\mathfrak{t})^{16} }
\nonumber\\
&\times 
(1+\mathfrak{t}+3\mathfrak{t}^2+9\mathfrak{t}^3+22\mathfrak{t}^4+43\mathfrak{t}^5+85\mathfrak{t}^6+153\mathfrak{t}^7+273\mathfrak{t}^8+440\mathfrak{t}^9+680\mathfrak{t}^{10}
\nonumber\\
&+982\mathfrak{t}^{11}+1364\mathfrak{t}^{12}+1778\mathfrak{t}^{13}
+2225\mathfrak{t}^{14}+2633\mathfrak{t}^{15}+2981\mathfrak{t}^{16}
\nonumber\\
&+3187\mathfrak{t}^{17}+6548\mathfrak{t}^{18}+\textrm{palindrome}+\mathfrak{t}^{36}
).
\end{align}
This reproduces the Hilbert series for four $USp(2)$ instantons \cite{Bourget:2017tmt}. 

\subsubsection{$O(6)$ with 1 antisym. and 1 fund. ($N=3$, $\gamma=0$ $\epsilon=0$, $l=1$)}
One can further test the duality \eqref{OCSdual} between the $O(2N)$ gauge theory with adjoint hyper and one flavor and the $U(N)_{2}\times U(N)\times U(N)_{-2}$ quiver Chern-Simons theory (see section \ref{sec_qCS}). 
The flavored index of the $O(6)$ gauge theory with an adjoint hyper and one flavor is evaluated as 
\begin{align}
\label{O6++_A_2_findex}
&I^{\textrm{$O(6)+$asym. hyper$-[2]$}}(\zeta=+;\chi'=+;x,y)
\nonumber\\
&=1+\Biggl(
(2+x^2+x^{-2}+y^2+y^{-2})t^2
+t^{-2}
\Biggr)q^{1/2}
+(2x+2x^{-1})t^{-1}q^{3/4}+\cdots. 
\end{align}
Turning off $x$ and $y$, this becomes 
\begin{align}
\label{O6++_A_2_index}
&I^{\textrm{$O(6)+$asym. hyper$-[2]$}}(\zeta=+;\chi'=+)
\nonumber\\
&=1+(6t^2+t^{-2})q^{1/2}+4t^{-1}q^{3/4}+(1+35t^4+4t^{-4})q
+(28t+4t^{-3})q^{5/4}
\nonumber\\
&+(162t^6-3t^2+33t^{-2}+7t^{-6})q^{3/2}
+\cdots
\end{align}
where we have evaluated the index up to $q^3$.  
In the Coulomb limit the index (\ref{O6++_A_2_index}) reduces to 
\begin{align}
    \label{HS_Sym3C2D1}
&\mathcal{I}^{\textrm{$O(6)+$asym. hyper$-[2]$}(C)}(\mathfrak{t})
=\frac{1-\mathfrak{t}^2+\mathfrak{t}^4+\mathfrak{t}^6+3\mathfrak{t}^8-\mathfrak{t}^{10}+4\mathfrak{t}^{12}+\textrm{palindrome}+\mathfrak{t}^{24}}{(1+\mathfrak{t}^2)^3(1+\mathfrak{t}^2+\mathfrak{t}^4)^2(1-\mathfrak{t}^2+\mathfrak{t}^4)(1-\mathfrak{t}^2)^6}. 
\end{align}
As expected, this agrees with the Hilbert series for $\mathrm{Sym}^3(\mathbb{C}^2/\widehat{D}_1)$ $\cong$ $\mathrm{Sym}^3(\mathbb{C}^2/\mathbb{Z}_4)$. 
In the Higgs limit we get 
\begin{align}
\label{O6asymhyp1Higgs}
\mathcal{I}^{\textrm{$O(6)+$asym. hyper$-[2]$}(H)}(\mathfrak{t})=    1+6\mathfrak{t}^2+35\mathfrak{t}^4+162\mathfrak{t}^6+636\mathfrak{t}^8+2193\mathfrak{t}^{10}+\cdots
\end{align}
This agrees with the Higgs limit (\ref{21kHiggs_N3_tonly}) of the flavored indices of the $U(3)_{2}\times U(3)\times U(3)_{-2}$ quiver Chern-Simons theory which is expected to be dual to the $O(6)$ gauge theory with an adjoint hyper and one flavor.

\section{ABJ(M) theory}
\label{sec_abjm}
In this section we consider supersymmetric indices of ABJ(M) theories. As reviewed in section \ref{sec_brane} the $U(N)_k\times U(N)_{-k}$ ABJM theory is a 3d $\mathcal{N}=6$ supersymmetric gauge theory 
consisting of the $\mathcal{N}=2$ vector multiplet of $U(N)\times U(N)$ gauge group with opposite Chern-Simons levels $k$ and $-k$ 
and a
twisted bifundamental ($(\Box,\bar{\Box})$) hypermultiplet $(T,{\widetilde T})$ and a bifundamental ($(\bar{\Box},\Box)$) hypermultiplet $(H,{\widetilde H})$ \cite{Aharony:2008ug}. 
\footnote{The 3d $\mathcal{N}=6$ Chern-Simons matter theories are classified in \cite{Schnabl:2008wj,Tachikawa:2019dvq}. }
When $k=1,2$ the supersymmetry is enhanced to $\mathcal{N}=8$. 

The ABJ theory \cite{Hosomichi:2008jb, Aharony:2008gk} is a generalization of the ABJM theory 
whose gauge group is replaced by a product $U(N)_{k}\times U(M)_{-k}$ of unitary gauge groups with $N\neq M$ 
or a product $O(2N+\gamma)_{2k}\times USp(2M)_{-k}$ of orthogonal and symplectic gauge groups. 
The $U(N)_{k}\times U(M)_{-k}$ ABJ theory can be unitary SCFT when $|M-N|\le |k|$. While the general ortho-symplectic ABJ model has $\mathcal{N}=5$ supersymmetry, the $O(2)_{2k}\times USp(2M)_{-k}$ ABJ theory has enhanced $\mathcal{N}\ge 6$ supersymmetry.

\subsection{Moduli spaces and local operators}
The moduli space of the ABJ(M) theory is parametrized by the bifundamental hyper and twisted hypers that dress the monopole operators. 
There exist two sets of topological currents and magnetic fluxes $\{m_i^{(1)}\}$, $\{m_i^{(2)}\}$ 
corresponding to the two gauge groups. 
The dimension of the monopole operator in the $U(N)_k\times U(M)_{-k}$ ABJ(M) model is 
\begin{align}
\Delta(m_i^{(1)}, m_j^{(2)})&=
-\sum_{I=1}^2\sum_{i<j} |m_{i}^{(I)}-m_{j}^{(I)}|+ \sum_{i,j} |m^{(1)}_i-m^{(2)}_j|,
\end{align}
and that in the $O(2N+\gamma)_{2k}\times USp(2M)_{-k}$ ABJ model is 
\begin{align}
\Delta(m_i^{(1)}, m_j^{(2)})&=
-\sum_{i=1}^N|m_i^{(1)}| 
-2\sum_{i=1}^M|m_i^{(2)}| 
-\sum_{I=1}^2\sum_{i<j} 
(|m_{i}^{(I)}-m_{j}^{(I)}|
+|m_{i}^{(I)}+m_{j}^{(I)}|)
\nonumber\\
&
+\sum_{i,j} 
(|m^{(1)}_i-m^{(2)}_j|
+|m^{(1)}_i+m^{(2)}_j|). 
\end{align}
Since the monopole operators carry electric charges due to the CS coupling, 
they are not gauge invariant by themselves so that the vevs do no parametrize the moduli space 
but rather fixes the action of the residual gauge group. 

The moduli space of the $U(1)_k\times U(1)_{-k}$ ABJM theory is $\mathbb{C}^4/\mathbb{Z}_k$ 
and that for the non-Abelian $U(N)_k\times U(M)_{-k}$ ABJ(M) theory is the $\mathrm{min}(N,M)$-th symmetric product \eqref{moduliABJM} of $\mathbb{C}^4/\mathbb{Z}_k$ \cite{Aharony:2008ug,Aharony:2008gk}
\begin{align}
\label{M_abjm}
\mathcal{M}_{\textrm{$U(N)_k\times U(M)_{-k}$ ABJ(M)}}&=\mathrm{Sym}^{\mathrm{min}(N,M)}(\mathbb{C}^4/\mathbb{Z}_k). 
\end{align}
For $N> M$ the effective theory on the moduli space has an extra $U(N-M)_{k}$ CS theory. 
The moduli space of the $O(2+\gamma)_{2}\times USp(2)_{-1}$  is $\mathbb{C}^4/\widehat{D}_k$ 
and that of the $O(2N+\gamma)\times USp(2M)$ ABJ theory is the $\mathrm{min}(N,M)$-th symmetric product \eqref{moduliABJOSp} of $\mathbb{C}^4/\widehat{D}_k$ \cite{Hosomichi:2008jb, Aharony:2008gk}
\begin{align}
\label{M_abj}
\mathcal{M}_{\textrm{$O(2N+\gamma)_k\times USp(2M)_{-k}$ ABJ}}&=\mathrm{Sym}^{\mathrm{min}(N,M)}(\mathbb{C}^4/\widehat{D}_k). 
\end{align}
There appears an effective CS theory on the moduli space. 
For $2N+\gamma \ge M$ it is a pure $\mathcal{N}=3$ $O(2N+\gamma-2M)_{2k}$ CS theory. 
For $2N+\gamma \le 2M$ it is a pure $\mathcal{N}=3$ $USp(2M-2N)_{-k}$ $\times$ $O(\gamma)_{2k}$ CS theory. The ABJ theory has a duality \eqref{ABJdual} \cite{Aharony:2008gk}. 

In the presence of the CS coupling,  
the monopole operators carry electric charges. 
For example, the basic monopole $v^{+,0,\cdots,0}$ in the $U(N)$ gauge theory of level $k$, it carries $k$ units of electric charges and transform as $k$-th symmetric representation of the $U(N)$ gauge group. Since the electric charge of the gauge invariant operator should vanish, the monopole operators in the ABJ(M) theory are not gauge invariant by themselves so that they need to be dressed by the bifundamental hyper and twisted hypermultiplets. 
According to the Gauss law constraint, i.e. equations of motion of gauge field, the monopole operators with $m^{(1)}_i=m^{(2)}_i$ are counted by the Hilbert series  \cite{Cremonesi:2016nbo}. When one computes the supersymmetric indices, the milder condition $\sum_i m_i^{(1)}=\sum_i m_i^{(2)}$ holds \cite{Beratto:2021xmn}. For example, 
one finds gauge invariant dressed monopole operators in the ABJ(M) theory of the following forms:
\begin{align}
    \label{abjm_mono1}
    &v^{\{m^{(1)}=m\};\{m^{(2)}=m\}} \cdot (H)^{m k}_{\mathrm{Sym}},\qquad 
    v^{\{m^{(1)}_i=-m\};\{m^{(2)}=-m\}} \cdot (\widetilde{H})^{m k}_{\mathrm{Sym}},
    \\
    \label{abjm_mono2}
    &v^{\{m^{(1)}=m\};\{m^{(2)}=m\}} \cdot (\widetilde{T})^{m k}_{\mathrm{Sym}},
    \qquad 
    v^{\{m^{(1)}=-m\};\{m^{(2)}=-m\}} \cdot (T)^{m k}_{\mathrm{Sym}}, 
\end{align}
where $m>0$ for the $m_i^{(1)}=m_i^{(2)}$ sector. In the ABJ(M) theory with unitary gauge groups each of the dressed monopole operators (\ref{abjm_mono1}) and (\ref{abjm_mono2}) parametrizes the factor $\mathbb{C}^2/\mathbb{Z}_k$ $\subset$ $\mathbb{C}^4/\mathbb{Z}_k$ probed by M2-branes. Similarly, for the ortho-symplectic ABJ theory each of them parametrizes the factor $\mathbb{C}^2/\widehat{D}_k$ $\subset$ $\mathbb{C}^4/\widehat{D}_k$. Also there are gauge invariant monopole operators dressed by both of the hyper and twisted hypermultiplets as well as their  fermionic superpartners. We will see them in the expansions in the indices in the following analysis and find the mapping of these operators under the relevant dualities. 

\subsection{Indices}
The index of the ABJ(M) theory is computed in \cite{Kim:2009wb,Bashkirov:2011pt,Gang:2011xp,Honda:2012ik,Cheon:2012be,Agmon:2017lga,Tachikawa:2019dvq,Beratto:2021xmn} from the UV gauge theory. 
The ABJM index in the large $N$ limit is shown to agree with the index of the Kaluza-Klein modes in the holographic dual $AdS_4\times S^7/\mathbb{Z}_k$ \cite{Kim:2009wb}. 
The finite $N$ corrections are proposed as contributions of the wrapped M5-branes in the gravity side \cite{Arai:2020uwd}. 

In order to investigate further dualities and geometries, 
we consider the $\mathcal{N}=4$ index (\ref{INDEX_def}) by introducing global fugacity $t$ coupled to the generators of the R-symmetry and 
additional fugacities for the flavor symmetry and the topological symmetry which allow for several limits. 

The index of the $U(N)_{k}\times U(M)_{-k}$ ABJ(M) theory is given by
\begin{align}
\label{abjm_index}
&I^{\textrm{$U(N)_{k}\times U(M)_{-k}$ABJM}}(t,x,z,y;q)
\nonumber\\
&=\frac{1}{N!M!}
\sum_{m^{(1)}_1,\cdots,m^{(1)}_N, m^{(2)}_1,\cdots,m^{(2)}_M \in \mathbb{Z}}
\oint \prod_{i=1}^N \frac{ds^{(1)}_{i}}{2\pi is^{(1)}_i}(s_i^{(1)})^{km^{(1)}_i}
\oint \prod_{i=1}^M \frac{ds^{(2)}_{i}}{2\pi is^{(2)}_i}(s_i^{(2)})^{-km^{(2)}_i}
\nonumber\\
&\times 
 \prod_{i<j}^N (1-q^{\frac{|m^{(1)}_i-m^{(1)}_j|}{2}}s^{(1)\pm}_{i}s^{(1)\mp}_{j})
 \prod_{i<j}^M (1-q^{\frac{|m^{(2)}_i-m^{(2)}_j|}{2}}s^{(2)\pm}_{i}s^{(2)\mp}_{j})
\nonumber\\
&\times 
\prod_{i,j} 
\frac{(q^{\frac34+\frac{|m_i^{(1)}-m_j^{(2)}|}{2}}ts_i^{(1)\mp} s_j^{(2)\pm} z^{\mp};q)_{\infty}}
{(q^{\frac14+\frac{|m_i^{(1)}-m_j^{(2)}|}{2}}t^{-1} s_i^{(1)\pm} s_j^{(2)\mp}z^{\pm};q)_{\infty}} 
\frac{(q^{\frac34+\frac{|m_i^{(2)} - m_j^{(1)}|}{2}}t^{-1}s_i^{(2)\mp} s_j^{(1)\pm}x^{\mp};q)_{\infty}}
{(q^{\frac14+\frac{|m_i^{(2)} - m_j^{(1)}|}{2}}t s_i^{(2)\pm} s_j^{(1)\mp}x^{\pm};q)_{\infty}}
\nonumber\\
&\times 
q^{-\frac12\sum_{I=1}^2\sum_{i<j} |m_{i}^{(I)}-m_{j}^{(I)}|+\frac12 \sum_{i,j} |m^{(1)}_i-m^{(2)}_j| }
y^{\frac12 \sum_{I=1}^2 \sum_{i} m_i^{(I)}}.
\end{align}
Note that there are redundancies\footnote{
Related to the enhanced global symmetry, there are many other ways to remove the redundancy.
For example, we can also write \eqref{abjm_index} as
\begin{align}
I^{U(N)_k\times U(M)_{-k}\text{ABJM}}(t,x,z,y;q)=I^{U(N)_k\times U(M)_{-k}\text{ABJM}}((xz)^{\frac{1}{2}},(xz^{-1})^{\frac{1}{2}},(xz^{-1})^{\frac{1}{2}},t;q).
\end{align}
} in the parameter dependence of the index \eqref{abjm_index}.
For example, by rescaling $s^{(1)}_i$ one can absorb $y$ to $x,z$, that is,
\begin{align}
I^{U(N)_k\times U(M)_{-k}\text{ABJM}}(t,x,z,y;q)=I^{U(N)_k\times U(M)_{-k}\text{ABJM}}(t,xy^{\frac{1}{k}},zy^{-\frac{1}{k}},1;q).
\label{ABJMremoveredundancy}
\end{align}
As we will see in the subsequent sections, once we fix $y$ to unity, the index of the ABJM theory with $k=1$ coincides with the index of the $U(N)$ ADHM theory with one flavor where the two fugacities $x,z$ are directly identified with the same fugacities in the ADHM theory.
However, for the purpose of reading off the operators corresponding to each term, we would like to keep $y$ in the subsequent sections.

For $\gamma=1$ or $\chi=1$ the flavored index of the $SO(2N+\gamma)_{2k}\times USp(2M)_{-k}$ ABJ model is 
\footnote{
While we get a consistent flavored index (\ref{abjso2sp2k1+_findex}) of the $SO(2)_{2}\times USp(2)_{-1}$ with $(\chi,\zeta)=(+,+)$ which include the fugacities $x$ and $z$, 
we are not sure how they can be consistently introduced in the general indices of the  $SO(2N+\gamma)_{2k}\times USp(2M)_{-k}$ ABJ model. }
\begin{align*}
&I^{\textrm{$SO(2N+\gamma)_{2k}\times USp(2M)_{-k}$ABJM}}(t;\zeta;q)
\nonumber\\
&=\frac{1}{2^{N+\gamma-1}N!2^M M!}
\sum_{m^{(1)}_1,\cdots,m^{(1)}_N, m^{(2)}_1,\cdots,m^{(2)}_M \in \mathbb{Z}}
\oint \prod_{i=1}^N \frac{ds^{(1)}_{i}}{2\pi is^{(1)}_i}(s_i^{(1)})^{2km^{(1)}_i}
\oint \prod_{i=1}^M \frac{ds^{(2)}_{i}}{2\pi is^{(2)}_i}(s_i^{(2)})^{-2km^{(2)}_i}
\nonumber\\
&\times 
\prod_{i=1}^{N}(1-\chi q^{\frac{|m_i^{(1)}|}{2}} s_i^{(1)\pm})^{\gamma}
 \prod_{i<j}^N (1-q^{\frac{|m^{(1)}_i-m^{(1)}_j|}{2}}s^{(1)\pm}_{i}s^{(1)\mp}_{j})
 \prod_{i<j}^N (1-q^{\frac{|m^{(1)}_i+m^{(1)}_j|}{2}}s^{(1)\pm}_{i}s^{(1)\pm}_{j})
\nonumber\\
&\times 
\prod_{i=1}^{M}(1-q^{|m_i^{(2)}|} s_i^{(2)\pm 2})
 \prod_{i<j}^M (1-q^{\frac{|m^{(2)}_i-m^{(2)}_j|}{2}}s^{(2)\pm}_{i}s^{(2)\mp}_{j})
 \prod_{i<j}^M (1-q^{\frac{|m^{(2)}_i+m^{(2)}_j|}{2}}s^{(2)\pm}_{i}s^{(2)\pm}_{j})
\end{align*}
\begin{align}
\label{abj_index1}
&\times 
\prod_{i,j}
\frac{(q^{\frac34+\frac{|m_i^{(1)}-m_j^{(2)}|}{2}}t^{-1}s_i^{(1)\mp} s_j^{(2)\pm};q)_{\infty}}
{(q^{\frac14+\frac{|m_i^{(1)}-m_j^{(2)}|}{2}}t s_i^{(1)\pm} s_j^{(2)\mp};q)_{\infty}} 
\frac{(q^{\frac34+\frac{|m_i^{(1)}+m_j^{(2)}|}{2}}t^{-1}s_i^{(1)\mp} s_j^{(2)\mp};q)_{\infty}}
{(q^{\frac14+\frac{|m_i^{(1)}+m_j^{(2)}|}{2}}t s_i^{(1)\pm} s_j^{(2)\pm};q)_{\infty}} 
\nonumber\\
&\times 
\frac{(q^{\frac34+\frac{|m_i^{(2)} - m_j^{(1)}|}{2}}ts_i^{(2)\mp} s_j^{(1)\pm};q)_{\infty}}
{(q^{\frac14+\frac{|m_i^{(2)} - m_j^{(1)}|}{2}}t^{-1} s_i^{(2)\pm} s_j^{(1)\mp};q)_{\infty}}
\frac{(q^{\frac34+\frac{|m_i^{(2)} + m_j^{(1)}|}{2}}ts_i^{(2)\mp} s_j^{(1)\mp};q)_{\infty}}
{(q^{\frac14+\frac{|m_i^{(2)} + m_j^{(1)}|}{2}}t^{-1} s_i^{(2)\pm} s_j^{(1)\pm};q)_{\infty}}
\nonumber\\
&\times 
\left[
\frac{(\chi q^{\frac34+\frac{|m_j^{(2)}|}{2}}t^{-1}s_j^{(2)\mp};q)_{\infty}}
{(\chi q^{\frac14+\frac{|m_j^{(2)}|}{2}}t s_j^{(2)\pm};q)_{\infty}} 
\frac{(\chi q^{\frac34+\frac{|m_j^{(2)}|}{2}}t s_j^{(2)\mp};q)_{\infty}}
{(\chi q^{\frac14+\frac{|m_j^{(2)}|}{2}}t^{-1} s_j^{(2)\pm};q)_{\infty}} 
\right]^{\gamma}
\nonumber\\
&\times 
q^{-\frac{\gamma}{2}\sum_{i=1}^N|m_i^{(1)}| -\sum_{i=1}^M|m_i^{(2)}| 
-\frac12\sum_{I=1}^2\sum_{i<j} 
(|m_{i}^{(I)}-m_{j}^{(I)}|
+|m_{i}^{(I)}+m_{j}^{(I)}|)
+\frac12 \sum_{i,j} 
(|m^{(1)}_i-m^{(2)}_j|
+|m^{(1)}_i+m^{(2)}_j|)
}
\zeta^{\sum_{i=1}^N m_i^{(1)}}.
\end{align}

For $\gamma=0$ and $\chi=-1$ we have 
\begin{align*}
&I^{\textrm{$SO(2N+\gamma)_{2k}\times USp(2M)_{-k}$ABJM}}(t;q)
\nonumber\\
&
=\frac{1}{2^{N-1}(N-1)! 2^M M!}
\sum_{m^{(1)}_1,\cdots,m^{(1)}_{N-1}, m^{(2)}_1,\cdots,m^{(2)}_M \in \mathbb{Z}}
\oint \prod_{i=1}^{N-1} \frac{ds^{(1)}_{i}}{2\pi is^{(1)}_i}(s_i^{(1)})^{2km^{(1)}_i}
\oint \prod_{i=1}^M \frac{ds^{(2)}_{i}}{2\pi is^{(2)}_i}(s_i^{(2)})^{-2km^{(2)}_i}
\nonumber\\
&\times 
\prod_{i=1}^{N-1} (1-q^{|m_i^{(1)}|} s_i^{(1)\pm 2})
 \prod_{i<j}^{N-1} (1-q^{\frac{|m^{(1)}_i-m^{(1)}_j|}{2}}s^{(1)\pm}_{i}s^{(1)\mp}_{j})
 \prod_{i<j}^{N-1} (1-q^{\frac{|m^{(1)}_i+m^{(1)}_j|}{2}}s^{(1)\pm}_{i}s^{(1)\pm}_{j})
\nonumber\\
&\times 
\prod_{i=1}^{M}(1-q^{|m_i^{(2)}|} s_i^{(2)\pm 2})
 \prod_{i<j}^M (1-q^{\frac{|m^{(2)}_i-m^{(2)}_j|}{2}}s^{(2)\pm}_{i}s^{(2)\mp}_{j})
 \prod_{i<j}^M (1-q^{\frac{|m^{(2)}_i+m^{(2)}_j|}{2}}s^{(2)\pm}_{i}s^{(2)\pm}_{j})
\end{align*}
\begin{align}
\label{abj_index2}
&\prod_{i,j}
\frac{(q^{\frac34+\frac{|m_i^{(1)}-m_j^{(2)}|}{2}}t^{-1}s_i^{(1)\mp} s_j^{(2)\pm};q)_{\infty}}
{(q^{\frac14+\frac{|m_i^{(1)}-m_j^{(2)}|}{2}}t s_i^{(1)\pm} s_j^{(2)\mp};q)_{\infty}} 
\frac{(q^{\frac34+\frac{|m_i^{(1)}+m_j^{(2)}|}{2}}t^{-1}s_i^{(1)\mp} s_j^{(2)\mp};q)_{\infty}}
{(q^{\frac14+\frac{|m_i^{(1)}+m_j^{(2)}|}{2}}t s_i^{(1)\pm} s_j^{(2)\pm};q)_{\infty}} 
\nonumber\\
&\times 
\frac{(\mp q^{\frac34+\frac{|m_j^{(2)}|}{2}}t^{-1}s_j^{(2)\pm};q)_{\infty}}
{(\pm q^{\frac14+\frac{|m_j^{(2)}|}{2}}t s_j^{(2)\mp};q)_{\infty}} 
\frac{(\pm q^{\frac34+\frac{|m_j^{(2)}|}{2}}t^{-1}s_j^{(2)\pm};q)_{\infty}}
{(\mp q^{\frac14+\frac{|m_j^{(2)}|}{2}}t s_j^{(2)\mp};q)_{\infty}} 
\nonumber\\
&\times 
\frac{(q^{\frac34+\frac{|m_i^{(2)} - m_j^{(1)}|}{2}}ts_i^{(2)\mp} s_j^{(1)\pm};q)_{\infty}}
{(q^{\frac14+\frac{|m_i^{(2)} - m_j^{(1)}|}{2}}t^{-1} s_i^{(2)\pm} s_j^{(1)\mp};q)_{\infty}}
\frac{(q^{\frac34+\frac{|m_i^{(2)} + m_j^{(1)}|}{2}}ts_i^{(2)\mp} s_j^{(1)\mp};q)_{\infty}}
{(q^{\frac14+\frac{|m_i^{(2)} + m_j^{(1)}|}{2}}t^{-1} s_i^{(2)\pm} s_j^{(1)\pm};q)_{\infty}}
\nonumber\\
&\times 
\frac{(\pm q^{\frac34+\frac{|m_j^{(2)}|}{2}}t s_j^{(2)\mp}z^{\mp};q)_{\infty}}
{(\mp q^{\frac14+\frac{|m_j^{(2)}|}{2}}t^{-1} s_j^{(2)\pm};q)_{\infty}} 
\frac{(\mp q^{\frac34+\frac{|m_j^{(2)}|}{2}}t s_j^{(2)\mp};q)_{\infty}}
{(\pm q^{\frac14+\frac{|m_j^{(2)}|}{2}}t^{-1} s_j^{(2)\pm};q)_{\infty}} 
\nonumber\\
&\times 
q^{-\sum_{i=1}^{N-1}|m_i^{(1)}| -\sum_{i=1}^M|m_i^{(2)}| 
-\frac12\sum_{I=1}^2\sum_{i<j} 
(|m_{i}^{(I)}-m_{j}^{(I)}|
+|m_{i}^{(I)}+m_{j}^{(I)}|)
+\frac12 \sum_{i,j} 
(|m^{(1)}_i-m^{(2)}_j|
+|m^{(1)}_i+m^{(2)}_j|)
}
\nonumber\\
&\times 
\zeta^{\sum_{i=1}^{N-1} m_i^{(1)}}.
\end{align}

\subsubsection{$U(1)_{1}\times U(1)_{-1}$ ABJM $(N=M=1, k=1)$}
For $N=1$ the bare monopole $v^{m;m}$ has electric charges $(m,-m)$ due to the effect of the Chern-Simons level $k=1$. 
It can form a gauge invariant operator when dressed by the chiral multiplet
$T$ (resp. ${\widetilde T}$) of electric charges $(+1,-1)$ (resp. $(-1,+1)$) 
and the chiral multiplet $H$ (resp. ${\widetilde H}$) of electric charges $(-1,+1)$ (resp. $(+1,-1)$).

The flavored index of $U(1)_1\times U(1)_{-1}$ ABJM is calculated as
\begin{align}
\label{abjmu1k1_findex}
&I^{\textrm{$U(1)_{1}\times U(1)_{-1}$ABJM}}(t,x,y,z;q)
\nonumber\\
&=1+
\Biggl[ 
(\underbrace{xy}_{v^{1;1}H}
+\underbrace{x^{-1}y^{-1}}_{v^{-1;-1}{\widetilde H}} )t 
+(\underbrace{yz^{-1}}_{v^{1;1}{\widetilde T}}
+\underbrace{y^{-1}z}_{v^{-1;-1} T}
)t^{-1}
\Biggr]q^{1/4}
+\Biggl[
\underbrace{xz}_{TH}
+\underbrace{x^{-1}z^{-1}}_{{\widetilde T}{\widetilde H}}
+\underbrace{xy^2z^{-1}}_{v^{2;2}{\widetilde T}H}
+\underbrace{x^{-1}y^{-2} z}_{v^{-2;-2}T{\widetilde H}}
\nonumber\\
&
+
(\underbrace{1}_{H{\widetilde H}}
+\underbrace{x^2y^2}_{v^{2;2}H^2}
+\underbrace{x^{-2}y^{-2}}_{v^{-2;-2}{\widetilde H}^2} )t^2
+
(\underbrace{1}_{T{\widetilde T}}
+\underbrace{y^2z^{-2}}_{v^{2;2}{\widetilde T}^2}
+\underbrace{y^{-2} z^2}_{v^{-2;-2} T^2})t^{-2}
\Biggr]q^{1/2}
+
\Biggl[
\Bigl(
\underbrace{
x^{-2} y^{-1} z^{-1}
}_{v^{-1;-1}{\widetilde T}{\widetilde H}^2}
+
\underbrace{
x^2 y z
}_{v^{1;1}T H^2}
\nonumber\\
&
+
\underbrace{
x^2y^3 z^{-1}
}_{v^{3;3}{\widetilde T}H^2}
+
\underbrace{
x^{-2}y^{-3}z
}_{v^{-3;-3}T {\widetilde H}^2}
\Bigr)t
+
\Bigl(
\underbrace{xy^{-1}z^2}_{v^{-1;-1}T^2H}
+
\underbrace{x^{-1}yz^{-2}}_{v^{1;1}{\widetilde T}^2{\widetilde H}}
+
\underbrace{xy^3 z^{-2}}_{v^{3;3} {\widetilde T}^2H}
+
\underbrace{x^{-1}y^{-3} z^2}_{v^{-3;-3}T^2 {\widetilde H}}
\Bigr)t^{-1} 
+
\Bigl(
\underbrace{xy}_{v^{1;1}H^2 {\widetilde H}}
+
\underbrace{x^{-1}y^{-1}}_{v^{-1;-1}H{\widetilde H}^2}
\nonumber\\
&
+
\underbrace{x^3y^3}_{v^{3;3}H^3}
+
\underbrace{x^{-3}y^{-3}}_{v^{-3;-3}{\widetilde H}^3}
\Bigr)t^3
+
\Bigl(
\underbrace{y^{-1}z}_{v^{-1;-1}T^2 {\widetilde T}}
+
\underbrace{yz^{-1}}_{v^{1;1}T{\widetilde T}^2}
+
\underbrace{y^{-3}z^3}_{v^{-3;-3}T^3}
+
\underbrace{y^3 z^{-3}}_{v^{3;3}{\widetilde T}^3}
\Bigr)t^{-3}
\Biggl]q^{3/4}+\cdots.
\end{align}
Note that the terms of order $y^{\pm }z^{\mp}tq^{\frac{3}{4}}$, $x^{\pm 1}y^{\pm}t^{-1}q^{\frac{3}{4}}$ are absent due to the cancellation by the fermionic modes.
For example, at the order $y^{-1}ztq^{\frac{3}{4}}$ there are a bosonic contribution $TH{\widetilde H}$ and a fermionic contribution $\psi_T$, hence the total coefficient vanishes.
See table \eqref{220427_ABJMk1N1auxdres} in appendix \ref{220427_app_CSMauxdres}.
As we will discuss the dualities between the ADHM theory and the ABJM theory in subsection \ref{sec_dual_ABJ}, 
when we set $y=1$, the index \eqref{abjmu1k1_findex} coincides with the index (\ref{u1_1_findex}) for the ADHM $U(1)$ with one flavor.
Thus the Coulomb and Higgs limits of the indices lead to the Hilbert series (\ref{HS_C2}) of the geometry $\mathbb{C}^2$ probed by a single M2-brane.

\subsubsection{$U(2)_{1}\times U(2)_{-1}$ ABJM $(N=M=2, k=1)$}
The flavored index of the $U(2)_1\times U(2)_{-1}$ ABJM theory is given by
\begin{align}
&I^{\textrm{$U(2)_{1}\times U(2)_{-1}$ABJM}}(t,x,y,z;q)
\nonumber\\
&=1+
\Biggl[ 
(\underbrace{xy}_{v^{1,0;1,0}H^{(1)}}
+\underbrace{x^{-1}y^{-1}}_{v^{-1,0;-1,0}{\widetilde H}^{(1)}} )t 
+(\underbrace{y^{-1}z}_{v^{-1,0;-1,0}T^{(1)}}
+\underbrace{yz^{-1}}_{v^{1,0;1,0}{\widetilde T}^{(1)}}
)t^{-1}
\Biggr]q^{1/4}
\nonumber\\
&+\Biggl[
\underbrace{2xz}_{\substack{\text{Tr} TH,\\ v^{1,-1;1,-1} T^{(2)} H^{(1)}}}
+\underbrace{2x^{-1}z^{-1}}_{\substack{\text{Tr} {\widetilde T}{\widetilde H},\\ v^{1,-1;1,-1} {\widetilde T}^{(1)}{\widetilde H}^{(2)}}}
+\underbrace{2xy^2z^{-1}}_{\substack{v^{2,0;2,0}{\widetilde T}^{(1)}H^{(1)},\\ v^{1,1;1,1}\text{Tr} {\widetilde T}H}}
+\underbrace{2x^{-1}y^{-2} z}_{\substack{v^{-2,0;-2,0}T^{(1)}{\widetilde H}^{(1)},\\ v^{-1,-1;-1,-1} \text{Tr}T {\widetilde H}}}
+
(
\underbrace{2}_{\substack{\text{Tr} H{\widetilde H},\\ v^{1,-1;1,-1}H^{(1)}{\widetilde H}^{(2)}}}
\nonumber\\
&
+\underbrace{2x^2y^2}_{\substack{v^{2,0;2,0}H^{(1)2},\\ v^{1,1;1,1} \text{Tr} H^2}}
+\underbrace{2x^{-2}y^{-2}}_{\substack{v^{-2,0;-2,0}{\widetilde H}^{(1)2},\\ v^{-1,-1;-1,-1} \text{Tr}{\widetilde H}^2}}
)t^2
+
(
\underbrace{2}_{\substack{\text{Tr}T{\widetilde T}, \\ v^{1,-1;1,-1}T^{(2)}{\widetilde T}^{(1)}}}
+\underbrace{2y^{-2}z^2}_{\substack{v^{-2,0;-2,0}T^{(1)2},\\ v^{-1,-1;-1,-1} \text{Tr}T^2}}
+\underbrace{2y^2 z^{-2}}_{\substack{v^{2,0;2,0}{\widetilde T}^{(1)2},\\ v^{1,1;1,1} \text{Tr}{\widetilde T}^2}}
)t^{-2}
\Biggr]q^{1/2}
\nonumber\\
&+
\Biggl[
\Bigl(
\underbrace{
3x^{-2} y^{-1}z^{-1}
}_{\substack{
v^{-1,0;-1,0} {\widetilde T}^{(1)}\tilde{H}^{(1)2},\\
v^{-1,0;-1,0} {\widetilde T}^{(1)}{\widetilde T}^{(2)}{\widetilde H}^{(2)},\\
v^{-2,1;-2,1} {\widetilde T}^{(2)}{\widetilde H}^{(1)2}
}}
+
\underbrace{
3x^2 y z
}_{\substack{
v^{1,0;1,0}T^{(1)} H^{(1)2},\\ 
v^{1,0;1,0}T^{(2)} H^{(1)} H^{(2)},\\ 
v^{2,-1;2,-1}T^{(2)} H^{(1)2}
}}
+
\underbrace{
3x^2y^3z^{-1}
}_{\substack{
v^{3,0;3,0}{\widetilde T}^{(1)} H^{(1)2},\\ 
v^{2,1;2,1}{\widetilde T}^{(1)} H^{(1)} H^{(2)},\\
v^{2,1;2,1}{\widetilde T}^{(2)} H^{(1)2}
}}
+
\underbrace{
3x^{-2}y^{-3}z
}_{\substack{
v^{-3,0;-3,0}T^{(1)} {\widetilde H}^{(1)2},\\
v^{-2,-1;-2,-1}T^{(1)} {\widetilde H}^{(1)} {\widetilde H}^{(2)},\\
v^{-2,-1;-2,-1}T^{(2)} {\widetilde H}^{(1)2}
}}\nonumber \\
&+\underbrace{
3y^{-1}z
}_{\substack{
v^{-1,0;-1,0}T^{(1)}H^{(1)}{\widetilde H}^{(1)},\\
v^{-1,0;-1,0}\psi_{T^{(1)}}, \\
v^{-1,0;-1,0}T^{(1)}H^{(2)}{\widetilde H}^{(2)},\\
v^{-1,0;-1,0}T^{(2)}H^{(2)}{\widetilde H}^{(1)},\\
v^{-2,1;-2,1}T^{(1)}H^{(2)}{\widetilde H}^{(1)}
}}
+\underbrace{
3yz^{-1}
}_{\substack{
v^{1,0;1,0}{\widetilde T}^{(1)}H^{(1)}{\widetilde H}^{(1)},\\
v^{1,0;1,0}\psi_{{\widetilde T}^{(1)}},\\
v^{1,0;1,0}{\widetilde T}^{(1)}H^{(2)}{\widetilde H}^{(2)},\\
v^{1,0;1,0}{\widetilde T}^{(2)}H^{(1)}{\widetilde H}^{(2)},\\
v^{2,-1;2,-1}{\widetilde T}^{(1)}H^{(1)}{\widetilde H}^{(2)}
}}
\Bigr)t
+
\Bigl(
\underbrace{
3xy^{-1}z^2
}_{\substack{
v^{-1,0;-1,0}T^{(1)2}H^{(1)},\\
v^{-1,0;-1,0}T^{(1)}T^{(2)}H^{(2)},\\
v^{-2,1;-2,1}T^{(1)2}H^{(2)}
}}
+
\underbrace{
3x^{-1}yz^{-2}
}_{\substack{
v^{1,0;1,0}{\widetilde T}^{(1)2}{\widetilde H}^{(1)},\\
v^{1,0;1,0}{\widetilde T}^{(1)}{\widetilde T}^{(2)}{\widetilde H}^{(2)},\\
v^{2,-1;2,-1}{\widetilde T}^{(1)2}{\widetilde H}^{(2)}
}}
\nonumber\\
&
+
\underbrace{
3xy^3z^{-2}
}_{\substack{
v^{3,0;3,0}{\widetilde T}^{(1)2}H^{(1)},\\
v^{2,1;2,1}{\widetilde T}^{(1)}{\widetilde T}^{(2)}H^{(1)},\\
v^{2,1;2,1}{\widetilde T}^{(1)2}H^{(2)}
}}+
\underbrace{
3x^{-1}y^{-3}z^{2}
}_{\substack{
v^{-3,0;-3,0}T^{(1)2}{\widetilde H}^{(1)},\\
v^{-2,-1;-2,-1}T^{(1)}T^{(2)}{\widetilde H}^{(1)},\\
v^{-2,-1;-2,-1}T^{(1)2}{\widetilde H}^{(2)}
}}
+\underbrace{
3xy
}_{\substack{
v^{1,0;1,0}T^{(1)}{\widetilde T}^{(1)}H^{(1)},\\
v^{1,0;1,0}\psi_{H^{(1)}},\\
v^{1,0;1,0}T^{(2)}{\widetilde T}^{(2)}H^{(1)},\\
v^{1,0;1,0}T^{(2)}{\widetilde T}^{(1)}H^{(2)},\\
v^{2,-1;2,-1}T^{(2)}{\widetilde T}^{(1)}H^{(1)}
}}
+\underbrace{
3x^{-1}y^{-1}
}_{\substack{
v^{-1,0;-1,0}T^{(1)}{\widetilde T}^{(1)}{\widetilde H}^{(1)},\\
v^{-1,0;-1,0}\psi_{{\widetilde H}^{(1)}},\\
v^{-1,0;-1,0}T^{(2)}{\widetilde T}^{(2)}{\widetilde H}^{(1)},\\
v^{-1,0;-1,0}T^{(1)}{\widetilde T}^{(2)}{\widetilde H}^{(2)},\\
v^{-2,1;-2,1}T^{(1)}{\widetilde T}^{(2)}{\widetilde H}^{(1)}
}}
\Biggr)t^{-1} 
\nonumber\\
&
+
\Bigl(
\underbrace{3xy}_{\substack{
v^{1,0;1,0} H^{(1)2}{\widetilde H}^{(1)},\\
v^{1,0;1,0} H^{(1)} H^{(2)}{\widetilde H}^{(2)},\\
v^{2,-1;2,-1} H^{(1)2}{\widetilde H}^{(2)}
}}
+
\underbrace{3x^{-1}y^{-1}}_{\substack{
v^{-1,0;-1,0}H^{(1)}{\widetilde H}^{(1)2},\\
v^{-1,0;-1,0}H^{(2)}{\widetilde H}^{(1)}{\widetilde H}^{(2)},\\
v^{-2,1;-2,1}H^{(2)}{\widetilde H}^{(1)2}
}}
+
\underbrace{2x^3y^3}_{\substack{
v^{3,0;3,0} H^{(1)3},\\
v^{2,1;2,1} H^{(1)2} H^{(2)}
}}
+
\underbrace{2x^{-3}y^{-3}}_{\substack{
v^{-3,0;-3,0} {\widetilde H}^{(1)3},\\
v^{-2,-1;-2,-1} {\widetilde H}^{(1)2} {\widetilde H}^{(2)}
}}
\Bigr)t^3
\nonumber\\
&
+
\Bigl(
\underbrace{3y^{-1}z}_{
\substack{
v^{-1,0;-1,0} T^{(1)2}{\widetilde T}^{(1)},\\
v^{-1,0;-1,0} T^{(1)} T^{(2)}{\widetilde T}^{(2)},\\
v^{-2,1;-2,1} T^{(1)2}{\widetilde T}^{(2)}
}}
+
\underbrace{3yz^{-1}}
_{
\substack{
v^{1,0;1,0} {\widetilde T}^{(1)2}T^{(1)},\\
v^{1,0;1,0} {\widetilde T}^{(1)} {\widetilde T}^{(2)}T^{(2)},\\
v^{2,-1;2,-1} {\widetilde T}^{(1)2}T^{(2)}
}}
+
\underbrace{2y^{-3}z^3}
_{\substack{
v^{-3,0;-3,0} T^{(1)3},\\
v^{-2,-1;-2,-1} T^{(1)2} T^{(2)}
}}
+
\underbrace{2y^3 z^{-3}}
_{\substack{
v^{3,0;3,0} {\widetilde T}^{(1)3},\\
v^{2,1;2,1} {\widetilde T}^{(1)2} {\widetilde T}^{(2)}
}}
\Bigr)t^{-3}
\Biggr]q^{3/4}
+\cdots.
\label{abjmu2k1_findex}
\end{align}
The difference from the Abelian theory starts from the terms with $q^{1/2}$. 
The terms with $q^{3/4}t$ and $q^{3/4}t^{-1}$ count the mixed branch operators which contain the cancellations between 
the monopole operators dressed by the (twisted) hypers and the monopole operators dressed by the (twisted) hyperrinos (also see appendix \ref{u2k1ABJM_auxiliary}):
\begin{align}
v^{-1,0;-1,0}  \widetilde{T}^{(1)}H^{(1)}\widetilde{H}^{(1)}
    &\leftrightarrow v^{-1,0;-1,0}\psi_{T^{(1)}},\nonumber\\
v^{1,0;1,0}  \widetilde{T}^{(1)}H^{(1)}\widetilde{H}^{(1)}
    &\leftrightarrow v^{1,0;1,0}\psi_{\widetilde{T}^{(1)}},\nonumber\\
v^{1,0;1,0}  T^{(1)}\widetilde{T}^{(1)}H^{(1)}
    &\leftrightarrow v^{1,0;1,0}\psi_{H^{(1)}},\nonumber\\
v^{-1,0;-1,0}  T^{(1)}\widetilde{T}^{(1)}\widetilde{H}^{(1)}
    &\leftrightarrow v^{-1,0;-1,0}\psi_{\widetilde{H}^{(1)}}.
\end{align}
These cancellations also occur in the $U(1)_1\times U(1)_{-1}$ ABJM theory so that they do not show up in the expansion (\ref{abjmu1k1_findex}). However, in the $U(2)_1\times U(2)_{-1}$ ABJM theory additional three operators with the same charges appear so that we get the non-trivial terms $3y^{\pm}z^{\mp}q^{3/4}t$ and $3x^{\pm}y^{\pm}q^{3/4}t^{-1}$. 

Again there exist redundancies of the fugacities and they can be fixed by setting $y=1$ \eqref{ABJMremoveredundancy}. Then
the index (\ref{abjmu2k1_findex}) is equal to the index (\ref{u2_1_findex}) 
for the $U(2)$ ADHM theory with one flavor.
This is the simplest non-Abelian duality between the ADHM theory and ABJM theory. 
In the Coulomb and Higgs limit we find the Hilbert series (\ref{HS_Sym2C2}) for $\mathrm{Sym}^2 (\mathbb{C}^2)$.

\subsubsection{$U(3)_{1}\times U(3)_{-1}$ ABJM $(N=M=3, k=1)$}
Similarly, one can check that the flavored index of the $U(3)_1\times U(3)_{-1}$ ABJM theory 
coincides with the index (\ref{u3_1_findex}) of the $U(3)$ ADHM theory with a single flavor. 
We have evaluated the index up to $q^2$. 
The Coulomb and Higgs limits of the index again give rise to the Hilbert series (\ref{HS_Sym3C2}) of $\mathrm{Sym}^{3}(\mathbb{C}^2)$. 

\subsubsection{$U(1)_{k}\times U(1)_{-k}$ ABJM $(N=M=1, k\ge 2)$}
The flavored index of the $U(1)_2\times U(1)_{-2}$ is given by 
\begin{align}
\label{abjmu1k2_findex}
&I^{U(1)_{2}\times U(1)_{-2}\text{ABJM}}(t,x,y,z;q)\nonumber \\
&=1+ \Bigl[
x^{-1} z^{-1} + x yz^{-1} + x z + x^{-1}y^{-1}z
+ (1 + x^{-2} y^{-1} + x^2 y)t^2
+ (1 + yz^{-2}\nonumber \\
&\quad + y^{-1}z^2)t^{-2}
\Bigr]q^{\frac{1}{2}}
+ \Bigl[
-3  + x^{-2} z^{-2} + x^2 y^2z^{-2} + x^2 z^2 + x^{-2}y^{-2}z^2
+ (1 + x^{-4} y^{-2}\nonumber \\
&\quad + x^{-2} y^{-1} + x^2 y + x^4 y^2)t^4
+ (x^{-3}y^{-1}z^{-1} + x^3 y^2z^{-1} + x^{-3}y^{-2}z + x^3 y z)t^2\nonumber \\
&\quad + ( x^{-1}yz^{-3} + x y^2z^{-3} + x^{-1}y^{-2}z^3 + x y^{-1}z^3)t^{-2}
+ ( 1 + y^2z^{-4} + yz^{-2} + y^{-1}z^2\nonumber \\
&\quad + y^{-2}z^4)t^{-4}
\Bigr]q
+\cdots.
\end{align}
For $x=y=z=1$ we have
\begin{align}
\label{abjmu1k2_index}
&I^{\textrm{$U(1)_{2}\times U(1)_{-2}$ABJM}}(t,x=1,y=1,z=1;q)
\nonumber\\
&=1+(4+3t^2+3t^{-2})q^{1/2}
\nonumber\\
&+(1+5t^4+4t^2+4t^{-2}+5t^{-4})q
+(4+7t^6+4t^4+4t^{-4}+7t^{-6})q^{3/2}
\nonumber\\
&+(7+9t^8+4t^6+8t^2+8t^{-2}+4t^{-6}+9t^{-8})q^2+\cdots.
\end{align}
In the Coulomb and Higgs limits the index (\ref{abjmu1k2_index}) gives rise to the Hilbert series (\ref{HS_C2Z2}) 
for the singularity $\mathbb{C}^2/\mathbb{Z}_2$ probed by the M2-brane. 
The variant $(U(1)_{2}\times U(1)_{-2})/\mathbb{Z}_2$ is dual to the $U(1)_1\times U(1)_{-1}$ ABJM theory as their indices are the same. 

For the $U(1)_k\times U(1)_{-k}$ ABJM theory with $k=3,4$ we have the following flavored indices
\begin{align}
\label{abjmu1k3_findex}
&I^{U(1)_{3}\times U(1)_{-3}\text{ABJM}}(t,x,y,z;q)\nonumber \\
&=1
+ (x^{-1} z^{-1} + x z
 + t^{-2} + t^2)q^{\frac{1}{2}} 
+ \Bigl[(x^{-3} y^{-1} + x^3 y)t^3  + (x^2 yz^{-1} + x^{-2}y^{-1}z)t\nonumber \\
&\quad + (x yz^{-2} + x^{-1}y^{-1}z^2)t^{-1} + (yz^{-3} + y^{-1}z^3)t^{-3}\Bigr]q^{\frac{3}{4}}
+ (-3 + x^{-2} z^{-2}
+ x^2 z^2+ t^4\nonumber \\
&\quad + t^{-4} )q +\cdots,
\end{align}
and
\begin{align}
\label{abjmu1k4_findex}
&I^{U(1)_{4}\times U(1)_{-4}\text{ABJM}}(t,x,y,z;q)\nonumber \\
&=
1
+ (x^{-1} z^{-1} + x z + t^2 + t^{-2})q^{\frac{1}{2}} 
+ [-3  + x^{-2} z^{-2} + x^2 yz^{-2}
+ x^2 z^2 + x^{-2}y^{-1}z^2\nonumber \\
&\quad + (1 + x^{-4} y^{-1} + x^4 y)t^4 + (x^3 yz^{-1} + x^{-3}y^{-1}z)t^2  + (x yz^{-3} + x^{-1}y^{-1}z^3)t^{-2}\nonumber \\
&\quad + ( 1 + yz^{-4}
+ y^{-1}z^4)t^{-4}]q 
+\cdots.
\end{align}
For $x=y=z=1$ we have
\begin{align}
\label{abjmu1k3_index}
&I^{\textrm{$U(1)_{3}\times U(1)_{-3}$ABJM}}(t,x=1,y=1,z=1;q)
\nonumber\\
&=1+(2+t^2+t^{-2})q^{1/2}+(2t^3+2t+2t^{-1}+2t^{-3})q^{3/4}
\nonumber\\
&+(-1+t^4+t^{-4})q+(2t^5+2t^3+2t^{-3}+2t^{-5})q^{5/4}+\cdots, 
\end{align}
and
\begin{align}
\label{abjmu1k4_index}
&I^{\textrm{$U(1)_{4}\times U(1)_{-4}$ABJM}}(t,x=1,y=1,z=1;q)
\nonumber\\
&=1+(2+t^2+t^{-2})q^{1/2}+(1+3t^4+2t^2+2t^{-2}+3t^{-4})q
\nonumber\\
&+(2+3t^6+2t^4+2t^{-4}+3t^{-6})q^{3/2}+\cdots.
\end{align}
As expected, we find the Hilbert series (\ref{HS_C2Z3}) and (\ref{HS_C2Z4}) in the Coulomb and Higgs limits. 
For general $k$ we get the Hilbert series (\ref{HS_C2Zl}) for $\mathbb{C}^2/\mathbb{Z}_k$. 
We will see that these indices have closed expressions in section \ref{sec_Zk}. 

\subsubsection{$U(N)_{k}\times U(N)_{-k}$ ABJM $(N=M\ge2, k\ge2)$}
For more general the $U(N)_k\times U(N)_{-k}$ ABJM theory the indices reduce the $N$-th symmetric product 
$\mathrm{Sym}^N (\mathbb{C}^2/\mathbb{Z}_k)$ in the Coulomb and Higgs limits. 
For example, the flavored indices for $N=2$ and $k=2,4$ are given by
\begin{align}
\label{abjmu2k2_findex}
&I^{\textrm{$U(2)_{2}\times U(2)_{-2}$ABJM}}(t,x,y,z;q)
\nonumber\\
&=1+\Bigl(
xz+x^{-1}z^{-1}+xyz^{-1}+x^{-1}y^{-1}z
+(1+x^2y+x^{-2}y^{-1})t^2
\nonumber\\
&+(1+yz^{-2}+y^{-1}z^2)t^{-2}
\Bigr)q^{1/2}
+\Bigl(
1+2x^{-2}y^{-1}+2x^2y+3x^2z^2+3x^{-2}z^{-2}
\nonumber\\
&
+3x^2y^2z^{-2}+3x^{-2}y^{-2}z^2
+2yz^{-2}+2y^{-1}z^2
+(3+2x^4y^2+2x^{-4}y^{-2}+2x^2y+2x^{-2}y^{-1})t^4
\nonumber\\
&+(2xz+2x^{-1}z^{-1}+2x^3y^2z^{-1}+2x^{-3}y^{-2}z+2xyz^{-1}+2x^{-1}y^{-1}z
\nonumber\\
&+2x^3yz+2x^{-3}y^{-1}z^{-1}
)t^2+(2xz+2x^{-1}z^{-1}+2x^{-1}y^{-2}z^{3}+2xy^2z^{-3}
\nonumber\\
&+2x^{-1}y^{-1}z+2xyz^{-1}+2xy^{-1}z^{3}+2x^{-1}yz^{-3})t^{-2}\nonumber \\
&+(3+2y^2z^{-4}+2yz^{-2}+2y^{-1}z^2+2y^{-2}z^4)t^{-4}\Bigr)q
+\cdots,
\end{align}
and
\begin{align}
\label{abjmu2k4_findex}
&I^{\textrm{$U(2)_{4}\times U(2)_{-4}$ABJM}}(t,x,y,z;q)
\nonumber\\
&=1+\Bigl(
xz+x^{-1}z^{-1}+t^2+t^{-2}
\Bigr)q^{1/2}
+
\Bigl(
2x^2z^2+2x^{-2}z^{-2}+x^2yz^{-2}+x^{-2}y^{-1}z^2
\nonumber\\
&+(2+x^4y+x^{-4}y^{-1})t^4
+(xz+x^{-1}z^{-1}+x^3yz^{-1}+x^{-3}y^{-1}z)t^2
\nonumber\\
&+(2+yz^{-4}+y^{-1}z^4)t^{-4}
+(xz+x^{-1}z^{-1}+xyz^{-3}+x^{-1}y^{-1}z^{3})t^{-2}
\Bigr)q+\cdots.
\end{align}
When we set $x=z=y=1$, we have 
\begin{align}
\label{abjmu2k2_index}
&I^{\textrm{$U(2)_{2}\times U(2)_{-2}$ABJM}}(t,x=1,y=1,z=1;q)
\nonumber\\
&=1+(4+3t^2+3t^{-2})q^{1/2}
+(21+11t^4+11t^{-4}+16t^2+16t^{-2})q
\nonumber\\
&+(32+22t^6+22t^{-6}+36t^4+36t^{-4}+36t^2+36t^{-2})q^{3/2}
\nonumber\\
&+(53+45t^8+45^{-8}+64t^6+64t^{-6}+54t^4+54t^{-4}+48t^2+48t^{-2})q^2+\cdots, 
\end{align}
and
%
%
\begin{align}
\label{abjmu2k4_index}
&I^{\textrm{$U(2)_{4}\times U(2)_{-4}$ABJM}}(t,x=1,y=1,z=1;q)
\nonumber\\
&=1+(2+t^2+t^{-2})q^{1/2}
+(6+4t^4+4t^2+4t^{-2}+4t^{-4})q
\nonumber\\
&+2(1+t^2)(1+t^{-2})(3t^4-t^2+3-t^{-2}+3t^{-4})q^{3/2}
\nonumber\\
&+
(17+14t^8+16t^6+15t^4+12t^2+(t\rightarrow t^{-1}))q^2+\cdots.
\end{align}
The Coulomb and Higgs limits of (\ref{abjmu2k2_index}) and (\ref{abjmu2k4_index}) reproduce 
the Hilbert series (\ref{HS_Sym2C2Z2}) and (\ref{HS_Sym2C2Z4}). 

\subsubsection{$U(N+k)_{k}\times U(N)_{-k}$ ABJ}
Because of the duality (\ref{ABJdual}), the $U(N+k)_{k}\times U(N)_{-k}$ ABJ theory is equivalent to the $U(N)_{k}\times U(N)_{-k}$ ABJM theory. 
So one can check that the corresponding ABJ index agrees with the ABJM index. 

\subsubsection{$U(2)_{2}\times U(1)_{-2}$ ABJ $(N=2, M=1, k=2)$}
The simplest ABJ theory which is not equivalent to the ABJM is the $U(2)_{2}\times U(1)_{-2}$ ABJ model. 
The index is 
\begin{align}
\label{abju2u1_findex}
&I^{U(2)_2\times U(1)_{-2}\text{ ABJ}}\nonumber \\
&=1
+\Bigl[t^2 (1 + x^{-2}y^{-1} + x^2 y) + x^{-1}z^{-1} + x yz^{-1} + x z + x^{-1}y^{-1}z\nonumber \\
&\quad\quad + (1 + yz^{-2} + y^{-1}z^2)t^{-2}\Bigr]q^{\frac{1}{2}}\nonumber \\
&\quad + [-2 + t^4 (1 + x^{-4}y^{-2} + x^{-2}y^{-1} + x^2 y + x^4 y^2) + x^{-2}z^{-2} + x^2 y^2z^{-2} + x^2 z^2\nonumber \\
&\quad\quad + x^{-2}y^{-2}z^2 + t^2 (x^{-3}y^{-1}z^{-1} + x^3 y^2z^{-1} + x^{-3}y^{-2}z + x^3 y z)  + (x^{-1} y z^{-3} + x y^2z^{-3}\nonumber \\
&\quad\quad + x^{-1}y^{-2}z^3 + x y^{-1}z^3)t^{-2} + ( 1 + y^2z^{-4} + yz^{-2} + y^{-1}z^2 + y^{-2}z^4)t^{-4}]q+\cdots.
\end{align}
Since the bare monopole has two units of an electric charge due to the Chern-Simons coupling of $k=2$, it can form the gauge invariant operators when dressed by quadratic polynomials in the  charged matter fields. 
The terms $q^{1/2}t^2$ and $q^{1/2}t^2x^{\pm2}y^{\pm}$ are contributed from the operators $H\widetilde{H}$, $v^{1;1} H^{2}$ and $v^{-1;-1}\widetilde{H}^2$. 
The terms $q^{1/2}t^{-2}$ and $q^{1/2}t^{-2}z^{\mp2}y^{\pm}$ correspond to the operators $T\widetilde{T}$, $v^{1;1} \widetilde{T}^{2}$ and $v^{-1;-1}T^2$. 
The terms $xz^{\pm}$ and $x^{\pm}y^{\pm}z^{\mp}$ count the operators $HT$, $\widetilde{HT}$, $v^{1;1}H\widetilde{T}$ and $v^{-1;-1}\widetilde{H}T$. 

For $x=y=z=1$ we have
\begin{align}
\label{abju2u1k2_index}
&I^{\textrm{$U(2)_{2}\times U(1)_{-2}$ABJ}}(t,x=1,y=1,z=1;q)
\nonumber\\
&=1+(4+3t^{2}+3t^{-2})q^{1/2}
+(2+4t^2+5t^4+4t^{-2}+5t^{-4})q
\nonumber\\
&+(-t^2+4t^4+7t^6-t^{-2}+4t^{-4}+7t^{-6})q^{3/2}
\nonumber\\
&+(15+12t^2+4t^6+9t^8+12t^{-2}+4t^{-6}+9t^{-8})q^2
\nonumber\\
&+(-20-8t^2+8t^4+4t^8+11t^{10}-8t^{-2}+8t^{-4}+4t^{-8}+11t^{-10})q^{5/2}+\cdots.
\end{align}
Here we find that 
\begin{align}
&I^{\textrm{$U(2)_{1}\times U(2)_{-1}$ABJM}}(t,x,y,z;q)
\nonumber\\
&=I^{\textrm{$U(1)_{1}\times U(1)_{-1}$ABJM}}(t,x,y,z;q)
I^{\textrm{$U(2)_{2}\times U(1)_{-2}$ABJ}}(t,x,y^2,z;q).
\end{align}
The operators corresponding to the terms with $q^{1/4}$ in the $U(2)_{1}\times U(2)_{-1}$ ABJM theory map to those in the $U(1)_1\times U(1)_{-1}$ ABJM theory and those for the terms $q^{1/2}$ correspond to those in the $U(2)_2\times U(1)_{-2}$ ABJ and the $U(1)_{1}\times U(1)_{-1}$ ABJM theory. 
From the terms with $q^{3/4}$ we find the following operator map:
\begin{align}
\begin{array}{c|c|c|c}
\text{fugacity }(x=z=1)      &U(2)_1\times U(2)_{-1}                         &U(1)_1\times U(1)_{-1}&U(2)_2\times U(1)_{-2} \\ \hline 
y^3t^3q^{\frac{3}{4}}&v^{3,0;3,0}(H^{(1)})^3                                     &v^{3;3}H^3                &1\ \\
                     &v^{2,1;2,1}(H^{(1)})^2{\widetilde H}^{(2)}                 &v^{1;1}H                    &v^{1;1}H^2\\ \hline
                     yt^3q^{\frac{3}{4}}  &v^{1,0;1,0}(H^{(1)})^2{\widetilde H}^{(1)}                 &v^{1;1}H^2{\widetilde H}    &1\ \\
                     &v^{1,0;1,0}H^{(1)}{\widetilde H}^{(1)}H^{(2)}              &v^{1;1}H                    &H{\widetilde H}\\
                     &v^{2,-1}(H^{(1)})^2{\widetilde H}^{(2)}                &v^{-1;-1}{\widetilde H}  &v^{1;1}H^2\\ \hline
y^3tq^{\frac{3}{4}}  &v^{3,0;3,0}{\widetilde T}^{(1)}(H^{(1)})^2                 &v^{3;3}{\widetilde T}H^2  &1\\
                     &v^{2,1;2,1}{\widetilde T}^{(2)}(H^{(1)})^2                 &v^{1;1}{\widetilde T}       &v^{1;1}{\widetilde H}^2\\
                     &v^{2,1;2,1}{\widetilde T}^{(1)}H^{(1)}H^{(2)}              &v^{1;1}H                    &v^{1;1}{\widetilde T}H\\ \hline
ytq^{\frac{3}{4}}    &v^{2,-1;2,-1}T^{(2)}(H^{(1)})^2                             &v^{-1;-1}T               &v^{1;1}H^2\\
                     &v^{2,-1;2,-1}{\widetilde T}^{(1)}H^{(1)}{\widetilde H}^{(2)}&v^{-1;-1}{\widetilde H}  &v^{1;1}{\widetilde T}H\\
                     &v^{1,0;1,0}{\widetilde T}^{(1)}H^{(2)}{\widetilde H}^{(2)} &v^{1;1}{\widetilde T}       &H{\widetilde H}\\
                     &v^{1,0;1,0}{\widetilde T}^{(2)}H^{(1)}{\widetilde H}^{(2)} &v^{1;1}H                    &{\widetilde T}{\widetilde H}
\end{array}.
\end{align}
Note that the monopole operators in the $U(2)_{1}\times U(2)_{-1}$ ABJM theory can be dressed by two components of the matter fields as each of the $U(2)$ gauge group is broken to the $U(1)\times U(1)$. When they are only dressed by the first components with the superscript $ ^{(1)}$, they correspond to the dressed monopoles in the $U(1)_{1}\times U(1)_{-1}$ ABJM theory. Otherwise, they map to composite operators constructed from the operators in the $U(1)_{1}\times U(1)_{-1}$ ABJM theory and those of the $U(2)_{2}\times U(1)_{-2}$ ABJ theory. 

Hence we conjecture a duality
\begin{align}
&\textrm{$U(2)_{2}\times U(1)_{-2}$ ABJ $\otimes$ $U(1)_{1}\times U(1)_{-1}$ ABJM}
\nonumber\\
&\Leftrightarrow
\textrm{$U(2)_{1}\times U(2)_{-1}$ ABJM}.
\label{220427ABJABJMABJMconjecture}
\end{align}
In other words, the $U(2)_{1}\times U(2)_{-1}$ ABJM theory is factorized into 
a product theory of the $U(1)_{1}\times U(1)_{-1}$ ABJM theory and the $U(2)_{2}\times U(1)_{-2}$ ABJ theory. 
The $U(1)_{1}\times U(1)_{-1}$ ABJM is a free theory that describes the center of motion of a stack of M2-branes.

Note that \eqref{220427ABJABJMABJMconjecture} is also consistent with the exact values of $S^3$ partition function $Z_{S^3}^{U(1)_1\times U(1)_{-1}}=\frac{1}{4}$, $Z_{S^3}^{U(2)_2\times U(1)_{-2}}=\frac{1}{4\pi}$, $Z_{S^3}^{U(2)_1\times U(2)_{-1}}=\frac{1}{16\pi}$ computed in \cite{Hatsuda:2012hm,Honda:2014npa}. 
The $U(2)_{1}\times U(1)_{-1}$ ABJ captures an interacting sector of the two coincident M2-branes. 

In the Coulomb and Higgs limits the index (\ref{abju2u1k2_index}) becomes the Hilbert series (\ref{HS_C2Z2}) for $\mathbb{C}^2/\mathbb{Z}_2$. 

\subsubsection{$U(3)_{4}\times U(1)_{-4}$ ABJ $(N=3, M=1, k=4)$}
For the $U(3)_{4}\times U(1)_{-4}$ ABJ theory we have the flavored index 
\begin{align}
\label{abju3u1k4_findex}
&I^{\textrm{$U(3)_{4}\times U(1)_{-4}$ABJ}}(t,x,y,z;q)
\nonumber\\
&=1+(xz+x^{-1}z^{-1}+t^2+t^{-2})q^{1/2}
+\Bigl(
-2+x^2yz^{-2}+x^{-2}y^{-1}z^{2}
+x^2z^2+x^{-2}z^{-2}
\nonumber\\
&+(1+x^4y+x^{-4}y^{-1})t^4
+(x^3yz^{-1}+x^{-3}y^{-1}z)t^2
\nonumber\\
&
+(xyz^{-3}+x^{-1}y^{-1}z^3)t^{-2}
+(1+yz^{-4}+y^{-1}z^4)t^{-4}
\Bigr)q+\cdots.
\end{align}
When the fugacities $x$, $y$ and $z$ are turned off, it reduces to
\begin{align}
\label{abju3u1k4_index}
&I^{\textrm{$U(3)_{4}\times U(1)_{-4}$ABJ}}(t,x=1,y=1,z=1;q)
\nonumber\\
&=1+(2+t^2+t^{-2})q^{1/2}+(2+3t^4+2t^2+2t^{-2}+3t^{-4})q
\nonumber\\
&+(3t^6+2t^4-t^2+t^{-2}+2t^{-4}+3t^{-6})q^{3/2}+\cdots.
\end{align}
The Coulomb and Higgs limits of the index give rise to the Hilbert series (\ref{HS_C2Z4}) for 
$\mathbb{C}^2/\widehat{D}_1$. 

\subsubsection{$U(3)_{2}\times U(2)_{-2}$ ABJ $(N=3, N=2, k=2)$}
For the $U(3)_2\times U(2)_{-2}$ ABJ theory we find the flavored index 
\begin{align}
\label{abju3u2k2_findex}
&I^{\textrm{$U(3)_{2}\times U(2)_{-2}$ABJ}}(t,x,y,z;q)
\nonumber\\
&=1+\Biggl(
xz+x^{-1}z^{-1}+xyz^{-1}+x^{-1}y^{-1}z+
(1+x^2y+x^{-2}y^{-1})t^2
+(1+y^{-1}z^2+yz^{-2})t^{-2}
\Biggr)q^{1/2}
\nonumber\\
&+\Bigl(
1+3(x^2z^2+x^{-2}z^{-2}+x^2y^2z^{-2}+x^{-2}y^{-2}z^2)
+2(x^2y+x^{-2}y^{-1}+y^{-1}z^2+yz^{-2})
\nonumber\\
&+(3+2x^4y^2+2x^{-4}y^{-2}+2x^2y+2x^{-2}y^{-1})t^4
+2(x^3yz+x^{-3}y^{-1}z^{-1}+x^3y^2z^{-1}+x^{-3}y^{-2}z
\nonumber\\
&+xyz^{-1}+x^{-1}y^{-1}z+xz+x^{-1}z^{-1})t^2
+2(xy^2z^{-3}+x^{-1}y^{-2}z^3+xy^{-1}z^3+x^{-1}yz^{-3}
\nonumber\\
&+xyz^{-1}+x^{-1}y^{-1}z+xz+x^{-1}z^{-1})t^{-2}
+(3+2y^{-2}z^4+2y^2z^{-4}+2y^{-1}z^2+2yz^{-2})t^{-4}
\Bigr)q+\cdots.
\end{align}
Setting $x$, $y$ and $z$ to unity, we get
\begin{align}
\label{abju3u2k2_index}
&I^{\textrm{$U(3)_{2}\times U(2)_{-2}$ABJ}}(t,x=1,y=1,z=1;q)
\nonumber\\
&=1+(4+3t^2+3t^{-2})q^{1/2}+(21+11t^4+16t^2+16t^{-2}+11t^{-4})q
\nonumber\\
&+(36+22t^6+36t^4+39t^2+36^{-4}+22t^{-6})q^{3/2}+\cdots.
\end{align}
Both of the Coulomb and Higgs limits of the index (\ref{abju3u2k2_index}) coincide with the Hilbert series (\ref{HS_Sym2C2Z2}) 
for $\mathrm{Sym}^2(\mathbb{C}^2/\mathbb{Z}_2)$. 
As discussed in section \ref{sec:typeIIAM}, this is dual to the $(SU(2)_{4}\times SU(2)_{-4})/\mathbb{Z}_2$ BLG theory. 

\subsubsection{$O(2)_{2}\times USp(2)_{-1}$ ABJ}
The flavored index of the $SO(2)_{2}\times USp(2)_{-1}$ ABJ theory with $(\zeta,\chi)=(+,+)$ coincides with the flavored index (\ref{abjmu1k2_findex}) of the $U(1)_{2}\times U(1)_{-2}$ ABJM theory for $x=z=y=1$. 
This implies the duality  \eqref{BMS1no2} \cite{Beratto:2021xmn}.
Thus it yields the Hilbert series (\ref{HS_C2Z2}) for $\mathbb{C}^2/\mathbb{Z}_2$ in the Coulomb and Higgs limits. 

The flavored indices with $(\zeta,\chi)=(+,-)$ and $(-,+)$ are given by 
\begin{align}
\label{abjso2sp2k1_index}
&I^{\textrm{$SO(2)_{2}\times USp(2)_{-1}$ABJ}}(t,\zeta=+,\chi=-;q)
\nonumber\\
&=I^{\textrm{$SO(2)_{2}\times USp(2)_{-1}$ABJ}}(t,\zeta=-,\chi=+;q)
\nonumber\\
&=1-(t^2+t^{-2})q^{1/2}+(1+t^4+t^{-4})q
-(t^6+t^{-6})q^{3/2}+(-1+t^8+t^{-8})q^2+\cdots.
\end{align}

The flavored index of the $O(2)_{2}\times USp(2)_{-1}$ ABJ theory 
that can be obtained by gauging the $\mathbb{Z}_2$ charge conjugation symmetry is equal to the flavored index (\ref{abjmu1k4_findex}) for the $U(1)_4\times U(1)_{-4}$ ABJM model for $x=y=z=1$. 
This is a consequence of the duality \eqref{OSpABJdualkeq1no1} \cite{Aharony:2008gk}. 
The gauge group $SO(2)\times USp(2)$ admits two families of theories 
whose indices involve the sum over the magnetic fluxes which take values in integers or half-integers (also see the discussion for the BLG theory in section \ref{sec_BLG}). 
The flavored index of the $(SO(2)_{2}\times USp(2)_{-1})/\mathbb{Z}_2$ ABJ theory in which the magnetic fluxes are summed over $\mathbb{Z}/2$ (i.e.~both integers and half-integers) is equal to the index (\ref{u1_1_index}) of the $U(1)$ ADHM theory with one flavor or equivalently to the index \ref{u1_1_findex}) of the $U(1)_1\times U(1)_{-1}$ ABJM theory for $x=z=y=1$. 
This implies the duality \eqref{BMS2no2} \cite{Beratto:2021xmn}.

We note that the refinement of the index of the $O(2)_{2}\times USp(2)_{-1}$ ABJ theory with additional fugacities $x$ and $z$ of the form
\begin{align}
\label{abjso2sp2k1+_findex}
&I^{\textrm{$SO(2)_{2}\times USp(2)_{-1}$ABJ}}(t,\zeta,\chi=+,x,z;q)
\nonumber\\
&=\frac12 \sum_{m^{(1)}, m^{(2)}\in \mathbb{Z}}\oint \frac{ds^{(1)}}{2\pi is^{(1)}}
\oint \frac{ds^{(2)}}{2\pi is^{(2)}}
(1-q^{|m^{(2)}|}{s^{(2)}}^{\pm 2})
{s^{(1)}}^{2m^{(1)}}{s^{(2)}}^{-2m^{(2)}}
\nonumber\\
&\times 
\frac{
(q^{\frac34+\frac{|m^{(1)}-m^{(2)}|}{2}} t^{-1}s^{(1)\mp} s^{(2)\pm}x^{\mp};q)_{\infty}
}
{
(q^{\frac14+\frac{|m^{(1)}-m^{(2)}|}{2}} ts^{(1)\pm} s^{(2)\mp}x^{\pm};q)_{\infty}
}
\frac{
(q^{\frac34+\frac{|m^{(1)}+m^{(2)}|}{2}} t^{-1}s^{(1)\mp} s^{(2)\mp}x^{\mp};q)_{\infty}
}
{
(q^{\frac14+\frac{|m^{(1)}+m^{(2)}|}{2}} ts^{(1)\pm} s^{(2)\pm}x^{\pm};q)_{\infty}
}
\nonumber\\
&\times \frac{
(q^{\frac34+\frac{|m^{(1)}-m^{(2)}|}{2}} ts^{(1)\mp} s^{(2)\pm}z^{\pm};q)_{\infty}
}
{
(q^{\frac14+\frac{|m^{(1)}-m^{(2)}|}{2}} t^{-1}s^{(1)\pm} s^{(2)\mp}z^{\mp};q)_{\infty}
}
\frac{
(q^{\frac34+\frac{|m^{(1)}+m^{(2)}|}{2}} ts^{(1)\mp} s^{(2)\mp}z^{\pm};q)_{\infty}
}
{
(q^{\frac14+\frac{|m^{(1)}+m^{(2)}|}{2}} t^{-1}s^{(1)\pm} s^{(2)\pm}z^{\mp};q)_{\infty}
}
\nonumber\\
&\times 
q^{-|m^{(2)}|+\frac{|m^{(1)}-m^{(2)}|}{2}+\frac{|m^{(1)}+m^{(2)}|}{2}}
\zeta^{m^{(1)}}
\end{align}
matches with the flavored index (\ref{abjmu1k2_findex}) for $y=1$ when $\zeta=+$. This generalizes the identity of the indices. 
Also for $\zeta=-$ we have
\begin{align}
\label{abjso2sp2k1-+_findex}
&I^{\textrm{$SO(2)_{2}\times USp(2)_{-1}$ABJ}}(t,\zeta=-,\chi=+,x,z;q)
\nonumber\\
&=1+
\Bigl(
xz+x^{-1}z^{-1}-xz^{-1}-x^{-1}z
+(1-x^2-x^{-2})t^2
+(1-z^2-z^{-2})t^{-2}
\Bigr)q^{1/2}
\nonumber\\
&+
\Bigl(
-3+x^2z^2+x^{-2}z^{-2}+x^2z^{-2}+x^{-2}z^2
+(1+x^4+x^{-4}-x^2-x^{-2})t^4
\nonumber\\
&+(x^3z^{-1}+x^{-3}z-x^3z-x^{-3}z^{-1})t^2
+(1+z^4+z^{-4}-z^2-z^{-2})t^{-4}
\nonumber\\
&+(xz^{-3}+x^{-1}z^3-xz^3-x^{-1}z^{-3})t^{-2}
\Bigr)q+\cdots. 
\end{align}
We find that 
the average of the refinement (\ref{abjso2sp2k1-+_findex}) with $\zeta=+$ and that with $\zeta=-$ exactly coincides with the flavored index (\ref{abjmu1k4_findex}) for $y=1$. 

\subsubsection{$O(4)_{2}\times USp(2)_{-1}$ ABJ}
We can increase the rank of the orthogonal group by $1$ and consider $O(4)_{2}\times USp(2)_{-1}$ ABJ theory. This theory is dual to the $U(3)_4\times U(1)_{-4}$ ABJ theory due to the duality \eqref{OSpABJdualkeq1no2}. Indeed, the index of the $O(4)_{2}\times USp(2)_{-1}$ ABJ theory,
$\frac{1}{2}(I^{SO(4)_2\times USp(2)_{-1}\text{ABJ}}(t,\zeta=+,\chi=+;q)+I^{SO(4)_2\times USp(2)_{-1}\text{ABJ}}(t,\zeta=+,\chi=-;q))$,
agrees with the flavored index (\ref{abju3u1k4_findex}) for $x=z=y=1$. 

\subsubsection{$O(4)_{2}\times USp(4)_{-1}$ ABJ}
Lastly let us consider the $O(4)_{2}\times USp(4)_{-1}$ ABJ theory. 
For $\zeta=+$ the indices are evaluated as
\begin{align}
&
I^{\textrm{$SO(4)_{2}\times USp(4)_{-1}$ABJ}}(t,\zeta=+,\chi=+;q)
\nonumber\\
&=\frac{1}{32}\sum_{m^{(1)}_i, m^{(2)}_i\in \mathbb{Z}}
\prod_{I=1}^2 \oint \prod_{i=1}^2 \frac{ds^{(I)}_i}{2\pi i s^{(I)}_i} 
(1-q^{\frac{|m^{(1)}_1-m^{(1)}_2|}{2}} s^{(1)\pm}_1 s^{(1)\mp}_2)
(1-q^{\frac{|m^{(1)}_1+m^{(1)}_2|}{2}} s^{(1)\pm}_1 s^{(1)\pm}_2)
\nonumber\\
&\times \Bigl(\prod_i(1-q^{|m_i^{(2)}|} s^{(2)\pm 2}_i) \Bigr)
(1-q^{\frac{|m^{(2)}_1-m^{(2)}_2|}{2}} s^{(2)\pm}_1 s^{(2)\mp}_2)
(1-q^{\frac{|m^{(2)}_1+m^{(2)}_2|}{2}} s^{(2)\pm}_1 s^{(2)\pm}_2)
\Bigl(\prod_i{s^{(1)}_{i}}^{2m^{(1)}_i}
{s^{(2)}_{i}}^{-2m^{(2)}_i} \Bigr)
\nonumber\\
&\times 
\prod_{i,j} \frac{(q^{\frac34+\frac{|m^{(1)}_i-m^{(2)}_j|}{2}} t^{-1} s^{(1)\mp}_i s^{(2)\pm}_j;q)_{\infty}}
{(q^{\frac14+\frac{|m^{(1)}_i-m^{(2)}_j|}{2}} t s^{(1)\pm}_i s^{(2)\mp}_j;q)_{\infty}}
\frac{(q^{\frac34+\frac{|m^{(1)}_i+m^{(2)}_j|}{2}} t^{-1} s^{(1)\mp}_i s^{(2)\mp}_j;q)_{\infty}}
{(q^{\frac14+\frac{|m^{(1)}_i+m^{(2)}_j|}{2}} t s^{(1)\pm}_i s^{(2)\pm}_j;q)_{\infty}}
\nonumber\\
&\times 
\prod_{i,j} \frac{(q^{\frac34+\frac{|m^{(1)}_i-m^{(2)}_j|}{2}} t s^{(1)\mp}_i s^{(2)\pm}_j;q)_{\infty}}
{(q^{\frac14+\frac{|m^{(1)}_i-m^{(2)}_j|}{2}} t^{-1} s^{(1)\pm}_i s^{(2)\mp}_j;q)_{\infty}}
\frac{(q^{\frac34+\frac{|m^{(1)}_i+m^{(2)}_j|}{2}} t s^{(1)\mp}_i s^{(2)\mp}_j;q)_{\infty}}
{(q^{\frac14+\frac{|m^{(1)}_i+m^{(2)}_j|}{2}} t^{-1} s^{(1)\pm}_i s^{(2)\pm}_j;q)_{\infty}}
\nonumber\\
&\times 
q^{-\frac12 |m^{(1)}_1-m^{(1)}_2|-\frac12 |m^{(1)}_1+m^{(1)}_2|
-\sum_{i}|m^{(2)}_i|-\frac12 |m^{(2)}_1-m^{(2)}_2|-\frac12 |m^{(2)}_1+m^{(2)}_2|
+\frac12 \sum_{i,j}
\left(
|m^{(1)}_i-m^{(2)}_j|
+|m^{(1)}_i+m^{(2)}_j|
\right)
}, 
\end{align}
\begin{align}
&
I^{\textrm{$SO(4)_{2}\times USp(4)_{-1}$ABJ}}(t,\zeta=+,\chi=-;q)
\nonumber\\
&=
\frac{1}{16} \sum_{m^{(1)}, m^{(2)}_i \in \mathbb{Z}}
\oint \frac{ds^{(1)}}{2\pi is^{(1)}}
\prod_{i=1}^2 \frac{ds^{(2)}_i}{2\pi is^{(2)}_i}
(1-q^{|m^{(1)}|} s^{(1)\pm 2}) 
\nonumber\\
&\times 
\Bigl(\prod_i(1-q^{|m_i^{(2)}|} s^{(2)\pm 2}_i)
\Bigr)
(1-q^{\frac{|m^{(2)}_1-m^{(2)}_2|}{2}} s^{(2)\pm}_1 s^{(2)\mp}_2)
(1-q^{\frac{|m^{(2)}_1+m^{(2)}_2|}{2}} s^{(2)\pm}_1 s^{(2)\pm}_2)
{s^{(1)}}^{2m^{(1)}} \Bigl(\prod_i{s^{(2)}_i}^{-2m^{(2)}_i}\Bigr)
\nonumber\\
&\times 
\prod_{i=1}^2 
\frac{
(q^{\frac34+\frac{|m^{(1)}-m^{(2)}_i|}{2}} t^{-1} s^{(1)\mp} s^{(2)\pm}_i;q)_\infty
}
{(q^{\frac14+\frac{|m^{(1)}-m^{(2)}_i|}{2}} t s^{(1)\pm} s^{(2)\mp}_i;q)_\infty}
\frac{
(q^{\frac34+\frac{|m^{(1)}+m^{(2)}_i|}{2}} t^{-1} s^{(1)\mp} s^{(2)\mp}_i;q)_\infty
}
{(q^{\frac14+\frac{|m^{(1)}+m^{(2)}_i|}{2}} t s^{(1)\pm} s^{(2)\pm}_i;q)_\infty}
\nonumber\\
&\times 
\frac{
(\pm q^{\frac34+\frac{|m^{(2)}_i|}{2}} t^{-1} s^{(2)\mp}_i;q)_\infty
}
{(\mp q^{\frac14+\frac{|m^{(2)}_i|}{2}} t s^{(2)\pm}_i;q)_\infty}
\frac{
(\mp q^{\frac34+\frac{|m^{(2)}_i|}{2}} t^{-1} s^{(2)\mp}_i;q)_\infty
}
{(\pm q^{\frac14+\frac{|m^{(2)}_i|}{2}} t s^{(2)\pm}_i;q)_\infty}
\nonumber\\
&\times 
\frac{
(q^{\frac34+\frac{|m^{(1)}-m^{(2)}_i|}{2}} t s^{(1)\mp} s^{(2)\pm}_i;q)_\infty
}
{(q^{\frac14+\frac{|m^{(1)}-m^{(2)}_i|}{2}} t^{-1} s^{(1)\pm} s^{(2)\mp}_i;q)_\infty}
\frac{
(q^{\frac34+\frac{|m^{(1)}+m^{(2)}_i|}{2}} t s^{(1)\mp} s^{(2)\mp}_i;q)_\infty
}
{(q^{\frac14+\frac{|m^{(1)}+m^{(2)}_i|}{2}} t^{-1} s^{(1)\pm} s^{(2)\pm}_i;q)_\infty}
\nonumber\\
&\times 
\frac{
(\pm q^{\frac34+\frac{|m^{(2)}_i|}{2}} t s^{(2)\mp}_i;q)_\infty
}
{(\mp q^{\frac14+\frac{|m^{(2)}_i|}{2}} t^{-1} s^{(2)\pm}_i;q)_\infty}
\frac{
(\mp q^{\frac34+\frac{|m^{(2)}_i|}{2}} t s^{(2)\mp}_i;q)_\infty
}
{(\pm q^{\frac14+\frac{|m^{(2)}_i|}{2}} t^{-1} s^{(2)\pm}_i;q)_\infty}
\nonumber\\
&\times 
q^{-|m^{(1)}|-\frac12
\left(
|m^{(2)}_1 -m^{(2)}_2|
+|m^{(2)}_1 +m^{(2)}_2|
\right)
+\frac12 \sum_{i,j} 
\left(
|m^{(1)}-m^{(2)}_i|
+|m^{(1)}+m^{(2)}_i|
\right)
}. 
\end{align}
We get the indices
\begin{align}
\label{abjso4sp4k1_index1}
&I^{\textrm{$SO(4)_{2}\times USp(4)_{-1}$ABJ}}(t,\zeta=+,\chi=+;q)
\nonumber\\
&=1+(2+t^2+t^{-2})q^{1/2}+(13+7t^4+8t^2+8t^{-2}+7t^{-4})q
\nonumber\\
&-(1+t^2)(1+t^{-2})(10t^4-2t^2+11-2t^{-2}+10t^{-4})q^{3/2}+\cdots
,\\
\label{abjso4sp4k1_index2}
&I^{\textrm{$SO(4)_{2}\times USp(4)_{-1}$ABJ}}(t,\zeta=+,\chi=-;q)
\nonumber\\
&=1+(2+t^2+t^{-2})q^{1/2}
+(-1+t^4+t^{-4})q
\nonumber\\
&-(1+t^2)(1+t^{-2})(2t^4-2t^2+1-2t^{-2}+2t^{-4})q^{3/2}+\cdots.
\end{align}
The index for the $O(4)_{2}\times USp(4)_{-1}$ ABJ theory obtained 
from (\ref{abjso4sp4k1_index1}) and (\ref{abjso4sp4k1_index2}) 
by gauging the charge conjugation symmetry matches with the index (\ref{abjmu2k4_index}), which is consistent with the duality \eqref{OSpABJdualkeq1no1}. 
So both theories describe two M2-branes probing $\mathbb{C}^2/\mathbb{Z}_2$. 

\subsubsection{ADHM-ABJM dualities}
\label{sec_dual_ABJ}

As reviewed in section \ref{sec:typeIIB}, the $U(N)$ ADHM theory with one flavor is conjectured to be equivalent to the ABJM theory of CS level $k=1$ in the IR \cite{Kapustin:2010xq}. 
Comparing the flavored indices, e.g. (\ref{abjmu1k1_findex}) and (\ref{abjmu2k1_findex}) with 
(\ref{u1_1_findex}) and (\ref{u2_1_findex}), 
we see that the ADHM index (\ref{uN_l_index}) with one flavor 
and the ABJM index (\ref{abjm_index}) with $N=M$ and $k=1$ perfectly agree with each other 
by turning off the topological fugacity $y$ for the ABJM model. 
Consequently, we find the operator mapping under the duality. 
For the Abelian case it is given by
\begin{align}
\label{adhm_abjm_map}
\begin{array}{c|c}
\textrm{$U(1)$ ADHM with one flavor}&\textrm{$U(1)_{1}\times U(1)_{-1}$ABJM} \\ \hline 
X&v^{1;1}H \\
Y&v^{-1;-1}{\widetilde H} \\
XY&H{\widetilde H}\\
X^l Y^m&v^{l-m;l-m}H^l{\widetilde H}^m\\ \hline 
v^{1}&v^{-1;-1}T \\
v^{-1}&v^{1;1}{\widetilde T} \\
\varphi&T{\widetilde T}\\ \hline 
v^1X&TH\\
v^1Y&v^{-2;-2}T{\widetilde H}\\
v^{-1}X&v^{2;2}{\widetilde T}H\\
v^{-1}Y&{\widetilde T}{\widetilde H}\\
v^1\psi_{\varphi}&v^{-1;-1}\psi_T\\
v^{-1}\psi_{\varphi}&v^{1;1}\psi_{\widetilde T}\\
\psi_X&v^{1;1}\psi_H\\
\psi_Y&v^{-1;-1}\psi_{\widetilde H}
\end{array}.
\end{align}
The Higgs branch operators in the ADHM theory correspond to the monopole operators dressed by the bifundamental hypermultiplet in the ABJM theory. On the other hand, the Coulomb branch operators in the ADHM theory map to the monopole operators only dressed by the bifundamental twisted hypermultiplet in the ABJM theory. The remaining mixed branch operator in the ADHM theory are dual to the monopole operators dressed by the both hyper and twisted hypers in the ABJM model. 

In addition, the flavored indices allow us to find the mapping of the mixed branch operators which contain the fermionic operators. 
The fermion $\psi_{\varphi}$ is the superpartner of the vector multiplet scalar $\varphi$ and $\psi_{X}$ (resp. $\psi_{Y}$) is that of the adjoint hypermultiplet scalar fields $X$ (resp. $Y$) in the ADHM theory. The mapping of the fermionic operators are consistent with that of their bosonic partners. 

For $U(2)$ gauge group, by comparing \eqref{u2_1_findex} and \eqref{abjmu2k1_findex} we conjecture the following operator mapping:
\begin{align}
\begin{array}{c|c}
\textrm{$U(2)$ ADHM with one flavor}&\textrm{$U(2)_{1}\times U(2)_{-1}$ABJM}\\ \hline 
\text{Tr}X                          &v^{1,0;1,0}H^{(1)}\\
\text{Tr}Y                          &v^{-1,0;-1,0}{\widetilde H}^{(1)}\\
\text{Tr}(XY)                       &\text{Tr}(H{\widetilde H})\\
\text{Tr}X\text{Tr}Y                &v^{1,-1;1,-1}H^{(1)}{\widetilde H}^{(2)}\\
\text{Tr}X^2                        &v^{1,1;1,1}\text{Tr}(H^2)\\
(\text{Tr}X)^2                      &v^{2,0;2,0}(H^{(1)})^2\\ \hline
v^{1,0}                             &v^{-1,0;-1,0}T^{(1)}\\
v^{-1,0}                            &v^{1,0;1,0}{\widetilde T}^{(1)}\\
\text{Tr}\varphi                    &\text{Tr}(T{\widetilde T})\\ 
v^{1,-1}                            &v^{1,-1;1,-1}T^{(2)}{\widetilde T}^{(1)}\\
v^{2,0}                             &v^{-2,0;-2,0}(T^{(1)})^2\\
v^{1,1}                             &v^{-1,-1;-1,-1}\text{Tr}(T^2)\\ \hline
\varphi^{(1)}                       &T^{(1)}{\widetilde T}^{(1)}\\
X^{(2)}                             &v^{1,0;1,0}H^{(1)}\\
X^{(1)}                             &v^{0,1;0,1}H^{(2)}\\
Y^{(2)}                             &v^{-1,0;-1,0}{\widetilde H}^{(1)}\\
Y^{(1)}                             &v^{0,-1;0,-1}{\widetilde H}^{(2)}\\
v^{1,0}J^{(1)}I^{(1)}+v^{1,0}\psi_{\varphi^{(1)}}+v^{1,0}\psi_{\varphi^{(2)}}&v^{-1,0;-1,0}\psi_{T^{(1)}}\\
v^{-1,0}J^{(1)}I^{(1)}+v^{-1,0}\psi_{\varphi^{(1)}}+v^{-1,0}\psi_{\varphi^{(2)}}&v^{1,0;1,0}\psi_{{\widetilde T}^{(1)}}
\end{array}.
\end{align}
Similarly the Higgs (resp. Coulomb) branch operators in the non-Abelian $U(N)$ ADHM theory map to the monopole operators dressed by the bifundamental hypermultiplets (resp. twisted hypermultiplets) in the non-Abelian $U(N)_1\times U(N)_{-1}$ ABJM theory. Each of these local operators corresponds to the plane partition with trace $N$ or a pair of column-strict plane partitions of shape $\lambda=\{\lambda_i\}_{i=1}^N$ with $\sum_i \lambda_i=N$ \cite{Okazaki:2022sxo}. 

\subsection{Duality to discrete gauge theories 
}
\label{sec_Zk}
In this section we consider discrete gauge theories which are expected to describe an M2-brane. 
When a gauge group is discrete, a theory is the rank-zero theory so that it has no gauge fields but matter fields may carry non-trivial gauge charges.  

Consider a 3d $\mathcal{N}=4$ gauge theory of a discrete cyclic group $\mathbb{Z}_k$ with a hypermultiplet and a twisted hypermultiplet. 
Based on the argument in \cite{Kapustin:2014gua}, 
it is conjectured that we have the following duality:
\begin{align}
\label{Zk_dual}
\begin{array}{ccc}
\textrm{$\mathbb{Z}_k$ gauge theory}
&\Leftrightarrow&\textrm{$U(1)_k\times U(1)_{-k}$ ABJM theory}\\
\textrm{$+$ a hyper $(X,Y)$$+$ a twisted hyper $(T,\widetilde{T})$}&&\\
\end{array}.
\end{align}
In the following we explicitly demonstrate this by computing supersymmetric indices which precisely agree with each other. 


\subsubsection{$\mathbb{Z}_2$ $(k=2)$}
A simple example is the $\mathbb{Z}_2$ gauge theory. 
It can be viewed as a generalization of the $O(1)_{+}$ gauge theory discussed in subsection \ref{sec_o1}. 
The index takes the form
\begin{align}
\label{Z2_index}
&I^{\textrm{$\mathbb{Z}_{2}$-hyper$+$thyper}}(t;x,z;q)
\nonumber\\
&=
\frac{1}{2}
\left[
\frac{(q^{\frac34}t^{-1}x^{\mp};q)_{\infty}}
{(q^{\frac14}tx^{\pm};q)_{\infty}}
\frac{(q^{\frac34}tz^{\mp};q)_{\infty}}
{(q^{\frac14}t^{-1}z^{\pm};q)_{\infty}}
+
\frac{(-q^{\frac34}t^{-1}x^{\mp};q)_{\infty}}
{(-q^{\frac14}tx^{\pm};q)_{\infty}}
\frac{(-q^{\frac34}tz^{\mp};q)_{\infty}}
{(-q^{\frac14}t^{-1}z^{\pm};q)_{\infty}}
\right].
\end{align}
The index (\ref{Z2_index}) matches with the index (\ref{abjmu1k2_findex}) for the $U(1)_{2}\times U(1)_{-2}$ ABJM model. 

\subsubsection{$\mathbb{Z}_3$ $(k=3)$}
$\mathbb{Z}_3$ contains three one-dimensional irreps, 
which correspond to the mapping of generators to $1$, $\omega = e^{2\pi i/3}=-1/2+\sqrt{3}i/2$ and $\omega^2$. 
The index reads
\begin{align}
\label{Z3_index}
&I^{\textrm{$\mathbb{Z}_{3}$-hyper$+$thyper}}(t;x,z;q)
\nonumber\\
&=
\frac{1}{3}
\Biggl[
\frac{(q^{\frac34}t^{-1}x^{\mp};q)_{\infty}}
{(q^{\frac14}tx^{\pm};q)_{\infty}}
\frac{(q^{\frac34}tz^{\pm};q)_{\infty}}
{(q^{\frac14}t^{-1}z^{\pm};q)_{\infty}}
+
\frac{(q^{\frac34}t^{-1}\omega^{\mp}x^{\mp};q)_{\infty}}
{(q^{\frac14}t\omega^{\pm}x^{\pm};q)_{\infty}}
\frac{(q^{\frac34}t\omega^{\mp}z^{\mp};q)_{\infty}}
{(q^{\frac14}t^{-1}\omega^{\pm}z^{\pm};q)_{\infty}}
\nonumber\\
&
+
\frac{(q^{\frac34}t^{-1}\omega^{\mp 2}x^{\mp};q)_{\infty}}
{(q^{\frac14}t\omega^{\pm 2}x^{\pm};q)_{\infty}}
\frac{(q^{\frac34}t\omega^{\mp 2}z^{\mp};q)_{\infty}}
{(q^{\frac14}t^{-1}\omega^{\pm 2}z^{\pm};q)_{\infty}}
\Biggr].
\end{align}
Again the index (\ref{Z3_index}) agrees with the index (\ref{abjmu1k3_findex}) for the $U(1)_{3}\times U(1)_{-3}$ ABJM theory. 

\subsubsection{$\mathbb{Z}_k$ $(k\ge 4)$}
For $\mathbb{Z}_4\cong \widehat{D}_1$ we have generators $\pm 1$, $\pm i$. 
The index is 
\begin{align}
\label{Z4_index}
I^{\textrm{$\mathbb{Z}_{4}$-hyper$+$thyper}}(t;x,z;q)
&=
\frac{1}{4}\sum_{l=0}^{3}
\frac{(q^{\frac34}t^{-1}i^{\mp l}x^{\pm};q)_{\infty}}
{(q^{\frac14}t i^{\pm l}x^{\pm};q)_{\infty}}
\frac{(q^{\frac34}t i^{\mp l}z^{\pm};q)_{\infty}}
{(q^{\frac14}t^{-1} i^{\pm l}z^{\pm};q)_{\infty}}. 
\end{align}
This is equal to the index (\ref{abjmu1k4_findex}) for the $U(1)_{4}\times U(1)_{-4}$ ABJM theory. 
Also we find that 
\begin{align}
\label{Z4half_index}
&I^{\textrm{$SO(2)_{2}\times USp(2)_{-1}$ABJ}}(t,\zeta=+,\chi=-;q)
\nonumber\\
&=
\frac{1}{2}
\left[
\frac{(q^{\frac34}t^{-1}i^{\mp};q)_{\infty}}
{(q^{\frac14}ti^{\pm};q)_{\infty}}
\frac{(q^{\frac34}ti^{\mp};q)_{\infty}}
{(q^{\frac14}t^{-1}i^{\pm};q)_{\infty}}
+
\frac{(-q^{\frac34}t^{-1}i^{\mp};q)_{\infty}}
{(-q^{\frac14}ti^{\pm};q)_{\infty}}
\frac{(-q^{\frac34}ti^{\mp};q)_{\infty}}
{(-q^{\frac14}t^{-1}i^{\pm};q)_{\infty}}
\right].
\end{align}

For general $k$ we have the index
\begin{align}
\label{Zk_index}
I^{\textrm{$\mathbb{Z}_{k}$-hyper$+$thyper}}(t;x,z;q)
&=
\frac{1}{k}\sum_{l=0}^{k-1}
\frac{(q^{\frac34}t^{-1}\omega^{\mp l}x^{\pm};q)_{\infty}}
{(q^{\frac14}t\omega^{\pm l}x^{\pm};q)_{\infty}}
\frac{(q^{\frac34}t\omega^{\mp l}z^{\pm};q)_{\infty}}
{(q^{\frac14}t^{-1}\omega^{\pm l}z^{\pm};q)_{\infty}},
\end{align}
where $\omega=e^{2\pi i/k}$. 
We confirm that the index (\ref{Zk_index}) agrees with the index for the $U(1)_k\times U(1)_{-k}$ ABJM theory when the fugacity $y$ is turned off. 

From the matching of indices we find the map of operators under the proposed duality (\ref{Zk_dual}): 
\begin{align}
\label{Zk_abjm_map}
\begin{array}{c|c}
\textrm{$Z_k$ gauge theory}&\textrm{$U(1)_{k}\times U(1)_{-k}$ABJM} \\ \hline 
H&v^{1;1}H \\
{\widetilde H}&v^{-1;-1}{\widetilde H} \\
H{\widetilde H}&H{\widetilde H}\\
H^l {\widetilde H}^m&v^{l-m;l-m}H^l{\widetilde H}^m\\ \hline 
T&v^{-1;-1}T \\
{\widetilde T}&v^{1;1}{\widetilde T} \\
T{\widetilde T}&T{\widetilde T}\\ \hline 
HT&TH\\
{\widetilde H}T&v^{-2;-2}T{\widetilde H}\\
H{\widetilde T}&v^{2;2}{\widetilde T}H\\
{\widetilde H}{\widetilde T}&{\widetilde T}{\widetilde H}\\
\end{array}.
\end{align}
The operators in the $\mathbb{Z}_k$ gauge theory are simply obtained from those in the $U(1)_k\times U(1)_{-k}$ ABJM theory by stripping off the monopole operators. 

\section{$\mathcal{N}=4$ quiver CS theories}
\label{sec_qCS}
We investigate a class of 3d $\mathcal{N}=4$ circular quiver Chern-Simons matter theories which describe M2-branes, which were discussed in section \ref{sec_brane}. 
In the following we focus on the $\mathcal{N}=4$ quiver CS theories which are conjecturally dual to the ADHM as in Figure \ref{fig:ADHM_mirror} and their cousin theories. 

\subsection{Moduli spaces and local operators}

The dimension of a monopole operator is given by
\begin{align}
\Delta(m^{(I)}_{i})
&=
-\sum_{I=1}^{l+1}\sum_{i<j} |m^{(I)}_i - m^{(I)}_j|
+\frac12 \sum_{I=1}^{l+1}\sum_{i,j}|m^{(I)}_i - m^{(I+1)}_j|. 
\end{align}
For the $\mathcal{N}=2$ vector multiplet with non-trivial CS level, 
the monopole operators carry electric charge so that they need to be dressed by matter fields to form gauge invariant operators. 
For the $\mathcal{N}=4$ vector multiplet with vanishing CS level, 
the monopole operators can form gauge invariant operators by themselves. 

Unlike the 3d $\mathcal{N}=4$ SYM theories coupled to the matter multiplets,  
the non-renormalization argument does not work for 3d $\mathcal{N}=4$ CS matter theories \cite{Gaiotto:2007qi}. 
So there are non-trivial quantum corrections to the moduli space of vacua. 

We examine the local operators by computing the supersymmetric indices and the moduli space of vacua by analyzing their  Coulomb and Higgs limits. 

\subsection{Indices}
The supersymmetric index of the quiver Chern-Simons theory in the right column in Figure \ref{fig:ADHM_mirror} with ranks $N_1,N_2,\cdots,N_{l+1}$ reads
\begin{align}
&I^{U(N_1)_k\times U(N_I)_0^{\otimes (l-1)}\times U(N_{l+1})_{-k}\text{quiver CS}}(t,x,y_I,z_I;q)\nonumber \\
&=\frac{1}{\prod_{I=1}^{l+1}N_I!}
\sum_{m_i^{(1)},\cdots,m_i^{(l+1)}\in\mathbb{Z}}
\prod_{I=1}^{l+1}y_I^{\sum_im_i^{(I)}}
\oint\prod_I\prod_{i=1}^{N_I}\frac{ds_i^{(I)}}{2\pi is_i^{(I)}}
\prod_{i=1}^{N_1}(s_i^{(1)})^{km_i^{(1)}}
\prod_{i=1}^{N_{l+1}}(s_i^{(l+1)})^{-km_i^{(l+1)}}\nonumber \\
&\quad \times \prod_{I=1}^{l+1}\prod_{i\neq j}^{N_I}\Bigl(1-q^{\frac{|m_i^{(I)}-m_j^{(I)}|}{2}}\frac{s_i^{(I)}}{s_j^{(I)}}\Bigr)
\prod_{I=2}^l\prod_{i,j=1}^{N_I}\frac{(q^{\frac{1}{2}+\frac{|m_i^{(I)}-m_j^{(I)}|}{2}}t^{-2}\frac{s_i^{(I)}}{s_j^{(I)}};q)_\infty}{(q^{\frac{1}{2}+\frac{|m_i^{(I)}-m_j^{(I)}|}{2}}t^2\frac{s_i^{(I)}}{s_j^{(I)}};q)_\infty}\nonumber \\
&\quad \times \prod_{I=1}^l\prod_{i=1}^{N_I}\prod_{j=1}^{N_{I+1}}\prod_{\pm}
\frac{(q^{\frac{3}{4}+\frac{|m_i^{(I)}-m_j^{(I+1)}|}{2}}t\Bigl(\frac{s_i^{(I)}}{s_j^{(I+1)}}\Bigr)^\pm z_I^\pm;q)_\infty}{(q^{\frac{1}{4}+\frac{|m_i^{(I)}-m_j^{(I+1)}|}{2}}t^{-1}\Bigl(\frac{s_i^{(I)}}{s_j^{(I+1)}}\Bigr)^\pm z_I^\pm;q)_\infty}\nonumber \\
&\quad \times \prod_{i=1}^{N_{l+1}}\prod_{j=1}^{N_1}\prod_{\pm}
\frac{(q^{\frac{3}{4}+\frac{|m_i^{(l+1)}-m_j^{(1)}|}{2}}t^{-1}\Bigl(\frac{s_i^{(l+1)}}{s_j^{(1)}}\Bigr)^\pm x^\pm;q)_\infty}{(q^{\frac{1}{4}+\frac{|m_i^{(l+1)}-m_j^{(1)}|}{2}}t\Bigl(\frac{s_i^{(l+1)}}{s_j^{(1)}}\Bigr)^\pm x^\pm;q)_\infty}\nonumber \\
&\quad \times q^{-\frac{1}{2}\sum_{I=1}^{l+1}\sum_{i<j}^{N_I}|m_i^{(I)}-m_j^{(I)}|+\frac{1}{4}\sum_{I=1}^{l+1}\sum_{i=1}^{N_I}\sum_{j=1}^{N_{I+1}}|m_i^{(I)}-m_j^{(I+1)}|}\nonumber \\
&\quad \times t^{-2\sum_{I=2}^l\sum_{i<j}^{N_I}|m_i^{(I)}-m_j^{(I)}|+\sum_{I=1}^l\sum_{i=1}^{N_I}\sum_{j=1}^{N_{I+1}}|m_i^{(I)}-m_j^{(I+1)}|-\sum_{i=1}^{N_{l+1}}\sum_{j=1}^{N_1}|m_i^{(l+1)}-m_j^{(1)}|}.
\label{220413_CSmatterindexnewnotation}
\end{align}
As in the case of the ABJM theory \eqref{ABJMremoveredundancy}, this index also has redundancies in the parameter dependence, which can be seen as follows.
First, by tracking the gauge indices, one can see that only the terms where the powers of $z_I$ are the same and the monopole charges satisfy $\sum_im_i^{(1)}=\sum_im_i^{(l+1)}$ can contribute to the full index.
Hence we conclude that the index depends on $z_1,z_2,\cdots,z_l$ and $y_1,y_{l+1}$ only through $z_1z_2\cdots z_l$ and $y_1y_{l+1}$.
Also, by rescaling the integration variable $s_i^{(1)}\rightarrow cs_i^{(1)}$, we find that the index is invariant under the change of parameters $(y_1,z_1,x)\rightarrow (c^k y_1,cz_1,c^{-1}x)$.
The redundancies can be fixed, for example, by imposing $\prod_{I=1}^{l+1}y_I=1$ and $y_{l+1}=z_2=z_3=\cdots=z_l=1$.
The parameters before and after imposing these constraints are related to each other as
\begin{align}
&{\cal I}^{U(N_1)_k\times U(N_I)_0^{\otimes (l-1)}\times U(N_{l+1})_{-k}\text{quiver CS}}(t,x,y_I,z_I;q)\nonumber \\
&\quad ={\cal I}^{U(N_1)_k\times U(N_I)_0^{\otimes (l-1)}\times U(N_{l+1})_{-k}\text{quiver CS}}(t,x',y_I',z_I';q),
\label{circularquiverCSredundancy}
\end{align}
with
\begin{align}
&x'=\biggl(\prod_{I=1}^{l+1}y_I\biggr)^{\frac{1}{k}}x,\quad y_1'=\biggl(\prod_{I=2}^ly_I\biggr)^{-1},\quad y_{l+1}'=1,\quad y_I'=y_I\quad (I=2,\cdots,l),\nonumber \\
&z_1'=\biggl(\prod_{I=1}^{l+1}y_I\biggr)^{-\frac{1}{k}}\biggl(\prod_{I=1}^lz_I\biggr),\quad z_I'=1\quad (I=2,\cdots,l).
\label{circularquiverCSfixredundancy}
\end{align}
Note that under the constraint $\prod_{I=1}^{l+1}y_I=1$ we can match $x,z_I,y_I$ with the fugacities in the dual ADHM theory \eqref{uN_l_index} without mixing between the Coulomb parameters ($z$) and the Higgs parameters ($x,y$) (see table \eqref{ADHM_qCS_fug}).

In the following sections we would like to keep these redundancies unfixed for the purpose of reading off the operators corresponding to each term.


\subsubsection{$U(1)_{1}\times U(1)_{0}\times U(1)_{-1}$}
In the first three subsections we consider circular quiver CS theories in Figure \ref{fig:ADHM_mirror}. Let us consider the $U(1)_1\times U(1)_0\times U(1)_{-1}$ CS theory with two bifundamental twisted hypermultiplets $(T_{1,2},{\widetilde T}_{1,2}),(T_{2,3},{\widetilde T}_{2,3})$ and a bifundamental hypermultiplet $(H_{3,1},{\widetilde H}_{3,1})$.
The flavored index \eqref{220413_CSmatterindexnewnotation} of this model to the order $q^{\frac{3}{4}}$ is given by
\begin{align}
&I^{U(1)_1\times U(1)_0\times U(1)_{-1}}(t,x,y_I,z_I;q)\nonumber \\
&=
1
+(
\underbrace{xy_1y_2y_3}_{v^{1;1;1}H_{3,1}}
+\underbrace{x^{-1}y_1^{-1}y_2^{-1}y_3^{-1}}_{v^{-1;-1;-1}{\widetilde H}_{3,1}}
)tq^{\frac{1}{4}}
+\Biggl[
(
\underbrace{2}_{\substack{\varphi^{(2)},\\ H_{3,1}{\widetilde H}_{3,1}}}
+\underbrace{x^2y_1^2y_2^2y_3^2}_{v^{2;2;2}H_{3,1}^2}
+\underbrace{x^{-2}y_1^{-2}y_2^{-2}y_3^{-2}}_{v^{-2;-2;-2}{\widetilde H}_{3,1}^2}
+\underbrace{y_2^{-1}}_{v^{0;-1;0}}
+\underbrace{y_2}_{v^{0;1;0}}
)t^2\nonumber \\
&+(
\underbrace{1}_{\substack{
\psi_{\varphi^{(2)}},\\
T_{1,2}{\widetilde T}_{1,2},\\
T_{2,3}{\widetilde T}_{2,3}
}}
+\underbrace{y_1y_2y_3z_1^{-1}z_2^{-1}}_{v^{1;1;1}{\widetilde T}_{1,2}{\widetilde T}_{2,3}}
+\underbrace{y_1^{-1}y_2^{-1}y_3^{-1}z_1z_2}_{v^{-1;-1;-1}T_{1,2}T_{2,3}}
)t^{-2}
\Biggr]q^{\frac{1}{2}}
+\Biggl[
(
\underbrace{xy_1^2y_2^2y_3^2z_1^{-1}z_2^{-1}}_{v^{2;2;2}{\widetilde T}_{1,2}{\widetilde T}_{2,3}H_{3,1}}
+\underbrace{x^{-1}y_1^{-2}y_2^{-2}y_3^{-2}z_1z_2}_{v^{-2;-2;-2}T_{1,2}T_{2,3}{\widetilde H}_{3,1}}\nonumber \\
&+\underbrace{x^{-1}z_1^{-1}z_2^{-1}}_{{\widetilde T}_{1,2}{\widetilde T}_{2,3}{\widetilde H}_{3,1}}
+\underbrace{xz_1z_2}_{T_{1,2}T_{2,3}H_{3,1}}
)t^{-1}
+(
\underbrace{2xy_1y_2y_3}_{\substack{
v^{1;1;1}H_{3,1}^2{\widetilde H}_{3,1},\\
v^{1;1;1}\varphi^{(2)}H_{3,1}
}}
+\underbrace{2x^{-1}y_1^{-1}y_2^{-1}y_3^{-1}}_{\substack{v^{-1;-1;-1}H_{3,1}{\widetilde H}_{3,1}^2\\
v^{-1;-1;-1}\varphi^{(2)}{\widetilde H}_{3,1}
}}
+\underbrace{x^{-3}y_1^{-3}y_2^{-3}y_3^{-3}}_{v^{-3;-3;-3}{\widetilde H}_{3,1}^3}
+\underbrace{x^3y_1^3y_2^3y_3^3}_{v^{3;3;3}H_{3,1}^3}\nonumber \\
&+\underbrace{xy_1y_3}_{v^{1;0;1}H_{3,1}}
+\underbrace{x^{-1}y_1^{-1}y_3^{-1}}_{v^{-1;0;-1}{\widetilde H}_{3,1}}
+\underbrace{xy_1y_2^2y_3}_{v^{1;2;1}H_{3,1}}
+\underbrace{x^{-1}y_1^{-1}y_2^{-2}y_3^{-1}}_{v^{-1;-2;-1}{\widetilde H}_{3,1}}
)t^3
\Biggr]q^{\frac{3}{4}}+\cdots.
\label{220413u1u1u1k+-0_findex}
\end{align}
The coefficient for the term $q^{1/2}t^{-2}$ will be contributed from 
the two bosonic operators $T_{1,2}{\widetilde T}_{1,2}$ and $T_{2,3}{\widetilde T}_{2,3}$ as well as a fermionic operator $\psi_{\varphi^{(2)}}$ that 
is the superpartner of $\varphi^{(2)}$.
This indicates that the bosonic operators consisting of bifundamental twisted hypers are not independent 
due to a constraint corresponding to the fermionic operator $\psi^{(2)}$.

The index (\ref{220413u1u1u1k+-0_findex}) coincides with the index (\ref{u1_2_findex}) for the $U(1)$ ADHM theory with two flavors, which agrees with the duality between the leftmost quiver and the rightmost quiver in Figure \ref{fig:ADHM_mirror}.



\subsubsection{$U(1)_{1}\times U(1)_0^{\otimes 2}\times U(1)_{-1}$}
Similarly, we get the index for the $U(1)_{1}\times U(1)_0\times U(1)_0\times U(1)_{-1}$ quiver CS theory 
\begin{align}
\label{u1u1u1u1k+-0_findex}
&I^{U(1)_{1}\times U(1)_0^{\otimes 2}\times U(1)_{-1}} (t,x,y_{I},z_{I};q)
\nonumber\\
&=1+(\underbrace{x\prod_{I=1}^{4} y_I}_{v^{1;1;1;1}H_{4,1}}
+\underbrace{x^{-1}\prod_{I=1}^4 y_I^{-1}}_{v^{-1;-1;-1;-1} {\widetilde H_{4,1}}})tq^{1/4}
+\Biggl[
(\underbrace{3}_{\substack{\varphi^{(2)},\\ \varphi^{(3)},\\ H_{4,1}{\widetilde H}_{4,1}}}
+\underbrace{x^2\prod_{I=1}^4y_I^2}_{v^{2;2;2;2}H_{4,1}^2}
+\underbrace{x^{-2}\prod_{I=1}y_I^{-2}}_{v^{-2;-2;-2;-2}{\widetilde H}_{4,1}^2}
\nonumber\\
&
+\underbrace{y_2 y_3}_{v^{0;1;1;0}}
+\underbrace{y_2^{-1}y_3^{-1}}_{v^{0;-1;-1;0}}
+\underbrace{y_3}_{v^{0;0;1;0}}
+\underbrace{y_3^{-1}}_{v^{0;0;-1;0}}
+\underbrace{y_2}_{v^{0;1;0;0}}
+\underbrace{y_2^{-1}}_{v^{0;-1;0;0}}
)t^2
+\underbrace{t^{-2}}_{
\substack{
\psi_{\varphi^{(2)}},\\
\psi_{\varphi^{(3)}},\\
T_{1,2}{\widetilde T}_{1,2},\\
T_{2,3}{\widetilde T}_{2,3},\\
T_{3,4}{\widetilde T}_{3,4}
}
}
\Biggr]q^{1/2}
\nonumber\\
&
+
(\underbrace{3x\prod_{I=1}^4 y_I}_{\substack{
v^{1;1;1;1}\varphi^{(2)}H_{4,1},\\ 
v^{1;1;1;1}\varphi^{(3)}H_{4,1},\\ 
v^{1;1;1;1} H_{4,1}^2 {\widetilde H_{4,1}}
}}
+\underbrace{3x^{-1}\prod_{I=1}^4y_I^{-1}}_{\substack{
v^{-1;-1;-1;-1}\varphi^{(2)}{\widetilde H}_{4,1},\\ 
v^{-1;-1;-1;-1}\varphi^{(3)}{\widetilde H}_{4,1},\\ 
v^{-1;-1;-1;-1}H_{4,1}^2{\widetilde H}_{4,1}
}
}
+\underbrace{x^{-3}\prod_{I=1}^4 y_I^{-3}}_{v^{-3;-3;-3;-3}{\widetilde H_{4,1}}^3}
+\underbrace{x^3\prod_{I=1}^4 y_I^{3}}_{v^{3;3;3;3}H_{4,1}^3}
\nonumber\\
&
+\underbrace{xy_1 y_4}_{v^{1;0;0;1}H_{4,1}}
+\underbrace{x^{-1}y_1^{-1} y_4^{-1}}_{v^{-1;0;0;-1}{\widetilde H}_{4,1}}
+\underbrace{xy_1 y_3y_4}_{v^{1;0;1;1}H_{4,1}}
+\underbrace{x^{-1}y_1^{-1} y_3^{-1}y_4^{-1}}_{v^{-1;0;-1;-1}{\widetilde H}_{4,1}}
+\underbrace{xy_1y_2^2y_3^2y_4}_{v^{1;2;2;1}H_{4,1}}
+\underbrace{x^{-1}y_1^{-1}y_2^{-2}y_3^{-2}y_4^{-1}}_{v^{-1;-2;-2;-1}{\widetilde H}_{4,1}}
\nonumber\\
&
+\underbrace{xy_1y_2^2y_3y_4}_{v^{1;2;1;1}H_{4,1}}
+\underbrace{x^{-1}y_1^{-1}y_2^{-2}y_3^{-1}y_4^{-1}}_{v^{-1;-2;-1;-1}{\widetilde H}_{4,1}}
+\underbrace{xy_1 y_2y_4}_{v^{1;1;0;1}H_{4,1}}
+\underbrace{x^{-1}y_1^{-1} y_2^{-1}y_4^{-1}}_{v^{-1;-1;0;-1}{\widetilde H}_{4,1}}
+\underbrace{xy_1y_2y_3^2y_4}_{v^{1;1;2;1}H_{4,1}}
+\underbrace{x^{-1}y_1^{-1}y_2^{-1}y_3^{-2}y_4^{-1}}_{v^{-1;-1;-2;-1}{\widetilde H}_{4,1}}
)t^{3}
\nonumber\\
&+(
\underbrace{z_1^{-1} z_2^{-1} z_3^{-1} \prod_{I}y_I}_{v^{1;1;1;1}{\widetilde T}_{1,2}{\widetilde T}_{2,3}{\widetilde T}_{3,4}}
+
\underbrace{z_1 z_2 z_3\prod_{I}y_I^{-2}}_{v^{-1;-1;-1;-1}T_{1,2}T_{2,3}T_{3,4}}
)t^{-3}
\Biggr]q^{3/4}+\cdots.
\end{align}
The index (\ref{u1u1u1u1k+-0_findex}) matches with the index (\ref{u1_3_findex}) for the $U(1)$ ADHM theory with three flavors.

\subsubsection{$U(2)_{1}\times U(2)_0\times U(2)_{-1}$}
We can increase the rank of the three unitary groups uniformly by $1$ and consider the $U(2)_{1}\times U(2)_0\times U(2)_{-1}$ quiver CS theory. The index precisely agrees with the index (\ref{u2_2_findex}) for the $U(2)$ ADHM theory with two flavors. 

\subsubsection{$U(1)_{2}\times U(1)_0\times U(2)_{-2}$}
\label{sec_USp112k2}
Let us then consider cases with different CS levels. The $U(1)_{2}\times U(1)_0\times U(2)_{-2}$ quiver CS theory is expected to have a $USp(2)$ dual given in \eqref{USpCSdual}. Indeed the index for the $U(1)_{2}\times U(1)_0\times U(2)_{-2}$ quiver CS theory agrees with the index (\ref{Sp2_S_2_findex}) for the $USp(2)$ gauge theory with an adjoint hyper and two fundamental half-hypers.
After fixing $\prod_{I=1}^3y_I=1$ we have
\begin{align}
&I^{U(1)_2\times U(1)_0\times U(2)_{-2}}(t,x,y_I,z_I;q)\nonumber \\
&=1
+ [t^{-2} + (2 + x^{-2} + x^2)t^2 ]q^{\frac{1}{2}} 
+ [(y_2^{-1} + x^2y_2^{-1} + y_2 + x^{-2}y_2)t^3 + (x^{-1}z_1^{-1}z_2^{-1}\nonumber \\
&\quad  + xz_1^{-1}z_2^{-1} + x^{-1}z_1 z_2 + x z_1 z_2)t^{-1}]q^{\frac{3}{4}} 
+ [-3 + (3 + x^{-4} + 2x^{-2} + 2 x^2 + x^4)t^4 + (1\nonumber \\
&\quad  + z_1^{-2} z_2^{-2} + z_1^2 z_2^2)t^{-4}]q 
+ [(2y_2^{-1} + x^{-2} y_2^{-1} + 2 x^2y_2^{-1} + x^4y_2^{-1} + 2 y_2 + x^{-4}y_2 + 2 x^{-2}y_2\nonumber \\
&\quad  + x^2 y_2)t^5 + (-y_2^{-1} - x^2y_2^{-1} - y_2 - x^{-2}y_2 + x^{-3} z_1^{-1} z_2^{-1} + x^3z_1^{-1} z_2^{-1} + x^{-3}z_1 z_2\nonumber \\
&\quad + x^3 z_1 z_2)t]q^{\frac{5}{4}} 
+\cdots,
\label{U1U1U2k+20-2_findex}
\end{align}
where we have set $y_1=y_2^{-1}y_3^{-1}$ so that $y_1y_2y_3=1$ which can be done without loss of generality due to the redundancy \eqref{circularquiverCSredundancy}.
The index \eqref{U1U1U2k+20-2_findex} agrees with the flavored index (\ref{Sp2_S_2_findex}) of the $USp(2)$ dual under the following identification of the fugacities:
\begin{align}
y_2&=xy^2,& z_1&=z_2=1. 
\end{align}
So the quiver CS theory can describe the M2-brane in $\mathbb{C}^2/\widehat{D}_1$. 

\subsubsection{$U(2)_{2}\times U(2)_0\times U(3)_{-2}$}
For $l=2,k=2,N_1=N_2=2,N_2=3$ we find the following index
\begin{align}
I^{U(2)_2\times U(2)_0\times U(3)_{-2}}(t,x,y_I,z_I;q)=1+[t^{-2}+(2+x^{-2}+x^2)t^2]q^{\frac{1}{2}}+\cdots,
\end{align}
where we have set $y_1=y_2^{-1}y_3^{-1}$.
This agrees with the index \eqref{Sp4_S_2_findex} for the $USp(4)$ theory with an adjoint hypermultiplet and two fundamental half-hypermultiplets, at least up to the order $q$.
This 
is again consistent with the duality \eqref{USpCSdual} between $U(N)_2\times U(N)_0\times U(N+1)_{-2}$ quiver Chern-Simons theory and the $USp(2N)$ theory with an adjoint hypermultiplet and two fundamental half-hypermultiplets.

\subsubsection{$U(1)_{2}\times U(1)_0\times U(1)_{-2}$}
Another interesting cases are the CS quiver theories involved in the duality \eqref{OCSdual}. Indeed we find that the index \eqref{220413_CSmatterindexnewnotation} for the $U(1)_{2}\times U(1)_0\times U(1)_{-2}$ quiver CS theory
\begin{align}
&I^{U(1)_2\times U(1)_0\times U(1)_{-2}}(t,x,y_I,z_I;q)\nonumber \\
&=1
+ [t^{-2} + (2 + x^{-2} + x^2 + y_2^{-1} + y_2)t^2]q^{\frac{1}{2}} 
+ (x^{-1} z_1^{-1} z_2^{-1}
+ xz_1^{-1}z_2^{-1} + x^{-1}z_1 z_2\nonumber \\
&\quad
+ x z_1 z_2) t^{-1} q^{\frac{3}{4}}
+ [-4 - y_2^{-1} - y_2 + (3 + x^{-4} + 2x^{-2} + 2 x^2 + x^4 + y_2^{-2} + 2y_2^{-1}\nonumber \\
&\quad  + x^{-2} y_2^{-1} + x^2y_2^{-1} + 2 y_2 + x^{-2}y_2 + x^2 y_2 + y_2^2)t^4  + (1 + z_1^{-2} z_2^{-2} + z_1^2 z_2^2)t^{-4}]q\nonumber \\
&\quad + (x^{-3} z_1^{-1} z_2^{-1} + x^3z_1^{-1} z_2^{-1} + x^{-3}z_1 z_2 + x^3 z_1 z_2) t q^{\frac{5}{4}}
+ \cdots
\end{align}
($y_1$ set to $y_2^{-1}y_3^{-1}$)
agrees (at least up to the order $q^{\frac{3}{2}}$)
with the index \eqref{O2++_A_2_indexFull} for the $O(2)$ gauge theory with an adjoint hyper and one flavor, with the following parameter identifications:
\begin{align}
x^{(\text{CS})}=x^{(O(2N))},\quad
z_1^{(\text{CS})}z_2^{(\text{CS})}=1,\quad
y_1^{(\text{CS})}y_3^{(\text{CS})}=(y^{(O(2N))})^{-2},\quad
y_2^{(\text{CS})}=(y^{(O(2N))})^2.
\label{O2Nantisym_212model_parameteridentification}
\end{align}
Thus the quiver CS theory describes a motion of M2-brane in $\mathbb{C}^2/\widehat{D}_1$. 

\subsubsection{$U(2)_{2}\times U(2)_0\times U(2)_{-2}$}
We can increase the ranks:
\begin{align}
&I^{U(2)_2\times U(2)_0\times U(2)_{-2}}(t,x,y_I,z_I;q)\nonumber \\
&=
1
+ [t^{-2} + t^2 (2 + x^{-2} + x^2 + y_2^{-1}+y_2)] q^{\frac{1}{2}}
+ (x^{-1} z_1^{-1} z_2^{-1} + xz_1^{-1}z_2^{-1} + x^{-1}z_1z_2\nonumber \\ 
&\quad  + x z_1z_2) t^{-1} q^{\frac{3}{4}} +\cdots,
\end{align}
($y_1$ set to $y_2^{-1}y_3^{-1}$) and we confirm that the index of the $U(2)_2\times U(2)_{0}\times U(2)_{-2}$ quiver CS theory matches with the index (\ref{O4++_A_2_index}) of the $O(4)$ gauge theory with an anti-symmetric hyper and one flavor with the parameter identification \eqref{O2Nantisym_212model_parameteridentification}, at least up to the order $q$.
We also confirm that the Coulomb limit of the two indices agree with \eqref{HS_Sym2C2Z4}
up to the order $\mathfrak{t}^{12}$,
and that the Higgs limit of the indices \eqref{O4asymhyp1Higgslimit},\eqref{220527_21modelHiggslimitsmalltexpansion} agree with each other
up to the order $\mathfrak{t}^{12}$.
These 
are again consistent with the conjectural duality \eqref{OCSdual} between 
the $O(2N)$ gauge theory with an anti-symmetric hyper and one flavor and the $U(N)_2\times U(N)_{0}\times U(N)_{-2}$ quiver CS theory.

\subsubsection{ADHM-CS dualities}

From the equivalence of the flavored indices we can derive the mapping of operators under the dualities 
between the ADHM theory and the circular quiver Chern-Simons theories given in Figure \ref{fig:ADHM_mirror}. 

The flavored index for the $U(1)$ ADHM theory with $l$ flavors 
and the flavored index for the $U(1)_{1}\times U(1)_0^{\otimes (l-1)}\times U(1)_{-1}$ CS theory agree with each other 
under the following fugacity map: 
\begin{align}
\label{ADHM_qCS_fug}
\begin{array}{c|c}
\text{$U(1)$ ADHM with $l$ flavors}&\text{$U(1)_{1}\times U(1)_0^{\otimes (l-1)}\times U(1)_{-1}$ CS theory} \\ \hline 
\text{$z^{\textrm{ADHM}}$ (topological sym.)}& z^{\text{CS}}_I (\text{flavor sym.~for $(T_{I,I+1},{\widetilde T}_{I,I+1})$}) \\ 
&{z^{\text{ADHM}}}^l=\prod_{I=1}^{l}{z^{\text{CS}}_I} \\ \hline
\text{$x^{\textrm{ADHM}}$ (flavor sym.~for $(X,Y)$)}& \text{$x^{\textrm{CS}}$ (flavor sym.~for $(H_{l+1,1},{\widetilde H}_{l+1,1})$)} \\ 
&x^{\text{ADHM}}=x^{\text{CS}} \\ \hline
\text{$y_{\alpha}^{\text{ADHM}}$ (flavor sym.~for $(I,J)$)} & \text{$y_I^{\text{CS}}$ (topological sym.)} \\
y_1^{\text{ADHM}}=1&\prod_{I=1}^{l+1}y_I^{\text{CS}}=1,\\
&y_\alpha^{\text{ADHM}}=\prod_{I=2}^{\alpha}y_I^{\text{CS}}\\
\end{array},
\end{align}
where we have distinguished the fugacities for the ADHM theory and those for the quiver CS theory by superscripts. 

As a consequence, we find the operator mapping under the duality between 
the $U(1)$ ADHM theory with $l$ flavors 
and the $U(1)_{1}\times U(1)_0^{\otimes (l-1)}\times U(1)_{-1}$ CS theory
\begin{align}
\label{ADHM_qCS_map1}
\begin{array}{c|c}
\textrm{$U(1)$ ADHM with $l$ flavors}&\textrm{$U(1)_{1}\times U(1)_0^{\otimes (l-1)}\times U(1)_{-1}$ CS theory} \\ \hline 
X^m&v^{m;m;\cdots;m}H_{3,1}^m \\
Y^m&v^{-m;-m;\cdots;-m}{\widetilde H}_{3,1}^m \\
XY,\quad \bigoplus_{\alpha(\neq l)} J_{\alpha}I_{\alpha}&H_{3,1}\tilde{H}_{3,1},\quad \bigoplus_{I=2}^l\varphi^{(I)} \\
J_\alpha I_\beta\quad (1<\alpha<\beta)&v^{0;\cdots;0;m^{(\alpha+1)}=1;\cdots;m^{(\beta)}=1;0;\cdots;0}\\
J_1I_\alpha\quad (\alpha>1)&v^{0;1;\cdots;m^{(\alpha)}=1;0;\cdots;0}\\
J_\alpha I_\beta\quad (\alpha>\beta>1)&v^{0;\cdots;0;m^{(\beta+1)}=-1;\cdots;m^{(\alpha)}=-1;0;\cdots;0}\\
J_\alpha I_1\quad (\alpha>1)&v^{0;-1;\cdots;m^{(\alpha)}=-1;0;\cdots;0}\\
v^{m}&v^{-m;-m;\cdots;-m}\prod_{I=1}^{l}T_{I,I+1} \\
v^{-m}&v^{m;m;\cdots;m}\prod_{I=1}^{l}{\widetilde T}_{I,I+1}  \\
\varphi&
\bigoplus_{I=1}^l
T_{I,I+1}{\widetilde T}_{I,I+1}
/
\bigoplus_{I=2}^{l}
\left[
\psi^{(I)}
\right]
 \\
\end{array},
\end{align}
In a similar manner as the mapping under the duality between the ADHM and ABJM theory discussed in section \ref{sec_dual_ABJ}, one can also generalize the map (\ref{ADHM_qCS_map1}) to the non-Abelian case. 

\subsection{Closed form expression for the Coulomb limit with general $k,N,l$}
\label{sec_coulombCSmatterfermigas}
We can also evaluate the Coulomb limit \eqref{220413_CSmatterindexnewnotation} of the supersymmetric index of the $U(N)_k\times U(N)_0^{\otimes (l-2)}\times U(N)_{-k}$ quiver Chern-Simons theory for general values of $k,N,l$, by the similar calculation as in the case of the $U(N)$ ADHM theory we considered in section \ref{ADHMCoulomballorder}.
We assume $k>0$, $|\mathfrak{t}|<1$, $|\prod_{I=1}^{l+1}y_I|=1$ and $|z_I|=1$.
First write the overall factor of $q$ and $t$ in terms of $\mathfrak{t}=q^{\frac{1}{4}}t^{-1}$ and $q$ as
\begin{align}
&q^{-\frac{1}{2}\sum_{I=1}^{l+1}\sum_{i<j}^N|m_i^{(I)}-m_j^{(I)}|+\frac{1}{4}\sum_{I=1}^{l+1}\sum_{i=1}^N\sum_{j=1}^N|m_i^{(I)}-m_j^{(I+1)}|}\nonumber \\
&\quad \times t^{-2\sum_{I=2}^l\sum_{i<j}^N|m_i^{(I)}-m_j^{(I)}|+\sum_{I=1}^l\sum_{i=1}^N\sum_{j=1}^N|m_i^{(I)}-m_j^{(I+1)}|-\sum_{i=1}^N\sum_{j=1}^N|m_i^{(l+1)}-m_j^{(1)}|}\nonumber \\
&=\mathfrak{t}^{
2\sum_{I=2}^l\sum_{i<j}^N|m_i^{(I)}-m_j^{(I)}|-\sum_{I=1}^l\sum_{i=1}^N\sum_{j=1}^N|m_i^{(I)}-m_j^{(I+1)}|+\sum_{i=1}^N\sum_{j=1}^N|m_i^{(l+1)}-m_j^{(1)}|
}\nonumber \\
&\quad \times q^{-\frac{1}{2}\sum_{I=1,l+1}\sum_{i<j}^N|m_i^{(I)}-m_j^{(I)}|-\sum_{I=2}^l\sum_{i<j}|m_i^{(I)}-m_j^{(I)}|+\frac{1}{2}\sum_{I=1}^l\sum_{i=1}^N\sum_{j=1}^N|m_i^{(I)}-m_j^{(I+1)}|
}.
\label{220430ICSMoverallqtinCoulomb}
\end{align}
We observe that the power of $q$ in \eqref{220430ICSMoverallqtinCoulomb} is positive semi-definite which vanishes if and only if all of the monopole charges $(m_1^{(I)},\cdots,m_N^{(I)})$ coincide up to permutations of $N$ charges for each $I$.
Hence the Coulomb limit of the index \eqref{220413_CSmatterindexnewnotation} simplififes as
\begin{align}
&{\cal I}^{U(N)_k\times U(N)_0^{\otimes (l-1)}\times U(N)_{-k}\text{quiver CS}\,(C)}\nonumber \\
&=
\lim_{\substack{\mathfrak{t}=q^{\frac{1}{4}}t^{-1}\text{: fixed}\\ q\rightarrow 0}}
I^{U(N)_k\times U(N)_0^{\otimes (l-1)}\times U(N)_{-k}\text{quiver CS}}\nonumber \\
&=\frac{1}{(N!)^{l+1}}
\sum_{m_i\in\mathbb{Z}}
r(m_i)^l
(\prod_{I=1}^{l+1}y_I)^{\sum_im_i}
\oint\prod_I\prod_{i=1}^N\frac{ds_i^{(I)}}{2\pi is_i^{(I)}}
\prod_{i=1}^N(s_i^{(1)})^{km_i}
\prod_{i=1}^N(s_i^{(l+1)})^{-km_i}\nonumber \\
&\quad \times \prod_{I=1}^{l+1}\prod_{i\neq j}^N
\lim_{\substack{\mathfrak{t}=q^{\frac{1}{4}}t^{-1}\text{: fixed}\\ q\rightarrow 0}}
\Bigl(1-q^{\frac{|m_i-m_j|}{2}}\frac{s_i^{(I)}}{s_j^{(I)}}\Bigr)
\prod_{I=2}^l\prod_{i,j=1}^N
\lim_{\substack{\mathfrak{t}=q^{\frac{1}{4}}t^{-1}\text{: fixed}\\ q\rightarrow 0}}
\frac{(q^{\frac{1}{2}+\frac{|m_i-m_j|}{2}}t^{-2}\frac{s_i^{(I)}}{s_j^{(I)}};q)_\infty}{(q^{\frac{1}{2}+\frac{|m_i-m_j|}{2}}t^2\frac{s_i^{(I)}}{s_j^{(I)}};q)_\infty}\nonumber \\
&\quad \times \prod_{I=1}^l\prod_{i,j}^N\prod_{\pm}
\lim_{\substack{\mathfrak{t}=q^{\frac{1}{4}}t^{-1}\text{: fixed}\\ q\rightarrow 0}}
\frac{(q^{\frac{3}{4}+\frac{|m_i-m_j|}{2}}t\Bigl(\frac{s_i^{(I)}}{s_j^{(I+1)}}\Bigr)^\pm z_I^\pm;q)_\infty}{(q^{\frac{1}{4}+\frac{|m_i-m_j|}{2}}t^{-1}\Bigl(\frac{s_i^{(I)}}{s_j^{(I+1)}}\Bigr)^\pm z_I^\pm;q)_\infty}\nonumber \\
&\quad \times \prod_{i,j}^N\prod_{\pm}
\lim_{\substack{\mathfrak{t}=q^{\frac{1}{4}}t^{-1}\text{: fixed}\\ q\rightarrow 0}}
\frac{(q^{\frac{3}{4}+\frac{|m_i-m_j|}{2}}t^{-1}\Bigl(\frac{s_i^{(l+1)}}{s_j^{(1)}}\Bigr)^\pm x^\pm;q)_\infty}{(q^{\frac{1}{4}+\frac{|m_i-m_j|}{2}}t\Bigl(\frac{s_i^{(l+1)}}{s_j^{(1)}}\Bigr)^\pm x^\pm;q)_\infty}.
\label{220430CSmatterCoulomb1}
\end{align}
where $r(m_i)$ is the number of permutations of $(m_1,m_2,\cdots,m_N)$.
If we label $m_i$ in the same way as we have done for the ADHM theory (see section \ref{ADHMCoulomballorder})
\begin{align}
m_i=(\cdots,\underbrace{-1,\cdots,-1}_{\nu_{-1}},\underbrace{0,\cdots,0}_{\nu_0},\underbrace{1,\cdots,1}_{\nu_1},\cdots,\underbrace{m,\cdots,m}_{\nu_m},\cdots)\quad (\text{up to permutation}),
\end{align}
then \eqref{220430CSmatterCoulomb1} can be written as
\begin{align}
&{\cal I}^{U(N)_k\times U(N)_0^{\otimes (l-1)}\times U(N)_{-k}\text{quiver CS}\,(C)}\nonumber \\
&=
\sum_{\substack{\nu_m\ge 0\\ (\sum_{m=-\infty}^\infty\nu_m=N)}}\frac{1}{\prod_{m=-\infty}^\infty(\nu_m!)^{l+1}}(\prod_{I=1}^{l+1}y_I)^{\sum_{m=-\infty}^\infty m\nu_m}
\prod_{m=-\infty}^\infty
\int\prod_{I=1}^{l+1}\prod_{i=1}^{\nu_m}\frac{ds_i^{(I)}}{2\pi is_i^{(I)}}\nonumber \\
&\quad \prod_{i=1}^{\nu_m}\Bigl(\frac{s_i^{(1)}}{s_i^{(l+1)}}\Bigr)^{km}
\prod_{I=1}^{l+1}\prod_{i\neq j}^{\nu_m}\Bigl(1-\frac{s_i^{(I)}}{s_j^{(I)}}\Bigr)
\prod_{I=2}^l\prod_{i,j}^{\nu_m}\Bigl(1-\mathfrak{t}^2\frac{s_i^{(I)}}{s_j^{(I)}}\Bigr)
\prod_{I=1}^l\prod_{i,j}^{\nu_m}\frac{1}{1-\mathfrak{t} z_I^{\pm 1}(\frac{s_i^{(I)}}{s_j^{(I+1)}})^{\pm 1}}.
\end{align}
Again, if we define the grand canonical sum we can remove the constraint of the summation:
\begin{align}
\Xi(\kappa)=\sum_{N=0}^\infty {\cal I}^{U(N)_k\times U(N)_0^{\otimes (l-1)}\times U(N)_{-k}\text{quiver CS}\,(C)}=\prod_{m=-\infty}^\infty{\widetilde \Xi}_{km}\Bigl(\kappa (\prod_{I=1}^{l+1}y_I)^m,z_I,\mathfrak{t}\Bigr),
\end{align}
where
\begin{align}
{\widetilde \Xi}_{k'}(\kappa,z_I,\mathfrak{t})&=\sum_{\nu=0}^\infty \kappa^\nu\Omega_{k',\nu}(\mathfrak{t},z_a)
\label{220518_Z1Higgsstart}
\end{align}
with
\begin{align}
\Omega_{k',\nu}(\mathfrak{t},z_I)&=
\frac{1}{(\nu!)^{l+1}}
\int\prod_{I=1}^{l+1}\prod_{i=1}^{\nu}\frac{ds_i^{(I)}}{2\pi is_i^{(I)}}
\prod_{i=1}^{\nu}(s_i^{(1)})^{k'}\prod_{i=1}^\nu(s_i^{(l+1)})^{-k'}
\prod_{I=1}^{l+1}\prod_{i\neq j}^{\nu}\Bigl(1-\frac{s_i^{(I)}}{s_j^{(I)}}\Bigr)\nonumber \\
&\quad \prod_{I=2}^l\prod_{i,j}^{\nu}\Bigl(1-\mathfrak{t}^2\frac{s_i^{(I)}}{s_j^{(I)}}\Bigr)
\prod_{I=1}^l\prod_{i,j}^{\nu}\frac{1}{1-\mathfrak{t} z_I^{\pm 1}(\frac{s_i^{(I)}}{s_j^{(I+1)}})^{\pm 1}}.
\end{align}
To evaluate $\Omega_{k',\nu}$ let us first write it as
\begin{align}
\Omega_{k',\nu}(\mathfrak{t},z_I)&=\frac{\prod_{I=1}^l(-\mathfrak{t}^{-1}z_I^{-1})^{\nu^2}}{(\nu!)^{l+1}}\int\prod_{I=1}^{l+1}\prod_{i=1}^\nu\frac{ds_i^{(I)}}{2\pi i}
\prod_i(s_i^{(1)})^{k'}
\prod_i(s_i^{(l+1)})^{-k'}\nonumber \\
&\quad \frac{\prod_{I=1}^{l+1}\prod_{i\neq j}(s_i^{(I)}-s_j^{(I)})\prod_{I=2}^l\prod_{i,j}(s_i^{(I)}-\mathfrak{t}^2s_j^{(I)})}{\prod_{i=I}^l\prod_{i,j}\prod_\pm(s_i^{(I)}-\mathfrak{t}^{\pm 1}z_I^{-1}s_j^{(I+1)})}.
\end{align}
This integration can be evaluated iteratively with respect to $I$ in the following way.
First let us suppose $k'\ge 0$.
The integrand has not pole at $s_i^{(1)}=0$, hence the integration over $s_i^{(I)}$ can be evaluated by picking the poles in $|s_i^{(I)}|<1$, which are
\begin{align}
s_i^{(1)}=\mathfrak{t}z_I^{-1}s_{\sigma(i)}^{(2)},
\end{align}
with any permutation $\sigma\in S_\nu$.
Since the integrand is symmetric in $(s_1^{(2)},\cdots,s_\nu^{(2)})$ the residue is independent of the choice of $\sigma$ and the summation over $\sigma\in S_\nu$ just gives an overall factor $\nu!$.
Evaluating the residue we end up with
\begin{align}
\Omega_{k',\nu}(\mathfrak{t},z_I)
&=(\mathfrak{t}z_1^{-1})^{k'\nu}\cdot \frac{\prod_{I=2}^l(-\mathfrak{t}^{-1}z_I^{-1})^{\nu^2}}{(\nu!)^l}\int\prod_{I=2}^{l+1}\prod_{i=1}^\nu\frac{ds_i^{(I)}}{2\pi i}
\prod_i(s_i^{(2)})^{k'}
\prod_i(s_i^{(l+1)})^{-k'}\nonumber \\
&\quad \frac{\prod_{I=2}^{l+1}\prod_{i\neq j}(s_i^{(I)}-s_j^{(I)})\prod_{I=3}^l\prod_{i,j}(s_i^{(I)}-\mathfrak{t}^2s_j^{(I)})}{\prod_{I=2}^l\prod_{i,j}\prod_\pm(s_i^{(I)}-\mathfrak{t}^{\pm 1}z_I^{-1}s_j^{(I+1)})}.
\end{align}
Repeating the same calculation we finally obtain
\begin{align}
\Omega_{k',\nu}(\mathfrak{t},z_I)=
\mathfrak{t}^{k'l\nu}\prod_{I=1}^lz_I^{-k'\nu}
\frac{1}{\nu!}\int\frac{ds_i^{(l+1)}}{2\pi i}\frac{\prod_{i\neq j}(s_i^{(l+1)}-s_j^{(l+1)})}{\prod_{i,j}(s_i^{(l+1)}-\mathfrak{t}^2s_j^{(l+1)})}.
\end{align}
Note that up to the overall factor $\mathfrak{t}^{k'l\nu}\prod_{I=1}^lz_I^{-k'\nu}$ the right-hand side coincides with $\Omega_\nu(\mathfrak{t})$ \eqref{220429Omeganu} introduced in the calculation of the Coulomb limit of the ADHM theory.
By performing the same calculation for $k'<0$ where we pick the poles $s_i^{(I)}=\mathfrak{t}^{-1}z_I^{-1}s_{\sigma(i)}^{(I+1)}$ ($\sigma\in S_\nu$) iteratively in $I$, we obtain
\begin{align}
\Omega_{k',\nu}(\mathfrak{t},z_I)=\mathfrak{t}^{|k'|l\nu}\prod_{I=1}^lz_I^{-k'\nu}\Omega_\nu(\mathfrak{t}).
\label{220518_Z1Higgsend}
\end{align}
Hence we conclude that
\begin{align}
{\cal I}^{U(N)_k\times U(N)_0^{\otimes (l-1)}\times U(N)_{-k}\text{quiver CS}\,(C)}(\mathfrak{t},y_I,z_I)
=
{\cal I}^{U(N) \text{ADHM-}[kl]\,(C)}\Bigl(\mathfrak{t},z=\Bigl(\prod_{i=1}^{l+1}y_I\prod_{I=1}^lz_I^{-k}\Bigr)^{\frac{1}{kl}}\Bigr).
\end{align}

\subsection{A simplification of Higgs limit with general $k,N,l$}
\label{sec_Higgssimplerform}
Lastly, let us also try to simplify the Higgs limit $q\rightarrow 0$ with $\mathfrak{t}=q^{\frac{1}{4}}t$ of the supersymmetric index of the Chern-Simons matter theory \eqref{220413_CSmatterindexnewnotation} ($k\ge 0$).
Here we consider only the cases with $N_{l+1}=N_1$.
We also assume $l\ge 2$, since for $l=1$ the Higgs limit manifestly coincides with the Coulomb limit which we have already treated in section \ref{sec_coulombCSmatterfermigas}, with $y_a\rightarrow y_a^{-1}$ and $x\leftrightarrow z$.
To treat the Higgs limit we write the overall factor of $q,t$ for each choice of the monopole charges as
\begin{align}
&q^{-\frac{1}{2}\sum_{I=1}^{l+1}\sum_{i<j}^N|m_i^{(I)}-m_j^{(I)}|+\frac{1}{4}\sum_{I=1}^{l+1}\sum_{i=1}^N\sum_{j=1}^N|m_i^{(I)}-m_j^{(I+1)}|}\nonumber \\
&\quad \times t^{
-2\sum_{I=2}^l\sum_{i<j}^N|m_i^{(I)}-m_j^{(I)}|+\sum_{I=1}^l\sum_{i=1}^N\sum_{j=1}^N|m_i^{(I)}-m_j^{(I+1)}|-\sum_{i=1}^N\sum_{j=1}^N|m_i^{(l+1)}-m_j^{(1)}|
}\nonumber \\
&=\mathfrak{t}^{
-2\sum_{I=2}^l\sum_{i<j}^N|m_i^{(I)}-m_j^{(I)}|+\sum_{I=1}^l\sum_{i=1}^N\sum_{j=1}^N|m_i^{(I)}-m_j^{(I+1)}|-\sum_{i=1}^N\sum_{j=1}^N|m_i^{(l+1)}-m_j^{(1)}|
}\nonumber \\
&\quad \times q^{-\frac{1}{2}\sum_{I=1,l+1}\sum_{i<j}|m_i^{(I)}-m_j^{(I)}|+\frac{1}{4}\sum_{i,j}|m_i^{(l+1)}-m_i^{(1)}|}.
\end{align}
The power of $q$ is positive semi-definite and vanishes if and only if $(m_1^{(1)},\cdots,m_N^{(1)})=(m_1^{(l+1)},\cdots,m_1^{(l+1)})$ up to permutation of the indices $i=1,\cdots,N$.
Hence the supersymmetric index simplifies in the Higgs limit as
\begin{align}
&{\cal I}^{U(N)_k\times U(N)_0^{\otimes (l-1)}\times U(N)_{-k}\text{ quiver CS}(H)}\nonumber \\
&=\lim_{\substack{\mathfrak{t}=q^{\frac{1}{4}}t\text{: fixed}\\
q\rightarrow 0}}
I^{U(N)_k\times U(N)_0^{\otimes (l-1)}\times U(N)_{-k}\text{ quiver CS}}\nonumber \\
&=\sum_{\substack{\nu_m^{(1)},\cdots,\nu_m^{(l)}\ge 0\\
(\sum_{m=-\infty}^\infty\nu_m^{(a)}=N_a)
}
}
\mathfrak{t}^{\epsilon_1(\nu_m^{(a)})}
\Biggl[\prod_{m=-\infty}^\infty (y_1y_{l+1})^{m\nu_m^{(1)}}
\prod_{a=2}^ly_a^{m\nu_m^{(a)}}\nonumber \\
&\quad \times \frac{1}{(\nu_m^{(1)}!)^2}\int
\Bigl(
\prod_{i=1}^{\nu_m^{(1)}}
\frac{ds_i^{(1)}}{2\pi is_i^{(1)}}
\frac{ds_i^{(l+1)}}{2\pi is_i^{(l+1)}}
(s_i^{(1)})^{km}
(s_i^{(l+1)})^{-km}
\Bigr)
\frac{
\prod_{i\neq j}^{\nu_m^{(1)}}
(1-\frac{s_i^{(1)}}{s_j^{(1)}})
(1-\frac{s_i^{(l+1)}}{s_j^{(l+1)}})
}{
\prod_{i,j=1}^{\nu_m^{(1)}}\prod_\pm(1-\mathfrak{t}(\frac{s_i^{(l+1)}}{s_j^{(1)}})^{\pm 1}x^{\pm 1})}\nonumber \\
&\quad \times \prod_{a=2}^l\frac{1}{\nu_m^{(a)}!}\int\prod_{i=1}^{\nu_m^{(a)}}\frac{ds_i^{(a)}}{2\pi is_i^{(a)}}\frac{\prod_{i\neq j}^{\nu_m^{(a)}}(1-\frac{s_i^{(a)}}{s_j^{(a)}})}{\prod_{i,j}^{\nu_m^{(a)}}(1-\mathfrak{t}^2\frac{s_i^{(a)}}{s_j^{(a)}})}
\Biggr],
\label{220518_Higgsmwritteninnu}
\end{align}
where we have labelled each monopole charge $(m_1^{(a)},\cdots,m_{N_a}^{(a)})$ with $\nu_m^{(a)}=\#\{m_i^{(a)}|m_i^{(a)}=m\}$, and defined $\epsilon_1$ as
\begin{align}
&\epsilon_1(\nu_m^{(a)})\nonumber \\
&=-2\sum_{I=2}^l\sum_{i<j}^{N_I}|m_i^{(I)}-m_j^{(I)}|+\sum_{I=1}^l\sum_{i=1}^{N_I}\sum_{j=1}^{N_{I+1}}|m_i^{(I)}-m_j^{(I+1)}|-\sum_{i=1}^{N_{I+1}}\sum_{j=1}^{N_I}|m_i^{(l+1)}-m_j^{(1)}|\biggr|_{m_i^{(l+1)}=m_i^{(1)}}\nonumber \\
&=-2\sum_{a=1}^l\sum_{m<m'}\nu_m^{(a)}\nu_{m'}^{(a)}|m-m'|+\sum_{a=1}^l\sum_{m,m'}\nu_m^{(a)}\nu_{m'}^{(a+1)}|m-m'|.
\end{align}
Here the $2\nu_m^{(1)}$ dimensional integration in the third line and the $\nu_m^{(a)}$ dimensional integrations in the fourth line of the final expression are what we have already calculated in \eqref{220518_Z1Higgsstart}-\eqref{220518_Z1Higgsend} and \eqref{220518_Z2inHiggsstart}-\eqref{220518_Z2inHiggsend}:
\begin{align}
&\frac{1}{(\nu!)^2}\int\prod_{i=1}^\nu\frac{ds_i}{2\pi is_i}\frac{ds'_i}{2\pi is'_i}s_i^{km}(s_i')^{-km}\frac{\prod_{i\neq j}(1-\frac{s_i}{s_j})(1-\frac{s_i'}{s_j'})}{\prod_{i,j,\pm}(1-\mathfrak{t}(\frac{s_i'}{s_j})^{\pm 1}x^{\pm 1})}
=\mathfrak{t}^{|km|\nu}x^{km\nu}\Omega_\nu(\mathfrak{t}),\nonumber \\
&\frac{1}{\nu!}\int\prod_{i=1}^\nu\frac{ds_i}{2\pi is_i}\frac{\prod_{i\neq j}(1-\frac{s_i}{s_j})}{\prod_{i,j}(1-\mathfrak{t}^2\frac{s_i}{s_j})}
=\Omega_\nu(\mathfrak{t}),
\end{align}
where $\Omega_\nu(\mathfrak{t})$ is given in \eqref{220518_Z2inHiggsend}.
Plugging these into \eqref{220518_Higgsmwritteninnu} we obtain
\begin{align}
&{\cal I}^{U(N)_k\times U(N)_0^{\otimes (l-1)}\times U(N)_{-k}\text{ quiver CS}(H)}\nonumber \\
&=\sum_{\substack{\nu_m^{(1)},\cdots,\nu_m^{(l)}\ge 0\\
(\sum_{m=-\infty}^\infty\nu_m^{(a)}=N_a)
}
}
\mathfrak{t}^{\epsilon_1(\nu_m^{(a)})}
\Biggl[\prod_{m=-\infty}^\infty \mathfrak{t}^{|km|\nu_m^{(1)}}(x^ky_1y_{l+1})^{m\nu_m^{(1)}}
\prod_{a=2}^ly_a^{m\nu_m^{(a)}}
\prod_{a=1}^l\Omega_{\nu_m^{(a)}}(\mathfrak{t})
\Biggr].
\label{220518_CSmatterHiggsdifficultsum}
\end{align}

When $l=2$ and $N_a=1$ we can perform the summation explicitly by relabelling $\nu_m^{(a)}$ in \eqref{220518_CSmatterHiggsdifficultsum} as $\nu_m^{(1)}=\delta_{m,m^{(1)}}$, $\nu_m^{(2)}=\delta_{m,m^{(1)}+n^{(2)}}$, as
\begin{align}
&{\cal I}^{U(1)_k\times U(1)_0\times U(1)_{-k}\text{ quiver CS}(H)}\nonumber \\
&=\Omega_1(\mathfrak{t})^2\sum_{m^{(1)}=-\infty}^\infty\mathfrak{t}^{k|m^{(1)}|}x^{km^{(1)}}\prod_{I=1}^3y_I^{m^{(1)}}
\sum_{n^{(2)}=-\infty}^\infty\mathfrak{t}^{2|n^{(2)}|}y_2^{n^{(2)}}\nonumber \\
&=
\frac{
(1-\mathfrak{t}^{2k})
(1+\mathfrak{t}^2)
}{
(1-\mathfrak{t}^2)\prod_{\pm}(1-x^{\pm k}\prod_{I=1}^3y_I^{\pm 1}\mathfrak{t}^k)(1-y_2^{\pm 1}\mathfrak{t}^2)
}.
\end{align}
In particular, if we set $k=1,2$ and $y_1=y_2^{-1}y_3^{-1}$ to fix the redundancy (see \eqref{circularquiverCSfixredundancy}), the result agree with the Higgs limit of the supersymmetric index of $U(1)$ ADHM theory \eqref{HS_1_sulinst} with $l=2$ and the Higgs limit of the supersymmetric index of $O(2)$ with one antisymmetric hypermultiplet and one fundamental hypermultiplet \eqref{HS_O2++_A_2} respectively.

Unfortunately, we are not able to perform the infinite sum over $\nu_m^{(a)}$ in \eqref{220518_CSmatterHiggsdifficultsum} explicitly for general $l\ge 2$ and $N_a$ due to the overall $\mathfrak{t}^{\epsilon_1(\nu_m^{(a)})}$ which does not factorize in $m,a$.
Nevertheless \eqref{220518_CSmatterHiggsdifficultsum} is useful for computing the small $\mathfrak{t}$ expansion of ${\cal I}^{(H)}$ to any finite order.
For example, for $l=2$ and $N_1=N_2=N_3=2$, by classifying the summations over $\nu_m^{(a)}$ into the following four types:
\begin{align}
&\text{(i):}\quad \nu_m^{(1)}=2\delta_{m,m^{(1)}},\quad \nu_{m^{(2)}}=2\delta_{m,m^{(2)}},\nonumber \\
&\text{(ii):}\quad \nu_m^{(1)}=2\delta_{m,m^{(1)}},\quad \nu_{m^{(2)}}=\delta_{m,m^{(2)}_1}+\delta_{m,m^{(2)}_2}\quad (m^{(2)}_1<m^{(2)}_2),\nonumber \\
&\text{(iii):}\quad \nu_m^{(1)}=\delta_{m,m^{(1)}_1}+\delta_{m,m^{(1)}_2},\quad \nu_m^{(2)}=2\delta_{m,m^{(2)}},\quad (m^{(1)}_1<m^{(1)}_2),\nonumber \\
&\text{(iv):}\quad \nu_m^{(1)}=\delta_{m,m^{(1)}_1}+\delta_{m,m^{(1)}_2},\quad \nu_m^{(2)}=\delta_{m,m^{(2)}_1}+\delta_{m,m^{(2)}_2},\quad (m^{(a)}_1<m^{(a)}_2),
\end{align}
we can write ${\cal I}^{U(2)_k\times U(2)_0\times U(2)_{-k}\text{ quiver CS}(H)}$ \eqref{220518_CSmatterHiggsdifficultsum} as
\begin{align}
&{\cal I}^{U(2)_k\times U(2)_0\times U(2)_{-k}\text{ quiver CS}(H)}\nonumber \\
&=
\Omega_2(\mathfrak{t})^2
\sum_{m^{(1)}=-\infty}^\infty\sum_{m^{(2)}=-\infty}^\infty
\mathfrak{t}^{8|m^{(1)}-m^{(2)}|+2|km^{(1)}|}
(x^ky_1y_3)^{2m^{(1)}}y_2^{2m^{(2)}}\nonumber \\
&\quad +\Omega_1(\mathfrak{t})^2\Omega_2(\mathfrak{t})\Bigl[
\sum_{m^{(1)}=-\infty}^\infty\sum_{m^{(2)}_1<m^{(2)}_2}^\infty
\mathfrak{t}^{-2|m^{(2)}_1-m^{(2)}_2|+4\sum_{i=1,2}|m^{(1)}-m^{(2)}_i|+2|km^{(1)}|}
(x^ky_1y_3)^{2m^{(1)}}\nonumber \\
&\quad\quad\quad\quad\quad\quad\quad\quad\quad\quad\quad\quad\times y_2^{m^{(2)}_1+m^{(2)}_2}\nonumber \\
&\quad\quad +
\sum_{m^{(1)}_1<m^{(1)}_2}^\infty\sum_{m^{(2)}=-\infty}^\infty
\mathfrak{t}^{-2|m^{(1)}_1-m^{(1)}_2|+4\sum_{i=1,2}(|m^{(1)}_i-m^{(2)}|+|km^{(1)}_i|)}
(x^ky_1y_3)^{m^{(1)}_1+m^{(1)}_2}y_2^{2m^{(2)}}
\Bigr]\nonumber \\
&\quad+\Omega_1(\mathfrak{t})^4
\sum_{m^{(1)}_1<m^{(1)}_2}^\infty\sum_{m^{(2)}_1<m^{(2)}_2}^\infty
\mathfrak{t}^{-2|m^{(1)}_1-m^{(1)}_2|-2|m^{(2)}_1-m^{(2)}_2|+2\sum_{i,j=1,2}|m^{(1)}_i-m^{(2)}_j|+\sum_{i=1,2}|km^{(1)}_i|}\nonumber \\
&\quad\quad\quad\quad\quad\quad\quad\quad\quad\quad\quad\quad \times (x^ky_1y_3)^{m^{(1)}_1+m^{(1)}_2}y_2^{m^{(2)}_1+m^{(2)}_2}.
\label{220527_Higgsl2N12N22N32final}
\end{align}
Now suppose we want to compute ${\cal I}^{U(2)_k\times U(2)_0\times U(2)_{-k}\text{ quiver CS}(H)}$ to the order $\mathfrak{t}^{p}$ with some $p$.
Since $\Omega_\nu(\mathfrak{t})$ \eqref{220518_Z2inHiggsend} only contains positive powers of $\mathfrak{t}$, we can truncate the summation over $(m^{(1)},m^{(2)})$, $(m^{(1)},m^{(2)}_i)$, $(m^{(1)}_i,m^{(2)})$, $(m^{(1)}_i,m^{(2)}_j)$ in \eqref{220527_Higgsl2N12N22N32final} so that the powers of $\mathfrak{t}$ written explicitly in \eqref{220527_Higgsl2N12N22N32final} are less than or equal to $p$.
For each $k$ and $p$, we observe that only a finite number of $(m^{(1)},m^{(2)})$, $(m^{(1)},m^{(2)}_i)$, $(m^{(1)}_i,m^{(2)})$, $(m^{(1)}_i,m^{(2)}_j)$ satisfies this condition (see Table \ref{220527_21modelHiggstruncate}).
\begin{table}
\begin{center}
\begin{tabular}{|c|c|c|c|c|c|c|c|}
\hline
$p\backslash k$&$1$         &$2$        &$3$        &$\cdots$\\ \hline
$1$            &$(0,0,0,1)$ &$(0,0,0,0)$&$(0,0,0,0)$&$\cdots$\\ \hline
$2$            &$(1,1,0,2)$ &$(0,1,0,1)$&$(0,1,0,0)$&$\cdots$\\ \hline
$3$            &$(1,1,1,3)$ &$(0,1,0,1)$&$(0,1,0,1)$&$\cdots$\\ \hline
$4$            &$(2,2,1,4)$ &$(1,2,1,2)$&$(0,2,0,1)$&$\cdots$\\ \hline
$5$            &$(2,2,2,5)$ &$(1,2,1,2)$&$(0,2,1,2)$&$\cdots$\\ \hline
$6$            &$(3,3,2,6)$ &$(1,3,1,3)$&$(1,3,1,2)$&$\cdots$\\ \hline
$7$            &$(3,3,3,7)$ &$(1,3,1,3)$&$(1,3,1,3)$&$\cdots$\\ \hline
$8$            &$(4,4,3,8)$ &$(2,4,2,4)$&$(1,4,1,3)$&$\cdots$\\ \hline
$9$            &$(4,4,4,9)$ &$(2,4,2,4)$&$(1,4,1,4)$&$\cdots$\\ \hline
$10$           &$(5,5,4,10)$&$(2,5,2,5)$&$(1,5,2,4)$&$\cdots$\\ \hline
$\vdots$       &$\vdots$   &$\vdots$   &$\vdots$   &\\ \hline
\end{tabular}
\caption{
Values of $(m_{max}^{\text{(i)}},m_{max}^{\text{(ii)}},m_{max}^{\text{(iii)}},m_{max}^{\text{(iv)}})$ such that
only $(m^{(1)},m^{(2)})$ with $|m^{(1)}|,|m^{(2)}|\le m_{max}^{\text{(i)}}$, $(m^{(1)},m^{(2)}_i)$ with $|m^{(1)}|,|m^{(2)}_i|\le m_{max}^{\text{(ii)}}$, $(m^{(1)}_i,m^{(2)})$ with $|m^{(1)}_i|,|m^{(2)}|\le m_{max}^{\text{(iii)}}$, and $(m^{(1)}_i,m^{(2)}_j)$ with $|m^{(1)}_i|,|m^{(2)}_j|\le m_{max}^{\text{(iv)}}$ contribute to ${\cal I}^{U(2)_k\times U(2)_0\times U(2)_{-k}\text{ quiver CS}(H)}$ \eqref{220527_Higgsl2N12N22N32final} expanded to the order $\mathfrak{t}^p$.
}
\label{220527_21modelHiggstruncate}
\end{center}
\end{table}
From \eqref{220527_Higgsl2N12N22N32final} (and table \ref{220527_21modelHiggstruncate} for $k=1,2,3$) we obtain (we have set $x=y_1=y_2=y_3=1$ for simplicity)
\begin{align}
{\cal I}^{U(2)_1\times U(2)_0\times U(2)_{-1}\text{ quiver CS}(H)}&=1 + 2 \mathfrak{t} + 9 \mathfrak{t}^2 + 22 \mathfrak{t}^3 + 55 \mathfrak{t}^4 + 116 \mathfrak{t}^5 + 242 \mathfrak{t}^6 + 448 \mathfrak{t}^7\nonumber \\
&\quad  + 820 \mathfrak{t}^8 + 1400 \mathfrak{t}^9 + 2334 \mathfrak{t}^{10}+\cdots,\nonumber \\
{\cal I}^{U(2)_2\times U(2)_0\times U(2)_{-2}\text{ quiver CS}(H)}&=1 + 6 \mathfrak{t}^2 + 35 \mathfrak{t}^4 + 131 \mathfrak{t}^6 + 427 \mathfrak{t}^8 + 1151 \mathfrak{t}^{10}+\cdots,\nonumber \\
{\cal I}^{U(2)_3\times U(2)_0\times U(2)_{-3}\text{ quiver CS}(H)}&=1 + 4 \mathfrak{t}^2 + 2 \mathfrak{t}^3 + 14 \mathfrak{t}^4 + 16 \mathfrak{t}^5 + 40 \mathfrak{t}^6 + 58 \mathfrak{t}^7 + 112 \mathfrak{t}^8\nonumber \\
&\quad  + 166 \mathfrak{t}^9 + 288 \mathfrak{t}^{10}+\cdots,\nonumber \\
{\cal I}^{U(2)_4\times U(2)_0\times U(2)_{-4}\text{ quiver CS}(H)}&=
1 + 4 \mathfrak{t}^2 + 16 \mathfrak{t}^4 + 51 \mathfrak{t}^6 + 143 \mathfrak{t}^8 + 350 \mathfrak{t}^{10},\nonumber \\
{\cal I}^{U(2)_5\times U(2)_0\times U(2)_{-5}\text{ quiver CS}(H)}&=
1 
+ 4 \mathfrak{t}^{2}
+ 14 \mathfrak{t}^{4}
+ 2 \mathfrak{t}^{5}
+ 35 \mathfrak{t}^{6}
+ 16 \mathfrak{t}^{7}
+ 80 \mathfrak{t}^{8}
+ 58 \mathfrak{t}^{9}\nonumber \\
&\quad + 163 \mathfrak{t}^{10}
+\cdots.
\label{220527_21modelHiggslimitsmalltexpansion}
\end{align}
The result for $k=1$ is consistent with the Higgs limit of the index of the $U(2)$ ADHM theory with $l=2$ in \eqref{HS_2_su2inst}.
From the results including even higher order corrections, we also guess the following closed form expressions for the ${\cal I}^{U(2)_k\times U(2)_0\times U(2)_{-k}\text{ quiver CS}(H)}$ indices with $k=2,3,4,5$:
\begin{align}
&{\cal I}^{U(2)_2\times U(2)_0\times U(2)_{-2}\text{ quiver CS}(H)}\nonumber \\
&=\frac{
1 
+ 2 \mathfrak{t}^{2}
+ 13 \mathfrak{t}^{4}
+ 15 \mathfrak{t}^{6}
+ 28 \mathfrak{t}^{8}
+ 15 \mathfrak{t}^{10}
+ 13 \mathfrak{t}^{12}
+ 2 \mathfrak{t}^{14}
+ \mathfrak{t}^{16}
}{
(1 - \mathfrak{t}^2)^4(1 - \mathfrak{t}^4)^4
}\quad (p_{th}=33),\nonumber \\
&{\cal I}^{U(2)_3\times U(2)_0\times U(2)_{-3}\text{ quiver CS}(H)}\nonumber \\
&=\frac{1}{(1 - \mathfrak{t}^2)(1 - \mathfrak{t}^3)(1 - \mathfrak{t}^4)^2(1 - \mathfrak{t}^5)^3(1 - \mathfrak{t}^6)}
(
1 
+ 3 \mathfrak{t}^{2}
+ \mathfrak{t}^{3}
+ 8 \mathfrak{t}^{4}
+ 8 \mathfrak{t}^{5}
+ 17 \mathfrak{t}^{6}
+ 21 \mathfrak{t}^{7}
+ 33 \mathfrak{t}^{8}\nonumber \\
&\quad + 35 \mathfrak{t}^{9}
+ 51 \mathfrak{t}^{10}
+ 49 \mathfrak{t}^{11}
+ 63 \mathfrak{t}^{12}
+ 54 \mathfrak{t}^{13}
+ 63 \mathfrak{t}^{14}
+ 49 \mathfrak{t}^{15}
+ 51 \mathfrak{t}^{16}
+ 35 \mathfrak{t}^{17}
+ 33 \mathfrak{t}^{18}
+ 21 \mathfrak{t}^{19}\nonumber \\
&\quad + 17 \mathfrak{t}^{20}
+ 8 \mathfrak{t}^{21}
+ 8 \mathfrak{t}^{22}
+ \mathfrak{t}^{23}
+ 3 \mathfrak{t}^{24}
+ \mathfrak{t}^{26}
)\quad (p_{th}=39),\nonumber \\
&{\cal I}^{U(2)_4\times U(2)_0\times U(2)_{-4}\text{ quiver CS}(H)}\nonumber \\
&=
\frac{1}{(1 - \mathfrak{t}^2)(1 - \mathfrak{t}^4)^4(1 - \mathfrak{t}^6)^3}
(
1 
+ 3 \mathfrak{t}^{2}
+ 8 \mathfrak{t}^{4} 
+ 20 \mathfrak{t}^{6}
+ 41 \mathfrak{t}^{8}
+ 61 \mathfrak{t}^{10}
+ 78 \mathfrak{t}^{12}
+ 84 \mathfrak{t}^{14}
+ 78 \mathfrak{t}^{16}\nonumber \\
&\quad + 61 \mathfrak{t}^{18}
+ 41 \mathfrak{t}^{20}
+ 20 \mathfrak{t}^{22}
+ 8 \mathfrak{t}^{24}
+ 3 \mathfrak{t}^{26}
+ \mathfrak{t}^{28}
)\quad (p_{th}=45),\nonumber \\
&{\cal I}^{U(2)_5\times U(2)_0\times U(2)_{-5}\text{ quiver CS}(H)}\nonumber \\
&=
\frac{1}{(1 - \mathfrak{t}^2)(1 - \mathfrak{t}^4)^2(1 - \mathfrak{t}^5)(1 - \mathfrak{t}^7)^3(1 - \mathfrak{t}^{10})}
(
1 
+ 3 \mathfrak{t}^{2}
+ 8 \mathfrak{t}^{4}
+ \mathfrak{t}^{5}
+ 15 \mathfrak{t}^{6}
+ 8 \mathfrak{t}^{7}
+ 26 \mathfrak{t}^{8}\nonumber \\
&\quad + 21 \mathfrak{t}^{9}
+ 41 \mathfrak{t}^{10}
+ 35 \mathfrak{t}^{11}
+ 63 \mathfrak{t}^{12}
+ 51 \mathfrak{t}^{13}
+ 87 \mathfrak{t}^{14}
+ 65 \mathfrak{t}^{15}
+ 105 \mathfrak{t}^{16}
+ 79 \mathfrak{t}^{17}
+ 111 \mathfrak{t}^{18}\nonumber \\
&\quad + 84 \mathfrak{t}^{19}
+ 111 \mathfrak{t}^{20}
+ 79 \mathfrak{t}^{21}
+ 105 \mathfrak{t}^{22}
+ 65 \mathfrak{t}^{23}
+ 87 \mathfrak{t}^{24}
+ 51 \mathfrak{t}^{25}
+ 63 \mathfrak{t}^{26}
+ 35 \mathfrak{t}^{27}
+ 41 \mathfrak{t}^{28}\nonumber \\
&\quad + 21 \mathfrak{t}^{29}
+ 26 \mathfrak{t}^{30}
+ 8 \mathfrak{t}^{31}
+ 15 \mathfrak{t}^{32}
+ \mathfrak{t}^{33}
+ 8 \mathfrak{t}^{34}
+ 3 \mathfrak{t}^{36}
+ \mathfrak{t}^{38}
)\quad (p_{th}=45),
\label{21modelHiggsN2closedconjecture}
\end{align}
where $p_{th}$ indicates that each expression is confirmed up to the order $\mathfrak{t}^{p_{th}+1}$.

In the same way we obtain for $N=3$ the following results for $x=y_1=y_2=y_3=1$
\begin{align}
&{\cal I}^{U(3)_1\times U(3)_0\times U(3)_{-1}\text{ quiver CS}(H)}\nonumber \\
&= 1 + 2 \mathfrak{t} + 9 \mathfrak{t}^2 + 26 \mathfrak{t}^3 + 73 \mathfrak{t}^4 + 178 \mathfrak{t}^5 + 430 \mathfrak{t}^6 + 940 \mathfrak{t}^7 + 
 1998 \mathfrak{t}^8 + 4008 \mathfrak{t}^9+\cdots,\nonumber \\
&{\cal I}^{U(3)_2\times U(3)_0\times U(3)_{-2}\text{ quiver CS}(H)}\nonumber \\
& = 1 + 6 \mathfrak{t}^{2} + 35 \mathfrak{t}^{4} + 162 \mathfrak{t}^{6} + 636 \mathfrak{t}^{8} + 2193 \mathfrak{t}^{10} + 6768 \mathfrak{t}^{12} + 18989 \mathfrak{t}^{14} + 49143 \mathfrak{t}^{16}\nonumber \\
&\quad  + 118565 \mathfrak{t}^{18}+\cdots,\nonumber \\
&{\cal I}^{U(3)_3\times U(3)_0\times U(3)_{-3}\text{ quiver CS}(H)}\nonumber \\
& = 1 + 4 \mathfrak{t}^{2} + 2 \mathfrak{t}^{3} + 14 \mathfrak{t}^{4} + 16 \mathfrak{t}^{5} + 45 \mathfrak{t}^{6} + 68 \mathfrak{t}^{7} + 144 \mathfrak{t}^{8} + 232 \mathfrak{t}^{9} + 438 \mathfrak{t}^{10} + 696 \mathfrak{t}^{11} + 1228 \mathfrak{t}^{12}\nonumber \\
&\quad + 1922 \mathfrak{t}^{13} + 3191 \mathfrak{t}^{14} + 4916 \mathfrak{t}^{15} + 7781 \mathfrak{t}^{16} + 11744 \mathfrak{t}^{17} + 17925 \mathfrak{t}^{18} + 26450 \mathfrak{t}^{19}+\cdots,\nonumber \\
&{\cal I}^{U(3)_4\times U(3)_0\times U(3)_{-4}\text{ quiver CS}(H)}\nonumber \\
& = 1 + 4 \mathfrak{t}^{2} + 16 \mathfrak{t}^{4} + 56 \mathfrak{t}^{6} + 173 \mathfrak{t}^{8} + 493 \mathfrak{t}^{10} + 1308 \mathfrak{t}^{12} + 3236 \mathfrak{t}^{14} + 7563 \mathfrak{t}^{16} + 16773 \mathfrak{t}^{18}+\cdots,\nonumber \\
&{\cal I}^{U(3)_5\times U(3)_0\times U(3)_{-5}\text{ quiver CS}(H)}\nonumber \\
& = 1 + 4 \mathfrak{t}^{2} + 14 \mathfrak{t}^{4} + 2 \mathfrak{t}^{5} + 40 \mathfrak{t}^{6} + 16 \mathfrak{t}^{7} + 100 \mathfrak{t}^{8} + 68 \mathfrak{t}^{9} + 232 \mathfrak{t}^{10} + 222 \mathfrak{t}^{11} + 523 \mathfrak{t}^{12} + 608 \mathfrak{t}^{13}\nonumber \\
&\quad  + 1157 \mathfrak{t}^{14} + 1478 \mathfrak{t}^{15} + 2509 \mathfrak{t}^{16} + 3310 \mathfrak{t}^{17} + 5281 \mathfrak{t}^{18} + 7014 \mathfrak{t}^{19}+\cdots,\nonumber \\
&{\cal I}^{U(3)_6\times U(3)_0\times U(3)_{-6}\text{ quiver CS}(H)}\nonumber \\
& = 1 + 4 \mathfrak{t}^{2} + 14 \mathfrak{t}^{4} + 42 \mathfrak{t}^{6} + 116 \mathfrak{t}^{8} + 295 \mathfrak{t}^{10} + 706 \mathfrak{t}^{12} + 1598 \mathfrak{t}^{14} + 3454 \mathfrak{t}^{16} + 7150 \mathfrak{t}^{18}+\cdots,
\label{21kHiggs_N3_tonly}
\end{align}
up to the order $\mathfrak{t}^{9+1}$ for $k=1$ and the order $\mathfrak{t}^{19+1}$ for $k=2,3,4,5,6$.

For $k=2$ and $N=2,3$, we obtain the following results before taking $x=y_1=y_2=y_3=1$:
\begin{align}
&{\cal I}^{U(2)_2\times U(2)_0\times U(2)_{-2}\text{ quiver CS}(H)}\nonumber \\
&=1
+ (2 + y_2^{-1} + y_2 + y_1^{-1}y_2^{-1}y_3^{-1}x^{-2} + x^2 y_1 y_2 y_3)\mathfrak{t}^2
+ [7 + y_2^{-2} + 3y_2^{-1} + 3 y_2+ y_2^2 \nonumber \\
&\quad  + (2 + 2y_2^{-2}+4y_2^{-1})y_1^{-1}y_3^{-1}x^{-2} + 2y_1^{-2}y_2^{-2}y_3^{-2}x^{-4} + 2 x^4 y_1^2 y_2^2 y_3^2\nonumber \\
&\quad  + (2 + 4 y_2 + 2 y_2^2)y_1 y_3 x^2]\mathfrak{t}^4
 + [15 + y_2^{-3} + 3y_2^{-2} + 10y_2^{-1} + 10 y_2 + 3 y_2^2 + y_2^3\nonumber \\
&\quad  + (3y_2^{-3} + 6y_2^{-2} + 3y_2^{-1})y_1^{-2}y_3^{-2}x^{-4} + ( 7 + 2y_2^{-3} + 7y_2^{-2} + 12y_2^{-1} + 2 y_2)y_1^{-1}y_3^{-1}x^{-2}\nonumber \\
&\quad  + 2y_1^{-3}y_2^{-3}y_3^{-3}x^{-6} + 2 x^6 y_1^3 y_2^3 y_3^3 + (7 + 2 y_2^{-2} + 12 y_2 + 7 y_2^2 + 2 y_2^3) y_1 y_3x^2\nonumber \\
&\quad  + (3 y_2  + 6 y_2^2 + 3 y_2^3)y_1^2 y_3^2x^4]\mathfrak{t}^6+\cdots,\nonumber \\
%
&{\cal I}^{U(3)_2\times U(3)_0\times U(3)_{-2}\text{ quiver CS}(H)}\nonumber \\
&=
1
+ (2 + y_2^{-1} + y_2 + y_1^{-1}y_2^{-1}y_3^{-1}x^{-2} + x^2 y_1 y_2 y_3)\mathfrak{t}^2
+ [7 + y_2^{-2} + 3y_2^{-1} + 3 y_2 + y_2^2\nonumber \\
&\quad  + (2 + 2y_2^{-2} + 4y_2^{-1})y_1^{-1}y_3^{-1}x^{-2} + 2y_1^{-2}y_2^{-2}y_3^{-2}x^{-4} + 2 x^4 y_1^2 y_2^2 y_3^2\nonumber \\
&\quad  + (2 + 4 y_2 + 2 y_2^2)y_1y_3x^2]\mathfrak{t}^4
+ [20 + y_2^{-3} + 3y_2^{-2} + 12y_2^{-1} + 12 y_2 + 3 y_2^2 + y_2^3\nonumber \\
&\quad  + (4y_2^{-3} + 8y_2^{-2} + 4y_2^{-1})y_1^{-2}y_3^{-2}x^{-4} + ( 8 + 2y_2^{-3} + 8y_2^{-2} + 16y_2^{-1} + 2 y_2)y_1^{-1}y_3^{-1}x^{-2}\nonumber \\
&\quad  + 3y_1^{-3}y_2^{-3}y_3^{-3}x^{-6} + 3 x^6 y_1^3 y_2^3 y_3^3 + (8 + 2 y_2^{-1} + 16 y_2 + 8 y_2^2 + 2 y_2^3 )y_1 y_3x^2\nonumber \\
&\quad + (4 y_2 + 8 y_2^2 + 4 y_2^3)y_1^2 y_3^2x^4 ]\mathfrak{t}^6
+\cdots,
\label{21kHiggs}
\end{align}
which are consistent with the indices of $O(2N)$ theory with an antisymmetric hypermultiplet and a fundamental hypermultiplets with $N=2,3$ \eqref{O4++_A_2_index},\eqref{O6++_A_2_findex}, with the parameter identification \eqref{O2Nantisym_212model_parameteridentification}.
\section{BLG theory}
\label{sec_BLG}
The BLG theories are 3d $\mathcal{N}=8$ Chern-Simons matter theories with $\mathfrak{so}(4)$ gauge algebra and Chern-Simons level $k\in \mathbb{Z}$ constructed in terms of Lie 3-algebra \cite{Bagger:2006sk,Bagger:2007jr,Bagger:2007vi,Gustavsson:2007vu,Gustavsson:2008dy}.  
They are two families of theories where one has gauge group $G=SU(2)\times SU(2)$ and the other has $G=(SU(2)\times SU(2))/\mathbb{Z}_2$ \cite{VanRaamsdonk:2008ft, Aharony:2008ug,Lambert:2010ji}. 

\subsection{Moduli spaces and local operators}
In the BLG model a bare monopole operator $v^{m^{(1)};m^{(2)}}$ has the conformal dimension 
\begin{align}
\Delta(m^{(1)}, m^{(2)})
&=-2|m^{(1)}|-2|m^{(2)}|+2|m^{(1)}-m^{(2)}|+2|m^{(1)}+m^{(2)}|,
\end{align}
where 
\begin{align}
\begin{cases}
m^{(1)}, m^{(2)}\in \mathbb{Z}&G=SU(2)\times SU(2)\cr
m^{(1)}, m^{(2)}\in \mathbb{Z}/2&G=(SU(2)\times SU(2))/\mathbb{Z}_2\cr
\end{cases}
\end{align}
 are the magnetic fluxes. 
 
 The moduli space is given by \eqref{moduliBLG1} or \eqref{moduliBLG2} \cite{Lambert:2008et,Distler:2008mk}
\begin{align}
\mathcal{M}_{\textrm{BLG}}&=
\begin{cases}
(\mathbb{C}^4\times \mathbb{C}^4)/\mathbb{D}_{4k}&G=SU(2)_{k}\times SU(2)_{-k}\cr
(\mathbb{C}^4\times \mathbb{C}^4)/\mathbb{D}_{2k}&G=SU(2)_{k}\times SU(2)_{-k}/\mathbb{Z}_2\cr
\end{cases}.
\end{align}
In particular, for $(SU(2)_{1}\times SU(2)_{-1})/\mathbb{Z}_2$, $SU(2)_{2}\times SU(2)_{-2}$ and $(SU(2)_{4}\times SU(2)_{-4})/\mathbb{Z}_2$ 
the moduli spaces are identified with 
\begin{align}
\mathcal{M}_{\textrm{$(SU(2)_{1}\times SU(2)_{-1})/\mathbb{Z}_2$ BLG}}
&=\mathrm{Sym}^2 (\mathbb{C}^4),\\
\mathcal{M}_{\textrm{$SU(2)_{2}\times SU(2)_{-2}$ BLG}}
&=\mathrm{Sym}^2 (\mathbb{C}^4/\mathbb{Z}_2),\\
\mathcal{M}_{\textrm{$(SU(2)_{4}\times SU(2)_{-4})/\mathbb{Z}_2$ BLG}}
&=\mathrm{Sym}^2 (\mathbb{C}^4/\mathbb{Z}_2),
\end{align}
which have the conjectural geometrical interpretation of two M2-branes. 
The difference between the $SU(2)_{2}\times SU(2)_{-2}$ and $(SU(2)_{4}\times SU(2)_{-4})/\mathbb{Z}_2$ BLG theories is expected to come from 
the absence or presence of discrete torsion for the background 4-form. 
 
\subsection{Indices}
The index of the BLG theory of level $k\in \mathbb{Z}$ is computed in \cite{Bashkirov:2011pt,Honda:2012ik,Agmon:2017lga,Tachikawa:2019dvq}. 
We find a simple equality which indicates a duality associated to the $SU(2)_1\times SU(2)_{-1}$ BLG theory. 
For completeness, we also show the known equalities of indices in our convention. 

In terms of the definition (\ref{INDEX_def}), the BLG index can be evaluated as 
\begin{align}
\label{blg_index}
&I^{\textrm{BLG}}(t,x,z;q)
\nonumber\\
&=\frac14 \sum_{m^{(1)}, m^{(2)}}
\prod_{I=1}^2 
\oint \frac{ds^{(I)}}{2\pi is^{(I)}}
(1-q^{|m^{(I)}|} s^{(I)\pm 2})
(s^{(1)})^{2km^{(1)}} (s^{(2)})^{-2km^{(2)}}
\nonumber\\
&\times 
\frac{
(q^{\frac34+\frac{|m^{(1)}-m^{(2)}|}{2}}t^{-1}s^{(1)\mp}s^{(2)\pm}x^{\mp};q)_{\infty} }
{(q^{\frac14+\frac{|m^{(1)}-m^{(2)}|}{2}}ts^{(1)\pm}s^{(2)\mp}x^{\pm};q)_{\infty} }
\frac{
(q^{\frac34+\frac{|m^{(1)}-m^{(2)}|}{2}}t^{-1}s^{(1)\mp}s^{(2)\pm}x^{\pm};q)_{\infty} }
{(q^{\frac14+\frac{|m^{(1)}-m^{(2)}|}{2}}ts^{(1)\pm}s^{(2)\mp}x^{\mp};q)_{\infty} }
\nonumber\\
&\times 
\frac{
(q^{\frac34+\frac{|m^{(1)}+m^{(2)}|}{2}}t^{-1}s^{(1)\mp}s^{(2)\mp}x^{\mp};q)_{\infty} }
{(q^{\frac14+\frac{|m^{(1)}+m^{(2)}|}{2}}ts^{(1)\pm}s^{(2)\pm}x^{\pm};q)_{\infty} }
\frac{
(q^{\frac34+\frac{|m^{(1)}+m^{(2)}|}{2}}t^{-1}s^{(1)\mp}s^{(2)\mp}x^{\pm};q)_{\infty} }
{(q^{\frac14+\frac{|m^{(1)}+m^{(2)}|}{2}}ts^{(1)\pm}s^{(2)\pm}x^{\mp};q)_{\infty} }
\nonumber\\
&\times 
\frac{
(q^{\frac34+\frac{|m^{(1)}-m^{(2)}|}{2}}ts^{(1)\mp}s^{(2)\pm}z^{\mp};q)_{\infty} }
{(q^{\frac14+\frac{|m^{(1)}-m^{(2)}|}{2}}t^{-1}s^{(1)\pm}s^{(2)\mp}z^{\pm};q)_{\infty} }
\frac{
(q^{\frac34+\frac{|m^{(1)}-m^{(2)}|}{2}}ts^{(1)\mp}s^{(2)\pm}z^{\pm};q)_{\infty} }
{(q^{\frac14+\frac{|m^{(1)}-m^{(2)}|}{2}}t^{-1}s^{(1)\pm}s^{(2)\mp}z^{\mp};q)_{\infty} }
\nonumber\\
&\times 
\frac{
(q^{\frac34+\frac{|m^{(1)}+m^{(2)}|}{2}}ts^{(1)\mp}s^{(2)\mp}z^{\mp};q)_{\infty} }
{(q^{\frac14+\frac{|m^{(1)}+m^{(2)}|}{2}}t^{-1}s^{(1)\pm}s^{(2)\pm}z^{\pm};q)_{\infty} }
\frac{
(q^{\frac34+\frac{|m^{(1)}+m^{(2)}|}{2}}ts^{(1)\mp}s^{(2)\mp}z^{\pm};q)_{\infty} }
{(q^{\frac14+\frac{|m^{(1)}+m^{(2)}|}{2}}t^{-1}s^{(1)\pm}s^{(2)\pm}z^{\mp};q)_{\infty} }
\nonumber\\
&\times 
q^{-|m^{(1)}|-|m^{(2)}|+|m^{(1)}-m^{(2)}|+|m^{(1)}+m^{(2)}| },
\end{align}
where the magnetic fluxes $m^{(1)}$ and $m^{(2)}$ are summed 
over integers and half-integers for the $SU(2)\times SU(2)$ BLG and the $(SU(2)\times SU(2))/\mathbb{Z}_2$ BLG respectively. 

\subsubsection{$SU(2)_{1}\times SU(2)_{-1}$ BLG}
The flavored index of the $SU(2)_{2}\times SU(2)_{-2}$ BLG theory is evaluated as
\begin{align}
\label{blgk1_findex}
&I^{\textrm{$SU(2)_{1}\times SU(2)_{-1}$ BLG}}(t,x,z;q)
\nonumber\\
&=1+2\Bigl(
(x+x^{-1})(z+z^{-1})+t^2(1+x^2+x^{-2})
+t^{-2}(1+z^2+z^{-2})
\Bigr)q^{1/2}
\nonumber\\
&+\Bigl(
1+4(x^2+x^{-2}+z^{2}+z^{-2})+5(x^2z^2+x^{-2}z^{-2}+5x^2z^{-2}+5x^{-2}z^2
\nonumber\\
&+t^4(5+3x^4+3x^{-4}+4x^2+4x^{-2})
+4t^2(x^3+x+x^{-1}+x^{-3})(z+z^{-1})
\nonumber\\
&+t^{-4}(5+3z^4+3z^{-4}+4z^2+4z^{-2})
+4t^{-2}(z^3+z+z^{-1}+z^{-3})(x+x^{-1})
\Bigr)q+\cdots.
\end{align}
When $x=z=1$, the index for the $SU(2)_{2}\times SU(2)_{-2}$ BLG theory reduces to
\begin{align}
\label{blgk1_index}
&I^{\textrm{$SU(2)_{1}\times SU(2)_{-1}$ BLG}}(t,x=1,z=1;q)
\nonumber\\
&=1+(8+6t^2+6t^{-2})q^{1/2}
+(37+19t^2+32t^2+32t^{-2}+19t^{-4})q
\nonumber\\
&+(64+44t^6+72t^4+70t^2+70t^{-2}+72t^{-4}+44t^{-6})q^{3/2}
\nonumber\\
&+(116+85t^8+128t^{6}+102t^4+104t^2+(t\rightarrow t^{-1}))q^2+\cdots.
\end{align}
Here we find that the flavored index \eqref{blgk1_findex} obeys a relation (see \eqref{abjmu1k2_findex}, \eqref{abju2u1_findex})
\begin{align}
&I^{\textrm{$SU(2)_{1}\times SU(2)_{-1}$ BLG}}(t;x;z;q)
\nonumber\\
&=
I^{\textrm{$U(2)_{2}\times U(1)_{-2}$ABJ}}(t,x,y=1,z;q)\times 
I^{\textrm{$U(1)_{2}\times U(1)_{-2}$ABJM}}(t,x,y=1,z;q). 
\end{align}
Note that the redundancies of the ABJ(M) indices are fixed by setting the topological fugacities to unity as in (\ref{ABJMremoveredundancy}). 
Accordingly, we conjecture a duality
\begin{align}
&\textrm{$SU(2)_1 \times SU(2)_{-1}$ BLG}
\nonumber\\
&\Leftrightarrow 
\textrm{$U(2)_{2}\times U(1)_{-2}$ ABJ} \otimes 
\textrm{$U(1)_{2}\times U(1)_{-2}$ ABJM}.
\end{align}

In the Coulomb and Higgs limits the index (\ref{blgk1_index}) reduces to  
\begin{align}
\mathcal{I}^{\textrm{$SU(2)_{1}\times SU(2)_{-1}$ BLG}(C)}(\mathfrak{t})&=\mathcal{I}^{\textrm{$SU(2)_{1}\times SU(2)_{-1}$ BLG}(H)}(\mathfrak{t})
=\frac{(1+\mathfrak{t}^2)^2}{(1-\mathfrak{t}^2)^4},
\end{align}
which agrees with \eqref{molienv2.D2n_rho2} with $n=2$. 
\subsubsection{$SU(2)_{2}\times SU(2)_{-2}$ BLG}
The flavored index for the $SU(2)_{2}\times SU(2)_{-2}$ BLG theory is equal to the flavored index (\ref{abjmu2k2_findex}). 
This demonstrates that they are dual to each other, mentioned in \eqref{BLGdual2} \cite{Lambert:2010ji,Bashkirov:2011pt}. 
They describe two M2-branes probing $\mathbb{C}^2/\mathbb{Z}_2$. 

\subsubsection{$SU(2)_{3}\times SU(2)_{-3}$ BLG}
For the $SU(2)_{3}\times SU(2)_{-3}$ BLG model we have the 
flavored index
\begin{align}
\label{blgk3_findex}
&I^{\textrm{$SU(2)_{3}\times SU(2)_{-3}$ BLG}}(t,x,z;q)
\nonumber\\
&=1+\Bigl(
xz+x^{-1}z^{-1}+xz^{-1}+x^{-1}z
+(1+x^2+x^{-2})t^2
+(1+z^2+z^{-2})t^{-2}
\Bigr)q^{1/2}
\nonumber\\
&+\Bigl(
x^2+x^{-2}+z^2+z^{-2}+2(x^2z^2+x^{-2}z^{-2}+x^2z^{-2}+x^{-2}z^2)
\nonumber\\
&+(2+x^4+x^{-4}+x^2+x^{-2})t^4
+(x^3z+x^{-3}z^{-1}+x^3z^{-1}+x^{-3}z
\nonumber\\
&+xz+x^{-1}z^{-1}+xz^{-1}+x^{-1}z
)t^2
+(xz^3+x^{-1}z^{-3}+xz^{-3}+x^{-1}z^3
\nonumber\\
&+xz+x^{-1}z^{-1}+xz^{-1}+x^{-1}z)t^{-2}
+(2+z^4+z^{-4}+z^2+z^{-2})t^{-4}
\Bigr)q+\cdots.
\end{align}
For $x=z=1$ it is simplified as
\begin{align}
\label{blgk3_index}
&I^{\textrm{$SU(2)_{3}\times SU(2)_{-3}$ BLG}}(t,x=1,z=1;q)
\nonumber\\
&=1+(4+3t^2+3t^{-2})q^{1/2}
+(12+6t^4+8t^2+8t^{-2}+6t^{-4})q
\nonumber\\
&+(24+17t^6+24t^4+27t^2+27t^{-2}+24t^{-4}+17t^{-6})q^{3/2}
+\cdots.
\end{align}
In the Coulomb and Higgs limits the index (\ref{blgk3_index}) gives
\begin{align}
\mathcal{I}^{\textrm{$SU(2)_{3}\times SU(2)_{-3}$ BLG}(C)}(\mathfrak{t})&=\mathcal{I}^{\textrm{$SU(2)_{3}\times SU(2)_{-3}$ BLG}(H)}(\mathfrak{t})
=\frac{1+\mathfrak{t}^2+\mathfrak{t}^4+6\mathfrak{t}^6+\mathfrak{t}^8+\mathfrak{t}^{10}+\mathfrak{t}^{12}}
{(1+\mathfrak{t}^2+\mathfrak{t}^4)^2 (1-\mathfrak{t}^2)^4},
\end{align}
which agrees with \eqref{molienv2.D2n_rho2} with $n=6$.

\subsubsection{$(SU(2)_{1}\times SU(2)_{-1})/\mathbb{Z}_2$ BLG}
The flavored index of the $(SU(2)_{1}\times SU(2)_{-1})/\mathbb{Z}_2$ BLG theory agrees with 
the flavored index (\ref{u2_1_findex}) for the $U(2)$ ADHM theory with one flavor or equivalently the $U(2)_1\times U(2)_{-1}$ ABJM theory \eqref{abjmu2k1_findex} with $y=1$.
This reflects the duality \eqref{BLGdual1} \cite{Lambert:2010ji,Bashkirov:2011pt}. 
They capture two M2-branes moving in $\mathbb{C}^2$. 

\subsubsection{$(SU(2)_{3}\times SU(2)_{-3})/\mathbb{Z}_2$ BLG}
For the $(SU(2)_{3}\times SU(2)_{-3})/\mathbb{Z}_2$ BLG model we have the flavored index 
\begin{align}
\label{blgk3z2_findex}
&I^{\textrm{$(SU(2)_{3}\times SU(2)_{-3})/\mathbb{Z}_2$ BLG}}(t,x,z;q)
\nonumber\\
&=1+\Bigl(
xz+x^{-1}z^{-1}+xz^{-1}+x^{-1}z
+(1+x^2+x^{-2})t^2
+(1+z^2+z^{-2})t^{-2}
\Bigr)q^{1/2}
\nonumber\\
&+\Bigl(
(x^3+x^{-3}+x+x^{-1})t^3
+(z+z^{-1}+x^2z+x^{-2}z^{-1}+x^{2}z^{-1}+x^{-2}z)t
\nonumber\\
&+(x+x^{-1}+xz^2+x^{-1}z^{-2}+xz^{-2}+x^{-1}z^2)t^{-1}
+(z^3+z^{-3}+z+z^{-1})t^{-3}
\Bigr)q^{3/4}+\cdots.
\end{align}
When the fugacities $x$ and $z$ are taken to unity, we have 
\begin{align}
\label{blgk3z2_index}
&I^{\textrm{$(SU(2)_{3}\times SU(2)_{-3})/\mathbb{Z}_2$ BLG}}(t,x=1,z=1;q)
\nonumber\\
&=1+(4+3t^2+3t^{-2})q^{1/2}
+(4t^3+6t+6t^{-1}+4t^{-3})q^{3/4}
\nonumber\\
&+(12+6t^4+8t^2+8t^{-2}+6t^{-4})q+\cdots.
\end{align}
The Coulomb and Higgs limits of the index (\ref{blgk3z2_index}) are
\begin{align}
&\mathcal{I}^{\textrm{$(SU(2)_{3}\times SU(2)_{-3})/\mathbb{Z}_2$ BLG}(C)}(\mathfrak{t})
=\mathcal{I}^{\textrm{$(SU(2)_{3}\times SU(2)_{-3})/\mathbb{Z}_2$ BLG}(H)}(\mathfrak{t})
\nonumber\\
&=\frac{1+\mathfrak{t}^2+2\mathfrak{t}^3+\mathfrak{t}^4+\mathfrak{t}^6}
{(1+\mathfrak{t})^2 (1+\mathfrak{t}+\mathfrak{t}^2)^2 (1-\mathfrak{t})^4},
\end{align}
which agrees with \eqref{molienv2.D2n_rho2} with $n=3$. 
From \eqref{u3_1_findex}, \eqref{abjmu1k1_findex} and \eqref{blgk3z2_findex}, 
we have 
\begin{align}
\label{blgk3z2_identity}
&I^{\textrm{$U(3)_{1}\times U(3)_{-1}$ABJM}}(t,x,1,z;q)
\nonumber\\
&=
I^{\textrm{$(SU(2)_{3}\times SU(2)_{-3})/\mathbb{Z}_2$ BLG}}(t,x,z;q)
\times 
I^{\textrm{$U(1)_{1}\times U(1)_{-1}$ABJM}}(t,x,1,z;q), 
\end{align}
which implies the duality \eqref{ACPdual}. 
This generalizes the identity of the indices in \cite{Agmon:2017lga} in such a way that (\ref{blgk3z2_identity}) reduces to it when $x=z=1$. 
While the $U(1)_{1}\times U(1)_{-1}$ ABJM describes a center of motion of three M2-branes, 
the $(SU(2)_3 \times SU(2)_{-3})/\mathbb{Z}_2$ BLG model describes an interacting sector \cite{Agmon:2017lga}. 

\subsubsection{$(SU(2)_{4}\times SU(2)_{-4})/\mathbb{Z}_2$ BLG}
The index of the $(SU(2)_{4}\times SU(2)_{-4})/\mathbb{Z}_{2}$ BLG theory coincides with the flavored index (\ref{abju3u2k2_findex}) of the $U(3)_{2}\times U(2)_{-2}$ ABJ theory for $y=1$ as we have the duality \eqref{BLGdual3} \cite{Bashkirov:2011pt}.
They capture two M2-branes probing $\mathbb{C}^2/\mathbb{Z}_2$ 
in the presence one unit of discrete torsion for 4-form flux. 


\subsection*{Acknowledgements}
The authors would like to thank Dongmin Gang, Naotaka Kubo, Kimyeong Lee, Sungjay Lee and Keita Nii for useful discussions and comments. 
The work of H.H. is supported in part by JSPS KAKENHI Grant Number JP18K13543. The work of T.O. was supported by KIAS Individual Grants (PG084301) at Korea Institute for Advanced Study. 

\appendix

\section{3d supersymmetric indices}
\label{app_notation}
The supersymmetric index of 3d supersymmetric field theory can be defined as a trace over the Hilbert space on $S^2$. 
We use the definition in \cite{Okazaki:2019ony} \footnote{This definition is also compatible with the half- and quarter-indices of 4d $\mathcal{N}=4$ SYM theory studied in \cite{Gaiotto:2019jvo} including the half-indices of 3d $\mathcal{N}=4$ gauge theories analyzed in \cite{Okazaki:2019bok,Okazaki:2020lfy}. } for the supersymmetric index of 3d $\mathcal{N}=4$ supersymmetric field theory
\begin{align}
\label{INDEX_def}
I(t,x;q)
&:={\Tr}_{\mathrm{Op}}(-1)^{F}q^{J+\frac{H+C}{4}}t^{H-C} x^{f},
\end{align}
as a trace over the cohomology of the preserved supercharges. 
Here we have introduced $F$ as the Fermion number operator, $J$ as the generator of the $U(1)_{J}$ rotational symmetry in the space-time, 
$H$ and $C$ as the Cartan generators of the $SU(2)_{H}$ and $SU(2)_{C}$ R-symmetry groups, 
$f$ as the Cartan generator of the global symmetry. 

The index can be calculated from the UV data via the localization \cite{Kim:2009wb,Imamura:2011su,Kapustin:2011jm}. 
It takes the following form: 
\begin{align}
&
I^{\textrm{3d $G$}}(t,x_H,x_C;q)
\nonumber\\
&=
\frac{1}{| \textrm{Weyl} (G)|}
\frac{
(q^{\frac12}t^2;q)_{\infty}^{\mathrm{rank}(G)}
}
{
(q^{\frac12}t^{-2};q)_{\infty}^{\mathrm{rank}(G)}
}
\sum_{m\in \mathrm{cochar}(G)}
\oint \prod_{\alpha\in \mathrm{roots} (G) }
\frac{ds}{2\pi is}
\frac{
\left(1-q^{\frac{|m\cdot \alpha|}{2}}s^{\alpha} \right)
\left(q^{\frac{1+|m\cdot \alpha|}{2}} t^2 s^{\alpha};q\right)_{\infty}
}
{
\left(q^{\frac{1+|m\cdot \alpha|}{2}} t^{-2} s^{\alpha};q\right)_{\infty}
}
\nonumber\\
&\times 
\prod_{\lambda \in {\bf R}}
\frac{
\left( q^{\frac34+\frac{|m\cdot \lambda|}{2}} t^{-1} s^{\lambda} x_{H};q \right)_{\infty}
}
{
\left( q^{\frac14+\frac{|m\cdot \lambda|}{2}} t s^{\lambda} x_{H};q \right)_{\infty}
}
q^{\frac{\Delta(m)}{2}}
\cdot 
t^{-2\Delta(m)}
\cdot 
x_{C}^{m}. 
\end{align}
The second line comes from the contribution of the $\mathcal{N}=4$ vector multiplet of gauge group $G$. 
The third line contains the contribution from the hypermultiplets transforming as representation ${\bf R}$ of the  gauge group $G$ as well as that from the monopole operators of dimension $\Delta(m)$ where $m\in \mathrm{cochar}(G)$ is a magnetic flux carried by the monopoles. 
The fugacities $x_H$ are coupled to the flavor symmetry, or the Higgs branch symmetry that rotates hypermultiplets. 
The fugacities $x_C$ are associated to the topological symmetry, or the Coulomb branch symmetry. 

The fugacity is fixed so that the power of $q$ is always strictly positive for a non-trivial local operator according to a unitarity bound. 
Therefore the index (\ref{INDEX_def}) is a formal power series in $q$ whose coefficients count the local operators as Laurent polynomials in the other fugacities. 

We have introduced the following notation by defining $q$-shifted factorial
\begin{align}
\label{qpoch_def}
(a;q)_{0}&:=1,\qquad
(a;q)_{n}:=\prod_{k=0}^{n-1}(1-aq^{k}),\qquad 
(q)_{n}:=\prod_{k=1}^{n}(1-q^{k}),\quad 
\quad  n\ge1,
\nonumber \\
(a;q)_{\infty}&:=\prod_{k=0}^{\infty}(1-aq^{k}),\qquad 
(q)_{\infty}:=\prod_{k=1}^{\infty} (1-q^k), 
\nonumber\\
(a^{\pm};q)_{\infty}&:=(a;q)_{\infty}(a^{-1};q)_{\infty},
\end{align}
where $a$ and $q$ are complex variables with $|q|<1$. 

The introduction of the fugacity $t$ for the R-charges allows us to study various aspects of the BPS local operators in the theories. 
For example, the 3d $\mathcal{N}=4$ index (\ref{INDEX_def}) can reduces to the Coulomb (resp. Higgs) branch Hilbert series in the Coulomb (resp. Higgs) limit \cite{Razamat:2014pta}
\begin{align}
\label{HS_lim}
\mathcal{I}^{(C)}(\mathfrak{t},x)=\lim_{\begin{smallmatrix}\mathfrak{t}=q^{1/4}t^{-1}:\textrm{fixed},\\q\rightarrow 0\end{smallmatrix}} I(t,x;q),\quad
\mathcal{I}^{(H)}(\mathfrak{t},x)=\lim_{\begin{smallmatrix}\mathfrak{t}=q^{1/4}t:\textrm{fixed},\\q\rightarrow 0\end{smallmatrix}} I(t,x;q). 
\end{align}
They can count the Coulomb (reps. Higgs) branch operators in the theory as generators of chiral rings of holomorphic functions on the Coulomb (resp. Higgs) branch. 
Also we can count the number of mixed branch operators from the coefficients of the term with $q^{n} t^m$ with $m\neq \pm 4n$.  

\section{Counting operator contents of indices by auxiliary dressing}
\label{app_auxdres}
\subsection{$U(N)$ ADHM theory with $l$ flavor}
One may consider the following integration
\begin{align}
&I^{U(N)\,\text{ADHM-}[l]}_{\text{aux.~dres}}(m_i)\nonumber \\
&=
\frac{1}{N!}\sum_{m_1,\cdots,m_N\in\mathbb{Z}}
\prod_{i=1}^N\frac{ds_i}{2\pi is_i}\prod_{i\neq j}^N\Bigl(1-q^{\frac{|m_i-m_j|}{2}}\frac{s_i}{s_j}A_{ij}\Bigr)\nonumber \\
&\quad \prod_{i,j=1}^N
\prod_{r=0}^\infty
\frac{
(1-q^{\frac{1}{2}+\frac{|m_i-m_j|}{2}+r}t^2\frac{s_i}{s_j}\partial^r(\psi_\varphi)_{ij})
}{
(1-q^{\frac{1}{2}+\frac{|m_i-m_j|}{2}+r}t^{-2}\frac{s_i}{s_j}\partial^r\varphi_{ij})
}\nonumber \\
&\quad \prod_{i,j=1}^N
\prod_{r=0}^\infty
\frac{
(1-q^{\frac{3}{4}+\frac{|m_i-m_j|}{2}+r}t^{-1}\frac{s_i}{s_j}x\partial^r(\psi_X)_{ij})
(1-q^{\frac{3}{4}+\frac{|m_i-m_j|}{2}+r}t^{-1}\frac{s_j}{s_i}x^{-1}\partial^r(\psi_Y)_{ji})
}{
(1-q^{\frac{1}{4}+\frac{|m_i-m_j|}{2}+r}t\frac{s_i}{s_j}x\partial^rX_{ij})
(1-q^{\frac{1}{4}+\frac{|m_i-m_j|}{2}+r}t\frac{s_j}{s_i}x^{-1}\partial^rY_{ji})
}\nonumber \\
&\quad \prod_{i=1}^N\prod_{\alpha=1}^l
\prod_{r=0}^\infty
\frac{
(1-q^{\frac{3}{4}+\frac{|m_i|}{2}+r}t^{-1}s_iy_\alpha \partial^r(\psi_{I^\alpha})_i)
(1-q^{\frac{3}{4}+\frac{|m_i|}{2}+r}t^{-1}s_i^{-1}y_\alpha^{-1} \partial^r(\psi_{J^\alpha})_i)
}{
(1-q^{\frac{1}{4}+\frac{|m_i|}{2}+r}ts_iy_\alpha \partial^rI^\alpha_i)
(1-q^{\frac{1}{4}+\frac{|m_i|}{2}+r}ts_i^{-1}y_\alpha^{-1}\partial^rJ^\alpha_i)
}\nonumber \\
&\quad q^{\sum_{i=1}^N\frac{l|m_i|}{4}}
t^{-l\sum_{i=1}^N|m_i|},
\label{220322_auxdres}
\end{align}
with $A_{ij}$, $\partial^r\varphi_{ij}$, $\partial^r(\psi_\varphi)_{ij}$, $\partial^rX_{ij}$, $\partial^r(\psi_X)_{ij}$, $\partial^rY_{ij}$, $\partial^r(\psi_Y)_{ij}$, $\partial^rI^\alpha_i$, $\partial^r(\psi_{I^\alpha})_i$, $\partial^rJ^\alpha_i$ and $\partial^r(\psi_{J^\alpha})_i$ a set of auxiliary parameters, instead of the supersymmetric index \eqref{uN_l_index}.
Though the auxiliary parameters are not allowed as the fugacities of the supersymmetric index, the integration \eqref{220322_auxdres} is useful to understand the operator content of each term in the full supersymmetric index.
For $l=1, N=1$ and to the order $q^{\frac{3}{4}}$ we obtain
\begin{align}
\begin{tabular}{|c|c|c|}
\hline
$m_i$&fugacity                     &auxiliary fugacity\\ \hline
$-3$ &$t^{-3}q^{\frac{3}{4}}$      &$1$\\ \hline
$-2$ &$t^{-2}q^{\frac{1}{2}}$      &$1$\\ \cline{2-3}
     &$xt^{-1}q^{\frac{3}{4}}$     &$X$\\ \cline{2-3}
     &$x^{-1}t^{-1}q^{\frac{3}{4}}$&$Y$\\ \hline
$-1$ &$t^{-1}q^{\frac{1}{4}}$      &$1$\\ \cline{2-3}
     &$xq^{\frac{1}{2}}$           &$X$\\ \cline{2-3}
     &$x^{-1}q^{\frac{1}{2}}$      &$Y$\\ \cline{2-3}
     &$t^{-3}q^{\frac{3}{4}}$      &$\varphi$\\ \cline{2-3}
     &$x^2tq^{\frac{3}{4}}$        &$X^2$\\ \cline{2-3}
     &$x^{-2}tq^{\frac{3}{4}}$     &$Y^2$\\ \cline{2-3}
     &$tq^{\frac{3}{4}}$           &$XY-\psi_\varphi$\\ \hline
$0$  &$1$                          &$1$\\ \cline{2-3}
     &$xtq^{\frac{1}{4}}$          &$X$\\ \cline{2-3}
     &$x^{-1}tq^{\frac{1}{4}}$     &$Y$\\ \cline{2-3}
     &$t^{-2}q^{\frac{1}{2}}$      &$\varphi$\\ \cline{2-3}
     &$t^2q^{\frac{1}{2}}$         &$IJ+XY-\psi_\varphi$\\ \cline{2-3}
     &$x^2t^2q^{\frac{1}{2}}$      &$X^2$\\ \cline{2-3}
     &$x^{-2}t^2q^{\frac{1}{2}}$   &$Y^2$\\ \cline{2-3}
     &$xt^{-1}q^{\frac{3}{4}}$     &$-\psi_X+\varphi X$ \\ \cline{2-3}
     &$x^{-1}t^{-1}q^{\frac{3}{4}}$&$-\psi_Y+\varphi Y$ \\ \cline{2-3}
     &$x^3t^3q^{\frac{3}{4}}$      &$X^3$ \\ \cline{2-3}
     &$x^{-3}t^3q^{\frac{3}{4}}$   &$Y^3$ \\ \cline{2-3}
     &$xt^3q^{\frac{3}{4}}$        &$IJX+X^2Y-\psi_\varphi X$ \\ \cline{2-3}
     &$x^{-1}t^3q^{\frac{3}{4}}$   &$IJY+XY^2-\psi_\varphi Y$ \\ \hline
$1$  &$t^{-1}q^{\frac{1}{4}}$      &$1$\\ \cline{2-3}
     &$xq^{\frac{1}{2}}$           &$X$\\ \cline{2-3}
     &$x^{-1}q^{\frac{1}{2}}$      &$Y$\\ \cline{2-3}
     &$t^{-3}q^{\frac{3}{4}}$      &$\varphi$\\ \cline{2-3}
     &$x^2tq^{\frac{3}{4}}$        &$X^2$\\ \cline{2-3}
     &$x^{-2}tq^{\frac{3}{4}}$     &$Y^2$\\ \cline{2-3}
     &$tq^{\frac{3}{4}}$           &$XY-\psi_\varphi$\\ \hline
$2$  &$t^{-2}q^{\frac{1}{2}}$      &$1$\\ \cline{2-3}
     &$xt^{-1}q^{\frac{3}{4}}$     &$X$\\ \cline{2-3}
     &$x^{-1}t^{-1}q^{\frac{3}{4}}$&$Y$\\ \hline
$3$  &$t^{-3}q^{\frac{3}{4}}$      &$1$\\ \hline
\end{tabular}.
\end{align}
This table reproduces the operator identification in \eqref{u1_1_findex}.
Note that the net coefficients of $tq^{\frac{3}{4}}$ at monopole charge $\pm 1$ and $x^{\pm 1}t^{-1}q^{\frac{3}{4}}$ at monopole charge $0$ in the table are zero, hence there are no corresponding terms in the supersymmetric index \eqref{u1_1_findex}.

For $l=2, N=1$ we obtain the following results:
\begin{align}
\begin{tabular}{|c|c|c|}
\hline
$m_i$&fugacity                             &auxiliary fugacity\\ \hline
$-1$ &$t^{-2}q^{\frac{1}{2}}$              &$1$\\ \cline{2-3}
     &$xt^{-1}q^{\frac{3}{4}}$             &$X$\\ \cline{2-3}
     &$x^{-1}t^{-1}q^{\frac{3}{4}}$        &$Y$\\ \hline
$0$  &$1$                                  &$1$\\ \cline{2-3}
     &$xtq^{\frac{1}{4}}$                  &$X$\\ \cline{2-3}
     &$x^{-1}tq^{\frac{1}{4}}$             &$Y$\\ \cline{2-3}
     &$t^{-2}q^{\frac{1}{2}}$              &$\varphi$\\ \cline{2-3}
     &$t^2q^{\frac{1}{2}}$                 &$\sum_\alpha I_\alpha J_\alpha+XY-\psi_\varphi$\\ \cline{2-3}
     &$x^2t^2q^{\frac{1}{2}}$              &$X^2$\\ \cline{2-3}
     &$y_\alpha y_\beta^{-1}t^2q^{\frac{1}{2}}$      &$I_\alpha J_\beta$\\ \cline{2-3}
     &$x^{-2}t^2q^{\frac{1}{2}}$           &$Y^2$\\ \cline{2-3}
     &$xt^{-1}q^{\frac{3}{4}}$             &$-\psi_X+\varphi X$\\ \cline{2-3}
     &$x^{-1}t^{-1}q^{\frac{3}{4}}$        &$-\psi_Y+\varphi Y$\\ \cline{2-3}
     &$x^3t^3q^{\frac{3}{4}}$              &$X^3$\\ \cline{2-3}
     &$x^{-3}t^3q^{\frac{3}{4}}$           &$Y^3$\\ \cline{2-3}
     &$xt^3q^{\frac{3}{4}}$                &$\sum_\alpha I_\alpha J_\alpha X+X^2Y-\psi_\varphi X$\\ \cline{2-3}
     &$xy_\alpha y_\beta^{-1}t^3q^{\frac{3}{4}}$     &$I_\alpha J_\beta X$\\ \cline{2-3}
     &$x^{-1}t^3q^{\frac{3}{4}}$           &$\sum_\alpha I_\alpha J_\alpha Y+XY^2-\psi_\varphi Y$\\ \cline{2-3}
     &$x^{-1}y_\alpha y_\beta^{-1}t^3q^{\frac{3}{4}}$&$I_\alpha J_\beta Y$\\ \hline
$1$  &$t^{-2}q^{\frac{1}{2}}$              &$1$\\ \cline{2-3}
     &$xt^{-1}q^{\frac{3}{4}}$             &$X$\\ \cline{2-3}
     &$x^{-1}t^{-1}q^{\frac{3}{4}}$        &$Y$\\ \hline
\end{tabular}.
\end{align}

For $l=3, N=1$ we obtain the following results:
\begin{align}
\begin{tabular}{|c|c|c|}
\hline
$m_i$&fugacity                                       &auxiliary fugacity\\ \hline
$-1$ &$t^{-3}q^{\frac{3}{4}}$                        &$1$\\ \hline
$0$  &$1$                                            &$1$\\ \cline{2-3}
     &$xtq^{\frac{1}{4}}$                            &$X$\\ \cline{2-3}
     &$x^{-1}tq^{\frac{1}{4}}$                       &$Y$\\ \cline{2-3}
     &$t^{-2}q^{\frac{1}{2}}$                        &$\varphi$\\ \cline{2-3}
     &$t^2q^{\frac{1}{2}}$                           &$\sum_\alpha I_\alpha J_\alpha+XY-\psi_\varphi$\\ \cline{2-3}
     &$x^2t^2q^{\frac{1}{2}}$                        &$X^2$\\ \cline{2-3}
     &$y_\alpha y_\beta^{-1}t^2q^{\frac{1}{2}}$      &$I_\alpha J_\beta$\\ \cline{2-3}
     &$x^{-2}t^2q^{\frac{1}{2}}$                     &$Y^2$\\ \cline{2-3}
     &$xt^{-1}q^{\frac{3}{4}}$                       &$-\psi_X+\varphi X$\\ \cline{2-3}
     &$x^{-1}t^{-1}q^{\frac{3}{4}}$                  &$-\psi_Y+\varphi Y$\\ \cline{2-3}
     &$x^3t^3q^{\frac{3}{4}}$                        &$X^3$\\ \cline{2-3}
     &$x^{-3}t^3q^{\frac{3}{4}}$                     &$Y^3$\\ \cline{2-3}
     &$xt^3q^{\frac{3}{4}}$                          &$\sum_\alpha I_\alpha J_\alpha X+X^2Y-\psi_\varphi X$\\ \cline{2-3}
     &$xy_\alpha y_\beta^{-1}t^3q^{\frac{3}{4}}$     &$I_\alpha J_\beta X$\\ \cline{2-3}
     &$x^{-1}t^3q^{\frac{3}{4}}$                     &$\sum_\alpha I_\alpha J_\alpha Y+XY^2-\psi_\varphi Y$\\ \cline{2-3}
     &$x^{-1}y_\alpha y_\beta^{-1}t^3q^{\frac{3}{4}}$&$I_\alpha J_\beta Y$\\ \hline
\end{tabular}.
\end{align}

For $l=4, N=1$ we obtain the following results:
\begin{align}
\begin{tabular}{|c|c|c|}
\hline
$m_i$&fugacity                                       &auxiliary fugacity\\ \hline
$0$  &$1$                                            &$1$\\ \cline{2-3}
     &$xtq^{\frac{1}{4}}$                            &$X$\\ \cline{2-3}
     &$x^{-1}tq^{\frac{1}{4}}$                       &$Y$\\ \cline{2-3}
     &$t^{-2}q^{\frac{1}{2}}$                        &$\varphi$\\ \cline{2-3}
     &$t^2q^{\frac{1}{2}}$                           &$\sum_\alpha I_\alpha J_\alpha+XY-\psi_\varphi$\\ \cline{2-3}
     &$x^2t^2q^{\frac{1}{2}}$                        &$X^2$\\ \cline{2-3}
     &$y_\alpha y_\beta^{-1}t^2q^{\frac{1}{2}}$      &$I_\alpha J_\beta$\\ \cline{2-3}
     &$x^{-2}t^2q^{\frac{1}{2}}$                     &$Y^2$\\ \cline{2-3}
     &$xt^{-1}q^{\frac{3}{4}}$                       &$-\psi_X+\varphi X$\\ \cline{2-3}
     &$x^{-1}t^{-1}q^{\frac{3}{4}}$                  &$-\psi_Y+\varphi Y$\\ \cline{2-3}
     &$x^3t^3q^{\frac{3}{4}}$                        &$X^3$\\ \cline{2-3}
     &$x^{-3}t^3q^{\frac{3}{4}}$                     &$Y^3$\\ \cline{2-3}
     &$xt^3q^{\frac{3}{4}}$                          &$\sum_\alpha I_\alpha J_\alpha X+X^2Y-\psi_\varphi X$\\ \cline{2-3}
     &$xy_\alpha y_\beta^{-1}t^3q^{\frac{3}{4}}$     &$I_\alpha J_\beta X$\\ \cline{2-3}
     &$x^{-1}t^3q^{\frac{3}{4}}$                     &$\sum_\alpha I_\alpha J_\alpha Y+XY^2-\psi_\varphi Y$\\ \cline{2-3}
     &$x^{-1}y_\alpha y_\beta^{-1}t^3q^{\frac{3}{4}}$&$I_\alpha J_\beta Y$\\ \hline
\end{tabular}.
\end{align}

\subsubsection{Results for higher ranks}
The following is the results for $l=1, N=2$.
Here we display only the contributions to $z^{\pm 1}tq^{\frac{3}{4}}$ and $x^{\pm 1}t^{-1}q^{\frac{3}{4}}$ in the mixed branch where there is a fermionic contribution ($\psi_{\varphi},\psi_X,\psi_Y,\psi_I,\psi_J$) from some of the monopole charge $m_i$.
\begin{align}
\begin{tabular}{|c|c|c|c|}
\hline
$m_i^{(a)}$ &fugacity                           &auxiliary fugacity                                           &gauge indices ignored\\ \hline
$(-1,0)$    &$tq^{\frac{3}{4}}$                 &$I_2J_2+\sum_{i}X_{ii}\sum_jY_{jj}-\sum_i(\psi_\varphi)_{ii}$&$IJ+4XY-2\psi_\varphi$\\ \hline
$(1,-1)$    &$xt^{-1}q^{\frac{3}{4}}$           &$\sum_iX_{ii}$                                               &$2X$\\ \cline{2-4}
            &$x^{-1}t^{-1}q^{\frac{3}{4}}$      &$\sum_iY_{ii}$                                               &$2Y$\\ \hline
$(0,0)$     &$xt^{-1}q^{\frac{3}{4}}$           &$\substack{(1+A_{12}A_{21})(-\sum_i(\psi_X)_{ii}+\sum_i\varphi_{ii}\sum_jX_{jj}+\varphi_{12}X_{21}\\
+\varphi_{21}X_{12})\\
-A_{21}(-(\psi_X)_{12}+\sum_i\varphi_{ii}X_{12}+\sum_i\varphi_{12}X_{ii})\\
-A_{12}(-(\psi_X)_{21}+\sum_i\varphi_{ii}X_{21}+\sum_i\varphi_{21}X_{ii})
}$                                                                                                            &$4\varphi X-2\psi_X$\\ \cline{2-4}
            &$x^{-1}t^{-1}q^{\frac{3}{4}}$      &$\substack{(1+A_{12}A_{21})(-\sum_i(\psi_Y)_{ii}+\sum_i\varphi_{ii}\sum_jY_{jj}+\varphi_{12}Y_{21}\\
+\varphi_{21}Y_{12})\\
-A_{21}(-(\psi_Y)_{12}+\sum_i\varphi_{ii}Y_{12}+\sum_i\varphi_{12}Y_{ii})\\
-A_{12}(-(\psi_Y)_{21}+\sum_i\varphi_{ii}Y_{21}+\sum_i\varphi_{21}Y_{ii})
}$                                                                                                            &$4\varphi Y-2\psi_Y$\\ \hline
$(1,0)$     &$tq^{\frac{3}{4}}$                 &$I_2J_2+\sum_iX_{ii}\sum_jY_{jj}-\sum_i(\psi_\varphi)_{ii}$  &$IJ+4XY-2\psi_\varphi$\\ \hline
\end{tabular}.
\label{ADHMl1N2auxdres}
\end{align}

The following is the results for $l=2, N=2$.
Here we display only the contributions to $x^{\pm 1}t^{-1}q^{\frac{3}{4}}$ in the mixed branch where there is a fermionic contribution ($\psi_{\varphi},\psi_X,\psi_Y,\psi_I,\psi_J$) from some of the monopole charge $m_i$.
\begin{align}
\begin{tabular}{|c|c|c|c|}
\hline
$m_i^{(a)}$ &fugacity                           &auxiliary fugacity&gauge indices ignored\\ \hline
$(0,0)$     &$xt^{-1}q^{\frac{3}{4}}$           &$\substack{
(1+A_{12}A_{21})(-\sum_i(\psi_X)_{ii}+\sum_{i,j}X_{ii}\varphi_{jj}+X_{12}\varphi_{21}\\
+X_{21}\varphi_{12})\\
-A_{21}(-(\psi_X)_{12}+X_{12}\sum_i\varphi_{ii}+\sum_iX_{ii}\varphi_{12})\\
-A_{12}(-(\psi_X)_{21}+X_{21}\sum_i\varphi_{ii}+\sum_iX_{ii}\varphi_{21})
}$& $4\varphi X-2\psi_X$\\ \cline{2-4}
&$x^{-1}t^{-1}q^{\frac{3}{4}}$           &$\substack{
(1+A_{12}A_{21})(-\sum_i(\psi_Y)_{ii}+\sum_{i,j}Y_{ii}\varphi_{jj}+Y_{12}\varphi_{21}\\
+Y_{21}\varphi_{12})\\
-A_{21}(-(\psi_Y)_{12}+Y_{12}\sum_i\varphi_{ii}+\sum_iY_{ii}\varphi_{12})\\
-A_{12}(-(\psi_Y)_{21}+Y_{21}\sum_i\varphi_{ii}+\sum_iY_{ii}\varphi_{21})
}$& $4\varphi Y-2\psi_Y$\\ \hline
\end{tabular}.
\end{align}

The following is the results for $l=4, N=2$.
Here we display only the contributions to $x^{\pm 1}t^{-1}q^{\frac{3}{4}}$ in the mixed branch where there is a fermionic contribution ($\psi_{\varphi},\psi_X,\psi_Y,\psi_I,\psi_J$) from some of the monopole charge $m_i$.
\begin{align}
\begin{tabular}{|c|c|c|c|}
\hline
$m_i^{(a)}$ &fugacity                           &auxiliary fugacity&gauge indices ignored\\ \hline
$(0,0)$     &$xt^{-1}q^{\frac{3}{4}}$           &$\substack{
(1+A_{12}A_{21})(-\sum_i(\psi_X)_{ii}+\sum_{i,j}X_{ii}\varphi_{jj}+X_{12}\varphi_{21}\\
+X_{21}\varphi_{12})\\
-A_{21}(-(\psi_X)_{12}+X_{12}\sum_i\varphi_{ii}+\sum_iX_{ii}\varphi_{12})\\
-A_{12}(-(\psi_X)_{21}+X_{21}\sum_i\varphi_{ii}+\sum_iX_{ii}\varphi_{21})
}$& $4\varphi X-2\psi_X$\\ \cline{2-4}
&$x^{-1}t^{-1}q^{\frac{3}{4}}$           &$\substack{
(1+A_{12}A_{21})(-\sum_i(\psi_Y)_{ii}+\sum_{i,j}Y_{ii}\varphi_{jj}+Y_{12}\varphi_{21}\\
+Y_{21}\varphi_{12})\\
-A_{21}(-(\psi_Y)_{12}+Y_{12}\sum_i\varphi_{ii}+\sum_iY_{ii}\varphi_{12})\\
-A_{12}(-(\psi_Y)_{21}+Y_{21}\sum_i\varphi_{ii}+\sum_iY_{ii}\varphi_{21})
}$& $4\varphi Y-2\psi_Y$\\ \hline
\end{tabular}.
\end{align}

The following is the results for $l=1, N=3$.
Here we display only the contributions to $tq^{\frac{3}{4}}$ and $x^{\pm 1}t^{-1}q^{\frac{3}{4}}$ in the mixed branch where there is a fermionic contribution ($\psi_{\varphi},\psi_X,\psi_Y,\psi_I,\psi_J$) from some of the monopole charge $m_i$.
\begin{align}
\begin{tabular}{|c|c|c|}
\hline
$m_i^{(a)}$   &fugacity                           &auxiliary fugacity (gauge indices ignored)\\ \hline
$(-1,0,0)$    &$tq^{\frac{3}{4}}$                 &$2IJ+10XY-4\psi_{\varphi}$\\ \hline
$(1,-1,0)$    &$xt^{-1}q^{\frac{3}{4}}$           &$3X$\\ \cline{2-3}
           &$x^{-1}t^{-1}q^{\frac{3}{4}}$      &$3Y$\\ \hline
$(0,0,0)$     &$xt^{-1}q^{\frac{3}{4}}$           &$-6\psi_{X}+12\varphi X$\\ \cline{2-3}
              &$x^{-1}t^{-1}q^{\frac{3}{4}}$      &$-6\psi_{Y}+12\varphi Y$\\ \hline
$(1,0,0)$     &$tq^{\frac{3}{4}}$                 &$2IJ+10XY-4\psi_{\varphi}$\\ \hline
\end{tabular}.
\label{ADHMl1N3auxdres}
\end{align}
Since with the full auxiliary dressing we have too many terms to write (in particular at $m_i=(0,0,0)$), here we have set $A_{ij}=1$ and also ignored all the gauge indices.

\subsection{$U(N)_k\times U(N)_0^{\otimes(l-1)}\times U(N)_{-k}$ Chern-Simons matter theory}
\label{220427_app_CSMauxdres}
For the $U(N)_k\times U(N)^{\otimes (l-1)}\times U(N)_{-k}$ supersymmetric Chern-Simons matter theory let us consider the following generalization of the contribution to the full index \eqref{220413_CSmatterindexnewnotation} from each monopole charge $m_i^{(I)}$:
\begin{align}
&I_{\text{aux.~dres}}^{U(N_1)_k\times U(N_I)_0^{\otimes (l-1)}\times U(N_{l+1})_{-k}}(m_i^{(I)})\nonumber \\
&=\oint\prod_I\prod_{i=1}^{N_I}\frac{ds_i^{(I)}}{2\pi is_i^{(I)}}
\prod_{i=1}^{N_1}(s_i^{(1)})^{km_i^{(1)}}
\prod_{i=1}^{N_{l+1}}(s_i^{(l+1)})^{-km_i^{(l+1)}}\nonumber \\
&\quad \times \prod_{I=1}^{l+1}\prod_{i\neq j}^{N_I}\Bigl(1-q^{\frac{|m_i^{(I)}-m_j^{(I)}|}{2}}\frac{s_i^{(I)}}{s_j^{(I)}}A_{ij}^{(I)}\Bigr)
\prod_{I=2}^l\prod_{i,j=1}^{N_I}\prod_{r=0}^\infty
\frac{
(1-q^{\frac{1}{2}+\frac{|m_i^{(I)}-m_j^{(I)}|}{2}+r}t^{-2}\frac{s_i^{(I)}}{s_j^{(I)}}\partial^r(\psi_{\varphi^{(I)}})_{ij})
}{
(1-q^{\frac{1}{2}+\frac{|m_i^{(I)}-m_j^{(I)}|}{2}+r}t^2\frac{s_i^{(I)}}{s_j^{(I)}}\partial^r\varphi^{(I)}_{ij})
}\nonumber \\
&\quad \times \prod_{I=1}^l\prod_{i=1}^{N_I}\prod_{j=1}^{N_{I+1}}
\prod_{r=0}^\infty
\frac{
(1-q^{\frac{3}{4}+\frac{|m_i^{(I)}-m_j^{(I+1)}|}{2}+r}t\frac{s_i^{(I)}}{s_j^{(I+1)}}z_I\partial^r(\psi_{T_{I,I+1}})_{ij})
}{
(1-q^{\frac{1}{4}+\frac{|m_i^{(I)}-m_j^{(I+1)}|}{2}+r}t^{-1}\frac{s_i^{(I)}}{s_j^{(I+1)}}z_I\partial^r(T_{I,I+1})_{ij})
}\nonumber \\
&\quad\quad\quad\quad\times\frac{(1-q^{\frac{3}{4}+\frac{|m_i^{(I)}-m_j^{(I+1)}|}{2}+r}t\frac{s_j^{(I+1)}}{s_i^{(I)}}z_I^{-1}\partial^r(\psi_{{\widetilde T}_{I,I+1}})_{ji})
}{(1-q^{\frac{1}{4}+\frac{|m_i^{(I)}-m_j^{(I+1)}|}{2}+r}t^{-1}\frac{s_j^{(I+1)}}{s_i^{(I)}}z_I^{-1}\partial^r({\widetilde T}_{I,I+1})_{ji})
}
\nonumber \\
&\quad \times \prod_{i=1}^{N_{l+1}}\prod_{j=1}^{N_1}
\prod_{r=0}^\infty
\frac{(1-q^{\frac{3}{4}+\frac{|m_i^{(l+1)}-m_j^{(1)}|}{2}+r}t^{-1}\frac{s_i^{(l+1)}}{s_j^{(1)}}x\partial^r(\psi_{H_{l+1,1}})_{ij})}{(1-q^{\frac{1}{4}+\frac{|m_i^{(l+1)}-m_j^{(1)}|}{2}+r}t\frac{s_i^{(l+1)}}{s_j^{(1)}}x\partial^r(H_{l+1,1})_{ij})}\nonumber \\
&\quad\quad\quad\quad\times\frac{(1-q^{\frac{3}{4}+\frac{|m_i^{(l+1)}-m_j^{(1)}|}{2}+r}t^{-1}\frac{s_j^{(1)}}{s_i^{(l+1)}}x^{-1}\partial^r(\psi_{{\widetilde H}_{l+1,1}})_{ji})}{(1-q^{\frac{1}{4}+\frac{|m_i^{(l+1)}-m_j^{(1)}|}{2}+r}t\frac{s_j^{(1)}}{s_i^{(l+1)}}x^{-1}\partial^r({\widetilde H}_{l+1,1})_{ji})}\nonumber \\
&\quad \times q^{-\frac{1}{2}\sum_{I=1}^{l+1}\sum_{i<j}^{N_I}|m_i^{(I)}-m_j^{(I)}|+\frac{1}{4}\sum_{I=1}^{l+1}\sum_{i=1}^{N_I}\sum_{j=1}^{N_{I+1}}|m_i^{(I)}-m_j^{(I+1)}|}\nonumber \\
&\quad \times t^{2\sum_{I=2}^l\sum_{i<j}^{N_I}|m_i^{(I)}-m_j^{(I)}|+\sum_{I=1}^l\sum_{i=1}^{N_I}\sum_{j=1}^{N_{I+1}}|m_i^{(I)}-m_j^{(I+1)}|-\sum_{i=1}^{N_{l+1}}\sum_{j=1}^{N_1}|m_i^{(l+1)}-m_j^{(1)}|},
\end{align}
with the auxiliary dressing parameters $A_{ij}^{(I)}$, $\partial^4\varphi_{ij}^{(I)}$, $\partial^r(\psi_{\varphi^{(I)}})_{ij}$, $\partial^r(T_{I,I+1})_{ij}$, $\partial^r(\psi_{T_{I,I+1}})_{ij}$, $\partial^r({\widetilde T}_{I,I+1})_{ij}$, $\partial^r(\psi_{{\widetilde T}_{I,I+1}})_{ij}$, $\partial^r(H_{l+1,1})_{ij}$, $\partial^r(\psi_{H_{l+1,1}})_{ij}$, $\partial^r({\widetilde H}_{l+1,1})_{ij}$ and $\partial^r(\psi_{{\widetilde H}_{l+1,1}})_{ij}$.
For $k=1, l=1, N_1=N_2=1$ and to the order $q^{\frac{3}{4}}$ we obtain
\begin{align}
\begin{tabular}{|c|c|c|}
\hline
$m_i^{(a)}$ &fugacity                           &auxiliary fugacity\\ \hline
$(-3;-3)$&$x^{-3}t^3q^{\frac{3}{4}}$            &${\widetilde H}^3$\\ \cline{2-3}
         &$x^{-2}ztq^{\frac{3}{4}}$             &$T{\widetilde H}^2$\\ \cline{2-3}
         &$x^{-1}z^2t^{-1}q^{\frac{3}{4}}$      &$T^2{\widetilde H}$\\ \cline{2-3}
         &$z^3t^{-3}q^{\frac{3}{4}}$            &$T^3$\\ \hline
$(-2;-2)$&$x^{-2}t^2q^{\frac{1}{2}}$            &${\widetilde H}^2$\\ \cline{2-3}
         &$x^{-1}zq^{\frac{1}{2}}$              &$T{\widetilde H}$\\ \cline{2-3}
         &$z^2t^{-2}q^{\frac{1}{2}}$            &$T^2$\\ \hline
$(-1;-1)$&$x^{-1}tq^{\frac{1}{4}}$              &${\widetilde H}$\\ \cline{2-3}
         &$zt^{-1}q^{\frac{1}{4}}$              &$T$\\ \cline{2-3}
         &$x^{-1}t^3q^{\frac{3}{4}}$            &$H{\widetilde H}^2$\\ \cline{2-3}
         &$zt^{-3}q^{\frac{3}{4}}$              &$T^2{\widetilde T}$\\ \cline{2-3}
         &$x^{-2}z^{-1}tq^{\frac{3}{4}}$        &${\widetilde T}{\widetilde H}^2$\\ \cline{2-3}
         &$ztq^{\frac{3}{4}}$                   &$TH{\widetilde H}-\psi_T$\\ \cline{2-3}
         &$x^{-1}t^{-1}q^{\frac{3}{4}}$         &$-\psi_{{\widetilde H}}+T{\widetilde T}{\widetilde H}$\\ \cline{2-3}
         &$xz^2t^{-1}q^{\frac{3}{4}}$           &$T^2H$\\ \hline
$(0;0)$  &$1$                                   &$1$\\ \cline{2-3}
         &$t^2q^{\frac{1}{2}}$                  &$H{\widetilde H}$\\ \cline{2-3}
         &$t^{-2}q^{\frac{1}{2}}$               &$T{\widetilde T}$\\ \cline{2-3}
         &$x^{-1}z^{-1}q^{\frac{1}{2}}$         &${\widetilde T}{\widetilde H}$\\ \cline{2-3}
         &$xzq^{\frac{1}{2}}$                   &$TH$\\ \hline
$(1;1)$  &$xtq^{\frac{1}{4}}$                   &$H$\\ \cline{2-3}
         &$z^{-1}t^{-1}q^{\frac{1}{4}}$         &${\widetilde T}$\\ \cline{2-3}
         &$xt^3q^{\frac{3}{4}}$                 &$H^2{\widetilde H}$\\ \cline{2-3}
         &$z^{-1}t^{-3}q^{\frac{3}{4}}$         &$T{\widetilde T}^2$\\ \cline{2-3}
         &$x^2ztq^{\frac{3}{4}}$                &$TH^2$\\ \cline{2-3}
         &$z^{-1}tq^{\frac{3}{4}}$              &${\widetilde T}H{\widetilde H}-\psi_{{\widetilde T}}$\\ \cline{2-3}
         &$xt^{-1}q^{\frac{3}{4}}$              &$-\psi_H+T{\widetilde T}H$\\ \cline{2-3}
         &$x^{-1}z^{-2}t^{-1}q^{\frac{3}{4}}$   &${\widetilde T}^2{\widetilde H}$\\ \hline
$(2;2)  $&$x^2t^2q^{\frac{1}{2}}$               &$H^2$\\ \cline{2-3}
         &$xz^{-1}q^{\frac{1}{2}}$              &${\widetilde T}H$\\ \cline{2-3}
         &$z^{-2}t^{-2}q^{\frac{1}{2}}$         &${\widetilde T}^2$\\ \hline
$(3;3)$  &$x^3t^3q^{\frac{3}{4}}$               &$H^3$\\ \cline{2-3}
         &$x^2z^{-1}tq^{\frac{3}{4}}$           &${\widetilde T}H^2$\\ \cline{2-3}
         &$xz^{-2}t^{-1}q^{\frac{3}{4}}$        &${\widetilde T}^2H$\\ \cline{2-3}
         &$z^{-3}t^{-3}q^{\frac{3}{4}}$         &${\widetilde T}^3$\\ \hline
\end{tabular}.
\label{220427_ABJMk1N1auxdres}
\end{align}

For $k=1, l=2, N_1=N_2=N_3=1$ and to the order $q^{\frac{3}{4}}$ we obtain
\begin{align}
\begin{tabular}{|c|c|c|}
\hline
$(m^{(1)},m^{(2)},m^{(3)})$&fugacity                           &auxiliary fugacity\\ \hline
$(-3;-3;-3)$&$x^{-3}t^3q^{\frac{3}{4}}$         &${\widetilde H}_{3,1}^3$\\ \hline
$(-2;-2;-2)$&$x^{-2}t^2q^{\frac{1}{2}}$         &${\widetilde H}_{3,1}^2$\\ \cline{2-3}
            &$z_1z_2x^{-1}t^{-1}q^{\frac{3}{4}}$&$T_{1,2}T_{2,3}{\widetilde H}_{3,1}$\\ \hline
$(-1;-2;-1)$&$x^{-1}t^3q^{\frac{3}{4}}$         &${\widetilde H}_{3,1}$\\ \hline
$(-1;-1;-1)$&$x^{-1}tq^{\frac{1}{4}}$           &${\widetilde H}_{3,1}$\\ \cline{2-3}
            &$z_1z_2t^{-2}q^{\frac{1}{2}}$      &$T_{1,2}T_{2,3}$\\ \cline{2-3}
            &$x^{-1}t^3q^{\frac{3}{4}}$         &$H{\widetilde H}_{3,1}^2+\varphi^{(2)}{\widetilde H}_{3,1}$\\ \cline{2-3}
            &$x^{-1}t^{-1}q^{\frac{3}{4}}$      &$-\psi_{{\widetilde H}_{3,1}}+T_{1,2}{\widetilde T}_{1,2}{\widetilde H}_{3,1}+T_{2,3}{\widetilde T}_{2,3}{\widetilde H}_{3,1}-\psi_{\varphi^{(2)}}{\widetilde H}_{3,1}$\\ \hline
$(-1;0;-1)$ &$x^{-1}t^3q^{\frac{3}{4}}$         &${\widetilde H}_{3,1}$\\ \hline
$(0;-1;0)$  &$t^2q^{\frac{1}{2}}$               &$1$\\ \hline
$(0;0;0)$   &$1$                                &$1$\\ \cline{2-3}
            &$z_1^{-1}z_2^{-1}x^{-1}t^{-1}q^{\frac{3}{4}}$&${\widetilde T}_{1,2}{\widetilde T}_{2,3}{\widetilde H}_{3,1}$\\ \cline{2-3}
            &$z_1z_2xt^{-1}q^{\frac{3}{4}}$     &$T_{1,2}T_{2,3}H_{3,1}$\\ \cline{2-3}
            &$t^2q^{\frac{1}{2}}$               &$H_{3,1}{\widetilde H}_{3,1}+\varphi^{(2)}$\\ \cline{2-3}
            &$t^{-2}q^{\frac{1}{2}}$            &$T_{1,2}{\widetilde T}_{1,2}+T_{2,3}{\widetilde T}_{2,3}-\psi_{\varphi^{(2)}}$\\ \hline
$(0;1;0)$   &$t^2q^{\frac{1}{2}}$               &$1$\\ \hline
$(1;0;1)$   &$xt^3q^{\frac{3}{4}}$              &$H_{3,1}$\\ \hline
$(1;1;1)$   &$xtq^{\frac{1}{4}}$                &$H_{3,1}$\\ \cline{2-3}
            &$z_1^{-1}z_2^{-1}t^{-2}q^{\frac{1}{2}}$&${\widetilde T}_{1,2}{\widetilde T}_{2,3}$\\ \cline{2-3}
            &$xt^3q^{\frac{3}{4}}$              &$H_{3,1}^2{\widetilde H}_{3,1}+\varphi^{(2)}H_{3,1}$\\ \cline{2-3}
            &$xt^{-1}q^{\frac{3}{4}}$           &$-\psi_{H_{3,1}}+T_{1,2}{\widetilde T}_{1,2}H_{3,1}+T_{2,3}{\widetilde T}_{2,3}H_{3,1}-\psi_{\varphi^{(2)}}H_{3,1}$\\ \hline
$(1;2;1)$   &$xt^3q^{\frac{3}{4}}$              &$H_{3,1}$\\ \hline
$(2;2;2)$   &$x^2t^2q^{\frac{1}{2}}$            &$H_{3,1}^2$\\ \cline{2-3}
            &$z_1^{-1}z_2^{-1}xt^{-1}q^{\frac{3}{4}}$&${\widetilde T}_{1,2}{\widetilde T}_{2,3}H_{3,1}$\\ \hline
$(3;3;3)$   &$x^3t^3q^{\frac{3}{4}}$            &$H_{3,1}^3$\\ \hline
\end{tabular}.
\end{align}
The operators contributing to each term of the full supersymmetric index \eqref{220413u1u1u1k+-0_findex} can also be read off from this table.
Note that the net coefficients of $x^{\pm 1}t^{-1}q^{\frac{3}{4}}$ at monopole charge $(\pm 1;\pm 1;\pm 1)$ in the table are zero, hence there are no corresponding terms in the supersymmetric index \eqref{220413u1u1u1k+-0_findex}.

For $k=1, l=3, N_1=N_2=N_3=N_4=1$ and to the order $q^{\frac{3}{4}}$ we obtain
\begin{align}
\begin{tabular}{|c|c|c|}
\hline
$m_i^{(a)}$ &fugacity                           &auxiliary fugacity\\ \hline
$(-3;-3;-3;-3)$&$x^{-3}t^3q^{\frac{3}{4}}$      &${\widetilde H}_{4,1}^3$\\ \hline
$(-2;-2;-2;-2)$&$x^{-2}t^2q^{\frac{1}{2}}$      &${\widetilde H}_{4,1}^2$\\ \cline{2-3}
$(-1;-2;-2;-1)$&$x^{-1}t^3q^{\frac{3}{4}}$      &${\widetilde H}_{4,1}$\\ \hline
$(-1;-2;-1;-1)$&$x^{-1}t^3q^{\frac{3}{4}}$      &${\widetilde H}_{4,1}$\\ \hline
$(-1;-1;-2;-1)$&$x^{-1}t^3q^{\frac{3}{4}}$      &${\widetilde H}_{4,1}$\\ \hline
$(-1;-1;-1;-1)$&$x^{-1}tq^{\frac{1}{4}}$        &${\widetilde H}_{4,1}$\\ \cline{2-3}
               &$z_1z_2z_3t^{-3}q^{\frac{3}{4}}$&$T_{1,2}T_{2,3}T_{3,4}$\\ \cline{2-3}
               &$x^{-1}t^3q^{\frac{3}{4}}$      &$H_{4,1}{\widetilde H}_{4,1}^2+{\widetilde H}(\varphi^{(2)}+\varphi^{(3)})$\\ \cline{2-3}
               &$x^{-1}t^{-1}q^{\frac{3}{4}}$   &$-\psi_{{\widetilde H}_{4,1}}+{\widetilde H}_{4,1}(T_{1,2}{\widetilde T}_{1,2}+T_{2,3}{\widetilde T}_{2,3}+T_{3,4}{\widetilde T}_{3,4}-\psi_{\varphi^{(2)}}-\psi_{\varphi^{(3)}})$\\ \hline
$(-1;-1;0;-1)$ &$x^{-1}t^3q^{\frac{3}{4}}$      &${\widetilde H}_{4,1}$\\ \hline
$(-1;0;-1;-1)$ &$x^{-1}t^3q^{\frac{3}{4}}$      &${\widetilde H}_{4,1}$\\ \hline
$(-1;0;0;-1)$  &$x^{-1}t^3q^{\frac{3}{4}}$      &${\widetilde H}_{4,1}$\\ \hline
$(0;-1;-1;0)$  &$t^2q^{\frac{1}{2}}$            &$1$\\ \hline
$(0;-1;0;0)$   &$t^2q^{\frac{1}{2}}$            &$1$\\ \hline
$(0;0;-1;0)$   &$t^2q^{\frac{1}{2}}$            &$1$\\ \hline
$(0;0;0;0)$    &$1$                             &$1$\\ \cline{2-3}
               &$t^2q^{\frac{1}{2}}$            &$H_{4,1}{\widetilde H}_{4,1}+\varphi^{(2)}+\varphi^{(3)}$\\ \cline{2-3}
               &$t^{-2}q^{\frac{1}{2}}$         &$T_{1,2}{\widetilde T}_{1,2}+T_{2,3}{\widetilde T}_{2,3}+T_{3,4}{\widetilde T}_{3,4}-\psi_{\varphi^{(2)}}-\psi_{\varphi^{(3)}}$\\ \hline
$(0;0;1;0)$    &$t^2q^{\frac{1}{2}}$            &$1$\\ \hline
$(0;1;0;0)$    &$t^2q^{\frac{1}{2}}$            &$1$\\ \hline
$(0;1;1;0)$    &$t^2q^{\frac{1}{2}}$            &$1$\\ \hline
$(1;0;0;1)$    &$xt^3q^{\frac{3}{4}}$           &$H_{4,1}$\\ \hline
$(1;0;1;1)$    &$xt^3q^{\frac{3}{4}}$           &$H_{4,1}$\\ \hline
$(1;1;0;1)$    &$xt^3q^{\frac{3}{4}}$           &$H_{4,1}$\\ \hline
$(1;1;1;1)$    &$xtq^{\frac{1}{4}}$             &$H_{4,1}$\\ \cline{2-3}
               &$z_1^{-1}z_2^{-1}z_3^{-1}t^{-3}q^{\frac{3}{4}}$&${\widetilde T}_{1,2}{\widetilde T}_{2,3}{\widetilde T}_{3,4}$\\ \cline{2-3}
               &$xt^3q^{\frac{3}{4}}$           &$H_{4,1}^2{\widetilde H}_{4,1}+H_{4,1}(\varphi^{(2)}+\varphi^{(3)})$\\ \cline{2-3}
               &$xt^{-1}q^{\frac{3}{4}}$        &$-\psi_{H_{4,1}}+H_{4,1}(T_{1,2}{\widetilde T}_{1,2}+T_{2,3}{\widetilde T}_{2,3}+T_{3,4}{\widetilde T}_{3,4}-\psi_{\varphi^{(2)}}-\psi_{\varphi^{(3)}})$\\ \hline
$(1;1;2;1)$    &$xt^3q^{\frac{3}{4}}$           &$H_{4,1}$\\ \hline
$(1;2;1;1)$    &$xt^3q^{\frac{3}{4}}$           &$H_{4,1}$\\ \hline
$(1;2;2;1)$    &$xt^3q^{\frac{3}{4}}$           &$H_{4,1}$\\ \hline
$(2;2;2;2)$    &$x^2t^2q^{\frac{1}{2}}$         &$H_{4,1}^2$\\ \hline
$(3;3;3;3)$    &$x^3t^3q^{\frac{3}{4}}$         &$H_{4,1}^3$\\ \hline
\end{tabular}.
\end{align}

\subsubsection{Results for higher ranks}
\label{u2k1ABJM_auxiliary}
The following is the results for $k=1, l=1, N_1=N_2=2$.
We call $T_{1,2},{\widetilde T}_{1,2},H_{2,1},{\widetilde H}_{2,1}$ respectively as $T,{\widetilde T},H,{\widetilde H}$.
Here we display only the contributions to $y_1^{\pm 1}y_2^{\pm 1}x^{\pm 1}t^{-1}q^{\frac{3}{4}}$ and $y_1^{\pm 1}y_2^{\pm 1}z_1^{\mp 1}tq^{\frac{3}{4}}$ in the mixed branch where there is a fermionic contribution ($\psi_{\varphi^{(a)}},\psi_T,\psi_{{\widetilde T}},\psi_H,\psi_{{\widetilde H}}$) from some of the monopole charge $(m_i^{(1)},m_i^{(2)})$.
\begin{align}
\begin{tabular}{|c|c|c|c|}
\hline
$m_i^{(a)}$ &fugacity                      &auxiliary fugacity                            &gauge indices ignored\\ \hline
$(-2,1;-2;1)$&$z_1tq^{\frac{3}{4}}$        &$H_{22}{\widetilde H}_{11}T_{11}$             &$TH{\widetilde H}$\\ \cline{2-4}
             &$x^{-1}t^{-1}q^{\frac{3}{4}}$&${\widetilde H}_{11}T_{11}{\widetilde T}_{22}$&$T{\widetilde T}{\widetilde H}$\\ \hline
$(-1,0;-1,0)$&$z_1tq^{\frac{3}{4}}$        &$\substack{H_{11}{\widetilde H}_{11}T_{11}+H_{22}{\widetilde H}_{22}T_{11}+H_{22}{\widetilde H}_{11}T_{22}\\
-(\psi_T)_{11}}$                                                                          &$3TH{\widetilde H}-\psi_T$\\ \cline{2-4}
             &$x^{-1}t^{-1}q^{\frac{3}{4}}$&$\substack{-(\psi_{{\widetilde H}})_{11}+{\widetilde H}_{11}T_{11}{\widetilde T}_{11}+{\widetilde H}_{22}T_{11}{\widetilde T}_{22}\\
+{\widetilde H}_{11}T_{22}{\widetilde T}_{22}}$                                           &$3T{\widetilde T}{\widetilde H}-\psi_{{\widetilde H}}$\\ \hline
$(-1,2;-1,2)$&$z_1^{-1}tq^{\frac{3}{4}}$   &$H_{22}{\widetilde H}_{11}{\widetilde T}_{22}$&${\widetilde T}H{\widetilde H}$\\ \cline{2-4}
             &$xt^{-1}q^{\frac{3}{4}}$     &$H_{22}T_{11}{\widetilde T}_{22}$             &$T{\widetilde T}H$\\ \hline
$(0,1;0,1)$  &$xt^{-1}q^{\frac{3}{4}}$     &$\substack{-(\psi_H)_{22}+H_{22}T_{11}{\widetilde T}_{11}+H_{11}T_{11}{\widetilde T}_{22}\\
+H_{22}T_{22}{\widetilde T}_{22}}$                                                        &$3T{\widetilde T}H-\psi_H$\\ \cline{2-4}
             &$z_1^{-1}tq^{\frac{3}{4}}$   &$\substack{H_{22}{\widetilde H}_{11}{\widetilde T}_{11}+H_{11}{\widetilde H}_{11}{\widetilde T}_{22}+H_{22}{\widetilde H}_{22}{\widetilde T}_{22}\\
-(\psi_{{\widetilde T}})_{22}}$                                                           &$3{\widetilde T}H{\widetilde H}-\psi_{{\widetilde T}}$\\ \hline
\end{tabular}.
\end{align}

\section{Hilbert series associated with dihedral groups}

Let $G$ be a finite group. We take a representation $\rho: G \to GL(m, \mathbb{C})$ and consider the action of $\rho(g)\;(g \in G)$ on complex coordinates of $\mathbb{C}^m$. Then the generating function for the number of independent polynomials invariant under the action $G$ can be computed by the Molien's formula,
\begin{equation}\label{molien}
\mathcal{I}(t;G,\rho)=\frac{1}{|G|}\sum_{g\in G}\frac{1}{\det(\text{id}-t\rho(g))},
\end{equation}
where $\text{id}$ is the $n\times n$ identity matrix. 

Let us apply the Molien's formula to some representations of dihedral groups. The dihedral group $\mathbb{D}_{2n}$ consists of $2n$ elements given by $\{1, r, \cdots, r^{n-1}, s, rs, \cdots, r^{n-1}s\}$ where $1$ is the identity element and $r, s$ satisfy 
\begin{equation}
    srs^{-1} = r^{-1}, \qquad r^n=1, \qquad s^2 = 1.
\end{equation}
For example a 2d representation of $r, s$ is given by
\begin{align}
    \rho_1(r)=\left(\begin{array}{cc}
    e^{\frac{2\pi i}{n}} & 0\\
    0 & e^{-\frac{2\pi i}{n}}
    \end{array}\right),\qquad
    \rho_1(s)=\left(\begin{array}{cc}
    0 & 1\\
    1 & 0
    \end{array}\right).
\end{align}
Then the application of the Molien's formula \eqref{molien} to the representation of the dihedral group yields
\begin{align}
    \mathcal{I}(t;\mathbb{D}_{2n},\rho_1)&=\frac{1}{2n}\left[\sum_{k=0}^{n-1}\det\left(
    \begin{array}{cc}
    1-te^{\frac{2\pi i k}{n}} & 0\\
    0 & 1 - te^{-\frac{2\pi i n}{k}}
    \end{array}\right)^{-1}+\sum_{k=0}^{n-1}\det\left(
    \begin{array}{cc}
    1 & -te^{\frac{2\pi i k}{n}}\\
    - te^{-\frac{2\pi i n}{k}} & 1
    \end{array}\right)^{-1}\right]\cr
    &=\frac{1}{2n}\left[\frac{n}{1-t^2} + \sum_{k=0}^{n-1}\frac{1}{(1-te^{\frac{2\pi i k}{n}})(1-te^{-\frac{2\pi i k}{n}})}\right]\cr
    &=\frac{1}{(1-t^2)(1-t^n)}.\label{molien.D2n_rho1}
\end{align}

We can also consider a 4d representation of $r, s$,
\begin{align}
    \rho_2(r)=\left(\begin{array}{cc|cc}
    e^{\frac{2\pi i}{n}} & 0 & 0 & 0\\
    0 & e^{-\frac{2\pi i}{n}} & 0 & 0\\\hline
    0 & 0 & e^{\frac{2\pi i}{n}} & 0\\
    0 & 0 & 0 & e^{-\frac{2\pi i}{n}}
    \end{array}\right),\qquad
    \rho_2(s)=\left(\begin{array}{cc|cc}
    0 & 1 & 0 & 0\\
    1 & 0 & 0 & 0\\\hline
    0 & 0 & 0 & 1\\
    0 & 0 & 1 & 0
    \end{array}\right).
\end{align}
The Molien's formula \eqref{molien} with this representation becomes 
\begin{equation}
    \mathcal{I}(t;\mathbb{D}_{2n}, \rho_2) = \frac{1}{2n}\left[\frac{n}{(1-t^2)^2} + \sum_{k=0}^{n-1}\frac{1}{(1-te^{\frac{2\pi i k}{n}})^2(1-te^{-\frac{2\pi i k}{n}})^2}\right].\label{molien.D2n_rho2}
\end{equation}
From the explicit evaluation of the expression \eqref{molien.D2n_rho2} for some orders of $t$ and some small $n$, we conjecture that \eqref{molien.D2n_rho2} can be also written as
\begin{equation}
    \mathcal{I}(t;\mathbb{D}_{2n}, \rho_2)=\frac{1+(n-1)t^n-(n-1)t^{n+2} - t^{2n+2}}{(1-t^2)^3(1-t^n)^2}.\label{molienv2.D2n_rho2}
\end{equation}

\bibliographystyle{utphys}
\bibliography{ref}
\end{document}